\documentclass[article]{aa} 

\usepackage{orcidlink}
\usepackage{caption}
\usepackage{longtable}
\usepackage{graphicx}
\usepackage{subcaption}
\usepackage{ulem}
\usepackage{natbib}
\usepackage{pdflscape}
\usepackage{txfonts}

\newcommand{\Px}{\ifmmode P_{\rm x}\else$P_{\rm x}$\fi}
\newcommand{\Po}{\ifmmode P_{\rm 1O}\else$P_{\rm 1O}$\fi}
\newcommand{\MS}{\ifmmode{\,}\else\thinspace\fi{\rm M}\ifmmode_{\odot}\else$_{\odot}$\fi}
\newcommand{\LS}{\ifmmode{\,}\else\thinspace\fi{\rm L}\ifmmode_{\odot}\else$_{\odot}$\fi}
\newcommand{\RS}{\ifmmode{\,}\else\thinspace\fi{\rm R}\ifmmode_{\odot}\else$_{\odot}$\fi}

\begin{document}
\title{Non-evolutionary effects on period change in Magellanic Cepheids.}
\subtitle{I. New binary systems revealed from Light Travel Time Effect}

\titlerunning{Non-evolutionary effects on period change in Magellanic Cepheids I}
\authorrunning{Rathour et al.}
\author{Rajeev Singh Rathour \inst{1}\orcidlink{0000-0002-7448-4285}, Gergely Hajdu \inst{1}\orcidlink{0000-0003-0594-9138}, Radosław Smolec \inst{1}\orcidlink{0000-0001-7217-4884}, Paulina Karczmarek \inst{2}\orcidlink{0000-0002-0136-0046}, Vincent Hocd\'e \inst{1}\orcidlink{0000-0002-3643-0366}, Oliwia Zi\'ołkowska \inst{1}\orcidlink{0000-0002-0696-2839}, Igor Soszy\'nski \inst{3}\orcidlink{0000-0002-7777-0842}, Andrzej Udalski \inst{3}\orcidlink{0000-0001-5207-5619}}

\institute{Nicolaus Copernicus Astronomical Centre, Polish Academy of Sciences, Bartycka 18, 00-716 Warszawa, Poland\\
\email{rajeevsr@camk.edu.pl}
\and
Departamento de Astronomía, Universidad de Concepción, Casilla 160-C, Concepción, Chile
\and
Astronomical Observatory, University of Warsaw, Aleje Ujazdowskie 4, Warsaw, 00-478, Poland
} 

\date{Received ??; accepted ??}

\abstract
{Period change studies offer a novel way to probe the evolution and dynamics of Cepheids. While evolutionary period changes have been well studied both observationally and theoretically, non-evolutionary period changes lack a systematic and quantitative description. Here, we deal with one such aspect of non-evolutionary period changes related to a crucial property, namely, the binarity-based nature of a Cepheid. With the advent of long-term photometry surveys covering Magellanic fields, the census of classical Cepheids in binary (or multiple) systems outside the Milky Way is timely. This may have implications for crucial aspects such as the period-luminosity relationship calibrations and our understanding of the nature of Cepheid companions.
}
{The overall objective is to have a quantitative understanding of the full picture of non-evolutionary period changes in Cepheids to develop a formalism to disentangle it from the secular evolutionary period change. In the first paper in the series, we aim to conduct a systematic search for non-evolutionary period changes to look for Cepheids in likely binary configurations and quantify their incidence rates in the Magellanic Clouds.
}
{We collected more than a decade-long time-series photometry from the publicly available, Optical Gravitational Lensing Experiment (OGLE) survey, with more than 7200 Cepheids altogether from the Large Magellanic Cloud (LMC) and Small Magellanic Cloud (SMC). Our sample contains both fundamental-mode and first-overtone mode Cepheids. Then, we calculate d the observed minus calculated ($O-C$) diagrams to reveal the light-travel time effect (LTTE). Finally, we calculated the minimum companion masses of the Cepheids and compared them with the predictions from Cepheid population synthesis results.
}
{In our search, out of an overall sample of more than 7200 Cepheids, we found 52 candidate Cepheid binary systems in the LMC (30 fundamental and 22 first-overtone mode) and 145 in the SMC (85 fundamental and 60 first-overtone mode). The majority of the sample is characterized by orbital periods of 2000-4000\,d and eccentricities of 0.2-0.5. Moreover, we report two candidates in each galaxy with the Cepheid likely existing with a giant companion. The incidence rate ratio for SMC to LMC calculated from our sample is in agreement with binary Cepheid population synthesis predictions. 
}
{In our attempt to quantify the non-evolutionary period change connected with the LTTE, our systematic search has enriched the Cepheid binary sample by a factor of about 2 in both galaxies. The future spectroscopic follow-up can confirm the binarity nature of our sample and constrain the orbital parameters.}

\maketitle

\section{Introduction}
\label{sec: Introduction}

Type I or classical Cepheids (hereafter referred to as Cepheids) have long been one of the benchmark stars for stellar evolution and pulsation studies. They are a class of variable stars with intermediate and high masses ($\sim3-13\MS$) and with pulsation periods of $\sim1-100$\,days. Cepheids have ages ranging between $10 - 100$ Myr and are reported to be pulsating in radial and non-radial modes \citep[e.g.,][]{Catelan-book,soszynski2008optical,soszynski2010optical,soszynski2011optical,Soszynski2015b}.

The pulsations in Cepheids are excited when they cross a region on the Hertzprung-Russell diagram, known as the instability strip (IS). These pulsations are driven by the $\kappa$ and $\gamma$ mechanisms, operating in the partially ionized helium and hydrogen zones \citep[e.g.,][]{Cox1980tsp..book.....C}. As these stars move across the IS, they experience overall changes in stellar radius (alongside effective temperature and luminosity), thereby changing the pulsation period. As a consequence, from the observational viewpoint, the measure of these period changes is a direct insight into the Cepheid evolution. However, Cepheid also exhibits period changes that are not related to its secular evolution, and thus of much shorter timescales ($\sim$100-10000 days). These period changes are classified as non-evolutionary and come in two types. The first kind manifests as periodic period changes and is a consequence of a Cepheid in a binary system. The second type is observed as short timescale irregularities in the period changes, hence the designation of irregular period changes. The occurrence of this kind of period change is still a puzzle with no consolidated mechanism proposed yet. In this work, we investigate the non-evolutionary period changes of the first kind, namely, those related to Cepheids in binary systems.

In the  Galactic sample, $60-80$ \% or more of Cepheids are likely to be part of a binary system, as observations indicate \citep{Szabados2003ASPC..298..237S,Evans2013AJ....146...93E,Kervella2019AA...623A.116K}. The well maintained Galactic binary Cepheids database\footnote{https://konkoly.hu/CEP/intro.html} contains $\sim$200 confirmed or suspected binary targets \citep{Szabados2003bIBVS.5394....1S}. However, the Magellanic Clouds' photometry and spectroscopic observations are way short of reproducing these statistics with 9 and 25 reported Cepheid binaries in the Small and Large Magellanic Clouds (SMC and LMC), respectively \citep{Szabados1983Ap&SS..96..185S,Pilecki2021ApJ...910..118P}. The significance of binary Cepheids is on two fronts. First, from a cosmological standpoint, Cepheids follow the well-known period-luminosity relationship \citep[Leavitt Law,][]{leavitt19081777,Leavitt1912HarCi.173....1L}, making them excellent extragalactic distance indicators \citep[e.g.,][and references therein]{Freedman2001ApJ...553...47F,Riess2021ApJ...908L...6R}. However, the presence of any companion can affect the measured brightness and color of the Cepheid \citep[e.g.,][]{Szabados2013MNRAS.430.2018S}, as a consequence that
may possibly affect the period-luminosity relationship \citep{SzabadosKlagyivik2012Ap&SS.341...99S,Pilecki2021ApJ...910..118P,Karczmarek2023ApJ...950..182K}. Second, binary Cepheids are significant to bridge the gap between stellar evolution and pulsation theory predictions. Nearly half a century has passed since the realization that the evolutionary masses of Cepheids are systematically higher (currently about $10-20$ \%) than the ones predicted by pulsation theory. This has led to a long-standing
enigma in astrophysics, known as the Cepheid mass discrepancy problem \citep{Stobie1969MNRAS.144..511S,Fricke1972ApJ...171..593F,Cox1980tsp..book.....C,Keller2008ApJ...677..483K,Bono2008}. The first purely dynamical mass of Cepheid (Polaris Aa) was presented by \cite{Evans2008AJ....136.1137E}, using \textit{Hubble Space Telescope (HST)} UV imaging observations. Thereafter, a landmark study by \cite{Pietrzynski2010Natur.468..542P} reported the dynamical mass of a Cepheid in an eclipsing binary system, OGLE-LMC-CEP-0227, to an accuracy of 1 \%, which endorsed the pulsation theory predictions and inspired revising stellar evolutionary models. Many more studies to measure dynamical masses of Cepheids in binary systems have been conducted thereafter \citep{Pietrzynski2011ApJ...742L..20P,Gieren2014ApJ...786...80G,Pilecki2015ApJ...806...29P,Pilecki2018ApJ...862...43P}. Therefore, Cepheids in binary systems (even better if eclipsing), can provide crucial and accurate knowledge in boosting the above frontiers, since we can no longer ignore the companion effects on avenues explored via Cepheids. 

Binary Cepheids are increasingly getting due attention, as solving for their orbital parameters can not only provide dynamical masses (and other physical properties) of the Cepheids, which has its crucial applications as described before, but also may contribute to better understanding of orbital evolution of the binary. Recently, a number of population synthesis studies have tried to approach Cepheids in binary systems \citep{Neilson2015AA...574A...2N,Karczmarek2022ApJ...930...65K,Karczmarek2023ApJ...950..182K}. The predicted minimum orbital period of Cepheid in a binary system is $200-300$ days for the Galactic Cepheid population. This prediction ensures that the binaries evolve without interacting when one of the components goes through the red giant branch phase. However, this does not apply to first-crossing Cepheids that are still in the sub-giant phase and may have shorter binary orbital periods, as observationally confirmed for OGLE-LMC-CEP-1347 \citep{Pilecki2022ApJ...940L..48P}. It is a double-mode Cepheid (first- and second-overtone modes) with a very short, 59-day orbit with a less massive companion (mass ratio $q=0.55$). 

Searching Cepheid-bearing binary systems is traditionally done via spectroscopic studies looking for radial velocity variation. \cite{Evans1992ApJ...384..220E} compiled several systems via direct imaging also for binaries containing luminous companions. Recently, a technique of searching Cepheids and RR Lyrae stars in binary and multiple systems using proper motion anomalies due to orbiting companions was demonstrated by \cite{Kervella2019AA...623A.116K,Kervellab2019AA...623A.117K}. Eclipsing binaries have also been reported using light curve modeling and radial velocity measurements \citep[e.g.][]{Pietrzynski2011ApJ...742L..20P,Pilecki2015ApJ...806...29P}. However, while effective and accurate, the spectroscopic methods are limited in the volume of binaries that can be searched for. Alternatively, there is a photometry-based method to significantly improve the number of binaries that can be searched. In binary systems, while the stars are orbiting around each other, there are apparent changes in the pulsation period as a consequence of the distance variation between Cepheid (or any regular variable) and the observer, known as the light-travel time effect \citep[LTTE,][]{Irwin1952ApJ...116..211I,Irwin1959AJ.....64..149I}. This can be noticed from the Cepheid's pulsational variability manifesting periodic phase shifts, and is most commonly measured using the technique of observed minus calculated ($O-C$) diagrams \citep[e.g.,][]{Sterken2005ASPC..335....3S}.

For more than two decades the Optical Gravitational Lensing Experiment \citep[OGLE,][]{Udalski1999AcA....49..223U,udalski2015ogle} has been continuing to collect \textit{V}- and \textit{I}-band photometry of selected Galactic fields and the Magellanic Clouds (MCs). The Cepheid catalog in both LMC and SMC has reached more than 99 \% in terms of completeness \citep{soszynski2008optical,soszynski2010optical,Soszynski2017aAcA....67..103S} and it covers fundamental-mode, overtone-mode, and even multi-mode objects. This inventory provides a valuable dataset to search for Cepheids in binary systems throughout the MCs. Therefore, our aim is to analyze $O-C$ diagrams of the MCs' single-mode Cepheids (fundamental or first-overtone mode) from the OGLE-III and OGLE-IV survey, to present a systematic investigation of Cepheids in binary system candidates and determine their incidence rate.

We describe the data in Section \ref{sec: Survey description} and present the analysis methodology in Section \ref{sec: Methodology}. Thereafter we discuss the final sample of binary candidates separately for SMC and LMC fields in Section \ref{sec: Results}. We present the distribution of the binary parameters of the two galaxies in Section \ref{sec: Discussion}. Finally, we conclude our findings in Section \ref{sec: Conclusions}.

\section{Data}
\label{sec: Survey description}

We analyzed the publicly available photometric data\footnote{\url{http://www.astrouw.edu.pl/ogle/}} from  OGLE-III and OGLE-IV \citep{soszynski2010optical,Soszynski2015a,Soszynski2017aAcA....67..103S} which provide the ideal sample for searching for binary candidates due to data homogeneity and a long temporal baseline. The span of observations for OGLE-III, which also contains OGLE-II data, is 1997-2009 and for OGLE-IV is 2010-present, hereby providing more than 20 years of data. The data are collected using the 1.3-m Warsaw telescope at the Las Campanas Observatory, Chile. 

The database contains photometry in the Johnson-Kron-Cousins \textit{V} and \textit{I} bands, however, we utilized the \textit{I}-band data only since it is more densely sampled than the \textit{V}-band data. The OGLE data is highly suitable for searching binary candidates with variable stars, as shown in several works \citep[e.g.,][]{Soszynski2009AcA....59....1S,soszynski2011optical,Hajdu2015MNRAS.449L.113H,Hajdu2021ApJ...915...50H}.

For the final list of our candidate binary systems, we obtained non-public data from the recent OGLE-IV extended survey (starting from August 12, 2022). Adding the new OGLE observations ($\sim$1.3 year) does improve the credibility of binary model fits.

\section{Methodology}
\label{sec: Methodology}

\subsection{Combining photometric data}
\label{ssec:combine}
Our analysis uses photometry from two phases of OGLE, with data obtained with different cameras between them. Therefore, as a first step, we combine the time-series of OGLE-III and OGLE-IV as follows. The \textit{I}-band data is converted from magnitudes to an arbitrary intensity scale. Then, a Fourier series is fit individually to both phased data, and the possible mean intensity difference is corrected. The order of the Fourier series is decided by the $A_{k}/\sigma(A_{k})>4$ empirical criterion, with ${A_{k}}$ and $\sigma(A_{k})$ being the amplitude and uncertainty of the $k$-th order term. This criterion is effective to prevent an overfitting of the light curves and typically results in higher ($\sim$10) order fits for fundamental-mode stars and lower ($\sim$6) order fits for first-overtone mode stars. The resulting combined homogeneous data set is then transformed back to the magnitude scale. Following a visual inspection, we chose not to use Cepheids with significant zero-point variations across observing seasons, as these affect the overall light curve and are not fit for our analysis.

\subsection{$O-C$ procedure applied to OGLE database}
An important tool for measuring and quantifying period changes is the well known observed-minus-calculated ($O-C$) diagram. A brief review of $O-C$ techniques is given in \cite{ Zhou1999PBeiO..33...17Z} and \cite{Sterken2005ASPC..335....3S}. The $O-C$ is estimated by recording the differences between specific points in the light curve (typically the maxima or minima) against the corresponding calculated (or expected) times, assuming a constant period. Such a method requires a good observing cadence around the specific points. Therefore, sampling of the OGLE survey is not suitable for utilizing this technique. However, there is an alternative method proposed by \citet{Hertzsprung1919AN....210...17H}, which uses a light curve template to measure phase shifts using the whole light curve to derive $O-C$ points. This technique has been used recently for period change studies in classical pulsators \citep[e.g.,][]{Hajdu2021ApJ...915...50H,Rodriguez-Segovia2022MNRAS.509.2885R}

Here we describe the procedure to compute the $O-C$ for the combined OGLE-III and OGLE-IV data. We started by performing a Fourier series fitting of suitable order (criteria same as in Sect.~\ref{ssec:combine}) on the combined OGLE-III+IV data. This serves as our template to compute "O" in the $O-C$. For "C", we used the linear ephemeris with a pulsation period derived for all data using our back-end scripts for the discrete Fourier transform (DFT) via the Kurtz algorithm \citep{Kurtz1985MNRAS.213..773K}. When phasing the data, we use this period and the starting observation timestamp as the phase folding reference epoch. The full data set for each Cepheid is divided into time bins corresponding to observing seasons by default. In order to resolve the period change effect due to binarity, it is pertinent to increase the resolution of the $O-C$ curve. Therefore, we further divided the seasons into finer time bins if the number of data points in each season exceeds a threshold of a minimum of 10 data points or the time span of the bin is larger than 160\,d. The individual phase-folded segments were then compared with the template, and $O-C$ points were derived as a phase shift by minimizing the scatter between the template and light curve folded within each time bin, using a non-linear least-squares technique. This provided us with 7212 $O-C$ diagrams covering both SMC and LMC samples.

Using linear and parabolic fits we used the Akaike information criterion (AIC) \citep{Akaike1974ANL},  Bayesian information criterion (BIC) \citep{Schwarz1978_10.1214/aos/1176344136} and chi-square statistics on the data to determine goodness of fit along with analyzing the resultant residuals \citep[Anderson-Darling and Ljung-Box tests, ][respectively]{Anderson&Darling10.1214/aoms/1177729437,Ljung&Box10.1093/biomet/65.2.297}
to filter the sample into three categories: (1) flat $O-C$, (2) parabolic trend and (3) variation of a third kind (neither flat nor parabolic). This third category sample was semi-automatically  divided further into two categories (3a) periodic $O-C$ trend and (3b) irregular shape $O-C$ trend. After visual confirmation of periodic structures in the $O-C$, we identified relevant candidates which were later fitted with a binary model to confirm good candidates (described in the following section). Full details of the classification method along with a sample of irregular period change candidates (performed as a parallel analysis) will be presented in a separate paper (Rathour et al., in prep.).

\subsection{Modeling the Light-Travel Time Effect}
Below, we describe the analysis of the Light-Travel Time Effect (LTTE), applied to our Cepheid inventory. A phase modulation of the light curve that is periodic in nature, is a signature of the LTTE. This is also reflected in the $O-C$ diagram as a repeating structure. Due to secular evolution, the linear period change effect, which manifests as a parabolic trend, is also overlaid on top of the LTTE. Hence, we use a composite LTTE and quadratic model presented below:

\begin{equation}
    \label{eq:lte}
    z(t) = \left[ a\sin i \frac{1-e^2}{1+e \cos(\nu)} \sin (\nu + \omega)\right] + \left[c_0 + c_1  t + c_2  t^2\right]\,.
\end{equation}

The first term is the LTTE model \citep{Irwin1952ApJ...116..211I} with $a\sin i$ as projected semi-major axis of the orbit (where $i$ is the inclination), $e$ denotes eccentricity of the orbit, $\nu$ is the true anomaly and $\omega$ denotes the argument of the periastron. The second term represents the quadratic function of time to model the linear period change with $c_0$, $c_1$, and $c_2$ being the coefficients. Here, $c_0$ and $c_1$ are the correction terms in reference epoch and pulsation period whereas the linear period-change rate (PCR) is calculated using the $c_2$ coefficient and pulsation period $P_\mathrm{pulsation}$  via the equation

\begin{eqnarray}
\label{eq:PCR_equation}
\mathrm{PCR} = 2 \cdot P_\mathrm{pulsation} \cdot c_2\,.
\end{eqnarray}

The true anomaly, $\nu$, and the mean anomaly, $M$, are calculated from the following relations: 

\begin{eqnarray}
\label{eq:true_anomaly}
\cos \nu = \frac{\cos E - e}{1-e \cos E},
\end{eqnarray}

\begin{eqnarray}
\label{eq:eccentric_anomaly}
M = E - e\sin E = \frac{2\pi}{P_\mathrm{orbital}} \left(t-T_0\right),
\end{eqnarray}

\noindent where $E$, $P_\mathrm{orbital}$ and $T_0$ represent eccentric anomaly, orbital
period and time of periastron passage, respectively.

Our procedure, similarly to \cite{Hajdu2021ApJ...915...50H}, is iterative. After we fit eq.~\eqref{eq:lte} to the obtained $O-C$ diagram, we subtracted the LTTE+quadratic effect from the data, which refines the phased light curve. This less scattered light curve was employed again to obtain a new template (typically resulting in a Fourier series of higher order) as well as the new pulsation period. Then we repeated the $O-C$ calculation procedure a second time using the new template. For estimating the uncertainties on the improved $O-C$ points, we used a bootstrapping method. We drew random samples from individual time bins with replacement, which were used to obtain $O-C$ points with a total of 500 iterations. The standard deviation of the resulting $O-C$ with respect to the original $O-C$ value of the respective time bin was treated as the uncertainty on that $O-C$ value.

Finally, the iterative procedure results in an $O-C$ diagram for a Cepheid, which was then fitted again with the binary model giving the final orbital parameters [$P_\mathrm{orbital}$, $T_0$, $e$, $\omega$, $a\sin i$]. Using these parameters we calculated the derived parameters, namely the semi-amplitude of the expected radial
velocity variation, $K$, and a mass function, $f(m)$, using the relations:

\begin{eqnarray}
\label{eq:K}
K = \frac{2 \pi a \sin i}{P_\mathrm{orbital} \sqrt{1-e^2}},
\end{eqnarray}

\begin{eqnarray}
\label{eq:fm1}
f(m) = \frac{a^3 \sin^3 i}{P_\mathrm{orbital}\sqrt{1-e^2}}.
\end{eqnarray}

We adopted a Bayesian formalism in calculating the binary parameters and their errors using the EMCEE package \citep{Foreman-Mackey2013PASP..125..306F}. In cases with $O-C$ points having large errors resulting in larger errors on eccentricities, the procedure may return unreasonable $\omega$ values. To avoid such a situation, in the MCMC procedure, instead of $\omega$ and $e$ fitted as independent parameters, we used their Cartesian equivalents, $\sqrt{e}\sin\omega$ and $\sqrt{e}\cos\omega$. After the MCMC calculations, we recover back the eccentricity and argument
of periastron values \citep[for details, see][]{Hajdu2021ApJ...915...50H}. We used the default ensemble sampler provided within the EMCEE package and adopted the well tested sampling parameters given in \cite{Hajdu2021ApJ...915...50H}, which were used for a similar study aimed at RR Lyrae stars using the OGLE data. Relying upon their convergence and autocorrelation lengths computations, we used 200 walkers, 31000 steps, burn-in of 1000 steps and thinning by 300 steps. For a few dozen candidates, we had to rely on manual adjustment, for example (i) where we needed to inspect the seasonal light curve corresponding to a deviant $O-C$ point, (ii) in cases where we needed to increase and decrease the adopted period to refine the $O-C$ diagram in order to make it more symmetric, and (iii) in cases where we needed to limit priors of some of the binary parameters to improve their posterior distribution.

\begin{table*}
\caption{Sample of MC Cepheids with distribution with respect to pulsation mode and final binary candidates.}
\label{tab:data working sample}
\begin{tabular}{lccccc}
\hline
\hline
\textbf{Field} & \textbf{Pulsation} & \textbf{Stars with data in}  & \textbf{Stars with data in}  &  \textbf{Stars with data in}  &  \textbf{Binary candidates} \\
\textbf{} & \textbf{mode} & \textbf{OGLE-III}  & \textbf{OGLE-IV}  &  \textbf{ both OGLE-III+IV}  &  \\ 
\hline
\hline
\textbf{SMC}  & F &  2626  & 2739 & 2582  &  85 \\
              & 1O & 1644  & 1790 & 1617  & 60 \\
\hline
\textbf{LMC}& F &  1818   &  2430  & 1801 & 30 \\ 
        & 1O & 1238  &  1769 & 1212   & 22 \\ 
\hline
\hline
\textbf{Total} &  &  7326   &  8728  & 7212 & 197  \\ 
\hline
\end{tabular}
\end{table*}

\begin{figure*}[p]
\begin{center}

{\includegraphics[height=4.5cm,width=0.49\linewidth]{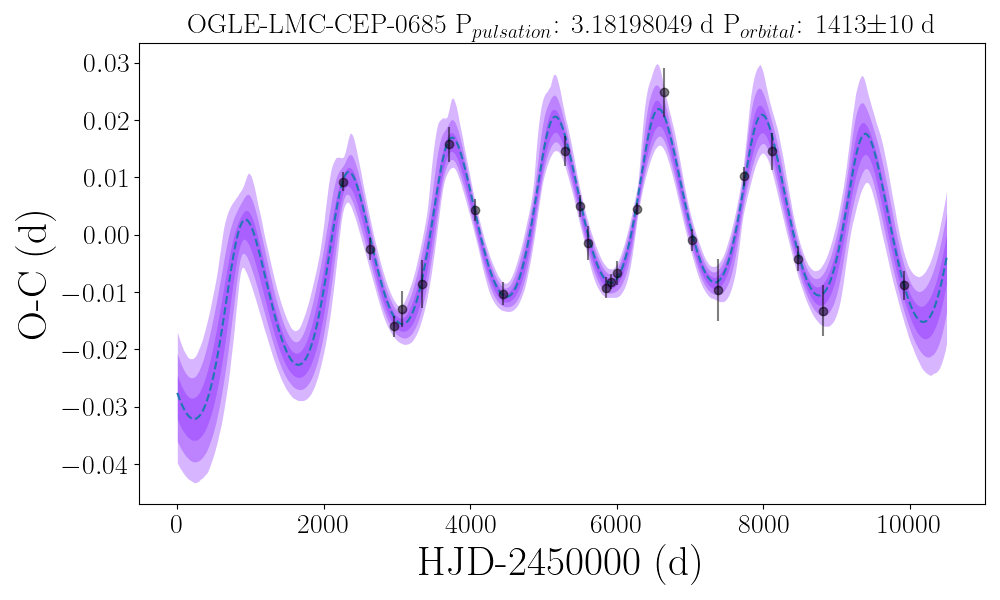}}
{\includegraphics[height=4.5cm,width=0.49\linewidth]{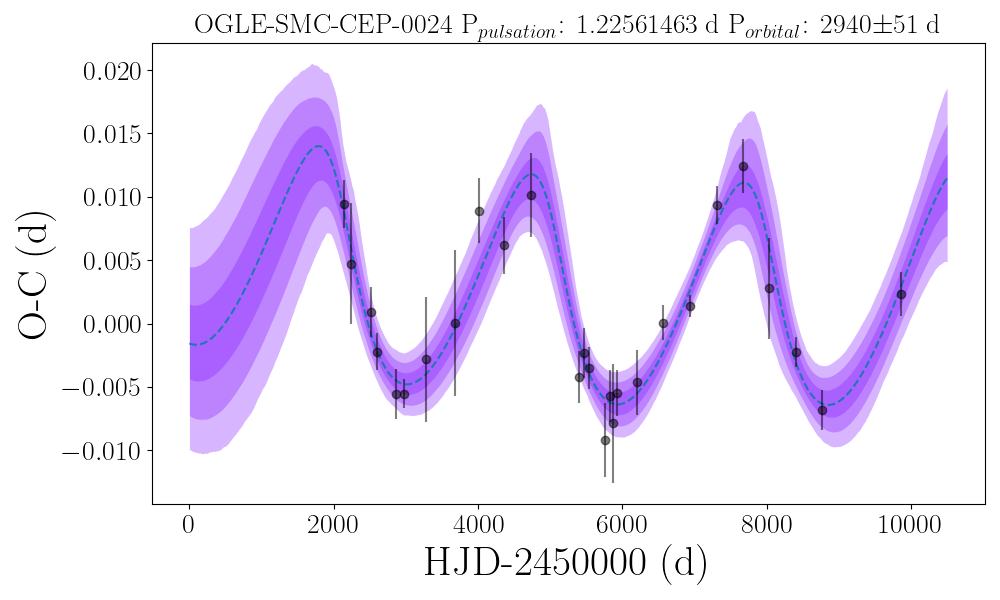}}
{\includegraphics[height=4.5cm,width=0.49\linewidth]{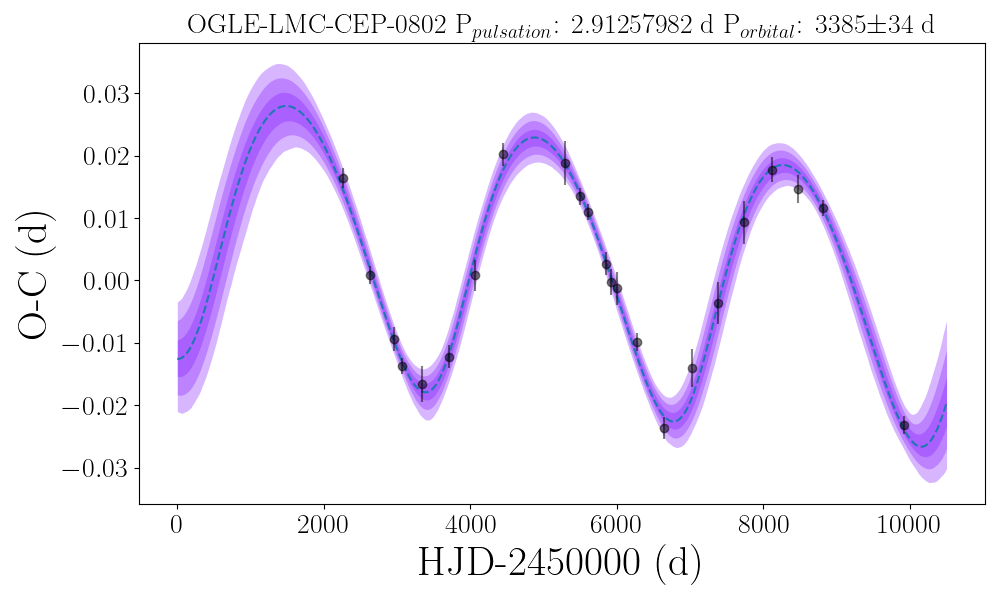}}
{\includegraphics[height=4.5cm,width=0.49\linewidth]{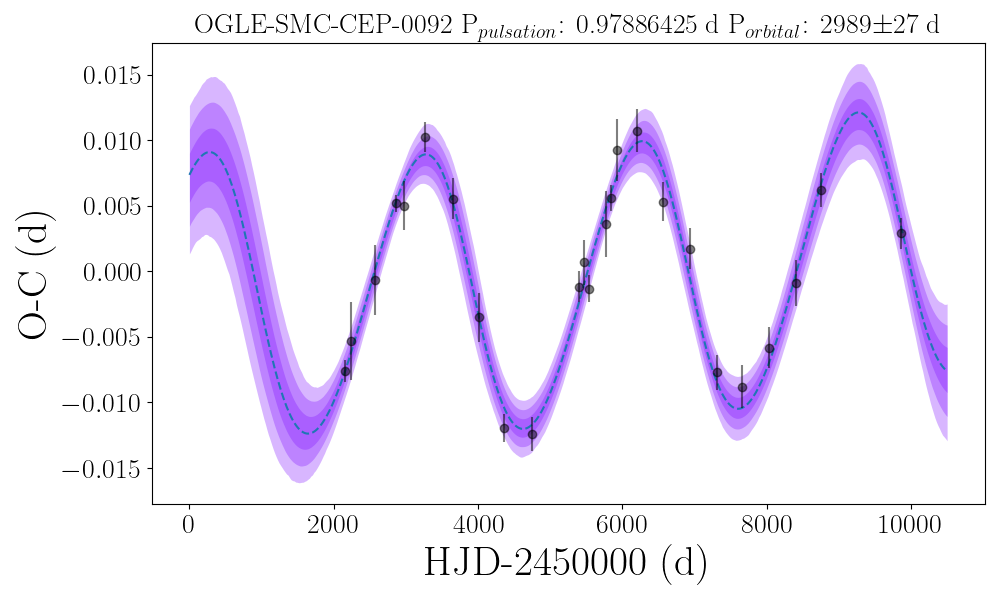}}
{\includegraphics[height=4.5cm,width=0.49\linewidth]{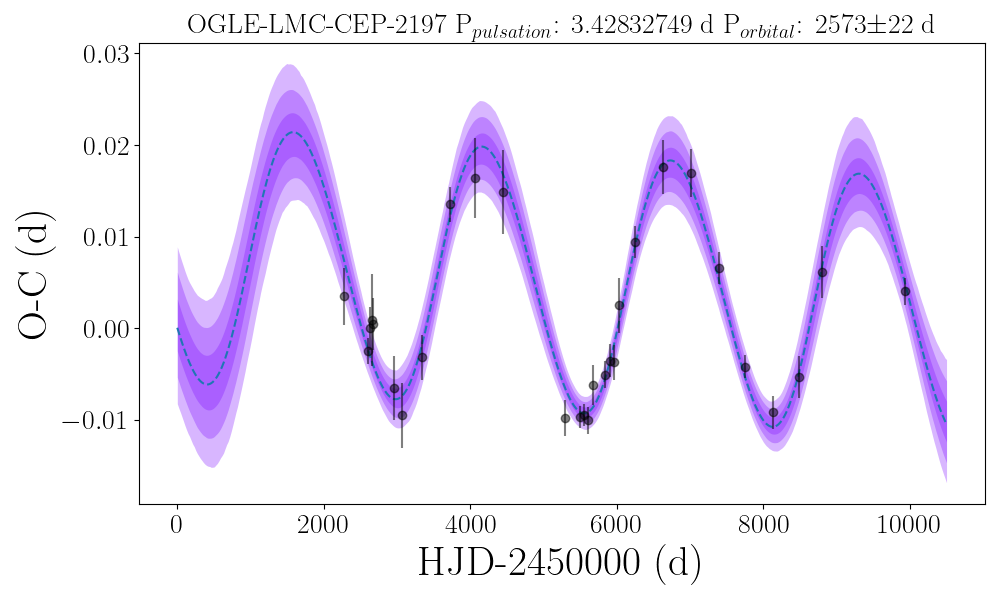}}
{\includegraphics[height=4.5cm,width=0.49\linewidth]{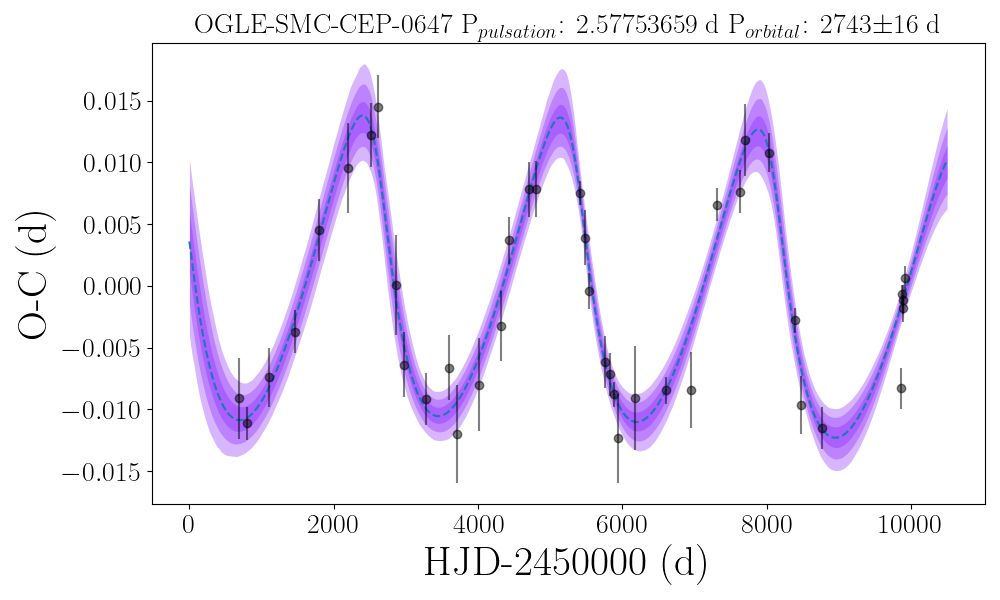}}
{\includegraphics[height=4.5cm,width=0.49\linewidth]{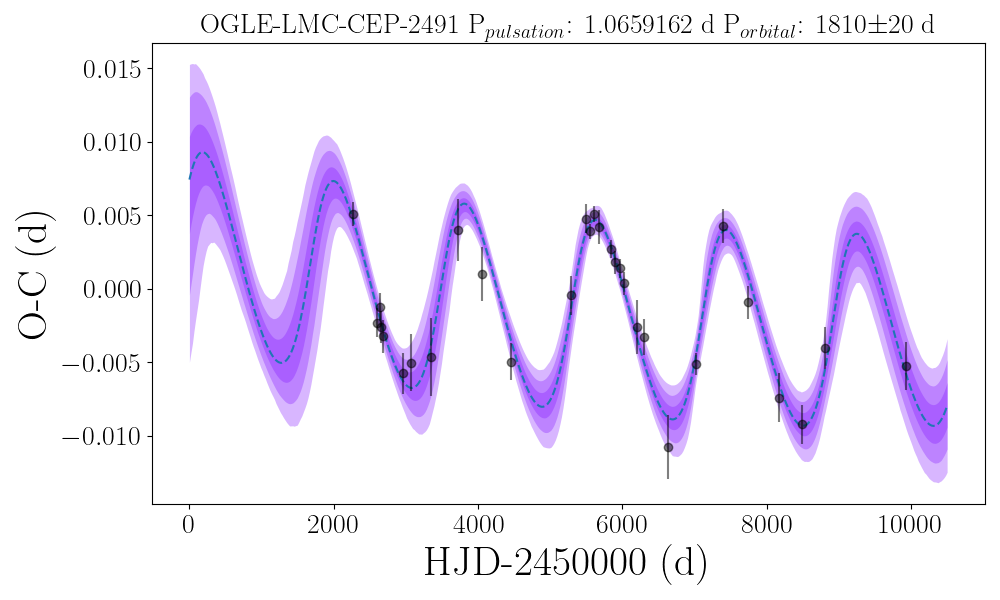}}
{\includegraphics[height=4.5cm,width=0.49\linewidth]{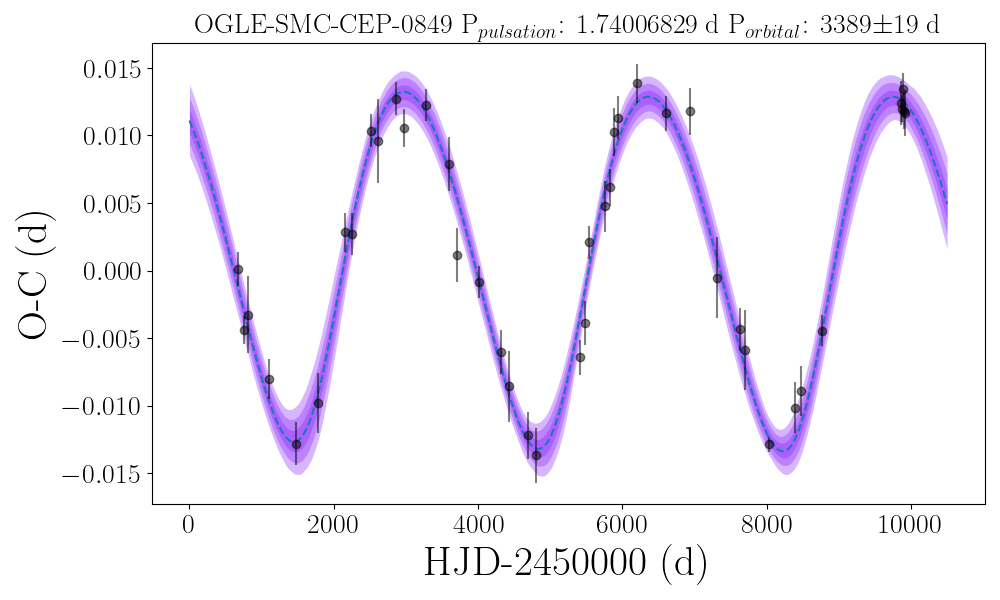}}
{\includegraphics[height=4.5cm,width=0.49\linewidth]{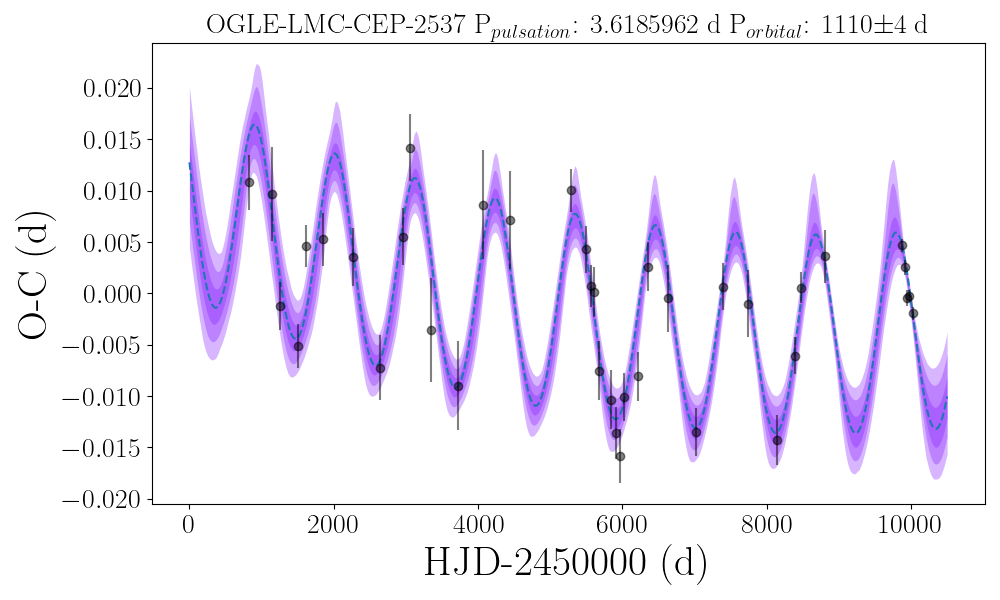}}
{\includegraphics[height=4.5cm,width=0.49\linewidth]{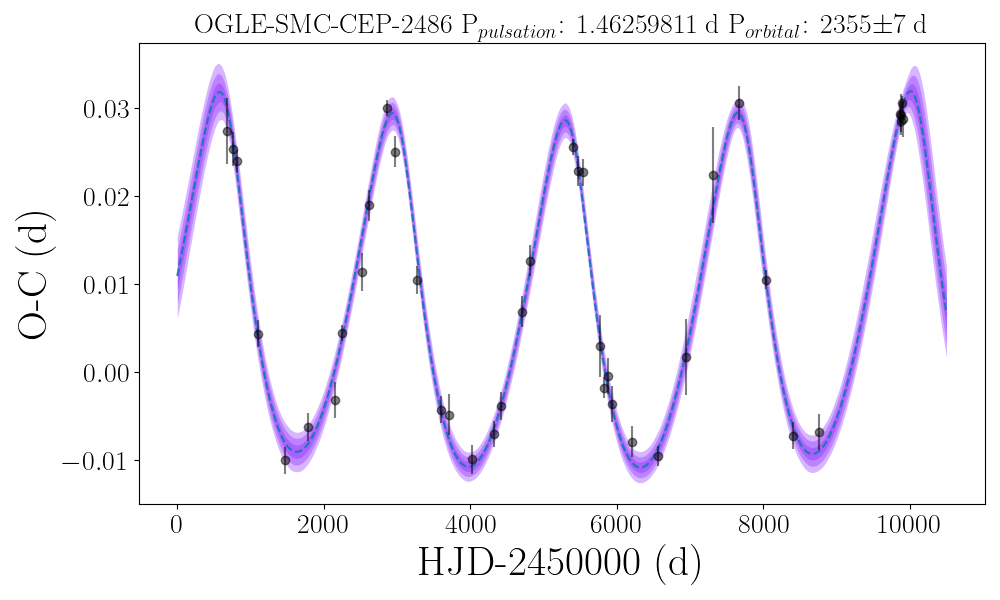}}

\caption{Examples of LMC fundamental-mode Cepheid binary candidate $O-C$ diagrams over-plotted with their MCMC binary fit solutions (left). The blue dashed line represents the orbital solution obtained from the median parameter values of the posterior distribution. The purple-shaded regions imply one, two, and three standard deviations of the resulting MCMC fit in increasing transparency order. Above each panel the OGLE-ID, adopted pulsation period to construct the $O-C$ diagram and the orbital period are shown. Same as above but for SMC fundamental-mode Cepheids (right). The remaining sample is presented in the appendix.}
\label{fig:ocplot_Fmode}
\end{center}
\end{figure*}

\begin{figure*}[p]
\begin{center}
{\includegraphics[height=4.5cm,width=0.49\linewidth]{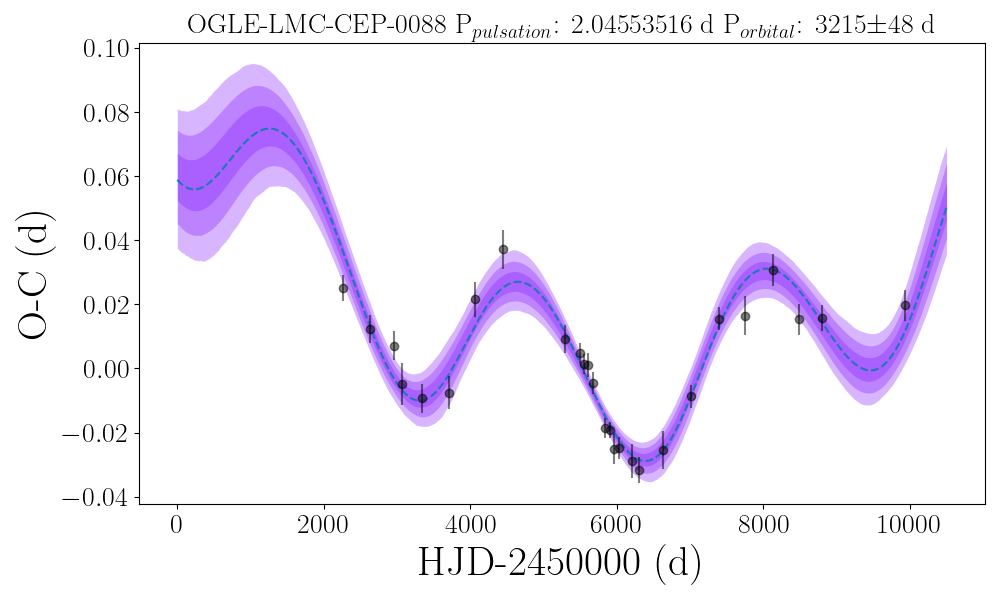}}
{\includegraphics[height=4.5cm,width=0.49\linewidth]{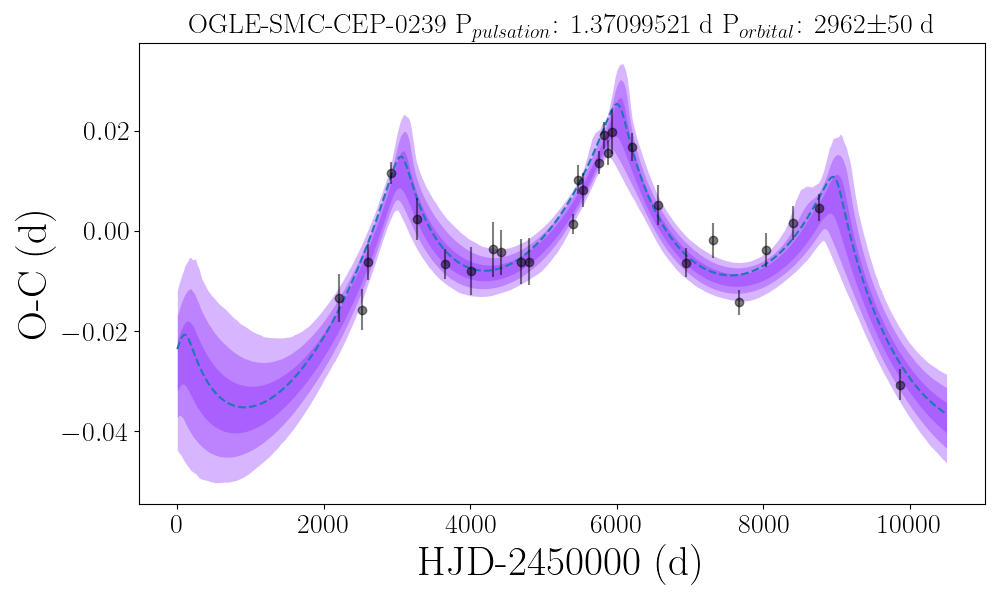}}
{\includegraphics[height=4.5cm,width=0.49\linewidth]{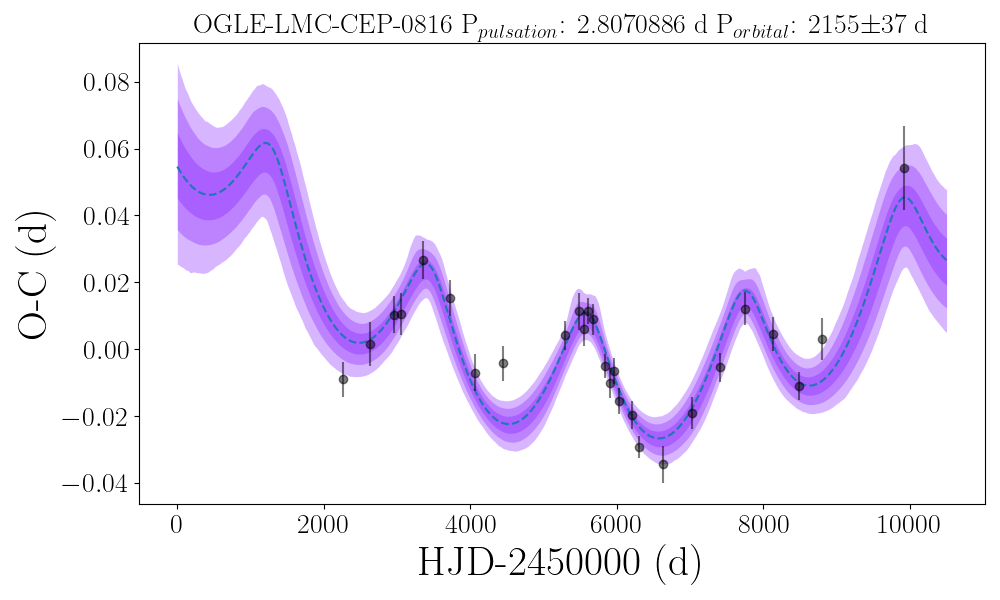}}
{\includegraphics[height=4.5cm,width=0.49\linewidth]{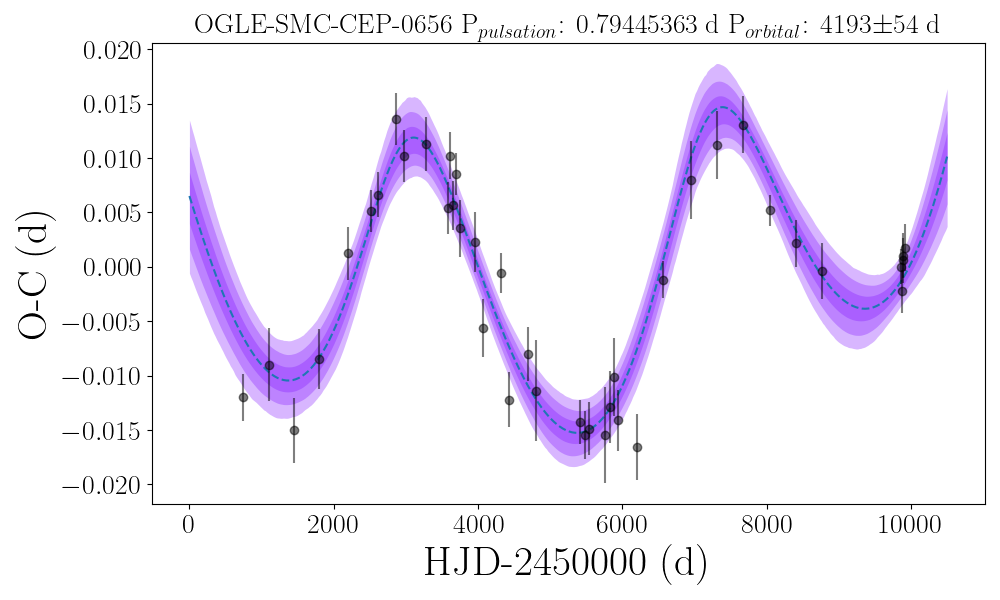}}
{\includegraphics[height=4.5cm,width=0.49\linewidth]{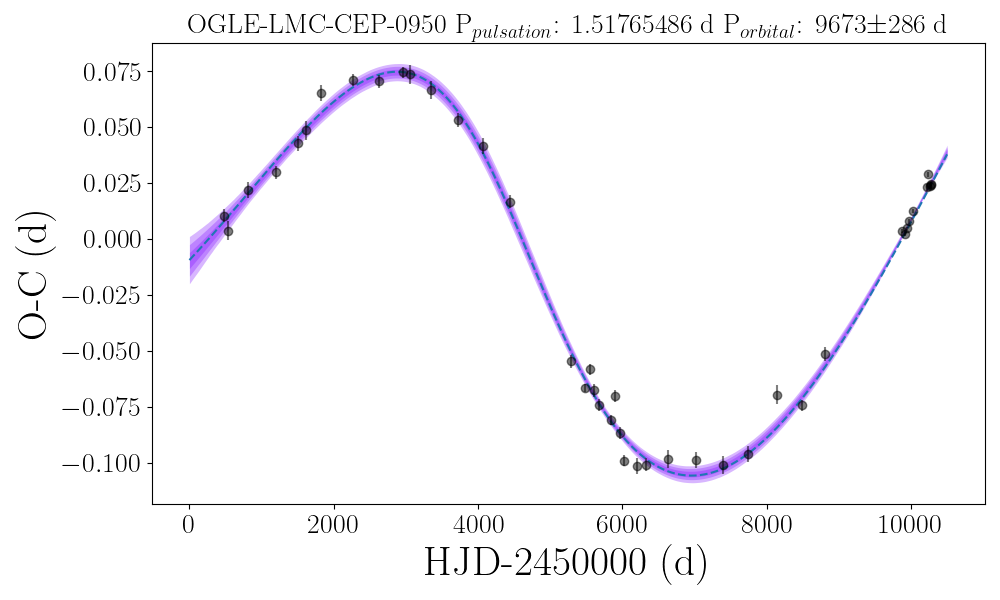}}
{\includegraphics[height=4.5cm,width=0.49\linewidth]{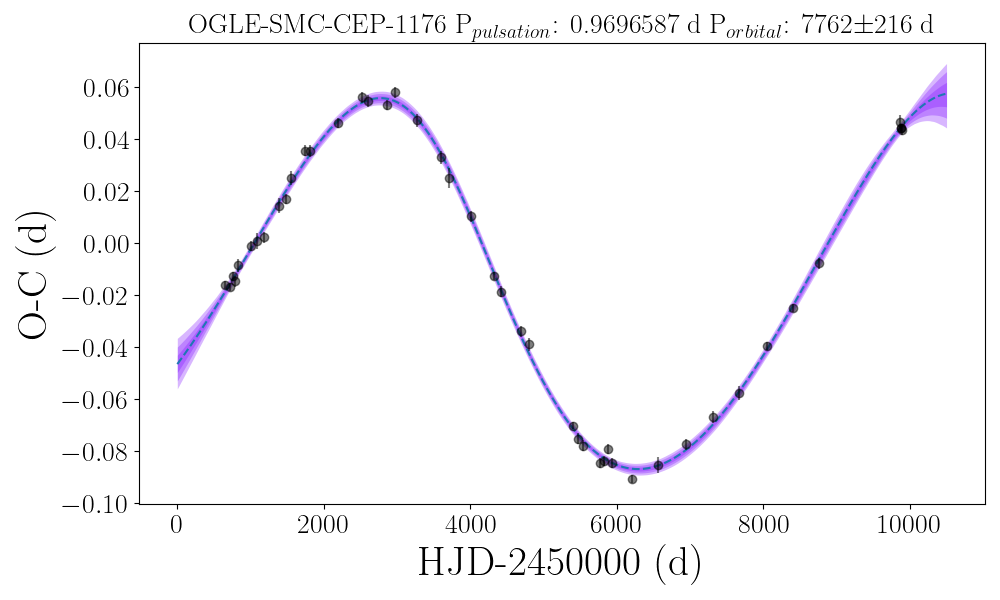}}
{\includegraphics[height=4.5cm,width=0.49\linewidth]{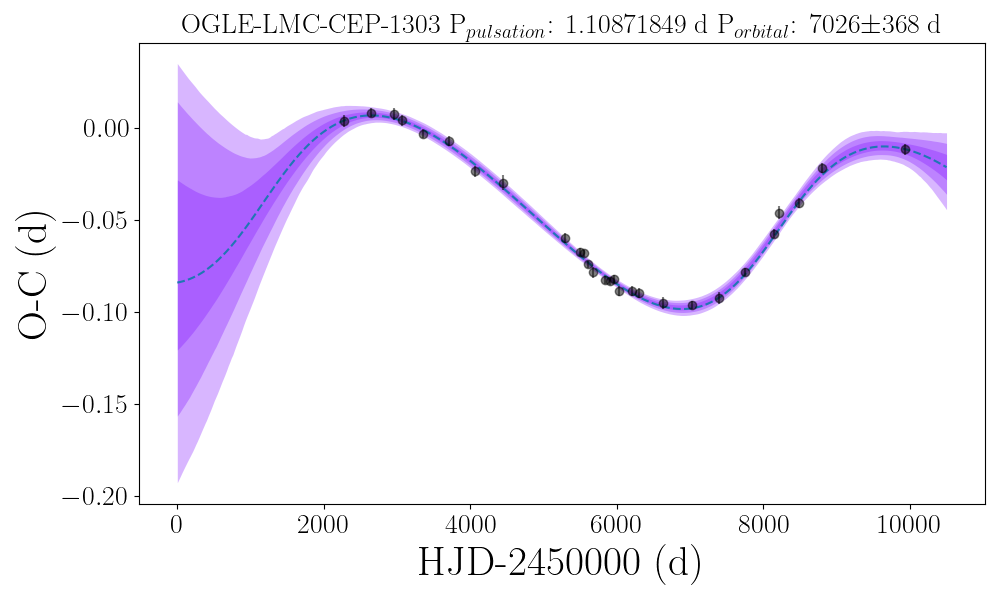}}
{\includegraphics[height=4.5cm,width=0.49\linewidth]{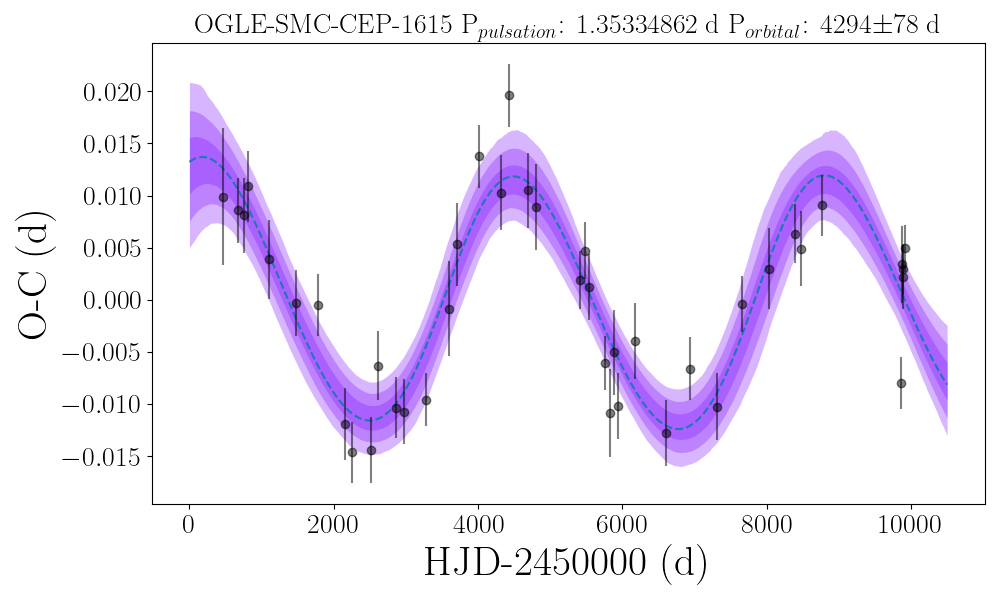}}
{\includegraphics[height=4.5cm,width=0.49\linewidth]{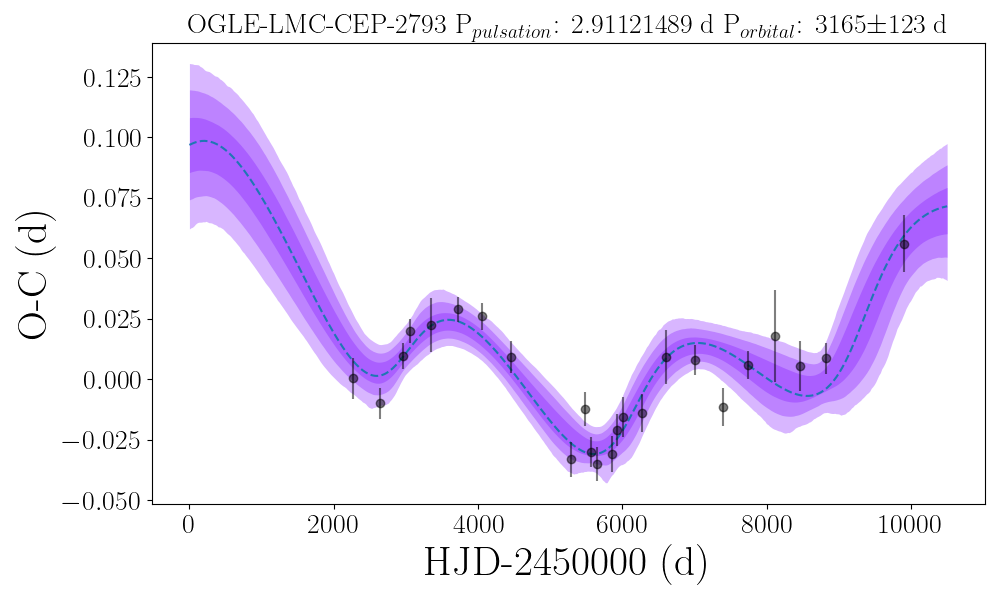}}
{\includegraphics[height=4.5cm,width=0.49\linewidth]{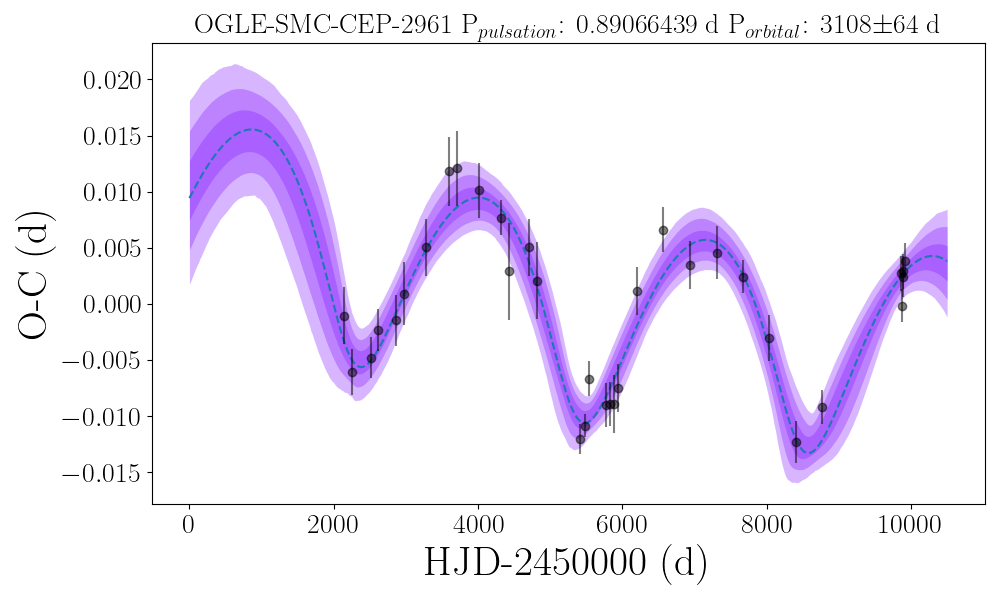}}
\caption{Examples of LMC first-overtone mode Cepheid binary candidate $O-C$ diagrams over-plotted with their MCMC binary fit solutions (left). The blue dashed line represents the orbital solution obtained from the median parameter values of the posterior distribution. The purple-shaded regions imply one, two, and three standard deviations of the resulting MCMC fit in increasing transparency order. Above each panel the OGLE-ID, adopted pulsation period to construct the $O-C$ diagram and the orbital period are shown. Same as above but for SMC first-overtone mode Cepheids (right). The remaining sample is presented in the appendix.}
\label{fig:ocplot_1Omode}
\end{center}
\end{figure*}

\section{Results}
\label{sec: Results}

From the possible candidate inventory of periodic variation, we ended up with 197 Cepheids (145 in SMC and 52 in LMC), for which we carefully inspect their resulting model fits, binary parameters and posterior distributions. To reach the above-mentioned final sample, many candidates were rejected if any of the following criteria were met: (i) strong amplitude changes in the raw photometry time series, (ii) a large gap (2-3 seasons) in the photometric data, (iii) an $O-C$ diagram that is shorter than one full cycle, thus preventing the periodic nature from being well established, (iv) significant $O-C$ point deviation from the resultant fit (more than ten points beyond 3-sigma in the residual), or (v) nonphysical or skewed distribution of fitted binary parameters in the posterior distribution. 

The case of a significant number of $O-C$ points deviating from periodic fit, as mentioned above, but classified as periodic, was more frequent in overtone-mode Cepheids in both the LMC and SMC. One likely nuisance factor could be the presence of non-linear period changes which can distort the original $O-C$ shape leading to misinterpretation. This is supported by earlier studies \citep{Berdnikov1997AstL...23..177B,Poleski2008AcA....58..313P,Evans2015MNRAS.446.4008E} that indicated first-overtone mode pulsation to be less stable and likely to show irregularities in $O-C$ diagrams.

The final sample consists of both strong candidates and marginal ones, and they are listed with their binary parameters in Tables~\ref{tab:LMC Binary list} and \ref{tab:SMC Binary list} for the LMC and SMC Cepheids, respectively. Candidates labeled as strong all have less than two \% uncertainty on the derived orbital period, an $O-C$ covering at least one full orbital cycle, and normal posterior distribution of all the fitted parameters. Marginal candidates do not satisfy either of these conditions, for instance: having a higher uncertainty on the orbital period, not enough coverage of the orbital cycle (mainly in longer orbital period binaries), or at least one unconstrained binary parameter in the posterior distribution or multiple $O-C$ points (more than 5) beyond the 3-sigma binary fit. In Figs.~\ref{fig:ocplot_Fmode} and \ref{fig:ocplot_1Omode} we present the gallery of a few exemplary $O-C$ diagrams from both galaxies for fundamental-mode and first-overtone candidates, respectively. The remaining $O-C$ diagrams for the candidates are provided in the appendix Figs. \ref{fig:appendix_ocplot_Fmode_LMC}, \ref{fig:appendix_ocplot_Fmode_SMC}, \ref{fig:appendix_ocplot_1Omode_LMC} and \ref{fig:appendix_ocplot_1Omode_SMC}. Each panel in the figure provides an $O-C$ diagram for the candidate with its OGLE ID, pulsation period (used to construct the $O-C$) and orbital period.

\subsection{Distribution of binary parameters}
\label{subsec: Distribution of binary parameters}
The distribution of the obtained binary parameters for the full sample is shown in Figs. \ref{fig:binary_params_Fmode} and \ref{fig:binary_params_1Omode} for fundamental-mode and first-overtone mode Cepheids, respectively.

\subsubsection{Fundamental-mode Cepheids}
\label{subsubsec: binary parameters distribution_Fmode}

Firstly we discuss fundamental-mode candidates in both LMC and SMC fields. Due to constraints on the total data span, our analysis will miss possible long orbital period ($P_\mathrm{orbital}>9500$\,d) binary candidates, forming the upper limit of detection. In the SMC, the orbital periods for fundamental-mode Cepheid binary candidates range between $500-7500$\,d, with a maximum at around $2000$\,d, from where the distribution falls off toward longer periods. In the LMC, they are distributed in the range of $500-4500$\,d, with a wide and flat maximum at around $3500$\,d. 

The eccentricity distribution for the SMC candidates shows multiple peaks with the vast majority of the stars having a value less than 0.7, whereas the LMC candidates show marginal peaks at around $0.15$ and $0.35$. The statistics are too low, however, to assign physical significance to these peaks. It is important to note that the LMC eccentricity distribution is roughly concentrated between $0.1-0.7$, whereas in the SMC, eccentricities go as high as $0.9$ with roughly half a dozen candidates with high eccentricities ($\geq0.8$). We note that the SMC sample is about three times larger, so the LMC distribution might not be large enough to draw conclusions. 

The logarithmic distribution of mass function shows asymmetrical behavior around the peak value in both SMC and LMC fundamental-mode Cepheids. For the SMC, the distribution roughly peaks at $-1.2$ whereas for the LMC the distribution peaks at a slightly higher value of $-0.3$. This may imply that high-mass binaries in more metal-rich environments result in higher initial mass ratios $q$. Moving towards higher mass function values, away from the peak, the SMC distribution decreases gradually, but for the LMC it falls off sharply, implying a narrower range of total masses for Cepheid binaries for fundamental-mode candidates.

\begin{figure*}
\begin{center}
{\includegraphics[height=9cm,width=1.\linewidth]{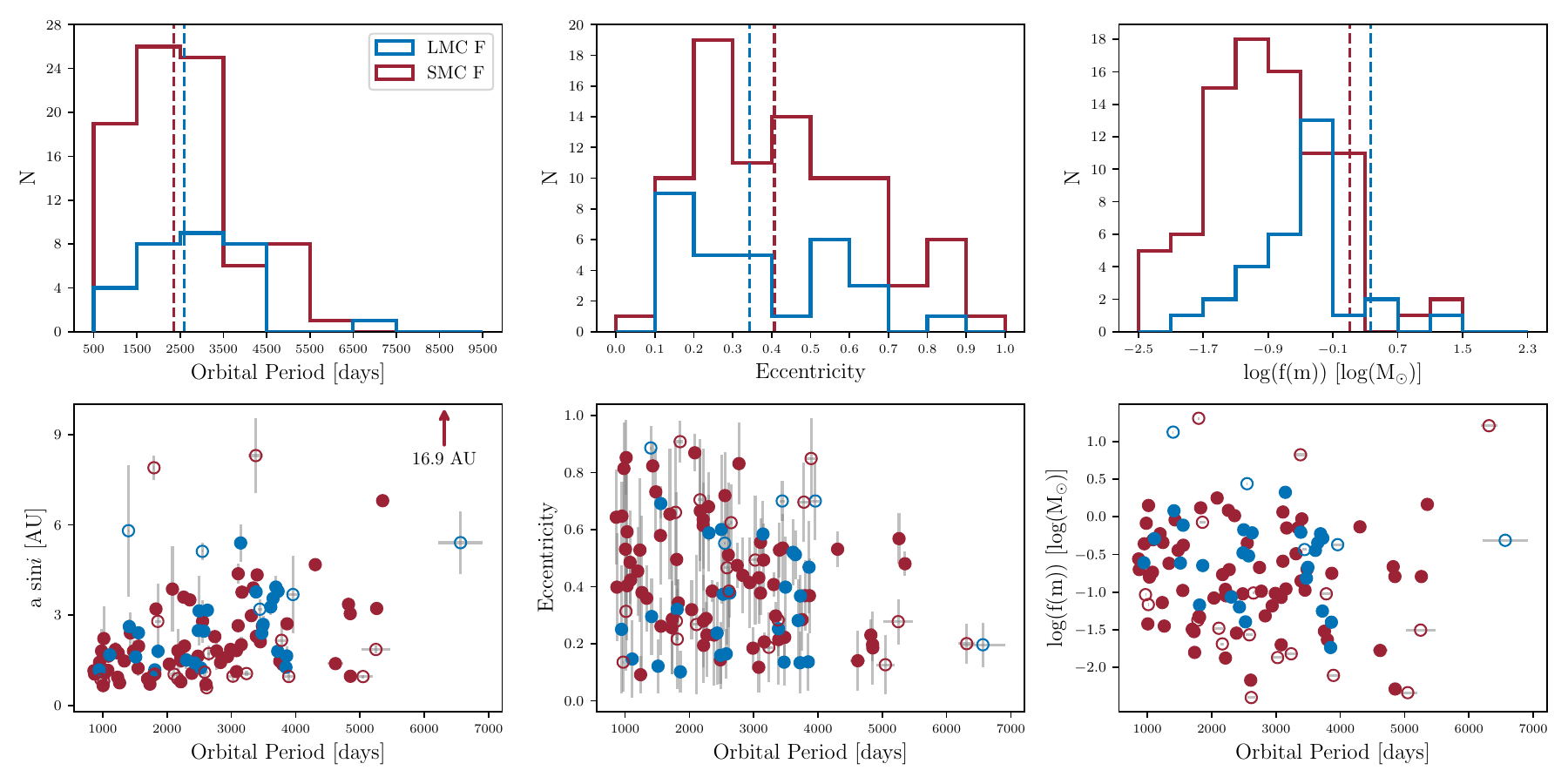}}
\caption{Distribution of orbital period, eccentricity and (logarithmic) mass functions of LMC and SMC fundamental-mode binary Cepheid candidates (shown at the top, from left to right). The dashed vertical line denotes the median value of the distribution. Distribution of the projected semimajor axes, eccentricities and (logarithmic) mass functions as a function of the orbital periods (bottom, from left to right). Filled blue and red dots denote LMC and SMC fundamental-mode strong candidates respectively. Empty symbols denote the marginal candidates.}
\label{fig:binary_params_Fmode}
\end{center}
\end{figure*}

The lower panels of Fig.~\ref{fig:binary_params_Fmode} show the distribution of three of the binary parameters: projected semi-major axis, eccentricity and logarithm of the mass function as a function of the orbital period. The majority of projected semi-major axis ($a\sin i$) values are within $6$\,AU. Only four candidates (three of them are marginal) extend beyond this limit, and all of them are from the SMC fundamental-mode sample. A weak relation of increasing $a\sin i$ value with increasing orbital period can be seen, at least for the LMC sample. The eccentricity values are uniformly distributed till $3500$\,d, and for longer orbital periods cases with $e>0.5$ are rare in both Magellanic Clouds. Similarly, the logarithm of mass function is well distributed across the orbital periods with no particular clustering of points or trends in either the LMC and the SMC.

\subsubsection{First-overtone mode Cepheids}
\label{subsubsec: binary parameters distribution_1Omode}

\begin{figure*}
\begin{center}
{\includegraphics[height=9cm,width=1.\linewidth]{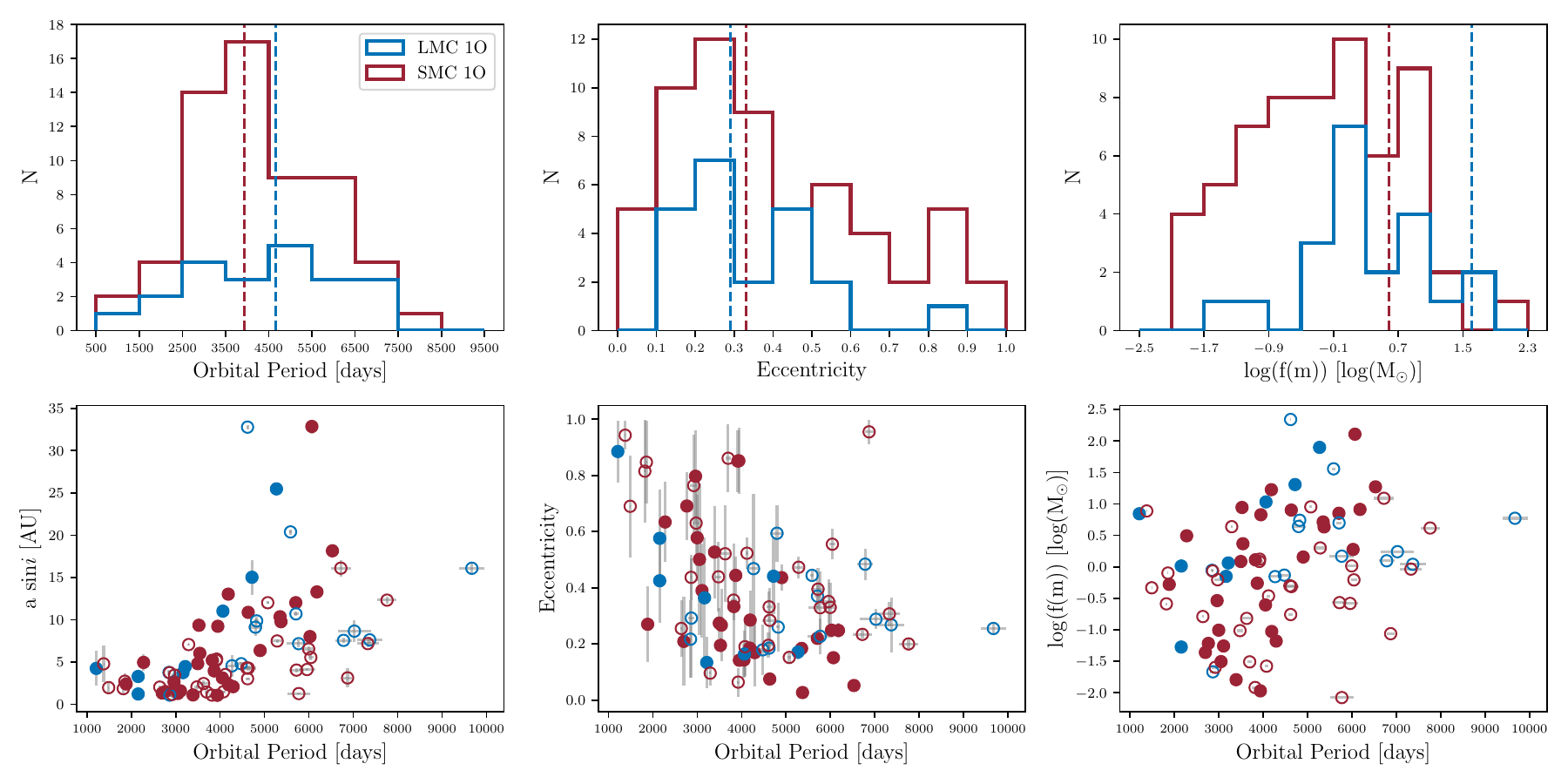}}
\caption{Same as Fig.~\ref{fig:binary_params_Fmode} but for first-overtone binary Cepheids candidates.} 
\label{fig:binary_params_1Omode}
\end{center}
\end{figure*}

The first-overtone Cepheid binary candidate parameters from the SMC and LMC are shown on Fig.~\ref{fig:binary_params_1Omode}. Compared to fundamental-mode candidates, the distribution of orbital periods extends to longer values ($\sim$8500\,d). Looking at the distribution of the orbital periods, the number of the LMC candidates is too low to draw conclusions. In the SMC, the maximum is at $\sim$3700\,d, from where the distribution shows an asymmetrical shape, which falls off steeply at lower orbital periods and is rather broad at the longer orbital periods. Considering eccentricity, both galaxies have the maximum at a roughly similar value of $\sim$0.2. A few candidates in the SMC do show much higher eccentricities (0.8 or more), similar to the SMC fundamental-mode candidates. The SMC distribution is broader, covering all possible eccentricity bins. In contrast, the majority of the LMC candidates have eccentricity below 0.3. However, as with the binary periods, the statistics for the LMC sample is too low to draw significant conclusions. The distribution of logarithm of the mass functions for the SMC candidates peaks at two values: $\log f(m)=0.1$ and $0.9$. Although having fewer first-overtone Cepheids in the LMC, its distribution also peaks at similar values. At lower values moving from the first peak ($\log f(m)=0.1$) the distribution is broader in the SMC and quite narrow in the LMC. However, any metallicity effect is inconclusive due to the small number of stars in the LMC sample.

For both galaxies, we notice a trend of increasing projected semi-major axis with increasing orbital period. In comparison to the previously discussed fundamental-mode candidates, the first-overtone sample extends to much higher $a\sin i$ values (up to $\sim$30\,AU). The eccentricities are uniformly distributed in the orbital period range of $2000-5500$\,d for both Magellanic Clouds. Below $\sim$2000\,d, eccentricities are higher than $0.6$, whereas above 5500\,d they are typically below $0.5$. The logarithm of the mass function shows a weak positive correlation with the orbital period in both LMC and SMC Cepheids, implying Cepheids with longer orbits have higher total mass of the binaries.

\subsection{Period change rate}
\label{subsec: Period change rate}
One important property of Cepheids (alongside
the binary parameters we determine in this work) is the period change rate. In our analysis, we fit an LTTE model along with an additional parabolic term (see Eq. \ref{eq:lte}). This latter term results from accounting for evolutionary changes in the pulsation period. This is calculated using Eq. \ref{eq:PCR_equation} reported under the PCR column in Tables~\ref{tab:LMC Binary list} and \ref{tab:SMC Binary list} in units d/Myr. A positive value of PCR implies a positive period change with a Cepheid showing secular evolution redwards on the HR diagram, whereas a negative value of PCR implies the opposite, namely, a blueward evolution indicating negative period change. In our combined sample we have 116 Cepheids with positive PCR and 81 Cepheids with negative PCR. The ratio of negative to positive PCR candidates is 0.7, which is close to the 0.8 value derived from the sample by \cite{Rodriguez-Segovia2022MNRAS.509.2885R}. This ratio is a proxy of the slower evolution of Cepheid in the third crossing, compared to the second crossing. Consequently, our reported PCR values are reasonably indicative of the long-term Cepheid evolution.

\subsection{Probing Cepheid companions}
\label{subsec: Searching for companions}

\subsubsection{Estimating mass}
\label{subsubsec: Estimating mass}

We estimated the minimum mass of Cepheid companions under certain assumptions, using the equation:
\begin{eqnarray}
\label{eq:companion_mass}
f(m) = \frac{m_{\text{c}}^3 \sin^3 i}{ (m_{\text{cep}} + m_{\text{c}})^{2}},
\end{eqnarray}
where $m_{\text{c}}$ and $m_{\text{cep}}$ are the masses of the companion and the Cepheid, respectively. If the mass function, inclination angle, and Cepheid mass are known, it allows us to estimate the companion mass. Since with the $O-C$ technique, we have no information regarding the inclination angle, for simplicity, we may assume $i = 90^{\circ}$ and then we can determine the "minimum" mass of the companion. In a similar study for RR~Lyrae stars, \cite{Hajdu2021ApJ...915...50H} assumed $m_{\text{RR}}=0.65\MS$ as a typical mass, since masses of RR~Lyrae stars lie in a narrow range. However, for Cepheids, the range is $3-13\MS$ and quite large to assume any typical value. To approximate the Cepheid mass, we use mass estimates provided by \cite{Groenewegen2023AA...676A.136G}, who studied fundamental-mode Magellanic Cloud Cepheids (142 in the LMC and 77 in the SMC) using spectral energy distributions (SEDs) and several empirical relations: period-luminosity (PL), period-radius (PR), and mass-luminosity (ML) relations. For each Cepheid, they calculated Cepheid mass using five different literature relations (see their appendix, Table B.1, and references therein). Finally, the adopted mass of a Cepheid was the median of the previous five estimates. From their sample, we used the Cepheid periods and corresponding mass estimates to construct a period-mass (PM) relation. Since their data extends for Cepheids with periods as high as 200\,d, and periods in our sample extend only up to 10\,d, we construct the relation using \cite{Groenewegen2023AA...676A.136G} data for pulsation periods less than 50\,d (to maintain Cepheids with statistically significant empirical relation). The period-mass (PM) relation estimated in log--log space is linear and is given as follows:

\begin{eqnarray}
\label{eq:logM-logP}
\log (M/\MS) = (0.368\pm0.022) + (0.352\pm0.018)\log P. 
\end{eqnarray}

Equation~\eqref{eq:logM-logP} for Cepheid mass is based on an average representation of literature relations transforming physical parameters to mass. As a consequence, the resulting estimate of Cepheid mass (and, hence, the minimum companion mass) is crude. Based on the dispersion of mass values for a given Cepheid, as given in \cite{Groenewegen2023AA...676A.136G}, we conclude that Cepheid masses derived with Eq.~\eqref{eq:logM-logP} are accurate to within $\pm1\MS$. 

Since \cite{Groenewegen2023AA...676A.136G} study was only limited to fundamental-mode Cepheids, therefore, for mass estimates of our overtone Cepheid sample, the pulsation periods are fundamentalized using the relation given by \cite{Pilecki2021ApJ...910..118P}:
\begin{eqnarray}
\label{eq:fundamentalized}
P_{\rm F} =  P_{\rm 1O} (1.418 + 0.115 \log P_{\rm 1O}),
\end{eqnarray}

\noindent where $P_{\rm F}$ and $P_{\rm 1O}$ are the fundamental-mode and first-overtone mode periods, respectively. 

\begin{figure}
\begin{center}
{\includegraphics[width=8cm]{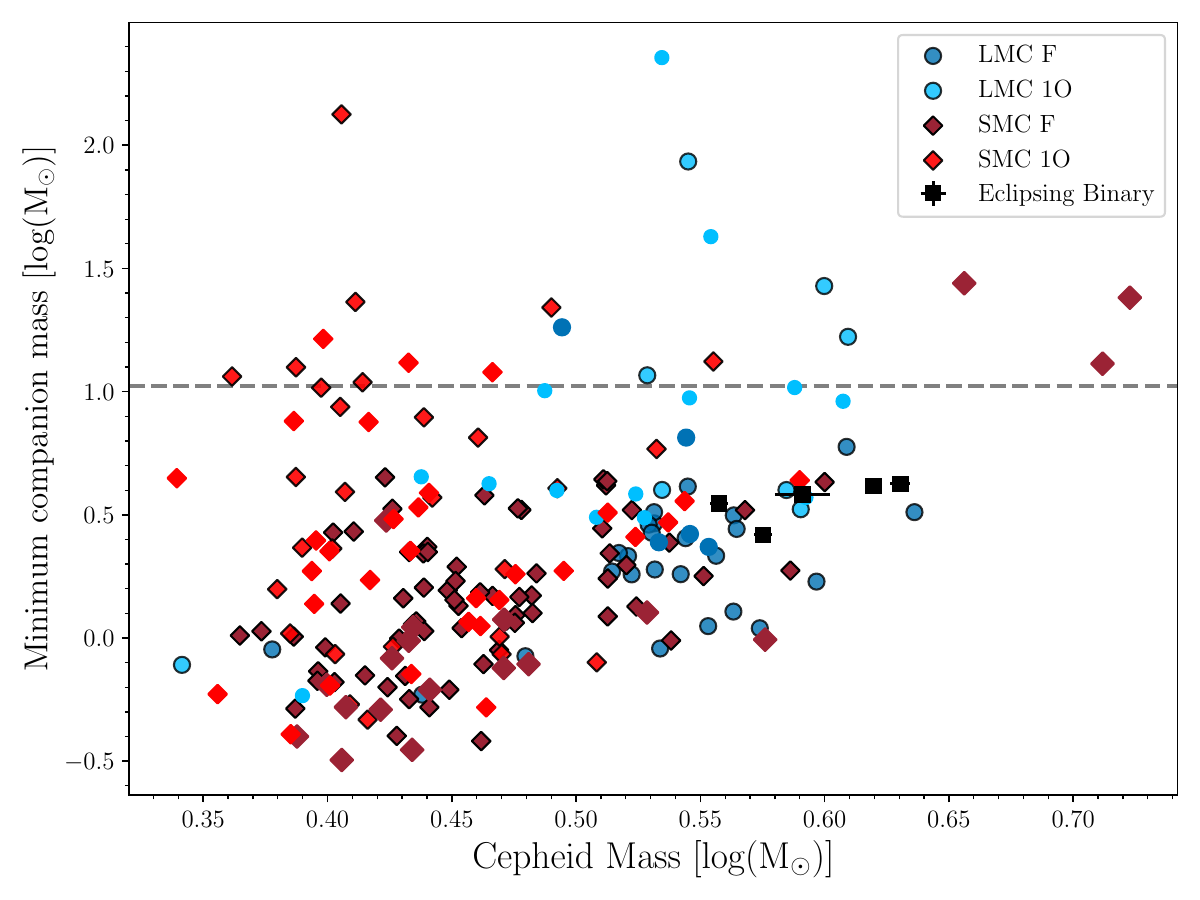}}
\caption{Distribution of estimated minimum companion masses as a function of derived Cepheid masses (see subsection \ref{subsec: Searching for companions}). The symbols represent Cepheids from the LMC F (blue circle), SMC F (brown diamond), LMC 1O (light blue circle) and SMC 1O (red diamond) samples. Symbols with black edges denote strong candidates whereas ones without denote marginal candidates. For comparison, we plot LMC Cepheid in eclipsing binaries (black filled square): OGLE-LMC-CEP-0227, OGLE-LMC-CEP-1718, OGLE-LMC-CEP-1812, OGLE-LMC-CEP-2532 and OGLE-LMC-CEP-4506 \citep{Pilecki2018ApJ...862...43P}. The dashed gray line denotes two times the maximum Cepheid mass in the overall sample.} 
\label{fig:Comp_mass}
\end{center}
\end{figure}

Figure~\ref{fig:Comp_mass} shows the above relations applied to our Cepheid sample. It shows, in the logarithmic space, the distribution of the estimated minimum companion mass, using Eq.~\eqref{eq:companion_mass}, and derived Cepheid mass, using the $P-M$ relation, Eq.~\eqref{eq:logM-logP}. For the SMC sample, the minimum companion masses are uniformly distributed up to the Cepheid mass of $\sim$ 3.2\MS, above which they tend to be systematically higher. For the LMC sample, there is a weak positive correlation between the minimum companion mass and the binary Cepheid mass. 

\begin{figure*}
\begin{center}
\begin{minipage}{0.95\linewidth}
\includegraphics[width=\linewidth]{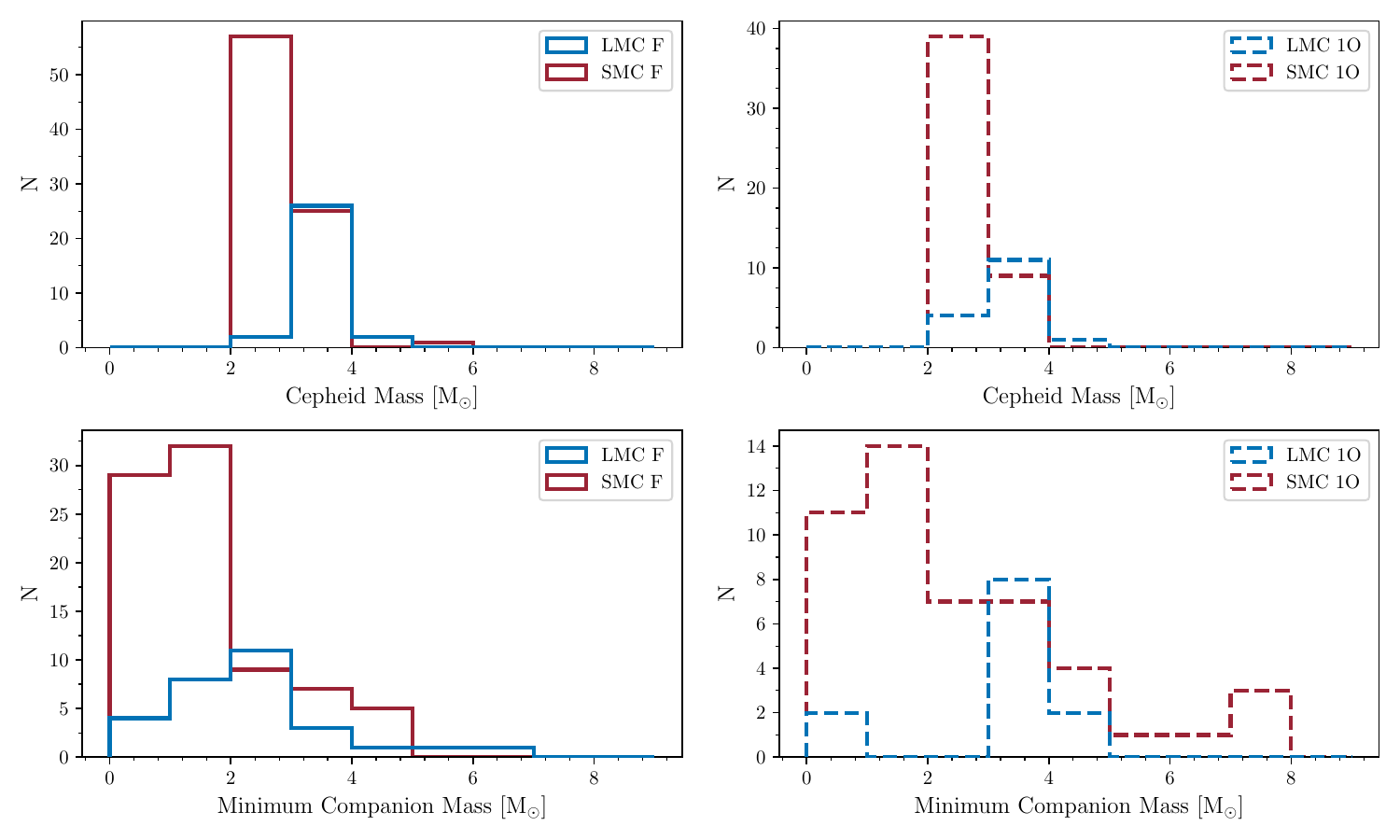}
\caption{Distribution of derived Cepheid mass (top) for the fundamental (left) and first-overtone (right) modes. The corresponding distributions of minimum companion masses (bottom).}
\label{fig:hist_CEP_COMP_mass}
\end{minipage}
\end{center}
\end{figure*}

In Figure~\ref{fig:hist_CEP_COMP_mass} we plot the distribution of both derived Cepheid mass and calculated minimum companion mass for each pulsation mode and both Magellanic Clouds. For this distribution, we exclude the peculiar mass function Cepheids (detailed in section \ref{subsec: High Mass function candidates}). The majority of the SMC fundamental-mode Cepheids of our sample have masses in the $2.5-3\MS$ range whereas in the LMC masses higher than $3\MS$ are observed; this is a consequence of different pulsation period distributions in both Clouds: the pulsation periods of SMC Cepheids are generally shorter than in the LMC. The distributions are similar for the first-overtone mode sample. For minimum companion masses, the majority of the SMC fundamental-mode candidates extend up to $2\MS$ with a tail extending up to $4\MS$. In the LMC, the distribution favors slightly higher minimum companion masses ($\sim1-3\MS$). In the first-overtone mode sample, the SMC candidates favor a broader range of companion masses (maximum at $2\MS$), extending until $\sim7.5\MS$. In the LMC, the majority of the sample lies in the $\sim3-5\MS$ range.

\subsubsection{Nature of the companions}
\label{subsubsec: Nature of the companion}
\begin{figure*}
\begin{center}
{\includegraphics[width=0.95\linewidth]{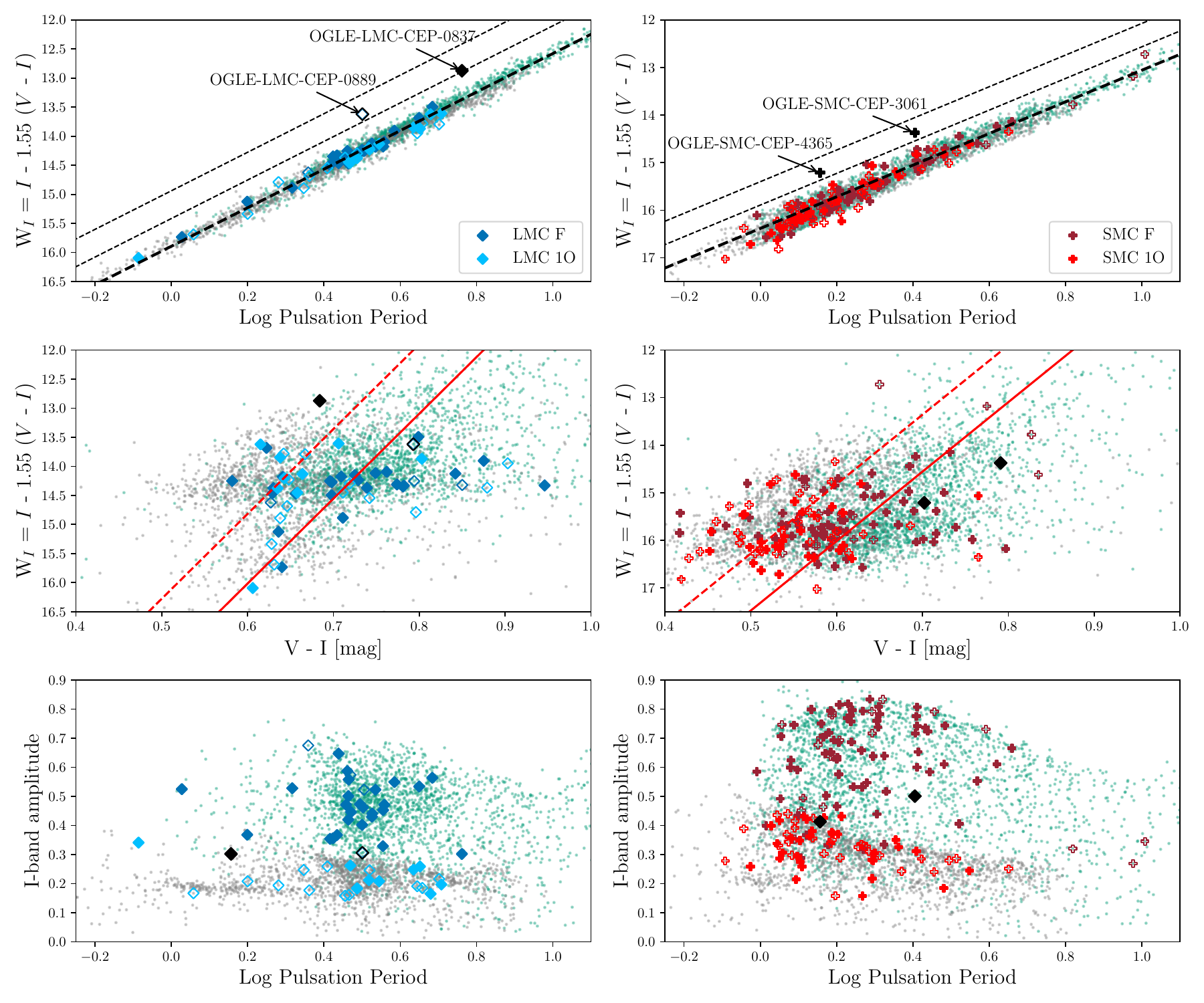}}
\caption{Cepheid Period-Luminosity ($P-L$) relation (top-left) using the reddening-free Wesenheit index for fundamental-mode (blue diamond symbol) and first-overtone mode (light blue diamond symbol) Cepheids from the LMC sample. The black solid line depicts the adopted Period-Wesenheit relation for the LMC whereas the black dashed lines indicate if Cepheids are 50\% and 100\% brighter than the P-L relation respectively. Color-magnitude diagram for the LMC (middle), where the solid red line denotes the color of a typical Cepheid, while the dashed red line denotes the total brightness of a Cepheid if it had a companion of similar color and luminosity. Distribution of I-band amplitude as a function of the logarithm of the pulsation period (bottom-left). The right three panels are the same as above but for the SMC. For both LMC and SMC fields, the candidate overbright Cepheids are marked with an arrow. Note: Filled symbols denote strong candidates, whereas empty symbols denote marginal candidates. The background green and gray points denote fundamental and first-overtone mode Cepheids from the full OGLE database respectively for LMC \citet{soszynski2008optical} and SMC \citet{soszynski2010optical}. The first-overtone candidates are ''fundamentalized'' using Eq.~\ref{eq:fundamentalized}.}
\label{fig:P-L relation}
\end{center}
\end{figure*}

To investigate further the nature of the companions, we utilized findings from a recent study by \cite{Pilecki2021ApJ...910..118P}, where the authors noted that Cepheids in binary systems with giant companion manifest as appearing overbright when compared to "typical" Cepheids. A typical Cepheid is either a Cepheid with a main-sequence companion or an isolated Cepheid, where there is negligible to no additional contribution to actual Cepheid's brightness, respectively. In the study by \cite{Pilecki2021ApJ...910..118P}, Cepheids with systematic offset from $P-L$ relationship were investigated to be mostly Cepheids with giant companions. Two fundamental-mode candidates from the LMC (OGLE-LMC-CEP-0837 and OGLE-LMC-CEP-0889) and two from the SMC (OGLE-SMC-CEP-3061 and OGLE-SMC-CEP-4365), show excess in their total observed brightness when compared to the respective Period-Wesenheit index from \cite{soszynski2008optical,soszynski2010optical} as seen in Fig.~\ref{fig:P-L relation}. These Cepheids (except OGLE-LMC-CEP-0837) appear redder in the color-magnitude diagram (see Fig.~\ref{fig:P-L relation} middle panels), in agreement with what was observed by \cite{Pilecki2021ApJ...910..118P} for their LMC sample. As shown previously, we can estimate the Cepheid mass and the minimum mass of its companion. Therefore, for these giant companions, using Eq.~\eqref{eq:companion_mass}, we estimate their minimum masses along with their primary Cepheid represented as $m_{\text{cep}}+m_{\text{c}}$, to be:  
\begin{align*}
\text{OGLE-LMC-CEP-0837:}\ 4.3\MS+3.2\MS\\
\text{OGLE-LMC-CEP-0889:}\ 3.5\MS+6.8\MS\\
\text{OGLE-SMC-CEP-3061:}\ 3.2\MS+2.8\MS\\
\text{OGLE-SMC-CEP-4365:}\ 2.7\MS+4.5\MS
\end{align*}

We note that these are not direct mass measurements and are dependent on the adopted empirical relation from Eq.~\eqref{eq:logM-logP} and derived parameters from the $O-C$ analysis. Our two LMC candidates were already reported as double-lined spectroscopic binaries (SB2 type) by \cite{Pilecki2021ApJ...910..118P}. We corroborate their findings regarding their binary nature via our independent, photometry-based method (LTTE analysis). The other two SMC candidates were not reported earlier. These targets could be observed spectroscopically to confirm if they show characteristics of double-lined binaries (SB2 type) in the composite spectra. This will eventually corroborate the binarity nature and aid in refining their orbital parameters.

\subsection{Low-amplitude variability or additional radial mode}
\label{subsec: Low amplitude variability}

Cepheids may show low-amplitude variability such as the presence of non-radial modes \citep[mainly in first-overtone stars,][]{Moskalik2008CoAst.157..343M,Moskalik2009MNRAS.394.1649M,soszynski2008optical,soszynski2010optical,Rathour2021MNRAS.505.5412R,Plachy2021ApJS..253...11P,Smolec2023MNRAS.519.4010S} and periodic modulation of the pulsation amplitude and/or phase \citep{Moskalik2009MNRAS.394.1649M,Soszynski2015b,Smolec2017MNRAS.468.4299S,Rathour2021MNRAS.505.5412R}. The latter is very common in another type of classical pulsators, RR~Lyrae stars, and is known as the Blazhko effect \citep{blazko1907mitteilung}, as also discussed, for example, in \citet{Kovacs2016CoKon.105...61K,Smolec2016pas..conf...22S,Netzel2023arXiv231014824N}. Such phenomena may be present in our sample and, particularly the possible periodic modulation of pulsation, may alter the light curve and, eventually, the LTTE-determined orbital parameters for our binary candidates. 

\subsubsection{Modulation and non-radial mode verification}
\label{subsec: Modulation and Non-radial mode verification}

\cite{Smolec2017MNRAS.468.4299S} searched through the OGLE Magellanic Cloud sample of fundamental-mode Cepheids to look for periodic modulation of the pulsation and compiled a list of 51 modulated Cepheids, 29 from the SMC and 22 from the LMC. Another recent study by \cite{Smolec2023MNRAS.519.4010S} reported periodic modulation in the LMC and SMC first-overtone OGLE samples. They reported 24 such candidates in the LMC and only 3 in the SMC. We cross-matched both our fundamental-mode and first-overtone samples and found no overlap with the above two reported studies, which was expected since the modulation periods ($\sim$10-100\,d) are much shorter than the ones captured by LTTE analysis. Hence, our Cepheid binary candidates are free from any likely low amplitude periodic modulation of pulsation.

We also check if our sample contains Cepheids with non-radial modes reported. We find an overlap in five SMC Cepheids (OGLE-SMC-CEP-1611, OGLE-SMC-CEP-1615, OGLE-SMC-CEP-2033, OGLE-SMC-CEP-2293 and OGLE-SMC-CEP-3097) and two LMC Cepheids (OGLE-LMC-CEP-0764 and OGLE-LMC-CEP-0839) which show periodicities with period ratios of $\Px/\Po\in (0.60,\, 0.65)$ co-existing with the radial first-overtone mode. Here, $\Px$ denotes the period of additional variability. In certain cases, instead of the above-mentioned period-ratio range, there can be additional periodicity at the sub-harmonic of $\nu_{x}$ (i.e. 1/$\Px$), which is the direct detection of the non-radial mode according to the theoretical framework proposed by \cite{Dziembowski2016CoKon.105...23D}. After cross-matching our first-overtone sample with a list of such candidates \citep{Smolec2023MNRAS.519.4010S}, we find an overlap of one SMC candidate (OGLE-SMC-CEP-3097) and two LMC Cepheids (OGLE-LMC-CEP-0764 and OGLE-LMC-CEP-0839) from our list. As a sanity check, namely, to inspect whether the low amplitude periodicities may affect the binary fits, we took the marginal LTTE candidate OGLE-LMC-CEP-0839 and performed a frequency analysis. Thus, we could identify the low-amplitude variability to eventually pre-whiten it from the frequency spectra. Then, we used the prewhitened data, with the additional variability filtered out, to obtain the $O-C$ diagram and re-estimate the binary fit and orbital parameters. Our analysis resulted in marginal differences in the final orbital parameters (well within the errors) implying that the low-amplitude variability indeed does not affect the LTTE signature in our sample.

\subsubsection{Additional radial mode verification}
\label{subsec: Additional radial mode verification}

Apart from the above-mentioned low-amplitude variability, the light curves can be distorted in the presence of another radial mode of pulsation. Therefore, as a check and to scrutinize our candidate binary sample, we check whether these are genuine single-mode Cepheids. \cite{Smolec2023MNRAS.519.4010S} provide a list of first-overtone Cepheids that likely have a presence of another radial mode of pulsation. We find an overlap with two Cepheids from the LMC only (OGLE-LMC-CEP-0839 and OGLE-LMC-CEP-1134). Both stars are candidate F+1O double-mode pulsators. This concludes that the majority of our samples are genuine single-mode Cepheids. OGLE-LMC-CEP-0839, as mentioned previously, also contains additional non-radial variability, whereas OGLE-LMC-CEP-1134 is classified as strong LTTE candidate in our first-overtone sample, so unlikely to be affected by the presence of the additional radial mode. We tested this hypothesis by pre-whitening the fundamental-mode signal in the frequency spectra and performing the $O-C$ analysis again to observe only a marginal difference in the fitted orbital parameters (within the errors). For OGLE-LMC-CEP-0839, we have already checked the negligible contribution of both the additional low-amplitude variability and the radial mode in LTTE analysis. In Tables~\ref{tab:LMC Binary list} and \ref{tab:SMC Binary list}, Cepheids are listed with their binary parameters, whereas the candidates with additional low amplitude variability are marked with $\ddag$ symbol, candidates with a sub-harmonic signal are marked with $\dagger$ symbol and candidates with additional radial mode are marked with $\ast$.

\subsection{Previously detected Cepheid binary candidates}
\label{subsec: Previously detected Cepheid binary candidates}

We compared the list of candidates in our study with earlier detections reported in the literature. In the LMC, OGLE discovered six Cepheids in eclipsing binary systems (OGLE-LMC-CEP-0227, OGLE-LMC-CEP-1718, OGLE-LMC-CEP-1812, OGLE-LMC-CEP-2532, OGLE-LMC-CEP-3782, and OGLE-LMC-CEP-4506 \citep{Udalski1999AcA....49..223U,soszynski2008optical,Udalski2015AcA....65..341U,Soszynski2012AcA....62..219S}. Then, OGLE-LMC-CEP-3782 and OGLE-LMC-CEP-4506 were not included in our original sample, since they lacked OGLE-III data (see our sample selection criteria in Sect.~\ref{ssec:combine}), but we also went on to analyse the available data for these stars.

In the automatic analysis of eclipsing binary stars, the $O-C$ reveals no strong periodic feature. This is due to the template getting distorted due to the eclipse feature getting incorporated into the phased light curve. Therefore, we manually re-analyzed $O-C$ diagrams of these candidates and detected LTTE in three candidates, OGLE-LMC-CEP-1812, OGLE-LMC-CEP-2532 and OGLE-LMC-CEP-4506. We report the orbital periods for these candidates: 554$\pm$1\,d, 788$\pm$4\,d, and 1671$\pm$176\,d, which are consistent with the ones derived from the eclipse timing: 551.776$\pm$0.003\,d, 800.419$\pm$0.009\,d, and 1550.354$\pm$0.009\,d, respectively \citep{Pilecki2018ApJ...862...43P}. For the remaining three candidates: OGLE-LMC-CEP-0227, OGLE-LMC-CEP-1718 and OGLE-LMC-CEP-3782, we still do not see any LTTE. The reason will be explained at the end of this section, along with similar SMC candidates.

From the SB2 candidate sample compiled by \cite{Pilecki2021ApJ...910..118P}, we could recover only two targets in our sample, OGLE-LMC-CEP-0837 and OGLE-LMC-CEP-0889. OGLE-LMC-CEP-0889 is not a spectroscopically confirmed SB2-type binary, but it does show an LTTE signature in the $O-C$. The orbital period we derived is $2569\pm12$\,d which is close to the value of $2533$\,d derived by \cite{Pilecki2021ApJ...910..118P}. However, their mass function estimate ($f(m)=1.83\MS$) is lower than ours ($f(m)=2.95\pm0.33\MS$). Another fundamental-mode candidate from \cite{Pilecki2021ApJ...910..118P}, OGLE-LMC-CEP-0837, is a spectroscopically confirmed SB2-type binary (see their tab.~3) and is also among our strong $O-C$ candidates. We derived an orbital period of $3637\pm62$\,d with an eccentricity of $0.2\pm0.1$. From our empirical relations in Sect.~\ref{subsec: Searching for companions}, we estimated the mass of the Cepheid and its companion to be $4.3\MS$ and $3.2\MS$, respectively.

In the SMC, the OGLE catalog contains five detections of classical Cepheids in eclipsing binary systems \citep[OGLE-SMC-CEP-0411, OGLE-SMC-CEP-1996, OGLE-SMC-CEP-2199, OGLE-SMC-CEP-3235, and OGLE-SMC-CEP-4795;][]{soszynski2010optical,Udalski2015AcA....65..341U}. It is important to note that none of the above SMC eclipsing binary Cepheids have been studied yet with radial velocity observations to obtain precise orbital parameters. We analyzed these stars manually, including OGLE-SMC-CEP-4795, for which only OGLE-IV data are available. After the manual analysis, we still did not see any LTTE in OGLE-SMC-CEP-1996 and OGLE-SMC-CEP-4795 that could corroborate the orbital periods reported from the analysis of the light curves. In OGLE-SMC-CEP-2199, we observed a LTTE signature corresponding to 1204$\pm$31\,d, which is in good agreement with the orbital period of 1210\,d estimated from photometric data. Lastly, OGLE-SMC-CEP-3235 is particularly interesting, as it not only shows a clear signature of LTTE in the $O-C$ diagram but also a single eclipse in $\sim$20-year OGLE baseline data, further validating its binarity nature. Our analysis recovered this first-overtone candidate with an orbital period of $7528\pm429$\,d, with other binary parameters presented in Table~\ref{tab:SMC Binary list}. \cite{Udalski2015AcA....65..341U} estimated the orbital period from the photometric data to be $\sim$4200\,d, which is discrepant from our value. This difference in the orbital period parameter has two reasons. First, our analysis uses data extended beyond 2015. Second, in \cite{Udalski2015AcA....65..341U}, their analysis was limited to fitting a sinusoidal fit to the $O-C$ data assuming a circular orbit ($e=0$) whereas we fit full binary model (including a period change term in Eq.\ref{eq:lte}) within a Bayesian framework.

In principle, all eclipsing binaries should show LTTE. However, to reveal LTTE in the $O-C$ diagram, the orbital period should be relatively long (and correspondingly the orbit should be sufficiently large) so that the amplitude of the $O-C$ variation is above the detection limit and the relatively coarse photometric data allow us to resolve the orbital variation. A total of six reported eclipsing binaries (three in each MC) from the literature did not show LTTEs when subject to analysis. A closer inspection reveals that their orbital periods are short, from 43\,d for OGLE-SMC-CEP-0411  to 412\,d for OGLE-LMC-CEP-1718. For such orbital periods, the amplitude of the $O-C$ produced by the LTTE clearly falls below the detection limit.

It is important to note that the majority of known Magellanic Cloud Cepheid eclipsing binaries have orbital periods less than 800\,d (8 out of 11 Cepheid). While these Cepheid binaries are valuable, searching for them and obtaining precise orbital parameters via radial velocity follow-ups and long-term monitoring is time-consuming and costly. Therefore, our sample is complementary to the known Magellanic Cloud Cepheid eclipsing binaries, as it extends from orbital periods starting above 600 days up to $\sim$9000 days. This shows that due to the longer temporal baseline of photometric observations, the $O-C$ technique is valuable for searching Cepheid binaries in the parameter space that is unattainable by other methods.

\subsection{Peculiar high-mass-function candidates}
\label{subsec: High Mass function candidates}

\begin{figure*}[p]
\begin{center}
{\includegraphics[height=4.5cm,width=0.49\linewidth]{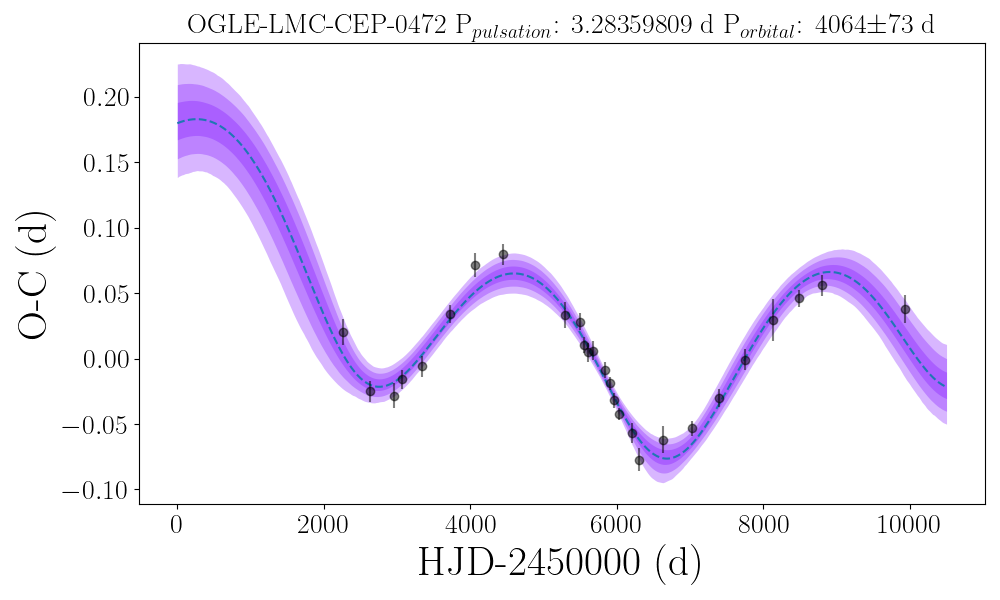}}
{\includegraphics[height=4.5cm,width=0.49\linewidth]{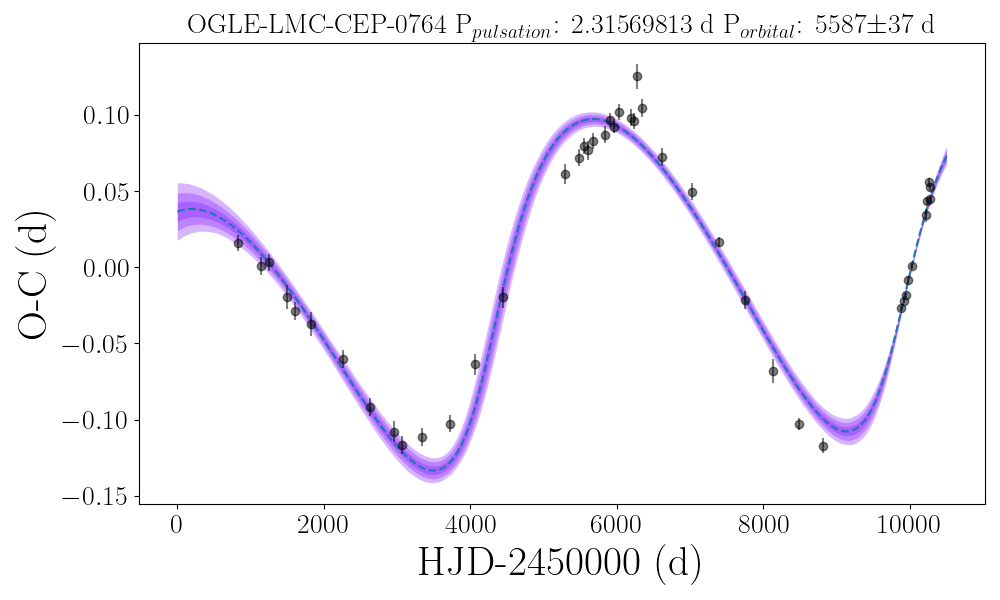}}
{\includegraphics[height=4.5cm,width=0.49\linewidth]{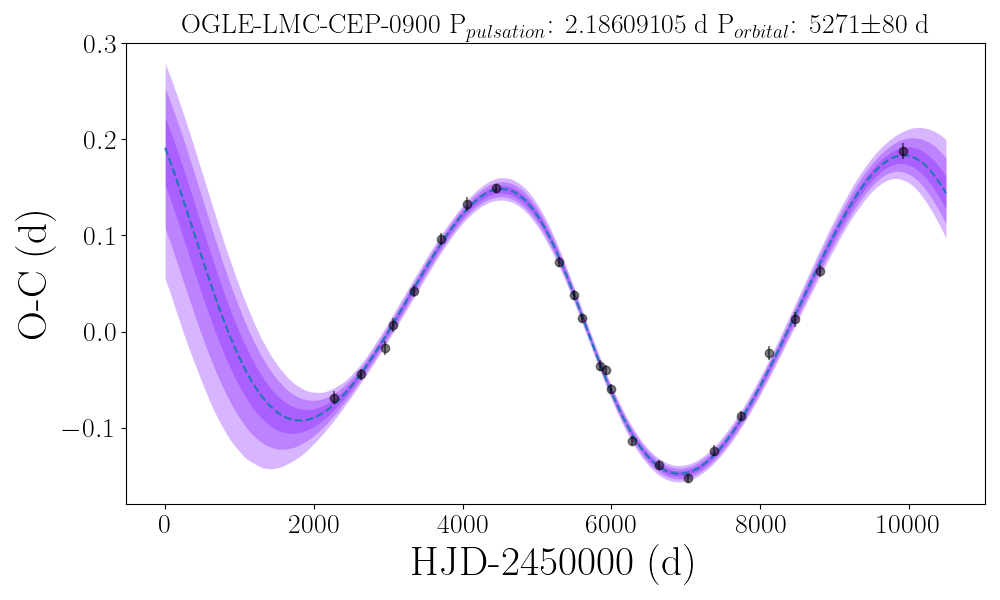}}
{\includegraphics[height=4.5cm,width=0.49\linewidth]{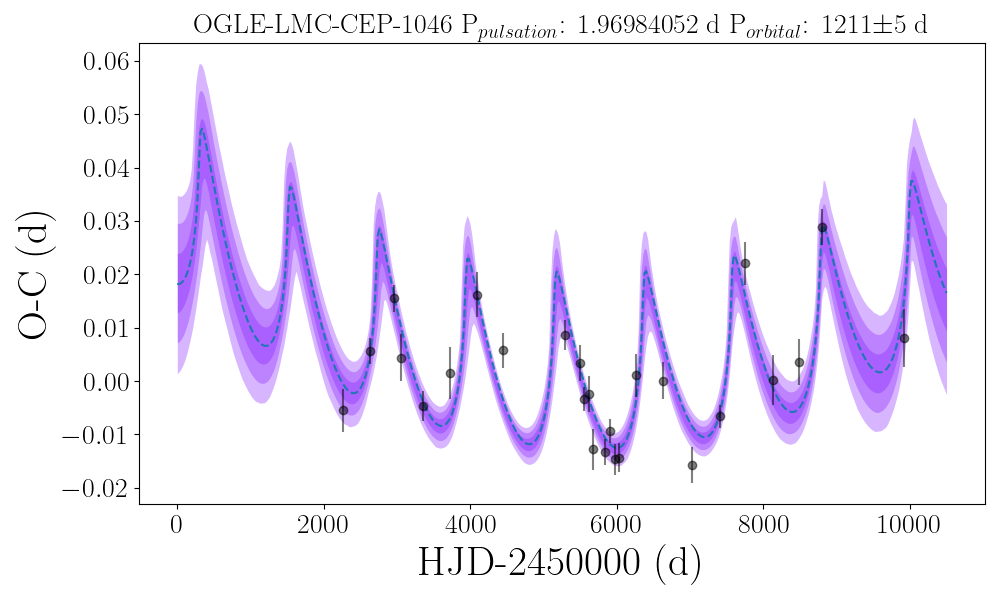}}
{\includegraphics[height=4.5cm,width=0.49\linewidth]{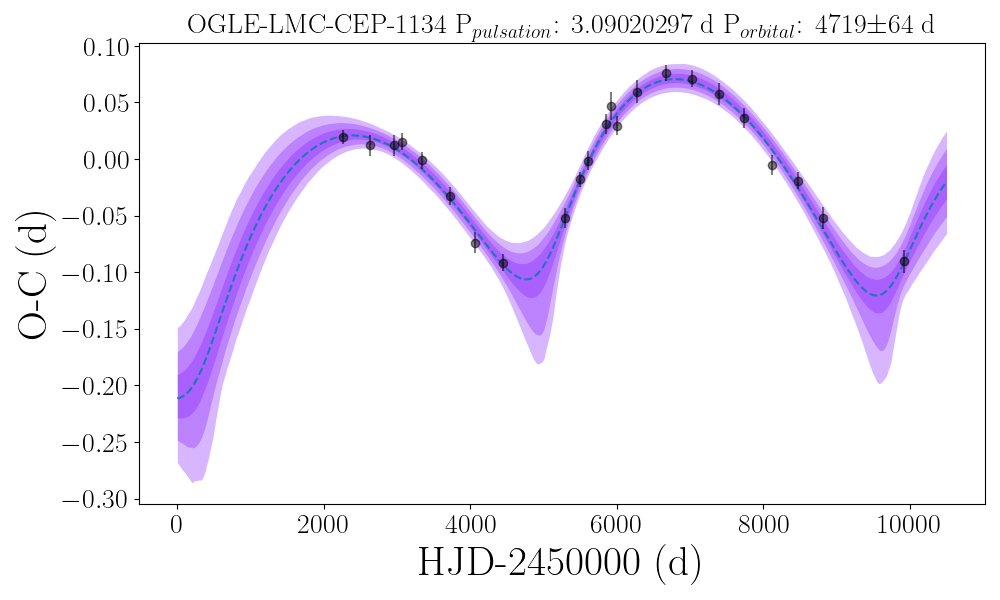}}
{\includegraphics[height=4.5cm,width=0.49\linewidth]{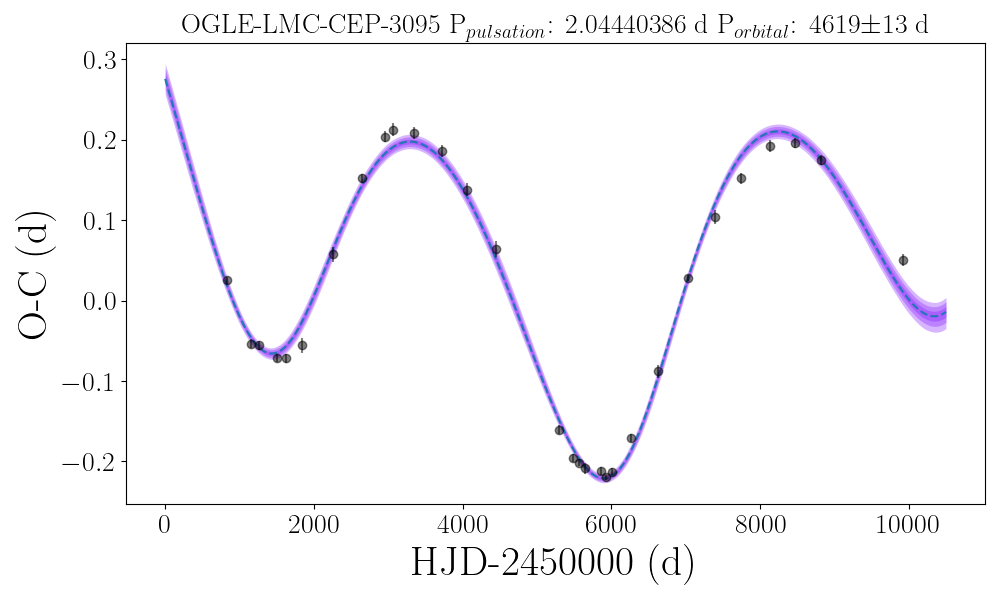}}
{\includegraphics[height=4.5cm,width=0.49\linewidth]{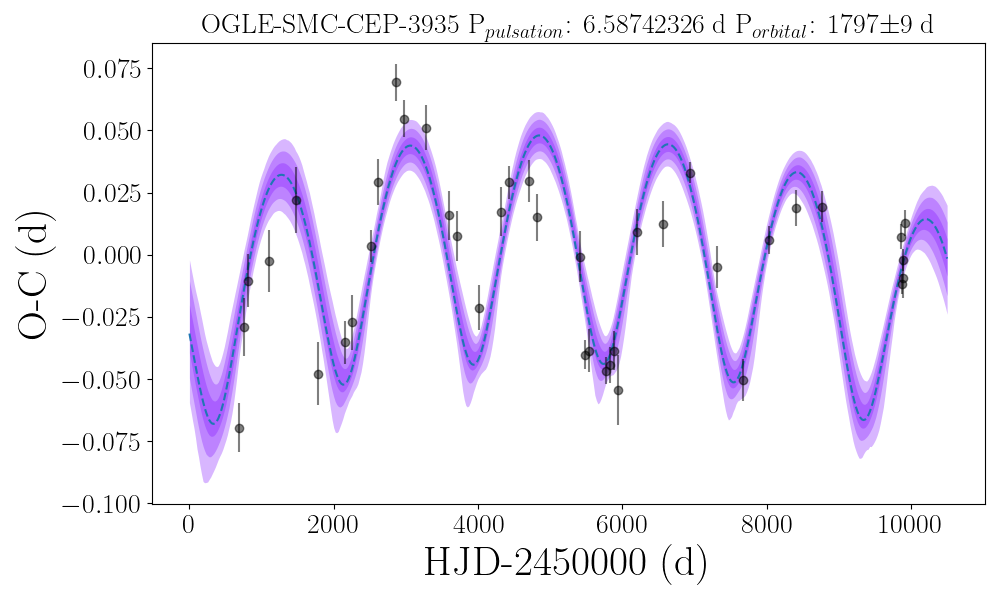}}
{\includegraphics[height=4.5cm,width=0.49\linewidth]{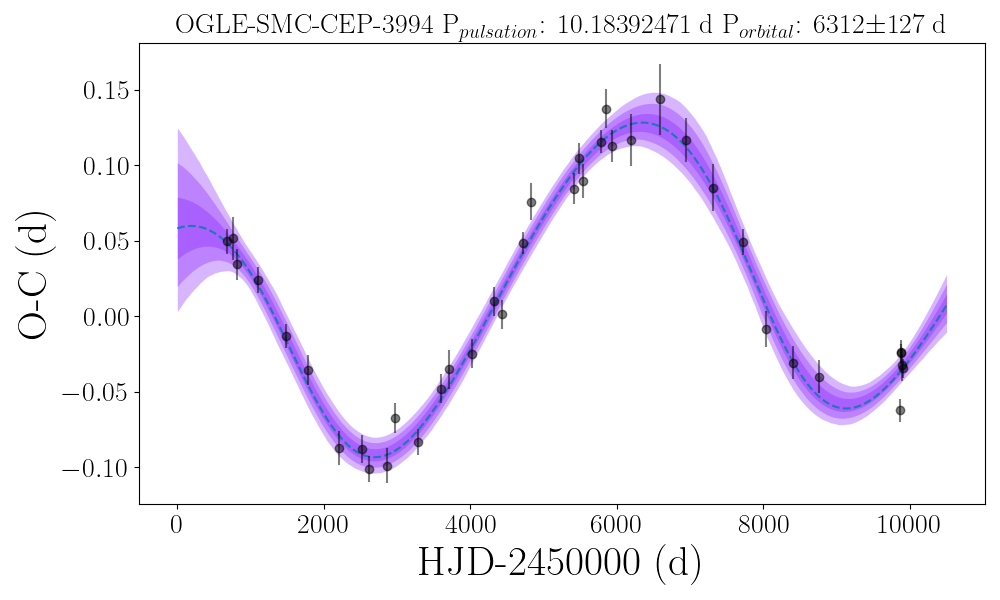}}
{\includegraphics[height=4.5cm,width=0.49\linewidth]{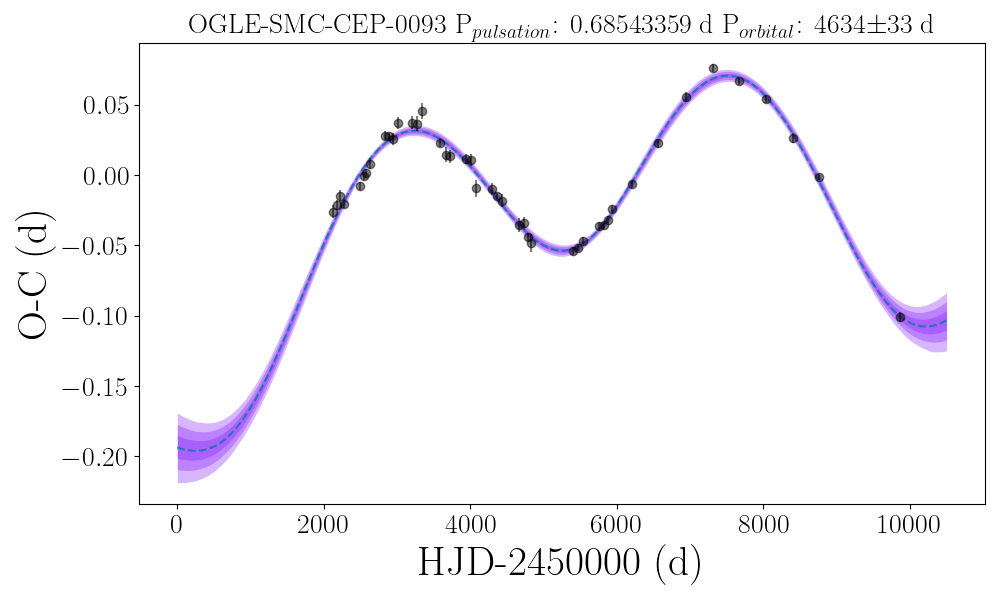}}
{\includegraphics[height=4.5cm,width=0.49\linewidth]{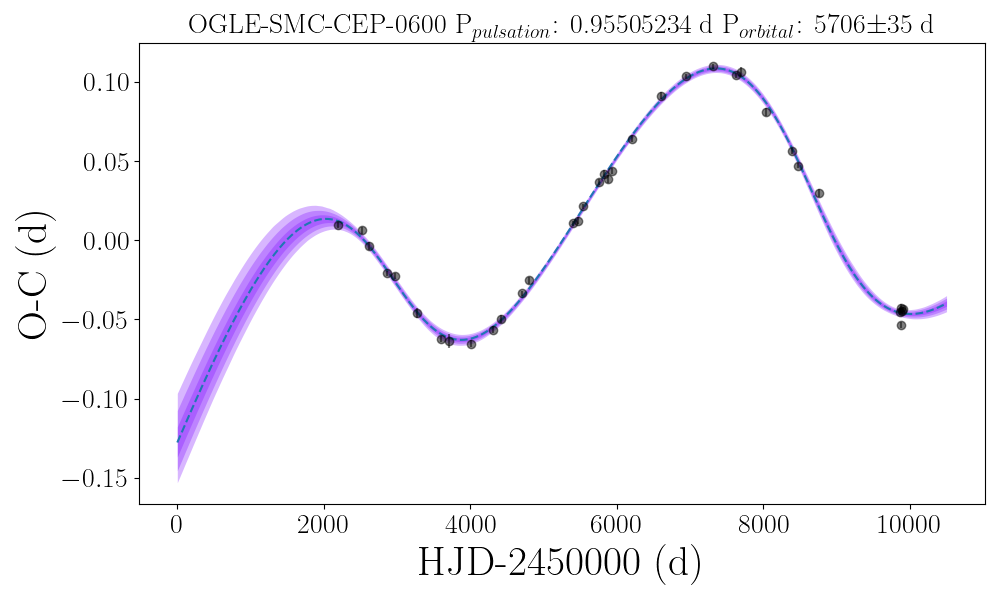}}

\caption{Examples of $O-C$ diagram for peculiar high-mass-function candidates. The remaining sample is presented in the appendix.}
\label{fig:Peculiar mass-function candidates}
\end{center}
\end{figure*}

Within our binary sample, an inspection of Tables~\ref{tab:LMC Binary list} and \ref{tab:SMC Binary list} reveals that for several candidates, the mass function is significantly larger than it is for the rest of the sample.

In Sect.~\ref{subsec: Searching for companions}, we describe how we estimated Cepheid mass from empirical period-mass relation for all candidates. Candidates for which mass function is twice as large as Cepheid's mass, namely, $f(m) \geq 2 \times m_\mathrm{cep}$, are classified as high-mass-function candidates. Using this criterion, there are 21 such Cepheids (LMC F mode: 1, LMC 1O mode: 6, SMC F mode: 2 and SMC 1O mode: 12). These candidates are marked with $\diamondsuit$ symbol in Tables~\ref{tab:LMC Binary list} and \ref{tab:SMC Binary list}. Their $O-C$ diagrams are compiled in Fig.~\ref{fig:Peculiar mass-function candidates} and Fig.~\ref{fig:appendix_Peculiar mass-function candidates}.

Out of the 21 candidates, when we calculate the minimum companion masses, the lowest mass is $8.7\MS$. Eleven targets have minimum companion masses in the $10-20\MS$ range, six in the $20-30\MS$ range, one in the $50-100\MS$ range, and two have even higher masses. These two extreme cases are OGLE-SMC-CEP-1267 and OGLE-LMC-CEP-3095 with derived minimum companion masses of $133.5\MS$ and $226.8\MS$, respectively. We recall that these are estimates and depend on an uncertain estimate of Cepheid's mass. These high values are difficult to explain in terms of a single Cepheid companion. Interestingly, the well-confirmed first-overtone mode Cepheid binary candidate OGLE-SMC-CEP-3235, discussed in the previous section, which shows eclipse as well as periodic $O-C$ variation \citep{Udalski2015AcA....65..341U}, is also one of the candidates in the peculiar high-mass-function category ($f(m)=12.32\pm1.43\MS$). This is one of the arguments in favor of not discarding all of these candidates straight away. We discuss possible explanations for these candidates below.

Firstly, stars can exist in more complex configurations than just a binary, for example, a hierarchical triple system with a tight massive inner binary and the third component being the Cepheid on a wide orbit. Theoretical population study by \cite{Dinnbier2022AA...659A.169D} validates that the fraction of Cepheids in multiple systems in significant. In principle, a companion may also be a black hole. \citep[e.g.][]{El-Badry2023MNRAS.521.4323E}. However, with the currently known mass distribution of black hole binaries \citep[e.g.][]{Bavera2023NatAs...7.1090B} it is not possible to reconcile the most extreme cases with the largest estimated minimum companion mass.

Secondly, the periodic $O-C$ may also arise due to a periodic modulation of pulsation phase intrinsic to Cepheid, such as the Blazhko effect in RR~Lyrae stars, but with pure phase modulation.

Thirdly, the $O-C$ diagrams may also be affected by the other category of non-evolutionary period changes. These are typically irregular, but we cannot exclude that on a shorter timescale, these changes may mimic periodic variations. Such changes may be the sole explanation for the observed $O-C$, and, consequently, a peculiar orbital solution, but may also be superimposed on a genuinely periodic $O-C$ due to LTTE in a binary system, distorting the determined orbital solutions and affecting the mass function. Some examples of candidates from our sample that may be affected by these non-linear period changes are OGLE-LMC-CEP-0889, OGLE-LMC-CEP-1454, OGLE-LMC-CEP-3095, OGLE-SMC-CEP-0340, and OGLE-SMC-CEP-3955 (see their $O-C$ diagrams). The proposed mechanisms for non-evolutionary period changes in classical pulsators are magnetic field \citep{Stothers1980PASP...92..475S}, short-term variations in the atmospheric structure \citep{Deasy1985MNRAS.212..395D}, changes in helium abundance gradients \citep{Cox1998ApJ...496..246C}, and semiconvection \citep{Sweigart1979AA....71...66S}. For Cepheids with longer pulsation periods, convection could play a significant role in the unstable behavior of the period \citep[e.g.][]{Neilson2014AA...563L...4N}.

The third explanation seems most probable. This is because first-overtone Cepheids are more prone to non-evolutionary period changes \citep[e.g.][]{Berdnikov1997AstL...23..177B, Poleski2008AcA....58..313P,Evans2015MNRAS.446.4008E}. In our high-mass-function sample, most of the candidates (18 out of 21), the Cepheid is indeed the first-overtone pulsator. The incidence rate and characteristics of irregular period changes (in particular, the dependence on pulsation mode) will be discussed in the next paper of the series (Rathour et al., in prep.). In that work, we will also systematically search for irregular period changes in Magellanic Cloud Cepheids.

Radial velocity observations will be the most straightforward way to confirm whether the observed variation in the $O-C$ is indeed due to LTTE in binary system. The longer time base of observations may allow us to more firmly conclude about the nature of $O-C$ variation in high-mass-function candidates.


\section{Discussion}
\label{sec: Discussion} 

\subsection{Incidence rates}
\label{subsec: Incidence rates}
We calculate the incidence rates of binary Cepheids, with respect to the working sample of Cepheids (mentioned in Table~\ref{tab:data working sample} column under stars with data in both OGLE-III+IV). Starting with the LMC, our binary candidate incidence rate for fundamental-mode Cepheids is $1.7\pm0.3$ \% and nearly the same for first-overtone Cepheids, namely, $1.8\pm0.4$ \%. In the SMC the incidence rate is higher, $3.3\pm0.4$ \% for fundamental-mode Cepheids and $3.7\pm0.5$ \% for first-overtone Cepheids. In a given galaxy, first-overtone Cepheids have a similar binarity detection fraction within the errors as compared to fundamental-mode Cepheids. However, when compared between the two galaxies, SMC incidence rates are about twice as high as the LMC ones. These incidence rates are clearly much lower than estimates for Galactic Cepheids, which are well above 50 percent \citep{Evans1992ApJ...384..220E,Szabados2003ASPC..298..237S,Evans2013AJ....146...93E,Kervella2019AA...623A.116K}.

There are multiple factors affecting the number of detected binary candidates. (a) The sample we analyze contains Cepheids with both OGLE-III and OGLE-IV data. In total, we discarded 1630 Cepheids (SMC F mode: 201; SMC 1O mode: 200; LMC F mode: 646; LMC 1O mode: 583) which had data in only one of the OGLE phases. It is likely we may miss some binary candidates just based on the sample quality cut. (b) The total time base including OGLE-III, OGLE-IV and OGLE-IV extended data is $\sim$10,000\,d. Any Cepheid binary candidate with an orbital period nearby or greater than this limit cannot be detected. Further observations of the OGLE survey in the future are necessary to recover those candidates. (c) Similarly, there is a detection limit on the lower end of orbital periods (below $\sim$600\,d), as we discussed in Section \ref{subsec: Previously detected Cepheid binary candidates} while searching for LTTE in eclipsing binary candidates. We could only establish LTTE for Cepheids with orbital periods longer than $\sim$600\,d. Hence, we would miss an important fraction of Cepheid binaries with orbital period less than this limit. (d) Irregular, non-evolutionary period changes may affect the $O-C$ diagrams (see Sect.~\ref{subsec: High Mass function candidates}). This is also supported by \cite{Pilecki2021ApJ...910..118P}, with some of their Cepheid $O-C$ diagrams showing quasiperiodic cyclic variation (marked as QPOC in that work). In extreme cases, real binary Cepheids will have a distorted $O-C$ diagram appearing semi-periodic or not periodic at all. (e) Finally, detectability is influenced by the inclination angle. The amplitude of LTTE signal measured from $O-C$ diagram is dependent on the inclination of the binary. In an ideal scenario at $90^{\circ}$ (edge-on configuration), the $O-C$ will have maximum amplitude and on the other hand at $0^{\circ}$ angle (face-on configuration) of the binary with respect to the observer, the $O-C$ will have no LTTE. Hence, assuming all possible inclination angles, the incidence rate of detected binaries will definitely be affected due to the ones missed in low inclination configuration binaries.

\subsection{Comparison with binary population synthesis models}
\label{subsec: Binarity fraction comparison with literature}

In order to understand the two-times higher incidence rate of binary Cepheid candidates in the SMC as compared to the LMC, we resorted to the synthetic populations of binary Cepheids \citep{Karczmarek2022ApJ...930...65K}. These populations consist of only fundamental-mode Cepheids in binary systems (i.e., the incidence rate of binary Cepheids is 100 percent) and show systematic differences in the distributions of luminosities, masses, pulsation periods, and so on, between the variants of different metallicities. We used these differences in an attempt to interpret the two-times higher incidence rate of binary candidates with fundamental-mode Cepheids in the SMC than in the LMC.

Firstly, the theory of stellar evolution predicts that the Cepheid blue loops are more pronounced in metal-poorer stars of a given mass than in their metal-richer counterparts \citep[see, e.g., Fig. 1 from ][]{Anderson_Richard2016AA...591A...8A}. This results in the SMC Cepheids crossing the instability strip (IS) during the blue loop at systematically lower masses, and therefore lower luminosities, than the LMC Cepheids. Indeed, synthetic SMC Cepheids (i.e. coming from the synthetic population, which reflects the average metallicity of the SMC Cepheids, $Z = 0.004$) have masses starting from $3.3\MS$ and a median mass of $3.7\MS$, while synthetic LMC Cepheids ($Z = 0.008$) have minimum and median masses of $3.8\MS$ and $5.0\MS$, respectively. The initial mass function \citep{kroupa03}, which was used to create both synthetic populations of LMC and SMC Cepheids, explains that low-mass stars are more abundant that massive ones; therefore, we can conclude that the SMC Cepheids should indeed be more numerous than the LMC Cepheids because they have systematically lower masses.

\begin{figure}
\begin{center}
\includegraphics[width=\linewidth]{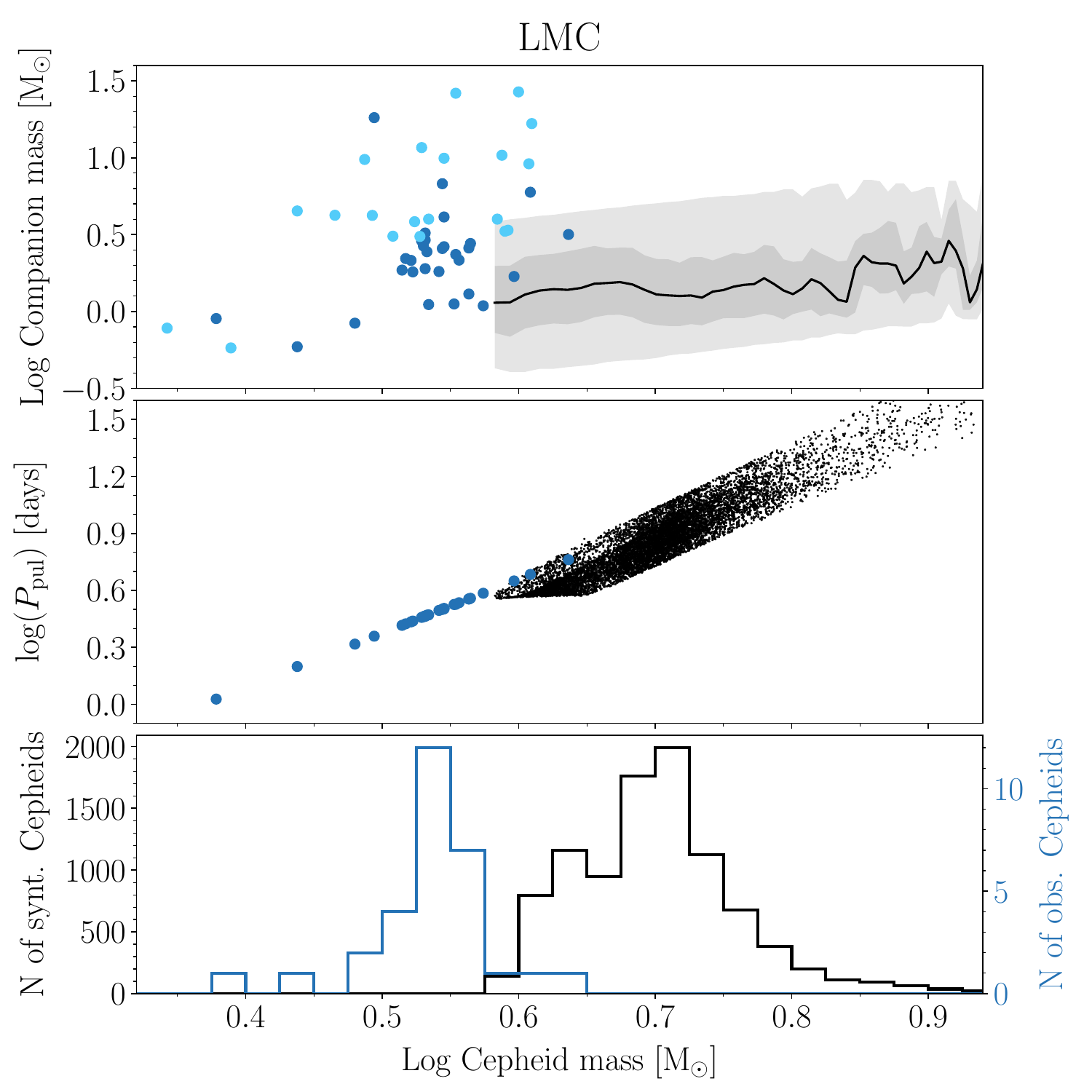}
\caption{Distribution of estimated minimum companion masses as a function of derived Cepheid masses in the LMC (top), represented by dark blue circles (fundamental-mode Cepheids) and light blue circles (first-overtone Cepheids). Black line indicates the median companion mass from the binary synthetic population \citep{Karczmarek2022ApJ...930...65K}, with half of the companions (1st to 3rd quartile) lying inside the dark-shaded area, and the entire sample enclosed by the light-shaded area. Relation between the Cepheid mass and pulsation period (middle) for observed fundamental-mode Cepheids (dark blue circles, derived from the inverse of Eq. \ref{eq:logM-logP}) and for synthetic ones (black dots). Mass distribution of synthetic LMC Cepheids (bottom), shown as a black histogram, which offers additional information about the number of systems that were used to calculate the median companion mass (in the top panel). For comparison the mass distribution of observed LMC fundamental-mode Cepheids is given in dark blue.}
\label{fig:syntpopLMC}
\end{center}
\end{figure}

\begin{figure}
\begin{center}
\includegraphics[width=\linewidth]{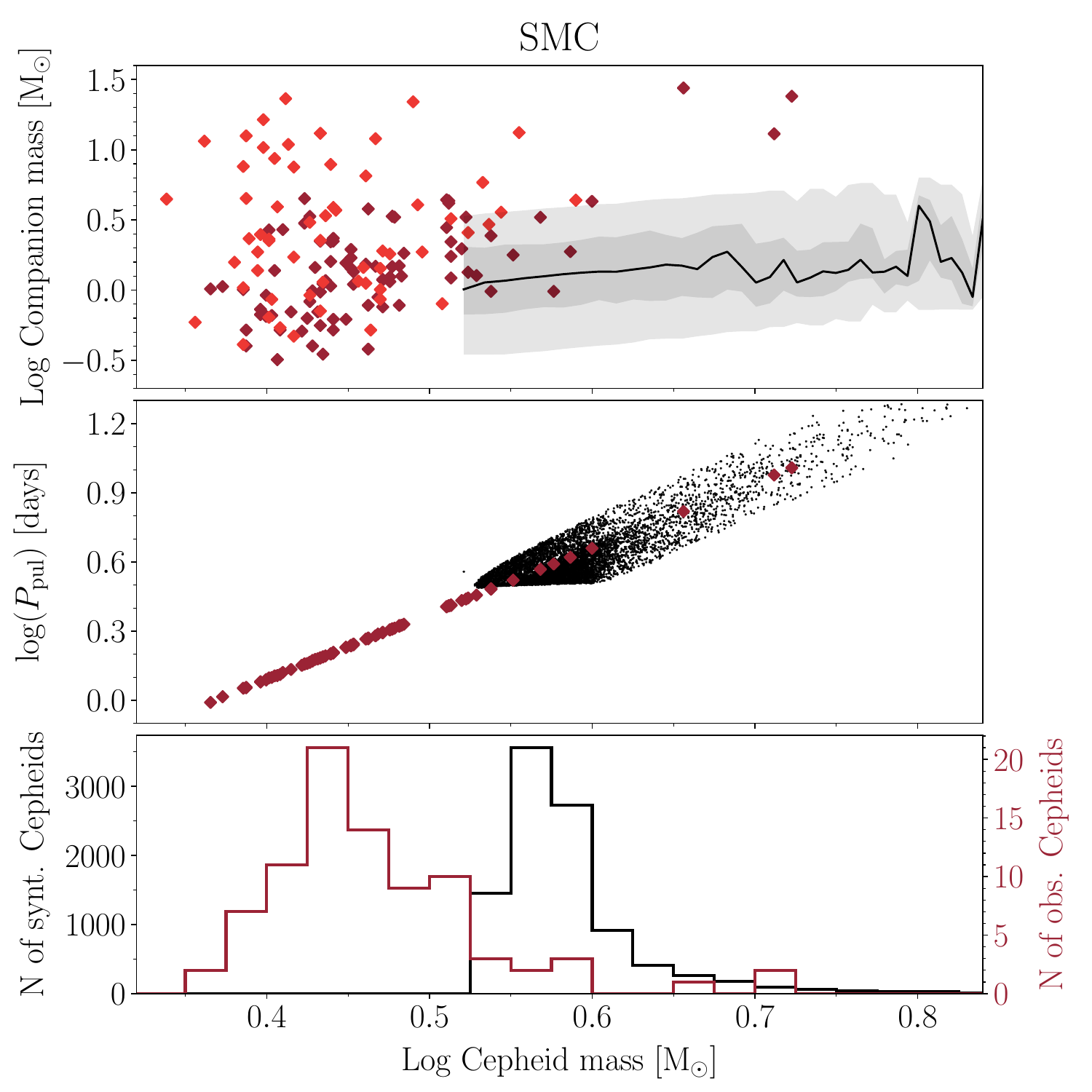}
\caption{Same as Fig \ref{fig:syntpopLMC}, but for the SMC Cepheid binary candidates: Dark red diamonds (fundamental-mode) and bright red diamonds (first-overtone), with the corresponding synthetic binary population.}
\label{fig:syntpopSMC}
\end{center}
\end{figure}

We notice that the aforementioned synthetic masses are systematically higher than the ones calculated for our sample. This discrepancy is clearly shown in Figs. \ref{fig:syntpopLMC} and \ref{fig:syntpopSMC} (bottom panels). Models of stellar evolution used in the binary population synthesis calculations fail to render Cepheid blue loops for lower-mass stars (below 3.3 and 3.8 $\MS$ for the SMC and LMC, respectively), even though Cepheids of even lower masses (derived from Eq. \ref{eq:logM-logP}) are evident in the observed data. The relative difference between the median mass of the synthetic and observed samples from the bottom panels of Figs. \ref{fig:syntpopLMC} and \ref{fig:syntpopSMC} is 26\% and 31\% for the SMC and LMC, respectively, which is consistent with the mass discrepancy phenomenon, existing between the mass derived from evolutionary models and other methods. Interestingly enough, the synthetic Cepheids from SMC and LMC alike follow the Period-Mass relation in Eq. (\ref{eq:logM-logP}), only starting at larger masses and pulsation periods, as shown in Figs. \ref{fig:syntpopLMC} and \ref{fig:syntpopSMC} (middle panels).

We also compared the time that synthetic Cepheids spend inside the instability strip (IS) in order to check the hypothesis that low-mass Cepheids spend more time inside the IS, because they evolve slower than the more massive ones. The median time that synthetic SMC Cepheids spend inside the IS is 4 and 6.3~Myr for the second and third IS crossing, respectively, whereas for the synthetic LMC Cepheids these times are 3.9 and 6 Myr. We regard these differences between crossing times for the SMC and LMC Cepheids as insignificant and insufficient to explain the twice higher incidence rate of SMC to LMC binary Cepheid candidates.

Another possibility might be different star formation rates (SFR) in the LMC and SMC. For example, a significantly larger SFR in the SMC about 200 Myr ago would result in a surplus of the 200 Myr-old SMC Cepheids today. Unfortunately, the star formation history used in the synthetic population approach was constructed from the pulsation period-age relation \citep{Bono2005ApJ...621..966B}; the number of Cepheids in a given pulsation period bin was translated to the SFR at a given age bin. This was, in turn, fed to the population synthesis code to induce the occurrence of Cepheids of a given age and pulsation period. This circular logic prevents us from using the synthetic population to establish the relevance of the SFR on the incident rates of SMC and LMC Cepheids in our sample.

As a next test, we compared the masses of Cepheids and their companions from our sample with the synthetic populations, as shown in Figs. \ref{fig:syntpopLMC} and \ref{fig:syntpopSMC}. The masses of our companions generally fall into the range of synthetic companion masses, with some exceptions, which are located above the upper limit for the synthetic companion mass. These exceptions are mostly overtone Cepheids (light-blue circle symbols in Fig. \ref{fig:syntpopLMC} and bright red diamond symbols in Fig. \ref{fig:syntpopSMC}), which endorses our hypothesis from Section \ref{subsec: High Mass function candidates} that such massive objects cannot exist within a framework of non-interacting stars. These more massive companions can be close binaries in hierarchical triples with the more distant Cepheids, mergers from a hierarchical triple, or an outcome of the mass gain, neither of which was explored in the population synthesis method \citep{Karczmarek2022ApJ...930...65K}. However, the recent theoretical studies on evolution of stars in stellar clusters hint at the possibility that significant percentage of Cepheids might be in multiple systems \citep{Dinnbier2022AA...659A.169D}, which sets an interesting course for future exploration.

Finally, we calculated the percentage of systems that could be recovered from the entire population of binary Cepheids using the LTTE method. We considered the following limitations, namely, the minimum and maximum values of our observed distributions of strong candidates among fundamental-mode Cepheids: (i) orbital period, (ii) logarithm of the mass function, and (iii) $a \sin(i)$ assuming isotropic inclination angle. We imposed them on the synthetic populations of binary Cepheids. Table \ref{tab:recovery_rate} presents the upper and lower limits, and the recovery rate of SMC and LMC fundamental-mode Cepheids for each limitation separately, and all combined. We were able to recover about 6.1 and 2.6 \% of the initial sample of SMC and LMC synthetic Cepheids, respectively. We notice that the percentages of individual limitations added together exceed 100 \%, which means that the recovery of some Cepheid binaries is affected by more than one limiting factor. The recovered incidence rates from the synthetic populations are about twice as large as the incidence rates from the empirical data, which can be explained on the basis of the Cepheid binarity fractions in the Magellanic Clouds. The binarity fractions in the synthetic populations are 100\% (all Cepheids are in binary systems), whereas the empirical binarity fractions of the SMC and LMC Cepheids are unknown. If the real binarity fractions were about 50\%, then the synthetic incidence rates should be halved, bringing them to a closer agreement with the empirical ones. The synthetic populations of binary Cepheids in the LMC and SMC were rendered using identical initial distributions; therefore, any differences between the SMC and LMC synthetic incidence rates arise from the limitations imposed on the orbital parameters (see Table. \ref{tab:recovery_rate}). This does not, however, explain why the limiting parameters have different ranges in the first place.

\begin{table}
    \centering
    \caption{Limitations imposed on the synthetic population of binary Cepheids and the recovery rates.}
    \label{tab:recovery_rate}
    \small
    \begin{tabular}{lcccc}
         \hline
         \hline
         & \multicolumn{2}{c}{SMC} & \multicolumn{2}{c}{LMC} \\
         Parameter & Limits & Recov. & Limits & Recov. \\
         \hline
         $P_\mathrm{orbital}$ [days] & [863, 5353] & 11.2\% & [945, 3897] & 8.1\% \\
         $\log f(m)$ [$\MS$] & [-2.29, 0.25] & 66.4\% & [-1.74, 0.33] &45.6\% \\
         $a \sin(i)$ [AU] & [0.66, 6.80] & 22.8\% & [1.18, 5.80] &15.2\% \\
         \hline
         \textbf{Combined} & --- & \textbf{6.1\%} &--- & \textbf{2.6\%}\\
         \hline
    \end{tabular}
\end{table}

Galactic field studies estimated binarity fractions to be as high as 80 \% \citep{Evans1992ApJ...384..220E,Szabados2003ASPC..298..237S,Evans2013AJ....146...93E,Kervella2019AA...623A.116K}, whereas population synthesis of LMC Cepheids implied the binarity fraction to be at least 55 \%, and likely much more \citep{Karczmarek2022ApJ...930...65K}. That study also predicts most of this fraction (up to 90 percent) to be dominated by hot main sequence companions, leaving only 3-5 percent for evolved giant star companions. Given the small sample statistic for our LMC candidates (52 out of roughly more than 3000 stars in the survey), it is clear why we did not detect any new overbright Cepheids (apart from previously established OGLE-LMC-CEP-0837 and OGLE-LMC-CEP-0889), which are likely tracers of Cepheids with giant companion origins. The SMC sample is three times as large as the LMC one, and does show new Cepheid candidates (OGLE-SMC-CEP-3061 and OGLE-SMC-CEP-4365) with giant companions (see Section \ref{subsec: Searching for companions}).

\section{Conclusions}
\label{sec: Conclusions} 

We have applied the modified Hertzsprung method to analyze period changes of classical Cepheids in the Magellanic Clouds based on OGLE data. This resulted in $O-C$ diagrams of more than 7200 Cepheids. Our systematic search for periodic $O-C$ features resulted in a sample of 197 candidates for binary Cepheids. The key findings are as follows:

\begin{itemize}
\item In the LMC, we found 52 candidates for Cepheid binary systems (30 fundamental and 22 first-overtone mode) and a roughly three times larger sample of 145 candidates in the SMC (85 fundamental and 60 first-overtone mode). The detection fraction of binary Cepheid candidates is as follows: LMC F 1.7 \%; LMC 1O 1.8 \%; SMC F 3.3\%; and SMC 1O 3.7\%. 

\item The Cepheid binary population synthesis results, when subject to the limits of our detected binary sample, provided a similar incidence rate ratio for SMC to LMC, as calculated from our observational sample, showing the agreement between theoretical predictions and observations.

\item Given that only a few spectroscopic detections of Cepheids in a binary system were reported, 25 in the LMC \citep{Szabados2012MNRAS.426.3148S,Pilecki2021ApJ...910..118P} and 9 in the SMC \citep{Szabados2012MNRAS.426.3148S}, our sample of Cepheid binary candidates is twice as large for the LMC and sixteen times as large for the SMC. We note that the sample compiled in this paper is a list of probable binary candidates and further follow-up is needed for verification.

\item From the SMC binary candidate sample, two fundamental-mode Cepheids, OGLE-SMC-CEP-3061 and OGLE-SMC-CEP-4365, are found to be in the class of overbright Cepheids \citep[see][]{Pilecki2021ApJ...910..118P}, which is likely a consequence of them being in a binary system with a giant companion. These two candidates are the very first detections of LTTE in overbright Cepheids in the SMC. Using empirical relations we estimate the Cepheid mass and minimum companion mass ($m_{\text{Cep}}+m_{\text{c}}$)  to be:
($3.2\MS+2.8\MS$) for OGLE-SMC-CEP-3061 and ($2.7\MS+4.5\MS$)  for OGLE-SMC-CEP-4365. In the LMC we report LTTE variation for two overbright Cepheids, OGLE-LMC-CEP-0837 and OGLE-LMC-CEP-0889, where the former is also a spectroscopically confirmed SB2-type binary system. For the remaining SB2-type systems in the LMC compiled by \cite{Pilecki2021ApJ...910..118P}, we do not see any significant LTTE variation.

\item Considering known literature Cepheids in eclipsing binary systems, from our $O-C$ analysis we confirm the LTTE signature for three candidates in the LMC (OGLE-LMC-CEP-1812, OGLE-LMC-CEP-2532 and OGLE-LMC-CEP-4506) and two in the SMC (OGLE-SMC-CEP-2199 and OGLE-SMC-CEP-3235). The reasons for no such signature for remaining eclipsing binary candidates are described in Sect.~\ref{subsec: Previously detected Cepheid binary candidates}.

\item We found 21 candidate binary systems with very high-mass-function values. From these, 85 \% come from the first-overtone mode sample, likely indicating a non-evolutionary period change phenomenon playing a role in observed the $O-C$ variation. Other explanations, however, including triple systems with a tight inner binary, or even black hole companions, cannot be excluded. To establish a better understanding of their nature, long-term monitoring and/or spectroscopic observations are needed.

\end{itemize}

Detailed spectroscopic follow-ups on at least the strong candidates may be able to accurately establish the binarity nature and lead to improved orbital parameters. A longer photometric baseline in the future will not only refine the proposed sample but also uncover more candidates in the longer orbital period regime. In addition, for a better characterization of non-evolutionary period changes in classical Cepheids, we will tackle the more complicated, second type of the non-evolutionary period change related to irregular $O-C$ diagrams (Rathour et al. in prep). The subject of our follow-up paper will be crucial for better understanding the biases and limitations in the study of binary candidates via the $O-C$ technique as well.

\begin{acknowledgements}
In tribute to late Krzysztof Belczyński, whose expert discussions on high-mass-function candidates have directly contributed to this article. We thank all the OGLE observers for their contribution to the collection of the photometric data over the decades. RSR, RS, VH and OZ are supported by the National Science Center, Poland, Sonata BIS project 2018/30/E/ST9/00598. GH and PK acknowledge grant support from the European Research Council (ERC) under the European Union's Horizon 2020 research and innovation program (grant agreement No. 695099). IS acknowledge support from the National Science Centre, Poland, grant no. 2022/45/B/ST9/00243.  For the purpose of Open Access, the author has applied a CC-BY public copyright license to any Author Accepted Manuscript (AAM) version arising from this submission. This research made use of the SIMBAD and VIZIER databases at CDS, Strasbourg (France) and the electronic bibliography maintained by the NASA/ADS system.
\end{acknowledgements}


\bibliographystyle{aa} 
\bibliography{aanda} 

\begin{appendix}

\section{Table with binary parameters}

\begin{table*}[ht]
\caption{List of LMC binary Cepheid candidates with their orbital parameters.}
\label{tab:LMC Binary list}
\scalebox{0.68}{
\begin{tabular}{lccccccccccc}
\hline
\hline
\textbf{Target} & \textbf{Fit order} & \textbf{P$_{\textbf{pulsation}}$} & \textbf{P$_{\textbf{orbital}}$} & \textbf{T$_{\textbf{0}}$} & \textbf{e} & \textbf{a sini} & \textbf{$\omega$} & \textbf{PCR} & \textbf{K} & \textbf{f(m)} & M$_{\textbf{c}}$ (M$_{\textbf{cep}}$)\\
{} & {} & {(d)} & {(d)} & {(d)} & {} & {AU} & {(deg)} & {(d Myr$^{-1}$)} & {(km s$^{-1}$)} & {(M$_{\odot}$)} & {(M$_{\odot}$)}\\
\hline
F mode &  &  &  &  &  &  &  &  & & & \\ 
Gold sample &  &  &  &  &  &  &  &  & & & \\ 
\hline
\textbf{0543} & 10  & 3.16340973 & 3649  $\pm$  60  & 9115  $\pm$  106 & 0.512  $\pm$  0.085 & 3.556  $\pm$  0.296 & -151  $\pm$  9 & -0.046  $\pm$  0.182 & 12.34   $\pm$  1.79  & 0.450	$\pm$	0.130    & 2.5(3.5) \\
\textbf{0631} & 10  & 3.58766239 & 2304  $\pm$  28  & 5370	$\pm$  89 & 0.588   $\pm$  0.134    & 1.505	 $\pm$ 0.107  & 129	 $\pm$  17 & +0.809	$\pm$  0.163  & 8.77   $\pm$  2.69  & 0.086	$\pm$	0.019   & 1.3(3.7) \\
\textbf{0637} & 6  & 2.60685284 & 945   $\pm$  10  & 2882  $\pm$  180 & 0.25   $\pm$  0.224 & 1.18   $\pm$  0.169 & 148   $\pm$  69 & +0.426   $\pm$  0.242 & 14.217  $\pm$  9.33  & 0.245	$\pm$	0.122    & 1.9(3.3) \\
\textbf{0662}  & 10  & 2.95839254 & 3861  $\pm$  69  & 8801	  $\pm$  257 & 0.468 $\pm$  0.161  & 1.645 $\pm$  0.307  & -13	 $\pm$  18 & +0.019   $\pm$  0.094  & 5.23   $\pm$  2.19   & 0.040	$\pm$	0.038    & 0.9(3.4) \\
\textbf{0685} & 9  & 3.18198049 & 1413  $\pm$  10  & 6471  $\pm$  124 & 0.295  $\pm$  0.135 & 2.62   $\pm$  0.243 & 58    $\pm$  32 & -1.355  $\pm$  0.282 & 21.103  $\pm$  3.23  & 1.201	$\pm$	0.384    & 4.1(3.5) \\
\textbf{0802} & 8  & 2.91257982 & 3385  $\pm$  34  & 7044  $\pm$  148 & 0.252  $\pm$  0.067 & 3.769  $\pm$  0.143 & -50   $\pm$  15 & +0.06    $\pm$  0.166 & 12.535  $\pm$  0.59  & 0.623	$\pm$	0.070    & 2.9(3.4)\\ 
\textbf{0837} & 10 & 5.77731047 &  3692  $\pm$	55 &	5408  $\pm$	338	& 0.281  $\pm$	0.143 &	3.932  $\pm$	0.379 &	-3	 $\pm$ 33	& -0.130	 $\pm$ 0.590 &	12.09	 $\pm$ 1.74	& 0.595	$\pm$	0.186 &  3.2(4.3)\\
\textbf{0952} & 6  & 3.12319326 & 3495  $\pm$  65  & 6312  $\pm$  264 & 0.398  $\pm$  0.177 & 2.692  $\pm$  0.293 & -38   $\pm$  28 & +1.011   $\pm$  0.437 & 9.086   $\pm$  2.12  & 0.213	$\pm$	0.083    & 1.8(3.5)\\
\textbf{1047} & 7  & 3.59110858  &  2498 $\pm$	54 &	4107 $\pm$	187 &	0.600 $\pm$	0.261 &	3.139 $\pm$	1.156 &	-177 $\pm$	25 &	0.112 $\pm$	0.504 &	17.26 $\pm$	16.66 &	0.672	$\pm$	1.771 & 3.2(3.7)\\
\textbf{1265} & 10 &  2.74481384 &   1860 $\pm$	19 &	5599 $\pm$	271 &	0.101 $\pm$	0.074 &	1.801 $\pm$	0.082 &	72 $\pm$	53 &	2.564 $\pm$	0.251 &	10.62 $\pm$	0.52 &	0.225	$\pm$	0.031 &  1.8(3.3)\\
\textbf{1856} & 10 & 2.07536078 & 2525  $\pm$  24  & 7335  $\pm$  164 & 0.373  $\pm$  0.13  & 1.244  $\pm$  0.108 & 44    $\pm$  24 & -0.041  $\pm$  0.059 & 5.794   $\pm$  0.8   & 0.040	$\pm$	0.011    &  0.8(3.0)\\
\textbf{2036} & 10 & 4.46453151 & 3473  $\pm$  25  & 7618  $\pm$  399 & 0.135  $\pm$  0.082 & 2.399  $\pm$  0.144 & 68    $\pm$  41 & -3.117  $\pm$  0.138 & 7.593   $\pm$  0.56  & 0.153	$\pm$	0.029    &  1.7(4.0)\\
\textbf{2039} & 9  & 1.58174298 & 3849  $\pm$  49  & 5628  $\pm$  516 & 0.136  $\pm$  0.1   & 1.268  $\pm$  0.055 & 57    $\pm$  48 & -0.06   $\pm$  0.034 & 3.639   $\pm$  0.19  & 0.018	$\pm$	0.003    &  0.6(2.7)\\
\textbf{2197} & 8  & 3.42832749 & 2573  $\pm$  22  & 6037  $\pm$  269 & 0.164  $\pm$  0.086 & 2.467  $\pm$  0.146 & -14   $\pm$  38 & +0.013   $\pm$  0.241 & 10.601  $\pm$  0.67  & 0.302	$\pm$	0.056     & 2.2(3.6)\\
\textbf{2295} & 10 & 2.91158082 & 1555  $\pm$  8   & 6856  $\pm$  25  & 0.691  $\pm$  0.052 & 2.412  $\pm$  0.113 & 33    $\pm$  7  & +0.063   $\pm$  0.188 & 23.317  $\pm$  2.38  & 0.772	$\pm$	0.111    & 3.2(3.4)\\
\textbf{2491} & 10 & 1.0659162  & 1810  $\pm$  20  & 5339  $\pm$  101 & 0.32   $\pm$  0.107 & 1.183  $\pm$  0.111 & 15    $\pm$  20 & +0.051   $\pm$  0.047 & 7.484   $\pm$  1.09  & 0.067	$\pm$	0.023    & 0.9(2.4)\\
\textbf{2537} & 10 & 3.6185962  & 1110  $\pm$  4   & 5357  $\pm$  160 & 0.146  $\pm$  0.135 & 1.68   $\pm$  0.113 & 92    $\pm$  53 & +0.462   $\pm$  0.16  & 16.729  $\pm$  1.71  & 0.513	$\pm$	0.107    & 2.8(3.7)\\
\textbf{2553} & 10 & 2.91710686 & 1513  $\pm$  7   & 6478  $\pm$  202 & 0.121  $\pm$  0.095 & 1.61   $\pm$  0.105 & -74   $\pm$  47 & -0.736  $\pm$  0.086 & 11.685  $\pm$  0.96  & 0.243	$\pm$	0.051    &  1.9(3.4)\\
\textbf{2634} & 6  & 3.35721485 & 2431  $\pm$  59  & 7436  $\pm$  424 & 0.237  $\pm$  0.18  & 1.411  $\pm$  0.197 & 36    $\pm$  61 & +1.816   $\pm$  0.357 & 6.541   $\pm$  4.71  & 0.063	$\pm$	0.031   &  1.1(3.6)\\
\textbf{2742} & 10 & 3.84564344 & 3721  $\pm$  65  & 4563  $\pm$  563 & 0.133  $\pm$  0.109 & 1.798  $\pm$  0.144 & -89   $\pm$  54 & -1.627  $\pm$  0.263 & 5.333   $\pm$  0.5   & 0.056	$\pm$	0.014    &   1.1(3.8)\\
\textbf{2829} & 9  & 2.87070846 & 2628  $\pm$  18  & 7564  $\pm$  95  & 0.377  $\pm$  0.086 & 3.163  $\pm$  0.185 & 10    $\pm$  12 & -0.844  $\pm$  0.17  & 14.121  $\pm$  1.34  & 0.611	$\pm$	0.111    &   2.9(3.4)\\
\textbf{3032} & 7  & 2.71904257 & 2486  $\pm$  43  & 5598  $\pm$  366 & 0.159  $\pm$  0.117 & 2.487  $\pm$  0.193 & 169   $\pm$  54 & +0.423   $\pm$  0.346 & 11.107  $\pm$  0.97  & 0.332	$\pm$	0.079    &  2.2(3.3)\\
\textbf{3133} & 9  & 4.83273655 & 3145  $\pm$  38  & 7082  $\pm$  62  & 0.584  $\pm$  0.117 & 5.396  $\pm$  0.621 & -42   $\pm$  7  & +5.401   $\pm$  0.548 & 22.813  $\pm$  7.72  & 2.112	$\pm$	1.008    &  6.0(4.1)\\
\textbf{3213} & 6  & 2.6542454  & 3613  $\pm$  56  & 7432  $\pm$  141 & 0.52   $\pm$  0.115 & 3.272  $\pm$  0.267 & 124   $\pm$  15 & +0.652   $\pm$  0.269 & 11.557  $\pm$  5.66  & 0.358	$\pm$	0.096    &  2.2(3.3)\\
\textbf{3298} & 10 & 2.89343793 & 3727  $\pm$  66  & 6260  $\pm$  206 & 0.367  $\pm$  0.075 & 3.793  $\pm$  0.137 & -60   $\pm$  17 & +5.915   $\pm$  0.236 & 11.865  $\pm$  0.82  & 0.520	$\pm$	0.061  &   2.7(3.4)\\
\hline
F mode &  &  &  &  &  &  &  &  & &  &\\ 
Bronze sample &  &  &  &  &  &  &  &  & &  &\\ 
\hline
0687  & 9  & 3.36220963 & 3444 $\pm$	36 &	5806 $\pm$	36 &	0.700 $\pm$	0.070 &	3.196 $\pm$	0.317  &	-34	$\pm$ 5	&  1.221 $\pm$	0.158	&  14.11 $\pm$	3.16 &	0.367	$\pm$	0.125 &  2.3(3.6)\\
0889 & 10 & 3.16841733 & 2549	$\pm$ 8	& 8061 $\pm$	25	&  0.552 $\pm$	0.079	&  5.118 $\pm$	0.292	&  164 $\pm$	3	&  +3.789 $\pm$	0.149	&  26.17 $\pm$	3.14	&  2.750	$\pm$	0.491 &  6.5(3.5)\\
2772 & 10 & 2.94871874 & 3958  $\pm$  74  & 6398  $\pm$  76  & 0.7    $\pm$  0.136 & 3.684  $\pm$  1.275 & 27    $\pm$  7  & -2.091  $\pm$  0.44  & 14.166  $\pm$  14.81 & 0.426	$\pm$	1.138    &   2.5(3.4)\\
2952 & 10 & 3.20001924 & 6564  $\pm$  347 & 3108  $\pm$  315 & 0.196  $\pm$  0.078 & 5.406  $\pm$  1.041 & 92    $\pm$  14 & +2.911   $\pm$  2.521 & 9.134   $\pm$  1.35  & 0.487	$\pm$	0.269     & 2.6(3.5)\\
3267 $\diamondsuit$ & 10 & 2.28513112 & 1400  $\pm$  10  & 4873  $\pm$  16  & 0.886  $\pm$  0.077 & 5.803  $\pm$  2.19  & -8    $\pm$  3  & +4.782   $\pm$  0.151 & 96.989  $\pm$  95.42 & 13.296	$\pm$	28.260   & 18.2(3.1)\\
\hline
1O mode &  &  &  &  &  &  &  &  & & & \\ 
Gold sample &  &  &  &  &  &  &  &  & & & \\ 
\hline
\textbf{0030} & 5 & 0.603548   & 2155 $\pm$  30  & 6874  $\pm$  115 & 0.576 $\pm$  0.173 & 1.23   $\pm$  0.278 & 22   $\pm$  18 & +0.332   $\pm$  0.042 & 7.536  $\pm$  4.72   & 0.053	$\pm$	0.059    & 0.8(2.2)\\
\textbf{0088} & 2 & 2.04553516 & 3215 $\pm$  48  & 6877  $\pm$  431 & 0.134 $\pm$  0.091 & 4.471  $\pm$  0.297 & -29  $\pm$  47 & +3.553   $\pm$  0.31  & 15.327 $\pm$  1.15   & 1.155	$\pm$	0.237    & 4.0(3.4)\\
\textbf{0472} $\diamondsuit$ & 4 & 3.28359809 & 4064 $\pm$  73  & 6466  $\pm$  362 & 0.164 $\pm$  0.084 & 11.026 $\pm$  0.465 & -111 $\pm$  32 & +8.127   $\pm$  1.201 & 29.935 $\pm$  1.66   & 10.790	$\pm$	1.444    & 16.7(4.1)\\
\textbf{0816} & 4 & 2.8070886  & 2155 $\pm$  37  & 5619  $\pm$  135 & 0.425 $\pm$  0.149 & 3.301  $\pm$  0.287 & 101  $\pm$  26 & +4.652   $\pm$  0.629 & 18.449 $\pm$  3.37   & 1.033	$\pm$	0.283    & 4.0(3.8)\\
\textbf{0900} $\diamondsuit$ & 2 & 2.18609105 & 5271 $\pm$  80  & 5447  $\pm$  131 & 0.171 $\pm$  0.027 & 25.476 $\pm$  0.468 & 159  $\pm$  9  & +4.771   $\pm$  0.849 & 53.338 $\pm$  0.93   & 79.301	$\pm$	3.584    & 85.9(3.5)\\
\textbf{1046} $\diamondsuit$& 3 & 1.96984052 & 1211 $\pm$  5   & 5147  $\pm$  26  & 0.885 $\pm$  0.111 & 4.253  $\pm$  2.061 & 29   $\pm$  11 & +1.347   $\pm$  0.285 & 82.487 $\pm$  140.27 & 6.999	$\pm$	27.987   & 11.7(3.4)\\
\textbf{1134} $\diamondsuit$ $\ast$& 3 & 3.09020297 & 4719 $\pm$  64  & 5017  $\pm$  126 & 0.441 $\pm$  0.134 & 15.031 $\pm$  2.047 & -56  $\pm$  10 & -5.971  $\pm$  1.177 & 38.541 $\pm$  12.85  & 20.344	$\pm$	12.133    & 26.8(4.0)\\
\textbf{2793} & 4 & 2.91121489 & 3165 $\pm$  123 & 6082  $\pm$  380 & 0.364 $\pm$  0.214 & 3.753  $\pm$  0.533 & -19  $\pm$  44 & +6.34    $\pm$  0.858 & 13.964 $\pm$  5.14   & 0.708	$\pm$	0.345    & 3.3(3.9)\\
\hline
1O mode &  &  &  &  &  &  &  &  & & &  \\ 
Bronze sample &  &  &  &  &  &  &  &  & & & \\ 
\hline
0134 & 4 & 3.24235812 & 4797 $\pm$  159 & 5436  $\pm$  174 & 0.594 $\pm$  0.101 & 9.125  $\pm$  0.998 & -31  $\pm$  15 & -14.237 $\pm$  1.46  & 25.659 $\pm$  5.71   & 4.387	$\pm$	1.622    & 9.1(4.1)\\
0315 & 3 & 1.73136384 & 4477 $\pm$  124 & 4283  $\pm$  438 & 0.179 $\pm$  0.09  & 4.81   $\pm$  0.234 & -79  $\pm$  35 & -3.291  $\pm$  0.382 & 11.907 $\pm$  0.66   & 0.740	$\pm$	0.103    & 3.1(3.2)\\
0764 $\diamondsuit$ $\ddag$ $\dagger$ & 5 & 2.31569813 & 5587 $\pm$	37 &	9807 $\pm$	31 &	0.444 $\pm$	0.022 &	20.391 $\pm$  0.318 &	-20	$\pm$ 2 &	-1.494 $\pm$ 0.317 &	44.29 $\pm$	1.14 &	36.209	$\pm$	1.692 &  42.6(3.6)\\
0839 $\ddag$ $\dagger$ $\ast$ & 4 & 2.19325821 & 5710 $\pm$	62 &	6987 $\pm$	106 &	0.370 $\pm$	0.041 &	10.701 $\pm$	0.270 &	-30 $\pm$	7 &	+3.669 $\pm$	0.339 &	21.92 $\pm$	0.91 &	5.007	$\pm$	0.395 &  9.4(3.5)\\
0924 & 6 & 2.94963963 &  2849 $\pm$	25 &	3756 $\pm$	316 &	0.217 $\pm$	0.138 &	3.772 $\pm$	0.302 &	-32 $\pm$	41 &	-0.450 $\pm$	0.263 &	14.82 $\pm$	1.66 &	0.883	$\pm$	0.220 &  3.7(3.9)\\
0950 & 5 & 1.51765486 & 9673 $\pm$	286	 & 4349 $\pm$	136 &	0.255 $\pm$	0.024 &	16.094 $\pm$	0.600 &	154 $\pm$	6 &	+1.323 $\pm$	0.225 &	18.72 $\pm$	0.28 &	5.928	$\pm$	0.347 &  10.1(3.1)\\
1147 & 3 & 1.56527096 & 6783 $\pm$ 	183 & 	4244 $\pm$ 	117 & 0.484 $\pm$ 	0.056 & 	7.568 $\pm$ 	0.317 & 	131 $\pm$ 	6 & 	+4.963 $\pm$ 	0.124 & 	13.82 $\pm$ 	1.11 & 	1.250	$\pm$	0.164 &   4.0(3.1)\\
1303 & 3 & 1.10871849 & 7026 $\pm$  368 & 1086  $\pm$  379 & 0.288 $\pm$  0.036 & 8.659  $\pm$  1.26  & -8   $\pm$  10 & -0.063  $\pm$  0.973 & 13.979 $\pm$  1.41   & 1.745	$\pm$	0.641    &   4.5(2.7)\\
1454 & 4 & 2.86627508 & 4824 $\pm$  38  & 9061  $\pm$  270 & 0.26  $\pm$  0.086 & 9.87   $\pm$  0.459 & -85  $\pm$  20 & -12.68  $\pm$  0.344 & 23.072 $\pm$  1.54   & 5.512	$\pm$	0.781    &   10.4(3.9)\\
1666 & 5 & 0.81963911 & 2868 $\pm$  73  & 6281  $\pm$  389 & 0.292 $\pm$  0.213 & 1.098  $\pm$  0.169 & 42   $\pm$  50 & -0.471  $\pm$  0.061 & 4.413  $\pm$  1.92   & 0.022	$\pm$	0.012    &  0.6(2.5)\\
2172 & 4 & 1.955718   & 4269 $\pm$  104 & 5071  $\pm$  318 & 0.468 $\pm$  0.265 & 4.559  $\pm$  1.322 & 102  $\pm$  28 & +3.369   $\pm$  0.339 & 13.046 $\pm$  66.73  & 0.701	$\pm$	1.008    &   3.1(3.4)\\
2967 & 2 & 1.91301    & 7371 $\pm$  291 & 2707  $\pm$  375 & 0.268 $\pm$  0.097 & 7.628  $\pm$  0.415 & -122 $\pm$  19 & -1.599  $\pm$  0.479 & 11.745 $\pm$  0.66   & 1.098	$\pm$	0.138    &   3.8(3.3)\\
3007 & 3 & 1.31733058 & 5770 $\pm$  280 & 10923 $\pm$  522 & 0.227 $\pm$  0.064 & 7.178  $\pm$  0.461 & 156  $\pm$  21 & +3.439   $\pm$  0.419 & 13.876 $\pm$  0.66   & 1.480	$\pm$	0.192     &  4.2(2.9)\\
3095 $\diamondsuit$ & 2 & 2.04440386 & 4619 $\pm$  13  & 6280  $\pm$  84  & 0.185 $\pm$  0.016 & 32.78  $\pm$  0.263 & -54  $\pm$  6  & +12.949  $\pm$  0.251 & 78.557 $\pm$  0.68   & 220.128	$\pm$	5.439    &   226.8(3.4)\\
\hline
\end{tabular}
}
\tablefoot{\textbf{Notes:} Strong candidates are marked in bold and constitute the gold sample. The remaining ones are the marginal candidates referred to as the bronze sample. Symbols denote: [$\diamondsuit$] peculiar mass function candidates; [$\ddag$] additional low amplitude variability; [$\dagger$] sub-harmonic signal; [$\ast$] additional radial mode.}
\end{table*}

\begin{table*}[ht]
\caption{Same as Table \ref{tab:LMC Binary list} but for SMC binary Cepheid candidates.}
\label{tab:SMC Binary list}
\scalebox{0.68}{
\begin{tabular}{lccccccccccc}
\hline
\hline
\textbf{Target} & \textbf{Fit order} & \textbf{P$_{\textbf{pulsation}}$} & \textbf{P$_{\textbf{orbital}}$} & \textbf{T$_{\textbf{0}}$} & \textbf{e} & \textbf{a sini} & \textbf{$\omega$} & \textbf{PCR} & \textbf{K} & \textbf{f(m)} & M$_{\textbf{c}}$ (M$_{\textbf{cep}}$)\\
{} & {} & {(d)} & {(d)} & {(d)} & {} & {AU} & {(deg)} & {(d Myr$^{-1}$)} & {(km s$^{-1}$)} & {(M$_{\odot}$)} & {(M$_{\odot}$)}\\
\hline
F mode &  &  &  &  &  &  &  &  & & & \\ 
Gold sample &  &  &  &  &  &  &  &  & & & \\ 
\hline

\textbf{0024} & 10 & 1.22561463  & 2940 $\pm$  51  & 5035 $\pm$  189 & 0.414 $\pm$  0.158 & 1.62   $\pm$  0.205 & 146  $\pm$  23  & +0.08   $\pm$  0.081 & 6.594  $\pm$  1.8   & 0.066	$\pm$	0.031   &  0.9(2.5)\\
\textbf{0092} & 10 & 0.97886425  & 2989 $\pm$  27  & 6831 $\pm$  265 & 0.184 $\pm$  0.075 & 1.86   $\pm$  0.086 & 167  $\pm$  31  & +0.047  $\pm$  0.05  & 6.893  $\pm$  0.39  & 0.096	$\pm$	0.014   &  1.0(2.3)\\
\textbf{0268} & 10 & 1.20302354  & 1006 $\pm$  15  & 6251 $\pm$  141 & 0.531 $\pm$  0.264 & 0.66   $\pm$  0.223 & -139 $\pm$  51  & +0.071  $\pm$  0.078 & 8.357  $\pm$  23.51 & 0.038	$\pm$	0.086    &  0.7(2.5)\\
\textbf{0349} & 10 & 1.47939779  & 2606 $\pm$  44  & 4389 $\pm$  233 & 0.511 $\pm$  0.254 & 0.702  $\pm$  0.147 & 78   $\pm$  34  & +0.029  $\pm$  0.044 & 3.411  $\pm$  10.31 & 0.007	$\pm$	0.006   &  0.4(2.7)\\
\textbf{0381} & 10 & 1.69160087  & 873  $\pm$  5   & 1473 $\pm$  87  & 0.398 $\pm$  0.188 & 1.044  $\pm$  0.173 & -158 $\pm$  35  & -0.665 $\pm$  0.063 & 14.199 $\pm$  5.61  & 0.199	$\pm$	0.129   &  1.6(2.8)\\
\textbf{0450} & 6  & 1.27595376  & 2169 $\pm$  36  & 7189 $\pm$  137 & 0.665 $\pm$  0.211 & 1.819  $\pm$  0.734 & 177  $\pm$  19  & -0.283 $\pm$  0.116 & 12.198 $\pm$  15.24 & 0.171	$\pm$	0.554   &  1.4(2.5)\\
\textbf{0586} & 10 & 1.90344226  & 1031 $\pm$  9   & 6324 $\pm$  69  & 0.592 $\pm$  0.202 & 1.102  $\pm$  0.234 & 89   $\pm$  30  & +0.015  $\pm$  0.064 & 14.378 $\pm$  32.72 & 0.168	$\pm$	0.195   &  1.5(2.9)\\
\textbf{0643} & 10 & 1.13272049  & 1734 $\pm$  27  & 5965 $\pm$  294 & 0.286 $\pm$  0.194 & 0.708  $\pm$  0.099 & -188 $\pm$  63  & +0.063  $\pm$  0.049 & 4.68   $\pm$  1.9   & 0.016	$\pm$	0.008   &   0.5(2.4)\\
\textbf{0647} & 10 & 2.57753659  & 2743 $\pm$  16  & 5378 $\pm$  58  & 0.474 $\pm$  0.079 & 2.287  $\pm$  0.129 & 145  $\pm$  8   & -0.1   $\pm$  0.076 & 10.293 $\pm$  1.02  & 0.212	$\pm$	0.037   &  1.7(3.3)\\
\textbf{0709} & 10 & 1.73208107  & 3311 $\pm$  14  & 5318 $\pm$  67  & 0.407 $\pm$  0.043 & 2.982  $\pm$  0.099 & 166  $\pm$  7   & +0.111  $\pm$  0.039 & 10.723 $\pm$  0.57  & 0.323	$\pm$	0.033   &   2.0(2.8)\\
\textbf{0771} & 10 & 1.84251862  & 1078 $\pm$  7   & 6235 $\pm$  80  & 0.485 $\pm$  0.183 & 1.17   $\pm$  0.153 & 145  $\pm$  26  & +0.042  $\pm$  0.052 & 13.388 $\pm$  4.09  & 0.184	$\pm$	0.088   &   1.5(2.9)\\
\textbf{0777} & 6  & 1.50491854  & 3867 $\pm$  64  & 5204 $\pm$  223 & 0.368 $\pm$  0.131 & 2.711  $\pm$  0.213 & -179 $\pm$  20  & +0.003  $\pm$  0.084 & 8.173  $\pm$  1.17  & 0.177	$\pm$	0.047   &   1.5(2.7)\\
\textbf{0849} & 10 & 1.74006829  & 3389 $\pm$  19  & 8597 $\pm$  142 & 0.214 $\pm$  0.057 & 2.297  $\pm$  0.068 & -40  $\pm$  16  & +0.019  $\pm$  0.04  & 7.548  $\pm$  0.3   & 0.141	$\pm$	0.013   &  1.4(2.8)\\
\textbf{0918} & 9  & 1.2558028   & 2385 $\pm$  35  & 5671 $\pm$  319 & 0.23  $\pm$  0.178 & 1.067  $\pm$  0.099 & -64  $\pm$  49  & +0.067  $\pm$  0.052 & 5.056  $\pm$  1.07  & 0.029	$\pm$	0.008   &  0.7(2.5)\\
\textbf{1146} & 9  & 2.02200462  & 3451 $\pm$  33  & 5958 $\pm$  91  & 0.535 $\pm$  0.12  & 2.115  $\pm$  0.151 & -87  $\pm$  11  & -0.03  $\pm$  0.084 & 7.897  $\pm$  1.68  & 0.106	$\pm$	0.024   &   1.2(3.0)\\
\textbf{1215} & 10 & 1.51245282  & 1725 $\pm$  18  & 6070 $\pm$  206 & 0.256 $\pm$  0.164 & 0.873  $\pm$  0.087 & -102 $\pm$  45  & -0.149 $\pm$  0.055 & 5.733  $\pm$  1.12  & 0.030	$\pm$	0.010   &   0.7(2.7)\\
\textbf{1245} & 6  & 2.11006384  & 5260 $\pm$  96  & 2811 $\pm$  138 & 0.568 $\pm$  0.089 & 3.221  $\pm$  0.162 & -94  $\pm$  10  & +0.112  $\pm$  0.118 & 8.091  $\pm$  1.2   & 0.161	$\pm$	0.026   &   1.5(3.0)\\
\textbf{1283} & 10 & 2.13681841  & 3106 $\pm$  20  & 7020 $\pm$  82  & 0.377 $\pm$  0.076 & 2.649  $\pm$  0.094 & 106  $\pm$  10  & +0.125  $\pm$  0.095 & 10.021 $\pm$  0.63  & 0.257	$\pm$	0.027   &   1.8(3.1)\\
\textbf{1348} & 10 & 1.60295544  & 1086 $\pm$  3   & 4779 $\pm$  38  & 0.423 $\pm$  0.083 & 1.637  $\pm$  0.103 & 154  $\pm$  14  & +0.025  $\pm$  0.037 & 18.082 $\pm$  1.86  & 0.495	$\pm$	0.096   &   0.4(2.4)\\
\textbf{1581} & 10 & 1.25120461  & 3165 $\pm$  10  & 7481 $\pm$  65  & 0.206 $\pm$  0.032 & 3.763  $\pm$  0.063 & 12   $\pm$  7   & +0.501  $\pm$  0.033 & 13.218 $\pm$  0.28  & 0.710	$\pm$	0.035   &   2.7(2.5)\\
\textbf{1645} & 10 & 1.69847602  & 3080 $\pm$  46  & 6336 $\pm$  181 & 0.431 $\pm$  0.161 & 1.124  $\pm$  0.132 & -38  $\pm$  21  & -0.062 $\pm$  0.076 & 4.393  $\pm$  1.08  & 0.020	$\pm$	0.008   &   0.6(2.8)\\
\textbf{1719} & 10 & 1.97438818  & 2482 $\pm$  19  & 5312 $\pm$  287 & 0.143 $\pm$  0.085 & 1.636  $\pm$  0.065 & 148  $\pm$  42  & +0.541  $\pm$  0.058 & 7.268  $\pm$  0.32  & 0.095	$\pm$	0.011   &   1.2(3.0)\\
\textbf{1795} & 10 & 2.05372202  & 1430 $\pm$  9   & 7606 $\pm$  34  & 0.823 $\pm$  0.083 & 2.406  $\pm$  0.686 & -171 $\pm$  5   & -0.622 $\pm$  0.053 & 32.18  $\pm$  23.82 & 0.909	$\pm$	1.414   &   3.3(3.0)\\
\textbf{1818} & 10 & 3.03115737  & 1558 $\pm$  9   & 5967 $\pm$  101 & 0.261 $\pm$  0.1   & 1.969  $\pm$  0.115 & -21  $\pm$  24  & -1.064 $\pm$  0.098 & 14.231 $\pm$  1.2   & 0.419	$\pm$	0.077   &   2.4(3.5)\\
\textbf{1868} & 10 & 1.52833063  & 4619 $\pm$  118 & 6097 $\pm$  703 & 0.14  $\pm$  0.104 & 1.39   $\pm$  0.073 & -48  $\pm$  53  & -0.091 $\pm$  0.061 & 3.322  $\pm$  0.23  & 0.017	$\pm$	0.003   &   0.6(2.7)\\
\textbf{1883} & 8  & 1.48845328  & 1231 $\pm$  11  & 6840 $\pm$  91  & 0.528 $\pm$  0.251 & 0.937  $\pm$  0.196 & 127  $\pm$  29  & +0.159  $\pm$  0.041 & 9.712  $\pm$  13.75 & 0.073	$\pm$	0.066    &   1.0(2.7)\\
\textbf{1921} & 10 & 1.93682713  & 2832 $\pm$  39  & 7603 $\pm$  160 & 0.439 $\pm$  0.134 & 1.427  $\pm$  0.157 & -29  $\pm$  22  & +0.137  $\pm$  0.066 & 6.092  $\pm$  1.29  & 0.048	$\pm$	0.018   &   0.9(2.9)\\
\textbf{1942} & 10 & 1.84741798  & 4852 $\pm$  92  & 6150 $\pm$  682 & 0.185 $\pm$  0.129 & 0.97   $\pm$  0.081 & 72   $\pm$  51  & +0.028  $\pm$  0.048 & 2.23   $\pm$  0.26  & 0.005	$\pm$	0.001   &   0.4(2.9)\\
\textbf{1960} & 10 & 4.16810092  & 3483 $\pm$  27  & 6349 $\pm$  238 & 0.222 $\pm$  0.07  & 2.637  $\pm$  0.124 & 37   $\pm$  25  & -0.563 $\pm$  0.189 & 8.454  $\pm$  0.49  & 0.201	$\pm$	0.029   &   1.9(3.9)\\
\textbf{2166} & 10 & 2.70734099  & 863  $\pm$  4   & 4115 $\pm$  31  & 0.643 $\pm$  0.166 & 1.155  $\pm$  0.213 & 127  $\pm$  15  & -0.814 $\pm$  0.093 & 18.925 $\pm$  12.96 & 0.275	$\pm$	0.225   &   2.0(3.3)\\
\textbf{2206} & 10 & 1.12805011  & 2298 $\pm$  24  & 5938 $\pm$  53  & 0.68  $\pm$  0.12  & 1.513  $\pm$  0.301 & -141 $\pm$  9   & +0.039  $\pm$  0.044 & 9.731  $\pm$  6.53  & 0.087	$\pm$	0.093    &   1.0(2.4)\\
\textbf{2358} & 10 & 1.72183113  & 4849 $\pm$  32  & 3942 $\pm$  202 & 0.197 $\pm$  0.059 & 3.049  $\pm$  0.085 & 52   $\pm$  14  & +0.08   $\pm$  0.043 & 6.983  $\pm$  0.25  & 0.161	$\pm$	0.014   &   1.4(2.8)\\
\textbf{2457} & 10 & 1.52821245  & 1241 $\pm$  5   & 5758 $\pm$  175 & 0.091 $\pm$  0.065 & 1.739  $\pm$  0.06  & 15   $\pm$  53  & -0.031 $\pm$  0.063 & 15.351 $\pm$  0.54  & 0.456	$\pm$	0.047   &   2.2(2.7)\\
\textbf{2486} & 10 & 1.46259811  & 2355 $\pm$  7   & 7773 $\pm$  34  & 0.384 $\pm$  0.037 & 3.505  $\pm$  0.066 & 122  $\pm$  5   & +0.156  $\pm$  0.049 & 17.535 $\pm$  0.48  & 1.035	$\pm$	0.058   &   3.4(2.7)\\
\textbf{2747} & 10 & 4.56171961  & 3107 $\pm$  15  & 7251 $\pm$  55  & 0.555 $\pm$  0.083 & 4.374  $\pm$  0.326 & -43  $\pm$  7   & +2.126  $\pm$  0.267 & 18.381 $\pm$  3.02  & 1.156	$\pm$	0.291   &   4.3(4.0)\\
\textbf{2833} & 8  & 1.75469245  & 3079 $\pm$  43  & 7081 $\pm$  431 & 0.117 $\pm$  0.089 & 1.821  $\pm$  0.083 & -55  $\pm$  49  & -0.06  $\pm$  0.086 & 6.503  $\pm$  0.35  & 0.085	$\pm$	0.012   &   1.1(2.8)\\
\textbf{2844} & 10 & 1.72678651  & 1336 $\pm$  6   & 8630 $\pm$  54  & 0.359 $\pm$  0.116 & 1.477  $\pm$  0.082 & 49   $\pm$  17  & -0.035 $\pm$  0.061 & 12.89  $\pm$  1.35  & 0.241	$\pm$	0.042    &   1.7(2.8)\\
\textbf{2849} & 8  & 1.58862022  & 4822 $\pm$  56  & 6968 $\pm$  193 & 0.23  $\pm$  0.054 & 3.358  $\pm$  0.11  & 4    $\pm$  13  & +0.24   $\pm$  0.056 & 7.789  $\pm$  0.35  & 0.217	$\pm$	0.022    &   1.6(2.8)\\
\textbf{2929} & 10 & 1.58662693  & 951  $\pm$  6   & 8456 $\pm$  34  & 0.647 $\pm$  0.121 & 1.438  $\pm$  0.298 & -18  $\pm$  10  & +0.031  $\pm$  0.053 & 21.5   $\pm$  10.41 & 0.438	$\pm$	0.441   &   2.2(2.8)\\
\textbf{2979} & 10 & 2.0186305   & 2221 $\pm$  20  & 7833 $\pm$  246 & 0.194 $\pm$  0.131 & 1.49   $\pm$  0.1   & 35   $\pm$  40  & -0.543 $\pm$  0.119 & 7.457  $\pm$  0.76  & 0.089	$\pm$	0.019   &   1.2(3.0)\\
\textbf{3061} & 10 & 2.54070246  & 1197 $\pm$  7   & 8733 $\pm$  64  & 0.454 $\pm$  0.156 & 1.857  $\pm$  0.253 & -171 $\pm$  21  & +0.162  $\pm$  0.161 & 18.886 $\pm$  5.19  & 0.595	$\pm$	0.307   &   2.8(3.2)\\
\textbf{3075} & 10 & 2.04232342  & 1032 $\pm$  6   & 2951 $\pm$  88  & 0.403 $\pm$  0.192 & 1.081  $\pm$  0.134 & 65   $\pm$  32  & -0.381 $\pm$  0.075 & 12.458 $\pm$  7.09  & 0.158	$\pm$	0.067   &   1.5(3.0)\\
\textbf{3087} & 10 & 1.55714085  & 1550 $\pm$  13  & 7913 $\pm$  82  & 0.579 $\pm$  0.173 & 1.239  $\pm$  0.253 & -130 $\pm$  18  & +0.103  $\pm$  0.05  & 10.593 $\pm$  6.93  & 0.105	$\pm$	0.103   &   1.2(2.7)\\
\textbf{3229} & 8  & 1.0361886   & 2771 $\pm$  41  & 5607 $\pm$  61  & 0.867 $\pm$  0.147 & 1.81   $\pm$  0.638 & 51   $\pm$  15  & +0.072  $\pm$  0.084 & 14.387 $\pm$  27    & 0.103	$\pm$	0.251   &   1.1(2.4)\\
\textbf{3308} & 10 & 1.36063995  & 1697 $\pm$  21  & 5793 $\pm$  106 & 0.654 $\pm$  0.23  & 0.884  $\pm$  0.182 & -89  $\pm$  30  & +0.07   $\pm$  0.067 & 7.405  $\pm$  23.88 & 0.032	$\pm$	0.035    &   0.7(2.6)\\
\textbf{3463} & 10 & 2.56622462  & 1828 $\pm$  8   & 8622 $\pm$  86  & 0.343 $\pm$  0.081 & 3.204  $\pm$  0.189 & -136 $\pm$  17  & -0.583 $\pm$  0.122 & 20.281 $\pm$  1.84  & 1.313	$\pm$	0.245   &   4.2(3.3)\\
\textbf{3466} & 10 & 1.85843731  & 1262 $\pm$  14  & 6069 $\pm$  173 & 0.379 $\pm$  0.269 & 0.752  $\pm$  0.141 & -73  $\pm$  52  & -0.079 $\pm$  0.094 & 7.024  $\pm$  7.4   & 0.035	$\pm$	0.028   &   0.8(2.9)\\
\textbf{3503} & 10 & 3.04494906  & 1802 $\pm$  32  & 6279 $\pm$  140 & 0.495 $\pm$  0.237 & 1.049  $\pm$  0.17  & 131  $\pm$  28  & -0.197 $\pm$  0.201 & 7.246  $\pm$  14.26 & 0.047	$\pm$	0.031   &   1.0(3.5)\\
\textbf{3556} & 9  & 3.69854923  & 4305 $\pm$  30  & 6375 $\pm$  65  & 0.531 $\pm$  0.064 & 4.676  $\pm$  0.219 & -59  $\pm$  6   & +5.489  $\pm$  0.182 & 13.932 $\pm$  1.34  & 0.735	$\pm$	0.107   &   3.3(3.7)\\
\textbf{3693} & 8  & 1.32095132  & 3341 $\pm$  16  & 6852 $\pm$  95  & 0.297 $\pm$  0.05  & 3.903  $\pm$  0.138 & -167 $\pm$  10  & -0.073 $\pm$  0.047 & 13.306 $\pm$  0.68  & 0.710	$\pm$	0.079     &  2.7(2.6)\\
\textbf{3838} & 10 & 1.61179355  & 2214 $\pm$  50  & 7392 $\pm$  180 & 0.635 $\pm$  0.248 & 0.787  $\pm$  0.185 & 63   $\pm$  29  & -0.01  $\pm$  0.076 & 5.015  $\pm$  18.86 & 0.013	$\pm$	0.016   &   0.5(2.8)\\
\textbf{3869} & 10 & 2.59092842  & 1478 $\pm$  7   & 6995 $\pm$  34  & 0.732 $\pm$  0.151 & 1.805  $\pm$  0.349 & 125  $\pm$  12  & -0.354 $\pm$  0.088 & 19.352 $\pm$  15.27 & 0.359	$\pm$	0.346   &   2.2(3.3)\\
\textbf{3881} & 10 & 1.44373436  & 3799 $\pm$  54  & 6932 $\pm$  179 & 0.368 $\pm$  0.124 & 1.361  $\pm$  0.068 & 96   $\pm$  17  & -0.124 $\pm$  0.053 & 4.192  $\pm$  0.42  & 0.023	$\pm$	0.004   &   0.6(2.7)\\
\textbf{3914} & 9  & 1.60525727  & 2553 $\pm$  18  & 6440 $\pm$  38  & 0.719 $\pm$  0.153 & 2.795  $\pm$  0.713 & -58  $\pm$  9   & +0.14   $\pm$  0.058 & 17.1   $\pm$  16.88 & 0.448	$\pm$	0.575   &   2.2(2.8)\\
\textbf{3929} & 10 & 2.57717076  & 2218 $\pm$  33  & 5891 $\pm$  262 & 0.279 $\pm$  0.161 & 1.498  $\pm$  0.131 & 138  $\pm$  45  & -0.207 $\pm$  0.186 & 7.677  $\pm$  1.35  & 0.091	$\pm$	0.025   &   1.2(3.3)\\
\textbf{4016} & 10 & 1.8631558   & 2259 $\pm$  9   & 6144 $\pm$  81  & 0.288 $\pm$  0.059 & 3.597  $\pm$  0.125 & 73   $\pm$  13  & -0.12  $\pm$  0.072 & 18.096 $\pm$  0.87  & 1.216	$\pm$	0.130      &   3.8(2.9)\\
\textbf{4111} & 10 & 2.77874082  & 3154 $\pm$  37  & 4978 $\pm$  134 & 0.493 $\pm$  0.129 & 2.016  $\pm$  0.267 & 173  $\pm$  14  & +0.086  $\pm$  0.121 & 7.97   $\pm$  2.4   & 0.110	$\pm$	0.060   &   1.3(3.3)\\
\textbf{4202} & 10 & 1.59026817  & 2035 $\pm$  14  & 6526 $\pm$  102 & 0.319 $\pm$  0.104 & 1.373  $\pm$  0.072 & 83   $\pm$  18  & -0.015 $\pm$  0.044 & 7.757  $\pm$  0.64  & 0.083	$\pm$	0.013   &   1.1(2.8)\\
\textbf{4218} & 10 & 2.03454707  & 3401 $\pm$  25  & 5601 $\pm$  50  & 0.528 $\pm$  0.097 & 4.336  $\pm$  0.178 & -102 $\pm$  6   & +0.506  $\pm$  0.109 & 16.279 $\pm$  2.14  & 0.937	$\pm$	0.128   &   3.4(3.0)\\
\textbf{4266} & 10 & 2.54817018  & 5353 $\pm$  76  & 2539 $\pm$  59  & 0.48  $\pm$  0.043 & 6.8    $\pm$  0.196 & 169  $\pm$  3   & +1.106  $\pm$  0.652 & 15.717 $\pm$  0.75  & 1.460	$\pm$	0.117   &  4.4(3.2)\\
\textbf{4289} & 8  & 3.31679034  & 2272 $\pm$  41  & 5884 $\pm$  263 & 0.23  $\pm$  0.148 & 1.973  $\pm$  0.164 & 36   $\pm$  41  & +1.298  $\pm$  0.516 & 9.772  $\pm$  1.24  & 0.199	$\pm$	0.051   &   1.8(3.6)\\
\textbf{4365} & 10 & 1.4343708   & 2084 $\pm$  17  & 5962 $\pm$  19  & 0.869 $\pm$  0.066 & 3.865  $\pm$  1.411 & -165 $\pm$  4   & +0.287  $\pm$  0.088 & 40.766 $\pm$  43.12 & 1.775	$\pm$	4.516   &   4.5(2.7)\\
\textbf{4400} & 9  & 2.57392249  & 1017 $\pm$  11  & 5135 $\pm$  35  & 0.852 $\pm$  0.131 & 2.222  $\pm$  1.062 & -167 $\pm$  8   & -0.459 $\pm$  0.17  & 45.306 $\pm$  61.53 & 1.412	$\pm$	4.826   &   4.3(3.3)\\
\textbf{4436} & 10 & 2.74656267  & 982  $\pm$  7   & 6326 $\pm$  39  & 0.814 $\pm$  0.161 & 1.809  $\pm$  0.742 & 133  $\pm$  18  & -0.321 $\pm$  0.222 & 35.063 $\pm$  73.84 & 0.820	$\pm$	2.510   &   3.3(3.3)\\
\textbf{4512} & 10 & 1.20054953  & 3753 $\pm$  68  & 6958 $\pm$  253 & 0.284 $\pm$  0.108 & 1.472  $\pm$  0.09  & 128  $\pm$  24  & +0.045  $\pm$  0.077 & 4.455  $\pm$  0.43  & 0.030	$\pm$	0.006   &   0.7(2.5)\\
\textbf{4571} & 10 & 2.1145914   & 2213 $\pm$  35  & 6867 $\pm$  114 & 0.614 $\pm$  0.151 & 1.588  $\pm$  0.363 & 11   $\pm$  15  & -0.431 $\pm$  0.146 & 9.861  $\pm$  5.3   & 0.109	$\pm$	0.119  &   1.3(3.0)\\

\hline
F mode &  &  &  &  &  &  &  &  & & & \\ 
Bronze sample &  &  &  &  &  &  &  &  & & & \\ 
\hline
0031 & 10 & 1.52691807  & 1014 $\pm$  9   & 5125 $\pm$  149 & 0.313 $\pm$  0.208 & 0.807  $\pm$  0.139 & -10  $\pm$  50  & +0.171  $\pm$  0.084 & 9.118  $\pm$  3.82  & 0.068	$\pm$	0.052   &   1.0(2.7)\\
0127 & 10 & 1.29405349  & 3236 $\pm$  72  & 6802 $\pm$  466 & 0.187 $\pm$  0.123 & 1.06   $\pm$  0.086 & 144  $\pm$  51  & +0.156  $\pm$  0.07  & 3.642  $\pm$  0.41  & 0.015	$\pm$	0.004   &   0.5(2.6)\\
0660 & 10 & 1.43903929  & 1856 $\pm$  46  & 5258 $\pm$  66  & 0.908 $\pm$  0.072 & 2.792  $\pm$  1.242 & -10  $\pm$  6   & +0.184  $\pm$  0.045 & 39.1   $\pm$  47.56 & 0.845	$\pm$	2.424   &   3.0(2.7)\\
1115 & 10 & 3.90037931  & 1784 $\pm$  21  & 6412 $\pm$  143 & 0.66  $\pm$  0.255 & 1.003  $\pm$  0.272 & 62   $\pm$  31  & -1.242 $\pm$  0.173 & 8.069  $\pm$  20.85 & 0.042	$\pm$	0.061   &   1.0(3.8)\\
1318 & 10 & 1.13801287  & 3893 $\pm$  103 & 5983 $\pm$  107 & 0.884 $\pm$  0.142 & 0.962  $\pm$  0.187 & 106  $\pm$  20  & -0.099 $\pm$  0.066 & 5.882  $\pm$  17.45 & 0.008	$\pm$	0.007    &   0.4(2.4)\\
1707 & 10 & 1.9624896   & 2649 $\pm$  18  & 7513 $\pm$  67  & 0.624 $\pm$  0.133 & 1.724  $\pm$  0.266 & 51   $\pm$  10  & -0.666 $\pm$  0.056 & 9.016  $\pm$  4.56  & 0.097	$\pm$	0.066    &   1.2(3.0)\\
2229 & 10 & 1.54470493  & 970  $\pm$  4   & 6747 $\pm$  204 & 0.135 $\pm$  0.135 & 0.867  $\pm$  0.09  & -3   $\pm$  108 & +0.013  $\pm$  0.046 & 9.893  $\pm$  1.36  & 0.092	$\pm$	0.030   &   1.1(2.7)\\
2504 & 10 & 1.27996851  & 2615 $\pm$  59  & 6072 $\pm$  308 & 0.384 $\pm$  0.236 & 0.59   $\pm$  0.122 & -45  $\pm$  44  & +0.001  $\pm$  0.036 & 2.661  $\pm$  3.05  & 0.004	$\pm$	0.004    &   0.3(2.6)\\
2989 & 10 & 1.96058204  & 5248 $\pm$  228 & 3975 $\pm$  176 & 0.277 $\pm$  0.079 & 1.859  $\pm$  0.132 & -107 $\pm$  12  & +0.807  $\pm$  0.302 & 4.032  $\pm$  0.21  & 0.031	$\pm$	0.005   &   0.8(3.0)\\
3053 & 10 & 1.54027597  & 5048 $\pm$  146 & 9785 $\pm$  724 & 0.126 $\pm$  0.103 & 0.96   $\pm$  0.059 & 14   $\pm$  52  & +0.09   $\pm$  0.058 & 2.095  $\pm$  0.15  & 0.005	$\pm$	0.001   &   0.4(2.7)\\
3247 & 10 & 1.46096693  & 1811 $\pm$  25  & 5972 $\pm$  282 & 0.216 $\pm$  0.176 & 1.045  $\pm$  0.14  & -134 $\pm$  57  & +0.108  $\pm$  0.112 & 6.483  $\pm$  2.3   & 0.046	$\pm$	0.026   &   0.8(2.7)\\
3618 & 10 & 1.61380046  & 2165 $\pm$  29  & 6384 $\pm$  86  & 0.704 $\pm$  0.214 & 0.896  $\pm$  0.188 & 95   $\pm$  19  & +0.131  $\pm$  0.037 & 6.314  $\pm$  22.91 & 0.021	$\pm$	0.017   &   0.6(2.8)\\
3664 & 10 & 2.09275118  & 2110 $\pm$  63  & 5566 $\pm$  347 & 0.267 $\pm$  0.245 & 1.036  $\pm$  0.191 & 98   $\pm$  60  & -0.384 $\pm$  0.173 & 5.595  $\pm$  5.95  & 0.033	$\pm$	0.025   &   0.8(3.0)\\
\hline
\end{tabular}}
\end{table*}

\begin{table*}[ht]
\ContinuedFloat
\caption{continued.}
\scalebox{0.68}{
\begin{tabular}{lccccccccccc}
\hline
\hline
\textbf{Target} & \textbf{Fit order} & \textbf{P$_{\textbf{pulsation}}$} & \textbf{P$_{\textbf{orbital}}$} & \textbf{T$_{\textbf{0}}$} & \textbf{e} & \textbf{a sini} & \textbf{$\omega$} & \textbf{PCR} & \textbf{K} & \textbf{f(m)} & M$_{\textbf{c}}$ (M$_{\textbf{cep}}$)\\
{} & {} & {(d)} & {(d)} & {(d)} & {} & {AU} & {(deg)} & {(d Myr$^{-1}$)} & {(km s$^{-1}$)} & {(M$_{\odot}$)} & {(M$_{\odot}$)}\\
\hline
F mode &  &  &  &  &  &  &  &  & & & \\ 
Bronze sample &  &  &  &  &  &  &  &  & & & \\ 
\hline
3799 & 10 & 2.85774356  & 3781 $\pm$  81  & 7254 $\pm$  90  & 0.696 $\pm$  0.14  & 2.162  $\pm$  0.671 & 11   $\pm$  9   & +0.247  $\pm$  0.158 & 8.665  $\pm$  7.08  & 0.095	$\pm$	0.182    &  1.3(3.4)\\
3935 $\diamondsuit$ & 6  & 6.58742326  & 1797 $\pm$  9   & 5721 $\pm$  120 & 0.279 $\pm$  0.086 & 7.894  $\pm$  0.407 & -93  $\pm$  24  & -5.736 $\pm$  0.96  & 49.812 $\pm$  3.7   & 20.312	$\pm$	3.232   &  27.5(4.5)\\
3994 $\diamondsuit$ & 6  & 10.18392471 & 6312 $\pm$  127 & 8123 $\pm$  365 & 0.2   $\pm$  0.07  & 16.922 $\pm$  0.524 & -156 $\pm$  20  & -8.069 $\pm$  5.302 & 29.712 $\pm$  1.26  & 16.175	$\pm$	1.485   &   24.1(5.3)\\
4064 & 7  & 1.23047345  & 2586 $\pm$  55  & 5220 $\pm$  244 & 0.466 $\pm$  0.271 & 1.106  $\pm$  0.199 & -112 $\pm$  35  & -0.219 $\pm$  0.141 & 5.265  $\pm$  11.38 & 0.027	$\pm$	0.020    &   0.7(2.5)\\
4355 & 10 & 1.41788388  & 3024 $\pm$  71  & 7103 $\pm$  243 & 0.493 $\pm$  0.225 & 0.973  $\pm$  0.205 & -52  $\pm$  31  & +0.053  $\pm$  0.093 & 4.023  $\pm$  28.09 & 0.014	$\pm$	0.013   &   0.5(2.6)\\
4456 & 4  & 9.47825523  & 3380 $\pm$  112 & 5087 $\pm$  539 & 0.219 $\pm$  0.221 & 8.297  $\pm$  1.253 & 132  $\pm$  58  & +41.56  $\pm$  7.423 & 27.711 $\pm$  11.09 & 6.656	$\pm$	3.653   &   13.0(5.2)\\
\hline
1O mode &  &  &  &  &  &  &  &  & & & \\ 
Gold sample &  &  &  &  &  &  &  &  & & & \\ 
\hline
\textbf{0093} $\diamondsuit$ & 4 & 0.68543359 & 4634 $\pm$  33  & 5595  $\pm$  270 & 0.075 $\pm$  0.024 & 10.881 $\pm$  0.146 & -67  $\pm$  20 & -2.132 $\pm$  0.103 & 25.619  $\pm$  0.36   & 8.001	$\pm$	0.319   &   11.5(2.3)\\
\textbf{0197} & 3 & 0.8843626  & 1884 $\pm$  18  & 5078  $\pm$  168 & 0.27  $\pm$  0.134 & 2.412  $\pm$  0.195 & -124 $\pm$  32 & -0.18  $\pm$  0.083 & 14.492  $\pm$  1.79   & 0.527	$\pm$	0.131   &   2.3(2.5)\\
\textbf{0239} & 3 & 1.37099521 & 2962 $\pm$  50  & 6035  $\pm$  101 & 0.797 $\pm$  0.163 & 2.68   $\pm$  0.632 & 106  $\pm$  20 & -1.436 $\pm$  0.145 & 16.322  $\pm$  33.51  & 0.291	$\pm$	0.314   &   1.9(3.0)\\
\textbf{0391} & 5 & 0.76872759 & 4053 $\pm$  56  & 6795  $\pm$  300 & 0.143 $\pm$  0.059 & 3.128  $\pm$  0.093 & -6   $\pm$  25 & +2.903  $\pm$  0.073 & 8.488   $\pm$  0.29   & 0.249	$\pm$	0.022   &   1.6(2.4)\\
\textbf{0600} $\diamondsuit$ & 4 & 0.95505234 & 5706 $\pm$  35  & 8708  $\pm$  57  & 0.22  $\pm$  0.021 & 12.026 $\pm$  0.115 & -172 $\pm$  3  & -2.152 $\pm$  0.115 & 23.49   $\pm$  0.3    & 7.120	$\pm$	0.198   &   10.9(2.6)\\
\textbf{0656} & 4 & 0.79445363 & 4193 $\pm$  54  & 7055  $\pm$  252 & 0.285 $\pm$  0.106 & 2.316  $\pm$  0.129 & 57   $\pm$  22 & +0.281  $\pm$  0.034 & 6.268   $\pm$  0.51   & 0.094	$\pm$	0.017   &   1.0(2.4)\\
\textbf{0744} $\diamondsuit$ & 3 & 0.85968448 & 3949 $\pm$  10  & 6588  $\pm$  130 & 0.142 $\pm$  0.02  & 9.236  $\pm$  0.134 & 59   $\pm$  11 & -4.436 $\pm$  0.058 & 25.709  $\pm$  0.42   & 6.739	$\pm$	0.301   &   10.4(2.5)\\
\textbf{1267} $\diamondsuit$ & 4 & 0.90520386 & 6068 $\pm$  62  & 9652  $\pm$  72  & 0.151 $\pm$  0.014 & 32.855 $\pm$  0.827 & -73  $\pm$  3  & +2.713  $\pm$  0.691 & 59.591  $\pm$  0.84   & 128.508	$\pm$	7.221   &   133.5(2.5)\\
\textbf{1468} & 3 & 0.91322521 & 4903 $\pm$  30  & 6775  $\pm$  86  & 0.436 $\pm$  0.047 & 6.372  $\pm$  0.174 & -77  $\pm$  6  & +2.237  $\pm$  0.042 & 15.711  $\pm$  0.73   & 1.436	$\pm$	0.118   &   3.9(2.6)\\
\textbf{1597} & 4 & 1.11620594 & 5373 $\pm$  19  & 8180  $\pm$  659 & 0.027 $\pm$  0.018 & 9.778  $\pm$  0.095 & -24  $\pm$  44 & +2.844  $\pm$  0.042 & 19.807  $\pm$  0.22   & 4.319	$\pm$	0.135   &   7.9(2.8)\\
\textbf{1611} $\ddag$& 4 & 1.7322348  & 3051 $\pm$  51  & 5805  $\pm$  254 & 0.501 $\pm$  0.269 & 1.297  $\pm$  0.381 & 157  $\pm$  30 & +0.052  $\pm$  0.071 & 5.315   $\pm$  5.15   & 0.031	$\pm$	0.051   &   0.8(3.2)\\
\textbf{1615} $\ddag$& 4 & 1.35334862 & 4294 $\pm$  78  & 3834  $\pm$  568 & 0.169 $\pm$  0.121 & 2.092  $\pm$  0.152 & 27   $\pm$  47 & +0.052  $\pm$  0.068 & 5.408   $\pm$  0.49   & 0.066	$\pm$	0.014   &   1.0(3.0)\\
\textbf{1871} & 5 & 1.03165542 & 2766 $\pm$  20  & 4443  $\pm$  54  & 0.691 $\pm$  0.121 & 1.518  $\pm$  0.264 & 157  $\pm$  7  & +0.049  $\pm$  0.03  & 8.225   $\pm$  3.66   & 0.061	$\pm$	0.044   &   0.9(2.7)\\
\textbf{1928} & 6 & 0.80609668 & 6026 $\pm$  22  & 3165  $\pm$  64  & 0.249 $\pm$  0.021 & 8.019  $\pm$  0.08  & -99  $\pm$  4  & +1.06   $\pm$  0.035 & 14.946  $\pm$  0.21   & 1.894	$\pm$	0.058    &   4.5(2.4)\\
\textbf{1985} & 8 & 1.14034791 & 3497 $\pm$  15  & 5018  $\pm$  65  & 0.274 $\pm$  0.054 & 4.817  $\pm$  0.113 & -108 $\pm$  6  & +5.698  $\pm$  0.054 & 15.586  $\pm$  0.53   & 1.219	$\pm$	0.084   &   3.7(2.8)\\
\textbf{2015} $\diamondsuit$ & 5 & 0.93782717 & 6531 $\pm$  23  & 2986  $\pm$  192 & 0.052 $\pm$  0.009 & 18.156 $\pm$  0.086 & 83   $\pm$  10 & +0.911  $\pm$  0.068 & 30.285  $\pm$  0.13   & 18.718	$\pm$	0.233   &   23.1(2.6)\\
\textbf{2127} $\diamondsuit$ & 4 & 0.90222227 & 5352 $\pm$  56  & 5165  $\pm$  107 & 0.184 $\pm$  0.023 & 10.365 $\pm$  0.139 & 141  $\pm$  7  & -0.484 $\pm$  0.097 & 21.432  $\pm$  0.3    & 5.184	$\pm$	0.184   &   8.7(2.5)\\
\textbf{2250} $\diamondsuit$ & 5 & 2.33089694 & 6182 $\pm$  59  & 4232  $\pm$  114 & 0.248 $\pm$  0.026 & 13.299 $\pm$  0.162 & -57  $\pm$  6  & +0.014  $\pm$  0.169 & 24.15   $\pm$  0.44   & 8.204	$\pm$	0.339   &   13.3(3.6)\\
\textbf{2357} & 5 & 0.81908458 & 3868 $\pm$  26  & 7090  $\pm$  78  & 0.444 $\pm$  0.066 & 3.955  $\pm$  0.166 & -75  $\pm$  7  & -0.127 $\pm$  0.031 & 12.408  $\pm$  0.94   & 0.551	$\pm$	0.072   &   2.3(2.5)\\
\textbf{2518} & 5 & 1.08540032 & 2999 $\pm$  43  & 5115  $\pm$  94  & 0.578 $\pm$  0.154 & 1.878  $\pm$  0.204 & -67  $\pm$  14 & -0.257 $\pm$  0.065 & 8.35    $\pm$  6.79   & 0.099	$\pm$	0.039   &   1.1(2.7)\\
\textbf{2961} & 4 & 0.89066439 & 3108 $\pm$  64  & 5294  $\pm$  206 & 0.39  $\pm$  0.167 & 1.593  $\pm$  0.12  & -120 $\pm$  25 & +0.077  $\pm$  0.043 & 6.005   $\pm$  1.12   & 0.055	$\pm$	0.015    &   0.9(2.5)\\
\textbf{2973} & 5 & 1.56733549 & 3825 $\pm$  22  & 7436  $\pm$  96  & 0.333 $\pm$  0.049 & 5.22   $\pm$  0.145 & 82   $\pm$  8  & +5.14   $\pm$  0.075 & 15.752  $\pm$  0.65   & 1.297	$\pm$	0.110   &   4.1(3.1)\\
\textbf{3097} $\ddag$ $\dagger$ & 2 & 2.0168275  & 3544 $\pm$  65  & 7447  $\pm$  396 & 0.267 $\pm$  0.128 & 6.038  $\pm$  0.426 & -93  $\pm$  39 & -5.169 $\pm$  0.418 & 19.294  $\pm$  2      & 2.337	$\pm$	0.500   &   5.9(3.4)\\
\textbf{3863} & 6 & 0.9672118  & 3938 $\pm$  59  & 6277  $\pm$  109 & 0.852 $\pm$  0.119 & 1.077  $\pm$  0.176 & 68   $\pm$  17 & +0.132  $\pm$  0.022 & 5.714   $\pm$  10.92  & 0.011	$\pm$	0.006   &   0.5(2.6)\\
\textbf{4075} $\diamondsuit$ & 5 & 0.8063606  & 3524 $\pm$  10  & 6742  $\pm$  74  & 0.195 $\pm$  0.026 & 9.359  $\pm$  0.126 & -132 $\pm$  7  & -0.235 $\pm$  0.059 & 29.46   $\pm$  0.47   & 8.803	$\pm$	0.353   &   12.6(2.4)\\
\textbf{4430} $\diamondsuit$ & 5 & 1.54368245 & 4185 $\pm$  22  & 6435  $\pm$  73  & 0.185 $\pm$  0.02  & 13.034 $\pm$  0.148 & -114 $\pm$  6  & +1.392  $\pm$  0.22  & 34.479  $\pm$  0.47   & 16.866	$\pm$	0.582   &   22.0(3.1)\\
\textbf{4496} & 5 & 1.36000613 & 2698 $\pm$  54  & 6815  $\pm$  403 & 0.209 $\pm$  0.158 & 1.338  $\pm$  0.14  & 100  $\pm$  54 & +1.622  $\pm$  0.125 & 5.559   $\pm$  1.48   & 0.044	$\pm$	0.015   &   0.9(3.0)\\
\textbf{4513} & 2 & 1.28093303 & 2277 $\pm$  33  & 11289 $\pm$  141 & 0.634 $\pm$  0.142 & 4.96   $\pm$  0.995 & -153 $\pm$  12 & +0.575  $\pm$  0.282 & 30.582  $\pm$  15.42  & 3.122	$\pm$	2.829   &   6.5(2.9)\\
\textbf{4524} & 5 & 0.92472577 & 3391 $\pm$  65  & 9390  $\pm$  193 & 0.527 $\pm$  0.164 & 1.116  $\pm$  0.12  & 104  $\pm$  24 & -0.181 $\pm$  0.058 & 4.218   $\pm$  5.01   & 0.016	$\pm$	0.006   &   0.5(2.6)\\
\hline
1O mode &  &  &  &  &  &  &  &  & & & \\ 
Bronze sample &  &  &  &  &  &  &  &  & & & \\ 

\hline
0180 & 4 & 0.97331376 & 5721 $\pm$  162 & 4163  $\pm$  138 & 0.395 $\pm$  0.076 & 4.057  $\pm$  0.274 & -44  $\pm$  10 & -0.059 $\pm$  0.261 & 8.392   $\pm$  0.62   & 0.271	$\pm$	0.047   &   1.7(2.6)\\
0337 & 4 & 1.13067702 & 3923 $\pm$  36  & 6969  $\pm$  584 & 0.064 $\pm$  0.051 & 5.351  $\pm$  0.181 & 99   $\pm$  53 & +3.717  $\pm$  0.125 & 14.89   $\pm$  0.54   & 1.327	$\pm$	0.138   &   3.9(2.8)\\
0340 & 4 & 0.80186187 & 3292 $\pm$  15  & 6600  $\pm$  236 & 0.096 $\pm$  0.045 & 7.073  $\pm$  0.141 & -77  $\pm$  25 & +2.52   $\pm$  0.08  & 23.504  $\pm$  0.48   & 4.356	$\pm$	0.258   &   7.6(2.4)\\
0536 & 4 & 1.03324391 & 2863 $\pm$  16  & 6304  $\pm$  105 & 0.437 $\pm$  0.08  & 3.755  $\pm$  0.21  & 62   $\pm$  14 & -1.484 $\pm$  0.06  & 15.85   $\pm$  1.6    & 0.861	$\pm$	0.150   &   3.0(2.7)\\
0775 & 4 & 1.59291308 & 5960 $\pm$  177 & 5003  $\pm$  202 & 0.351 $\pm$  0.08  & 4.122  $\pm$  0.184 & -27  $\pm$  11 & -1.525 $\pm$  0.1   & 8.036   $\pm$  0.5    & 0.263	$\pm$	0.035   &   1.9(3.1)\\
0792 & 4 & 1.25033041 & 3479 $\pm$  66  & 4879  $\pm$  194 & 0.439 $\pm$  0.125 & 2.063  $\pm$  0.248 & -3   $\pm$  19 & +0.456  $\pm$  0.084 & 7.175   $\pm$  1.53   & 0.097	$\pm$	0.041   &   1.2(2.9)\\
0859 & 4 & 0.87780105 & 4603 $\pm$  151 & 5616  $\pm$  209 & 0.195 $\pm$  0.045 & 4.311  $\pm$  0.172 & -106 $\pm$  16 & -1.14  $\pm$  0.282 & 10.366  $\pm$  0.24   & 0.501	$\pm$	0.035    &   2.3(2.5)\\
0885 & 5 & 0.83943679 & 4121 $\pm$  35  & 6469  $\pm$  69  & 0.523 $\pm$  0.056 & 3.532  $\pm$  0.146 & 22   $\pm$  5  & -0.328 $\pm$  0.025 & 10.94   $\pm$  0.86   & 0.346	$\pm$	0.044   &   1.9(2.5)\\
1103 & 5 & 1.08102444 & 3697 $\pm$  52  & 2338  $\pm$  101 & 0.861 $\pm$  0.121 & 1.469  $\pm$  0.623 & -23  $\pm$  9  & +0.629  $\pm$  0.042 & 8.469   $\pm$  11.4   & 0.031	$\pm$	0.089   &   0.7(2.7)\\
1176 & 5 & 0.9696587  & 7762 $\pm$  216 & 3710  $\pm$  141 & 0.2   $\pm$  0.02  & 12.326 $\pm$  0.418 & 141  $\pm$  8  & +0.142  $\pm$  0.118 & 17.641  $\pm$  0.2    & 4.150	$\pm$	0.205   &   7.5(2.6)\\
1498 & 5 & 1.27470408 & 2642 $\pm$  26  & 6664  $\pm$  182 & 0.255 $\pm$  0.101 & 2.041  $\pm$  0.085 & 41   $\pm$  24 & -0.131 $\pm$  0.05  & 8.705   $\pm$  0.53   & 0.163	$\pm$	0.020   &   1.5(2.9)\\
1626 & 6 & 0.84449431 & 4620 $\pm$  51  & 8472  $\pm$  94  & 0.332 $\pm$  0.054 & 3.036  $\pm$  0.08  & -123 $\pm$  7  & -1.524 $\pm$  0.047 & 7.58    $\pm$  0.28   & 0.175	$\pm$	0.013   &   1.4(2.5)\\
1951 & 6 & 1.30785364 & 3819 $\pm$  70  & 5645  $\pm$  471 & 0.356 $\pm$  0.202 & 1.099  $\pm$  0.133 & -87  $\pm$  49 & +0.313  $\pm$  0.061 & 3.356   $\pm$  1.11   & 0.012	$\pm$	0.005   &   0.5(2.9)\\
2027 & 5 & 0.79555168 & 5773 $\pm$  262 & 4265  $\pm$  299 & 0.329 $\pm$  0.13  & 1.28   $\pm$  0.088 & -67  $\pm$  21 & -0.072 $\pm$  0.067 & 2.558   $\pm$  0.29   & 0.008	$\pm$	0.002    &   0.4(2.4)\\
2033 $\ddag$& 5 & 1.40873535 & 1820 $\pm$  16  & 5652  $\pm$  55  & 0.815 $\pm$  0.183 & 1.86   $\pm$  0.57  & -115 $\pm$  18 & +0.421  $\pm$  0.061 & 19.343  $\pm$  46.99  & 0.259	$\pm$	0.451   &   1.8(3.0)\\
2035 & 5 & 0.84843745 & 6045 $\pm$  64  & 6823  $\pm$  81  & 0.555 $\pm$  0.054 & 5.546  $\pm$  0.322 & -160 $\pm$  4  & +2.785  $\pm$  0.067 & 11.985  $\pm$  1.2    & 0.622	$\pm$	0.111   &   2.5(2.5)\\
2117 & 5 & 2.9034422  & 3915 $\pm$  78  & 6845  $\pm$  124 & 0.851 $\pm$  0.11  & 5.205  $\pm$  0.723 & -78  $\pm$  16 & +0.572  $\pm$  0.3   & 27.589  $\pm$  73.2   & 1.221	$\pm$	0.659   &   4.4(3.9)\\
2125 & 6 & 2.16627716 & 7329 $\pm$  242 & 7399  $\pm$  162 & 0.307 $\pm$  0.044 & 7.186  $\pm$  0.47  & 20   $\pm$  8  & -4.874 $\pm$  0.6   & 11.238  $\pm$  0.44   & 0.923	$\pm$	0.130   &   3.6(3.5)\\
2293 $\ddag$& 5 & 1.28879424 & 6873 $\pm$  95  & 2986  $\pm$  88  & 0.955 $\pm$  0.044 & 3.134  $\pm$  1.102 & 35   $\pm$  12 & +0.092  $\pm$  0.082 & 16.796  $\pm$  20.2   & 0.087	$\pm$	0.182   &  1.1(2.9)\\
2328 & 5 & 0.59565295 & 5286 $\pm$  109 & 4734  $\pm$  112 & 0.473 $\pm$  0.039 & 7.497  $\pm$  0.218 & -67  $\pm$  8  & -0.373 $\pm$  0.076 & 17.46   $\pm$  0.79   & 2.003	$\pm$	0.170      &   4.5(2.2)\\
2491 & 3 & 2.07736033 & 2975 $\pm$  74  & 6379  $\pm$  182 & 0.63  $\pm$  0.154 & 3.461  $\pm$  0.693 & 39   $\pm$  22 & +1.903  $\pm$  0.387 & 16.287  $\pm$  16     & 0.626	$\pm$	0.558   &   2.9(3.4)\\
2499 & 5 & 1.78113087 & 1855 $\pm$  11  & 7980  $\pm$  49  & 0.896 $\pm$  0.147 & 2.742  $\pm$  0.921 & -58  $\pm$  21 & +0.928  $\pm$  0.125 & 36.859  $\pm$  148.97 & 0.799	$\pm$	1.636   &   3.2(3.3)\\
2980 $\diamondsuit$ & 4 & 1.07331802 & 5074 $\pm$  34  & 3478  $\pm$  101 & 0.152 $\pm$  0.019 & 12.021 $\pm$  0.14  & 130  $\pm$  7  & +6.592  $\pm$  0.176 & 26.076  $\pm$  0.28   & 8.998	$\pm$	0.268   &   13.1(2.7)\\
3023 & 5 & 0.87912243 & 4076 $\pm$  44  & 5489  $\pm$  447 & 0.19  $\pm$  0.105 & 1.488  $\pm$  0.076 & -97  $\pm$  40 & -0.391 $\pm$  0.025 & 4.067   $\pm$  0.26   & 0.026	$\pm$	0.004   &   0.6(2.5)\\
3159 & 3 & 1.10072567 & 5996 $\pm$  150 & 7596  $\pm$  180 & 0.33  $\pm$  0.088 & 6.551  $\pm$  0.323 & 23   $\pm$  10 & +1.885  $\pm$  0.278 & 12.545  $\pm$  0.84   & 1.038	$\pm$	0.132   &   3.4(2.7)\\
3235 $\diamondsuit$ & 4 & 0.86408921 & 6721 $\pm$  224 & 5153  $\pm$  85  & 0.234 $\pm$  0.016  & 16.106 $\pm$  0.965 & 45   $\pm$  5  & +1.367 $\pm$  0.643 & 26.795  $\pm$  0.81   & 12.322	$\pm$	1.425   &   16.4(2.5)\\
3955 & 5 & 1.07831811 & 1490 $\pm$  8   & 6053  $\pm$  49  & 0.69  $\pm$  0.182 & 1.98   $\pm$  0.784 & -139 $\pm$  13 & -0.146 $\pm$  0.028 & 19.985  $\pm$  26.34  & 0.466	$\pm$	1.318   &   2.3(2.7)\\
3991 $\diamondsuit$ & 5 & 1.32878771 & 1377 $\pm$  4   & 4332  $\pm$  16  & 0.96  $\pm$  0.052 & 4.793  $\pm$  2.13  & 24   $\pm$  9  & -1.252 $\pm$  0.093 & 134.705 $\pm$  177.92 & 7.744	$\pm$	21.732  &   12.0(2.9)\\
4013 & 4 & 1.35226337 & 3629 $\pm$  108 & 3771  $\pm$  165 & 0.521 $\pm$  0.173 & 2.462  $\pm$  0.338 & -121 $\pm$  19 & +2.142  $\pm$  0.203 & 8.666   $\pm$  5.53   & 0.152	$\pm$	0.081   &   1.4(3.0)\\
4245 & 5 & 0.66073172 & 2921 $\pm$  76  & 7671  $\pm$  130 & 0.764 $\pm$  0.18  & 1.175  $\pm$  0.37  & -131 $\pm$  18 & -0.837 $\pm$  0.048 & 6.778   $\pm$  9.98   & 0.025	$\pm$	0.049   &   0.6(2.3)\\
4277 & 3 & 1.91138938 & 4636 $\pm$  172 & 2327  $\pm$  332 & 0.284 $\pm$  0.114 & 4.284  $\pm$  0.322 & 157  $\pm$  23 & -4.136 $\pm$  0.658 & 10.476  $\pm$  1.07   & 0.487	$\pm$	0.106   &   2.6(3.3)\\
\hline
\end{tabular}}
\end{table*}

\section{$O-C$ curves for LMC F mode}

\begin{figure*}[ht!]
\begin{center}
{\includegraphics[height=4.5cm,width=0.49\linewidth]{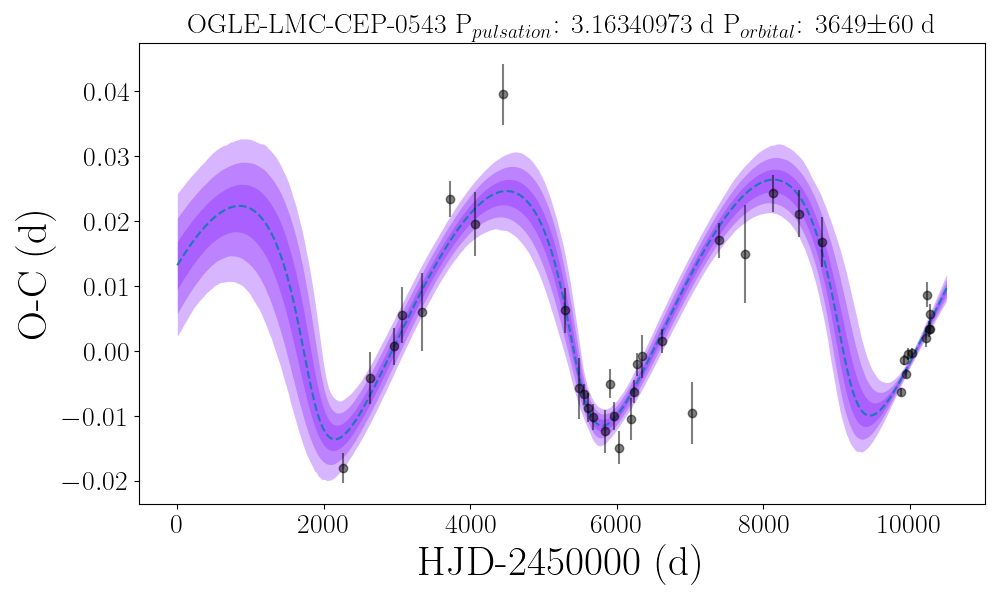}}
{\includegraphics[height=4.5cm,width=0.49\linewidth]{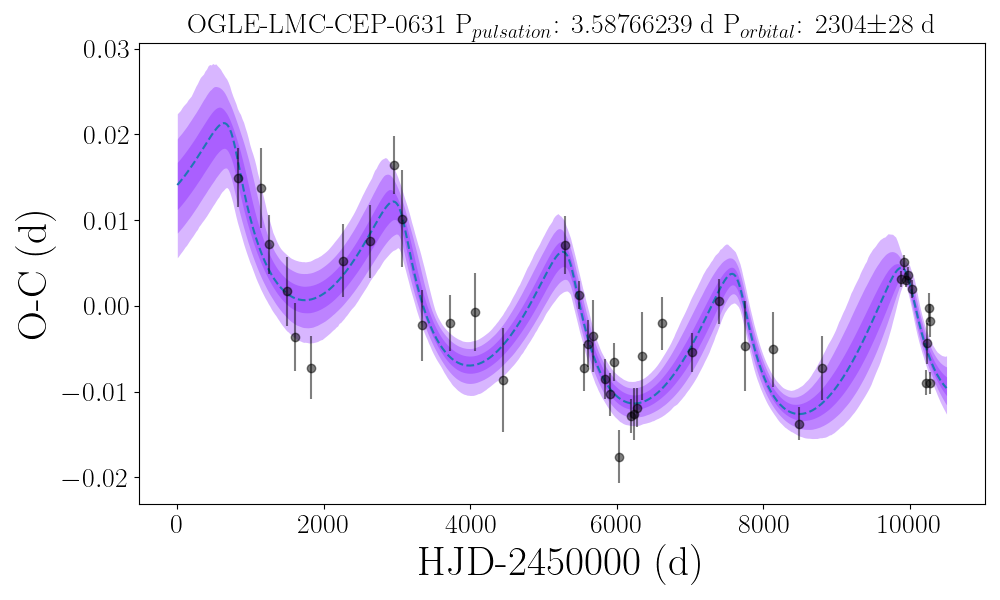}}
{\includegraphics[height=4.5cm,width=0.49\linewidth]{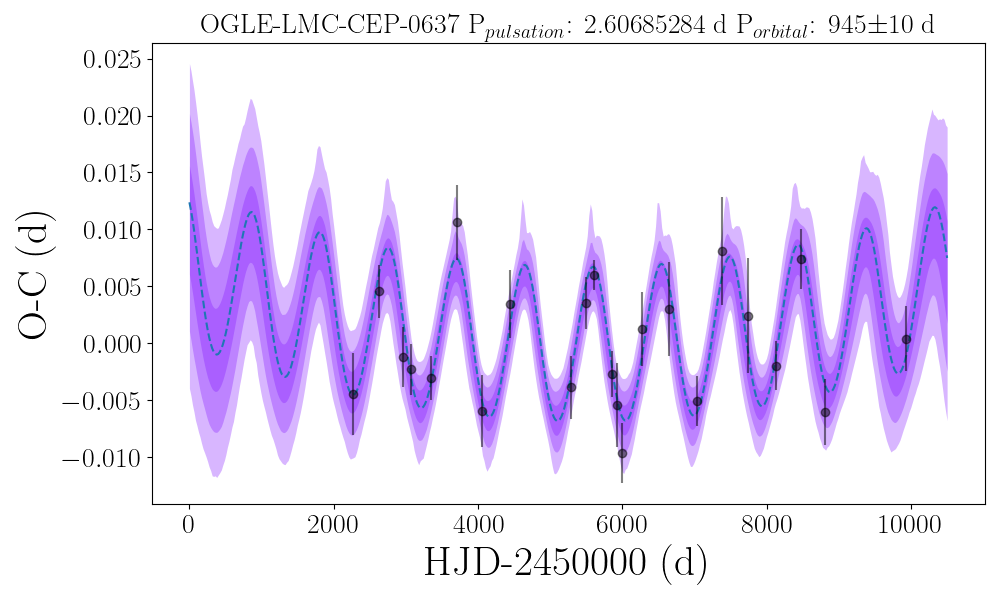}}
{\includegraphics[height=4.5cm,width=0.49\linewidth]{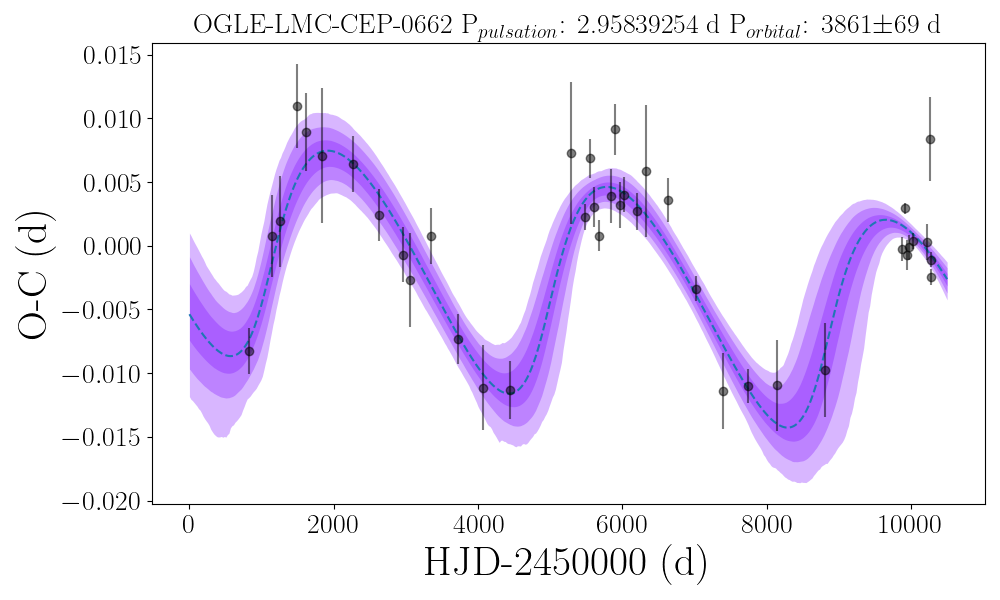}}
{\includegraphics[height=4.5cm,width=0.49\linewidth]{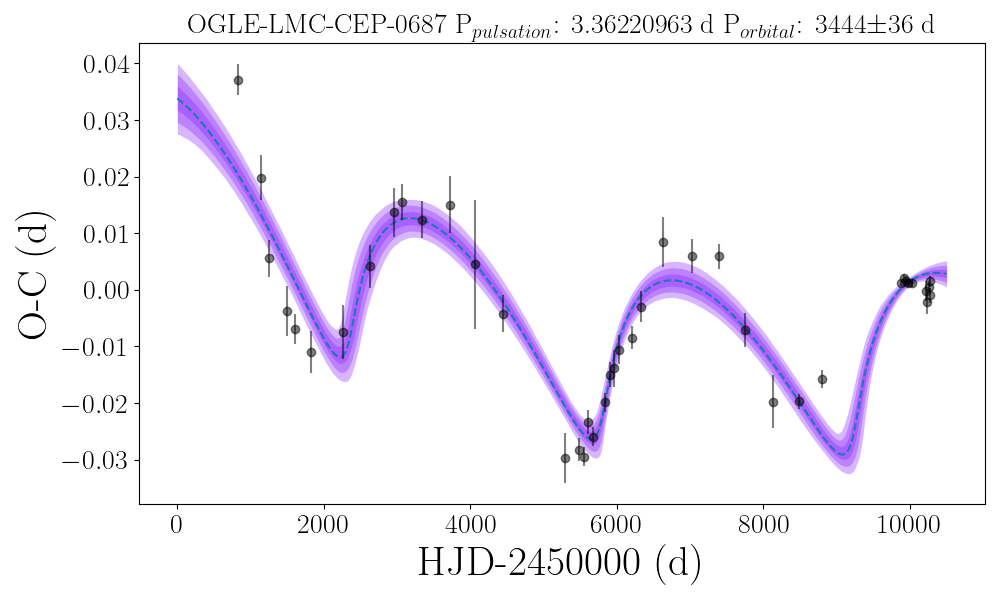}}
{\includegraphics[height=4.5cm,width=0.49\linewidth]{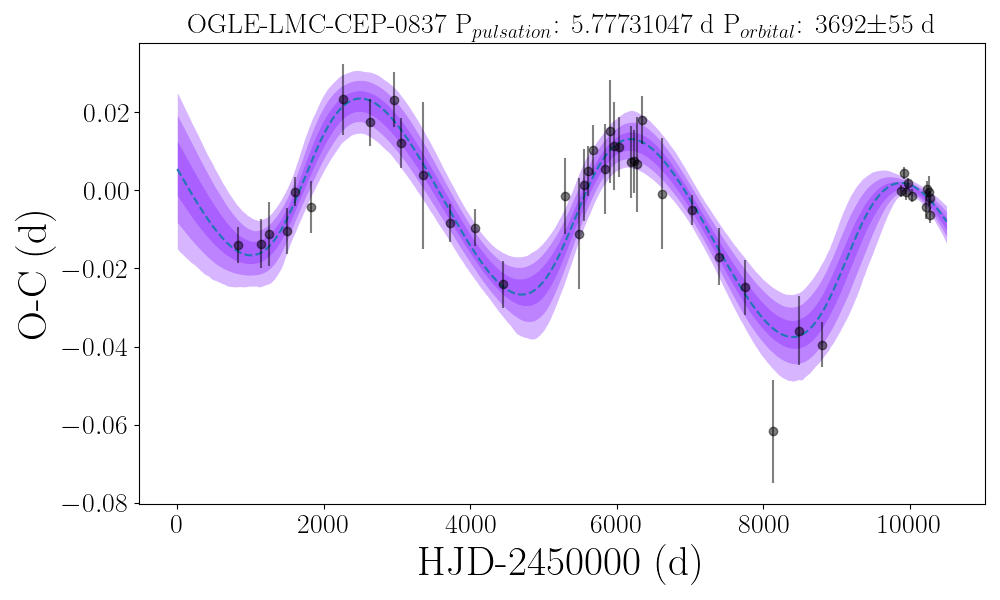}}
{\includegraphics[height=4.5cm,width=0.49\linewidth]{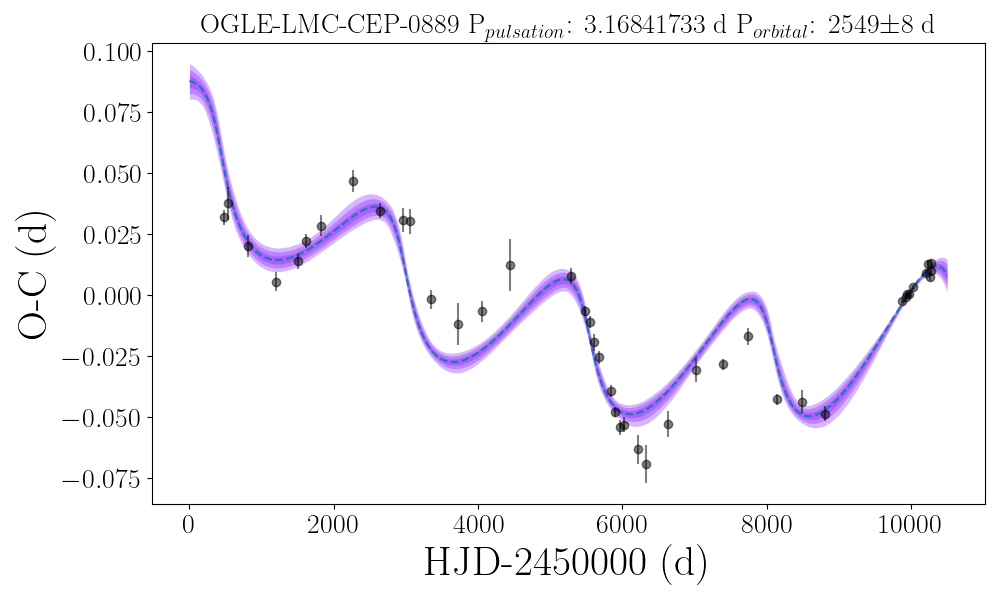}}
{\includegraphics[height=4.5cm,width=0.49\linewidth]{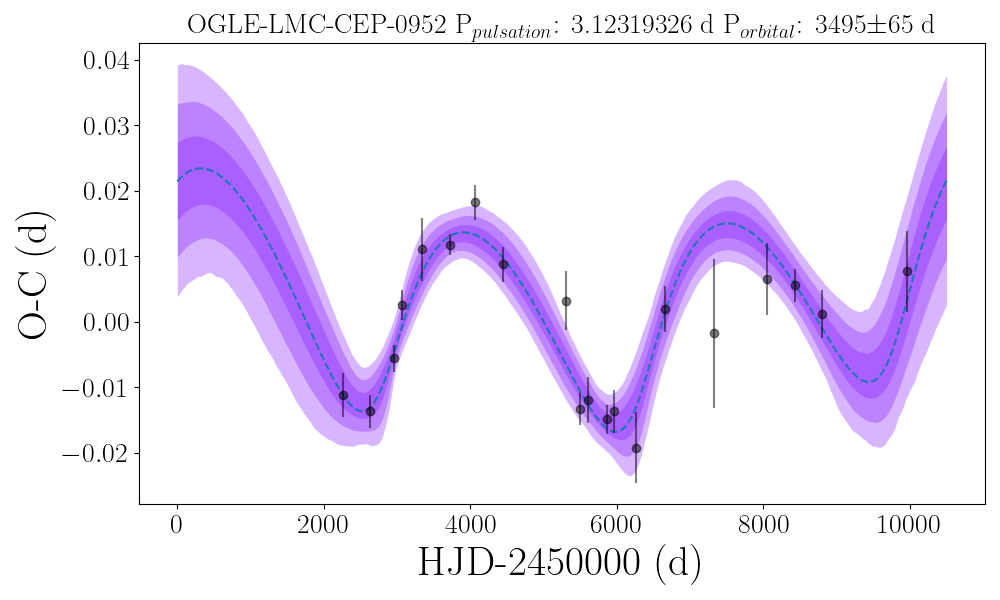}}
{\includegraphics[height=4.5cm,width=0.49\linewidth]{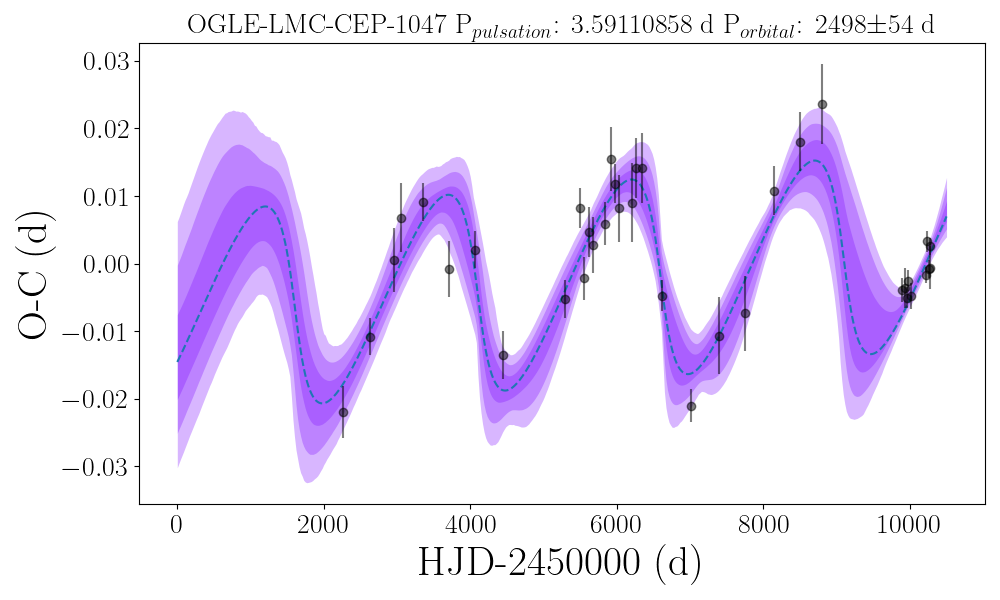}}
{\includegraphics[height=4.5cm,width=0.49\linewidth]{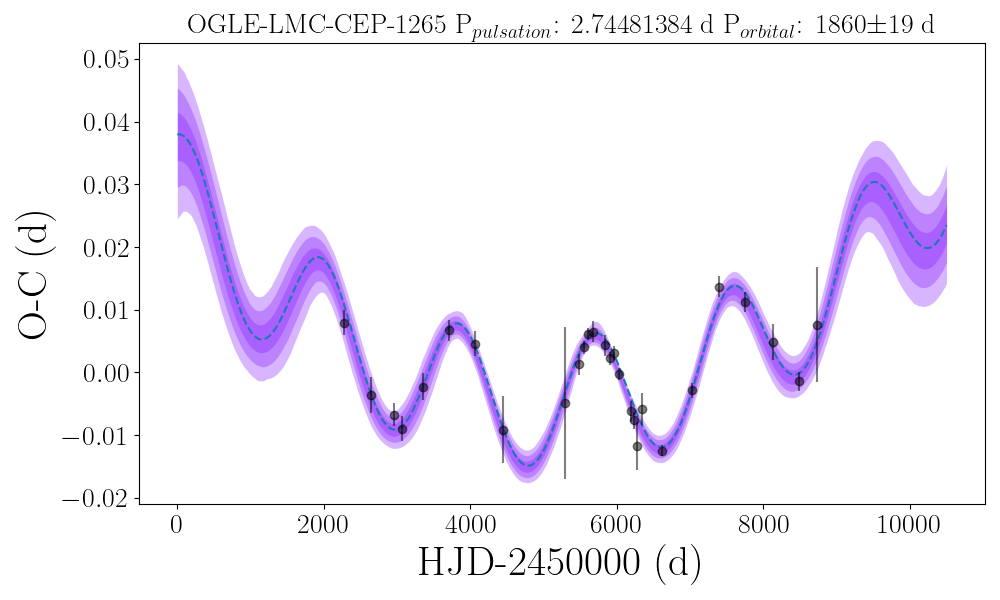}}
\caption{Remaining $O-C$ diagrams for LMC fundamental-mode Cepheid binary candidate sample}
\label{fig:appendix_ocplot_Fmode_LMC}
\end{center}
\end{figure*}

\begin{figure*}[ht!]
\ContinuedFloat
\begin{center}
{\includegraphics[height=4.5cm,width=0.49\linewidth]{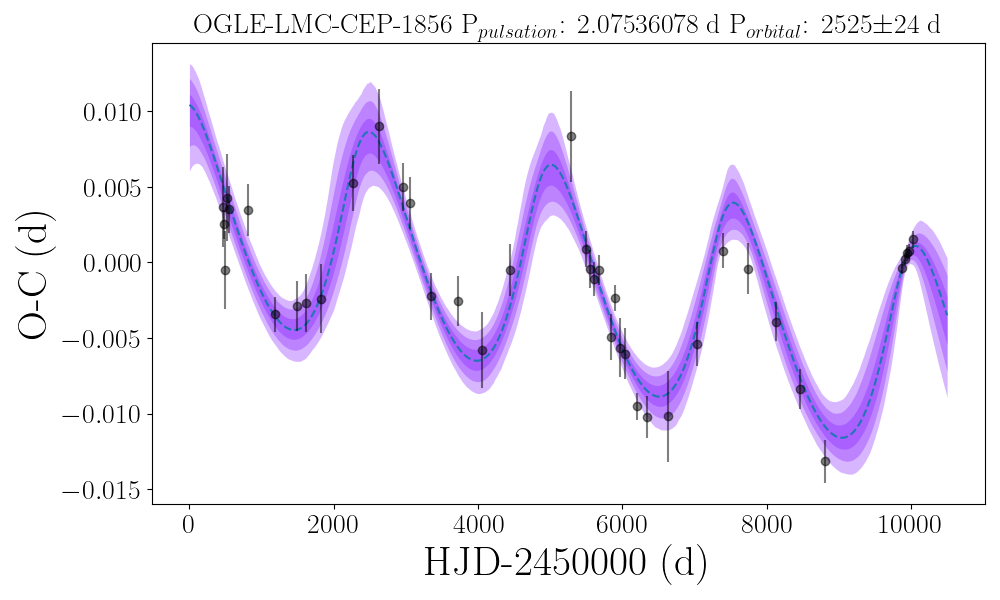}}
{\includegraphics[height=4.5cm,width=0.49\linewidth]{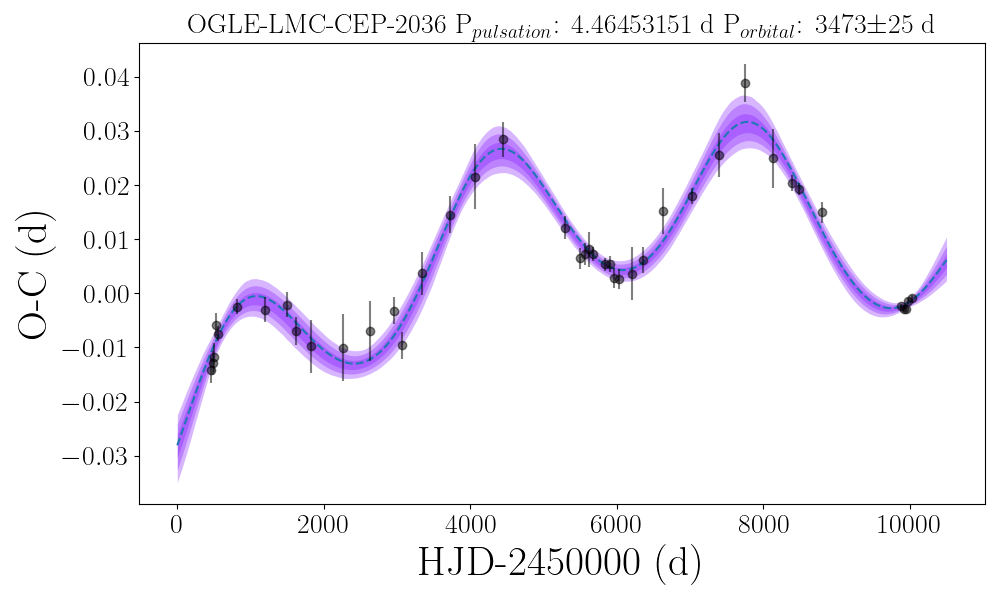}}
{\includegraphics[height=4.5cm,width=0.49\linewidth]{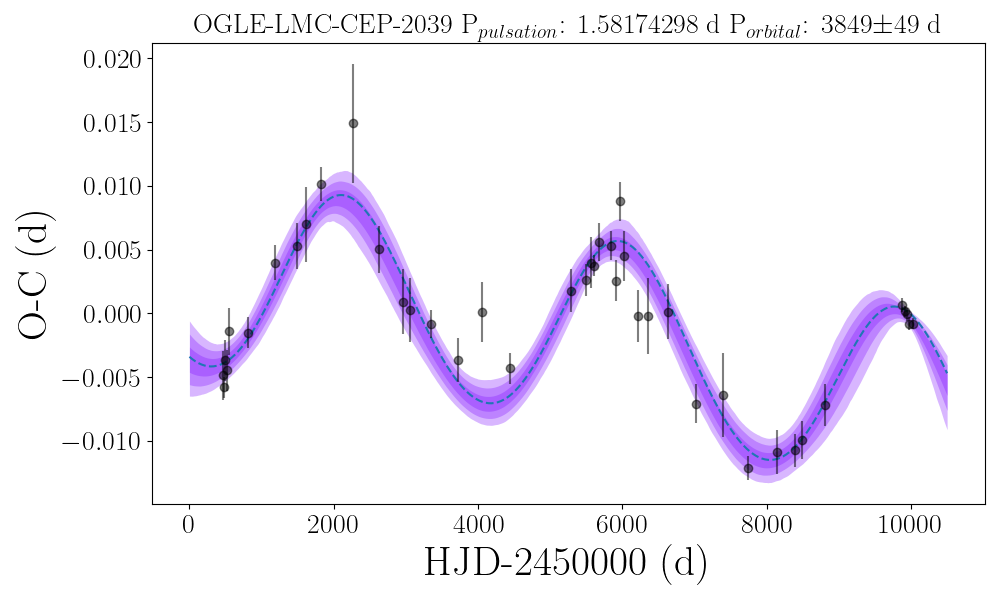}}
{\includegraphics[height=4.5cm,width=0.49\linewidth]{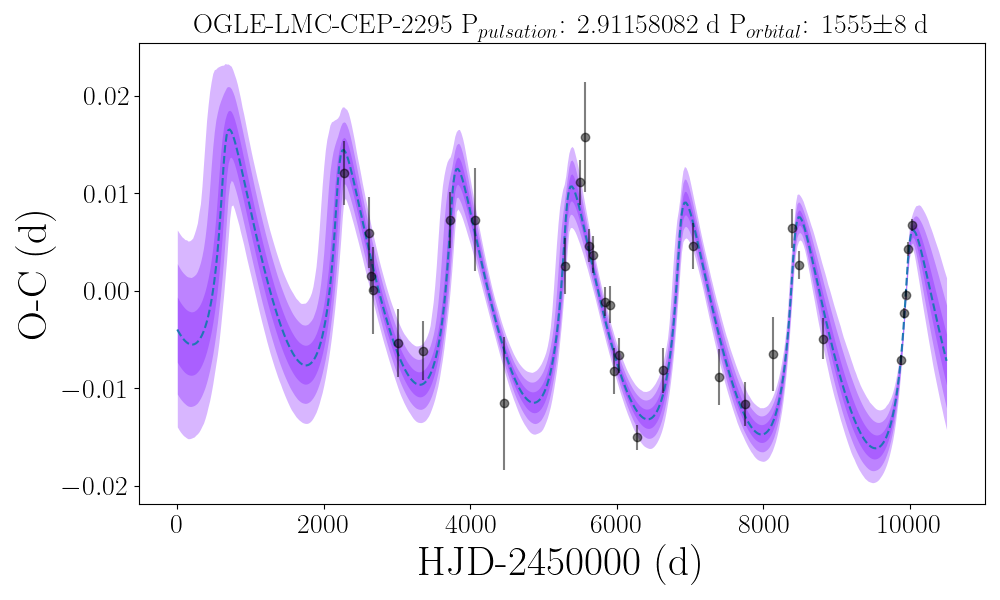}}
{\includegraphics[height=4.5cm,width=0.49\linewidth]{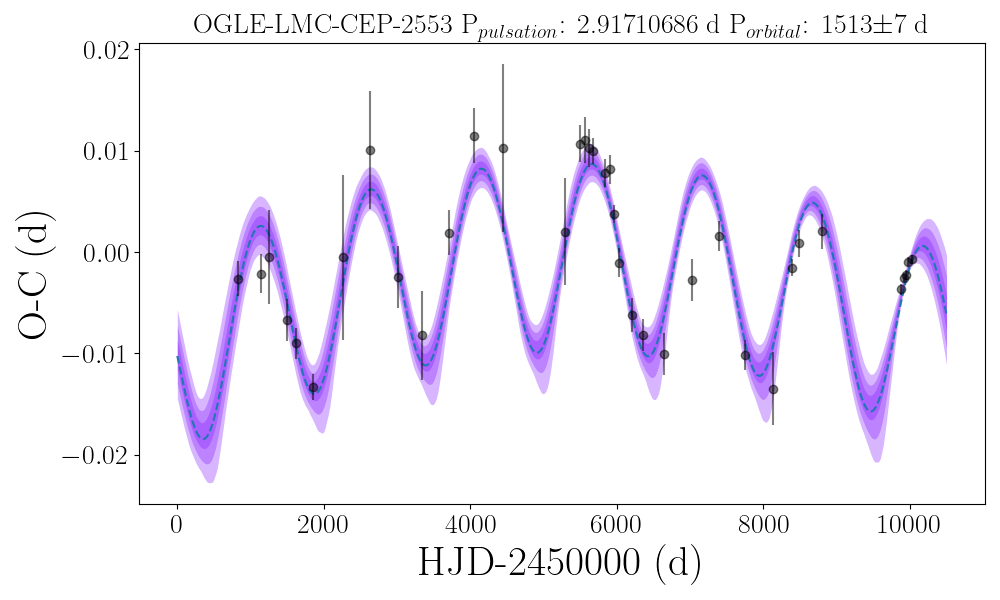}}
{\includegraphics[height=4.5cm,width=0.49\linewidth]{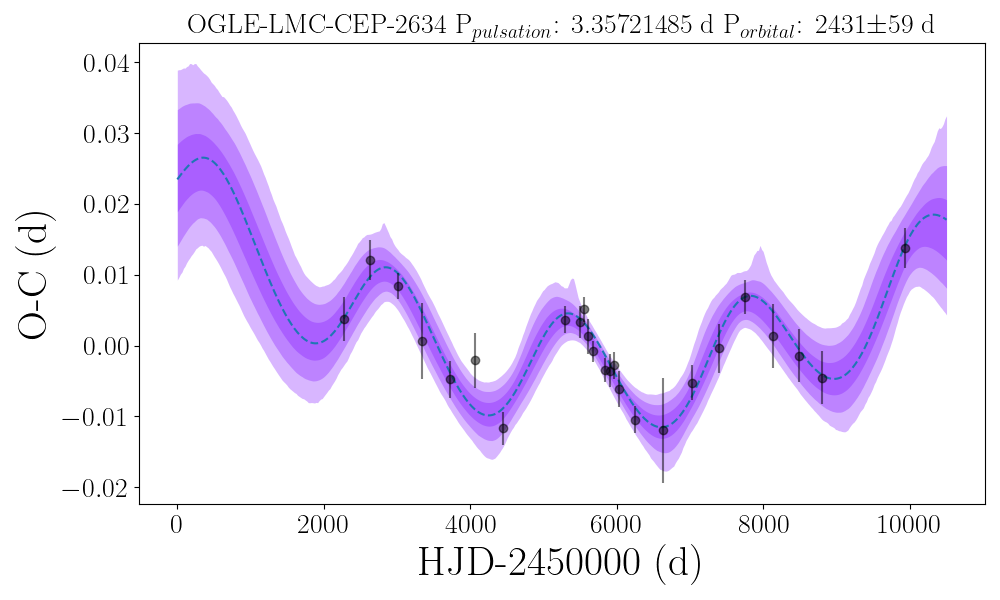}}
{\includegraphics[height=4.5cm,width=0.49\linewidth]{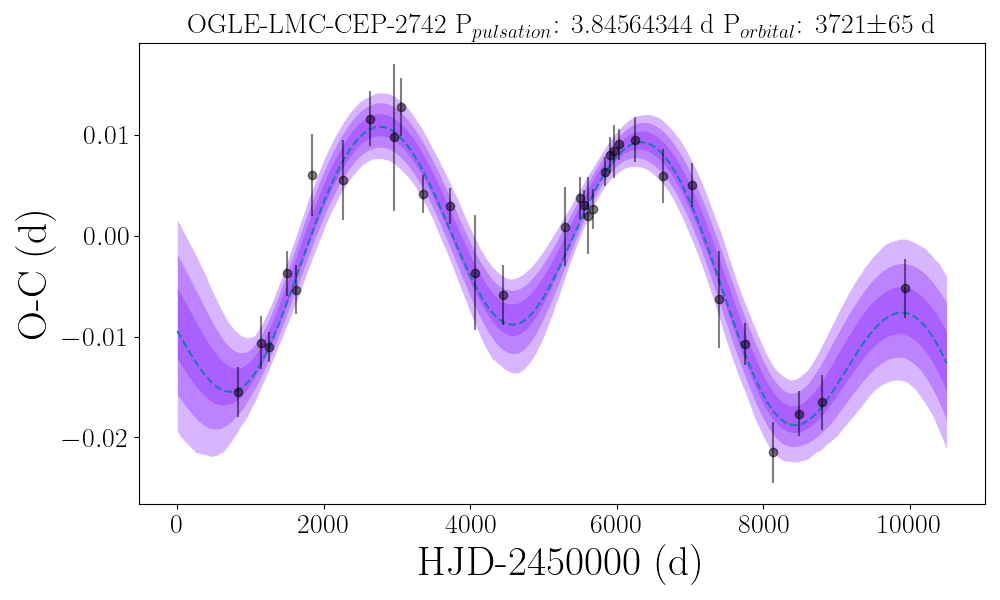}}
{\includegraphics[height=4.5cm,width=0.49\linewidth]{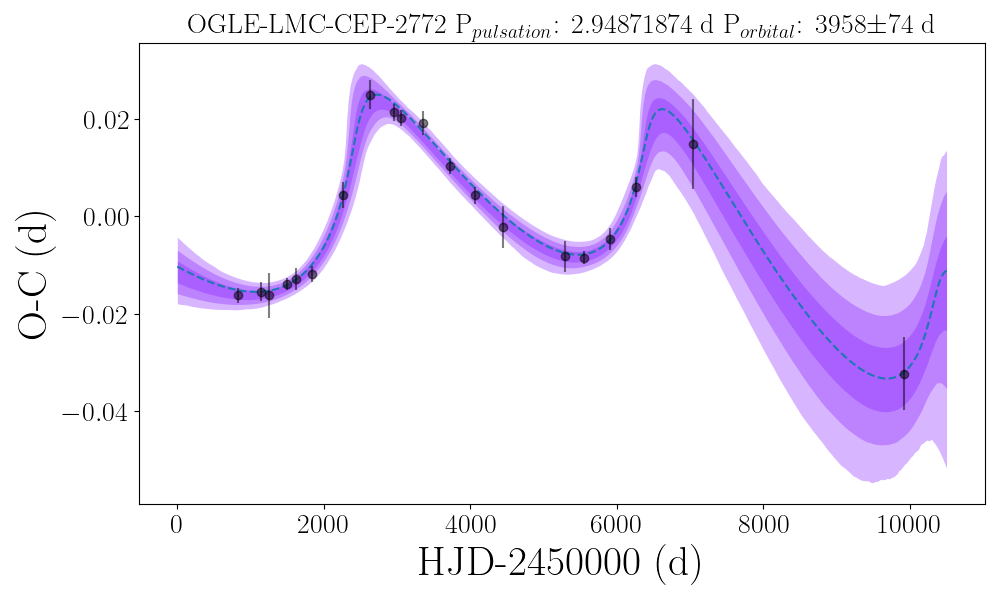}}
{\includegraphics[height=4.5cm,width=0.49\linewidth]{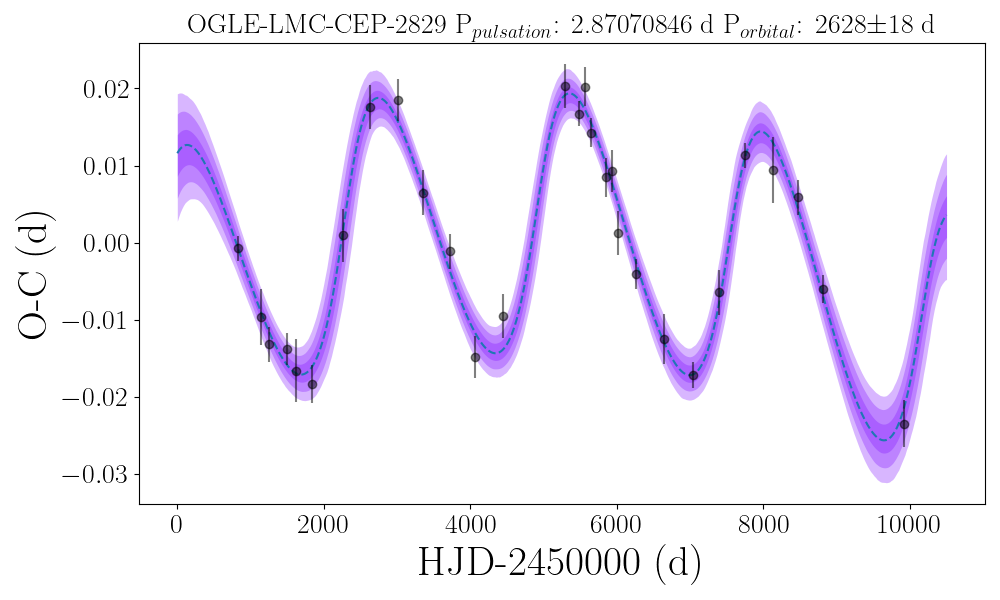}}
{\includegraphics[height=4.5cm,width=0.49\linewidth]{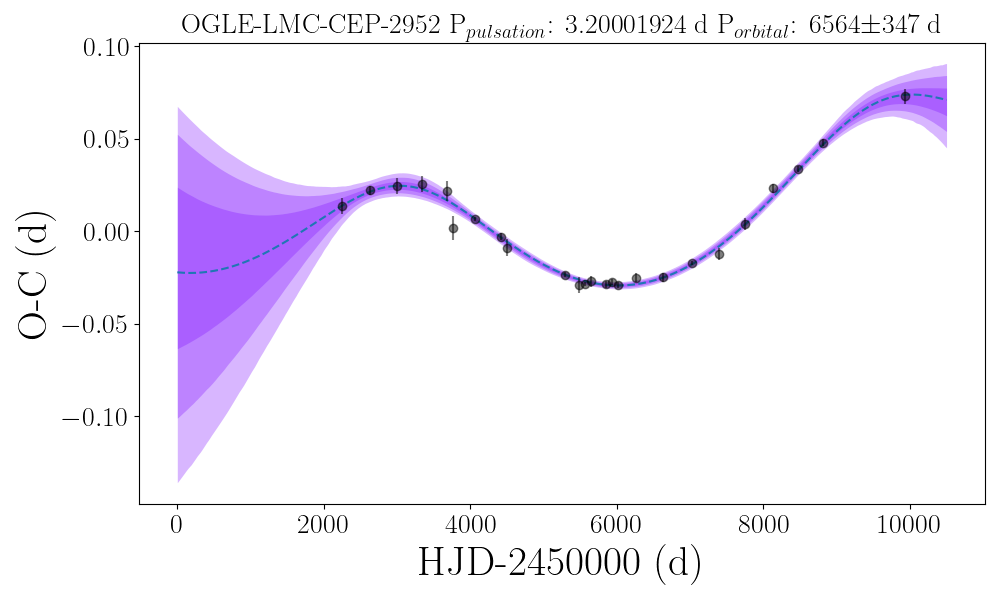}}
\caption{continued.}
\end{center}
\end{figure*}

\begin{figure*}[ht!]
\ContinuedFloat
\begin{center}
{\includegraphics[height=4.5cm,width=0.49\linewidth]{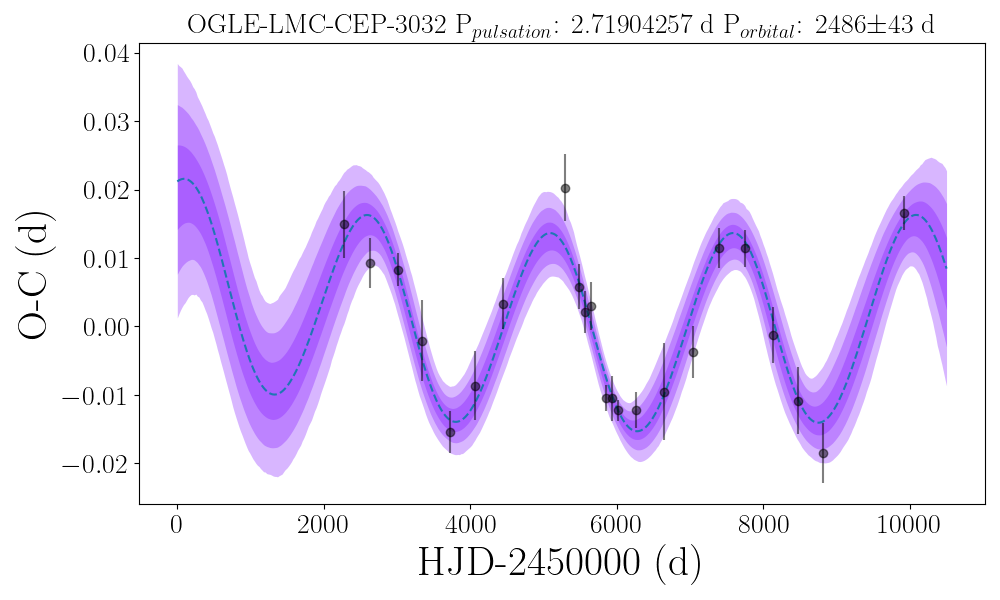}}
{\includegraphics[height=4.5cm,width=0.49\linewidth]{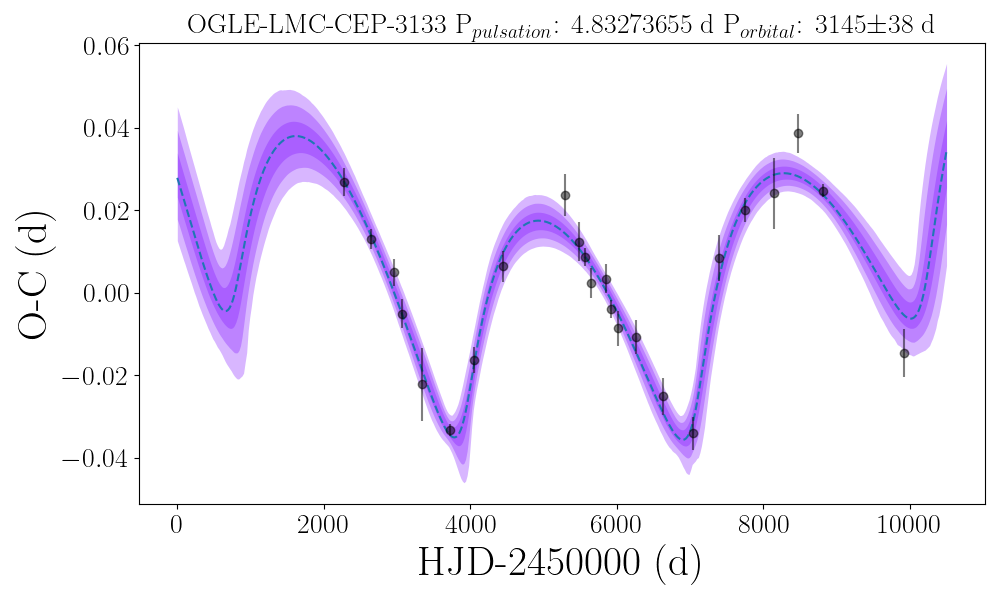}}
{\includegraphics[height=4.5cm,width=0.49\linewidth]{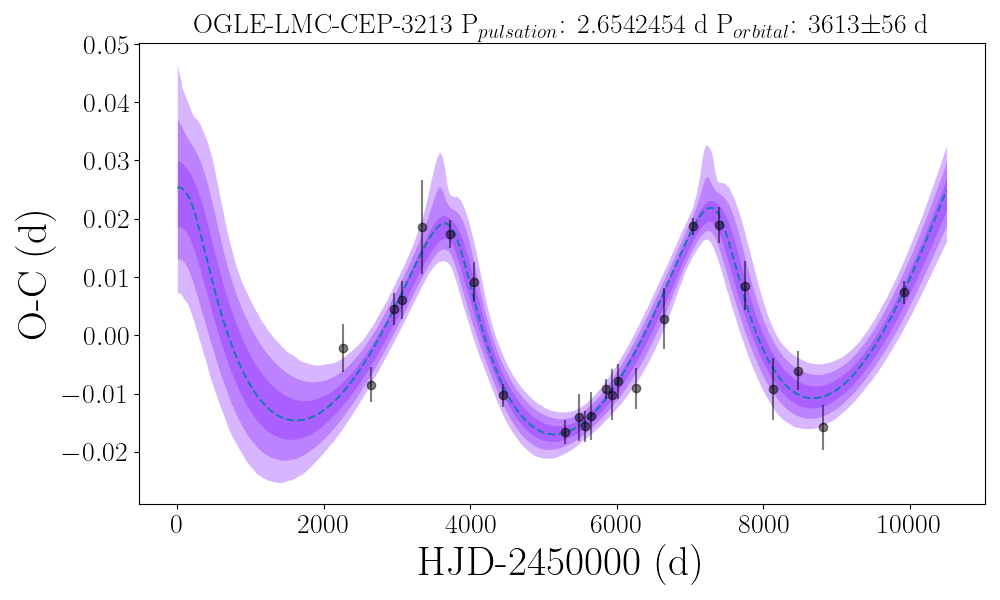}}
{\includegraphics[height=4.5cm,width=0.49\linewidth]{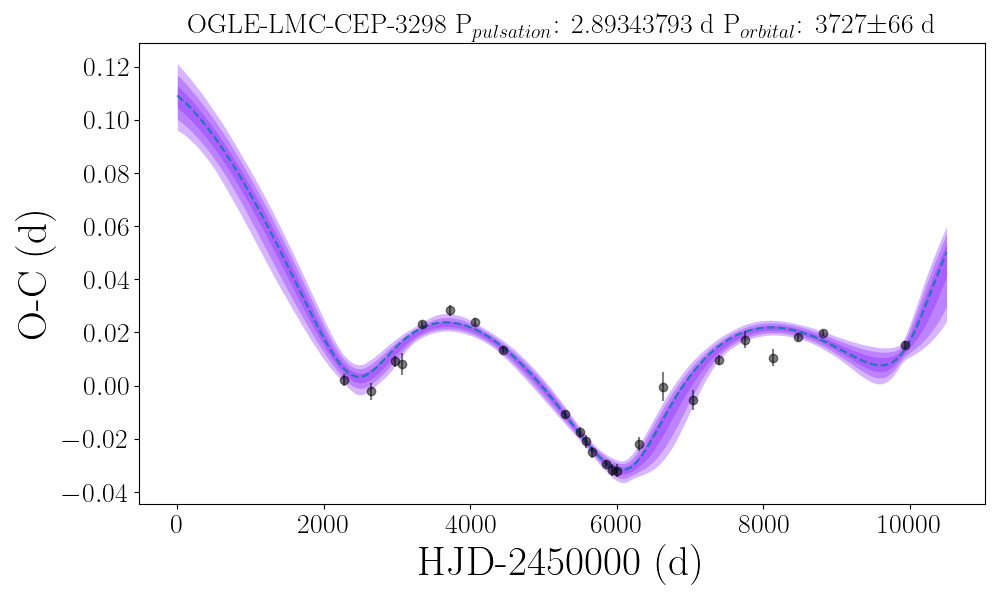}}
\caption{continued.}
\end{center}
\end{figure*}


\section{$O-C$ curves for LMC 1O mode}

\begin{figure*}[ht!]
\begin{center}
{\includegraphics[height=4.5cm,width=0.49\linewidth]{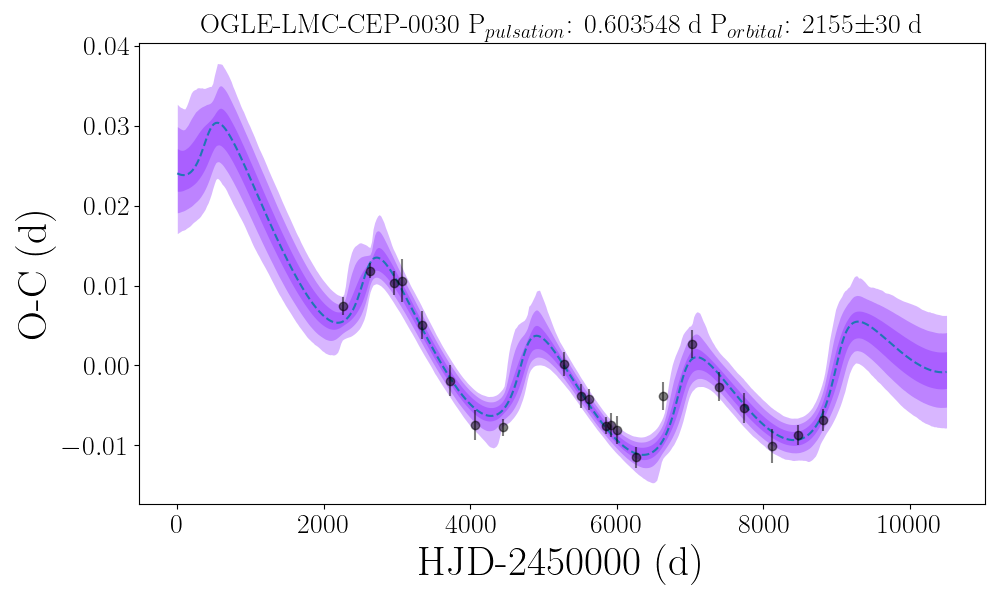}}
{\includegraphics[height=4.5cm,width=0.49\linewidth]{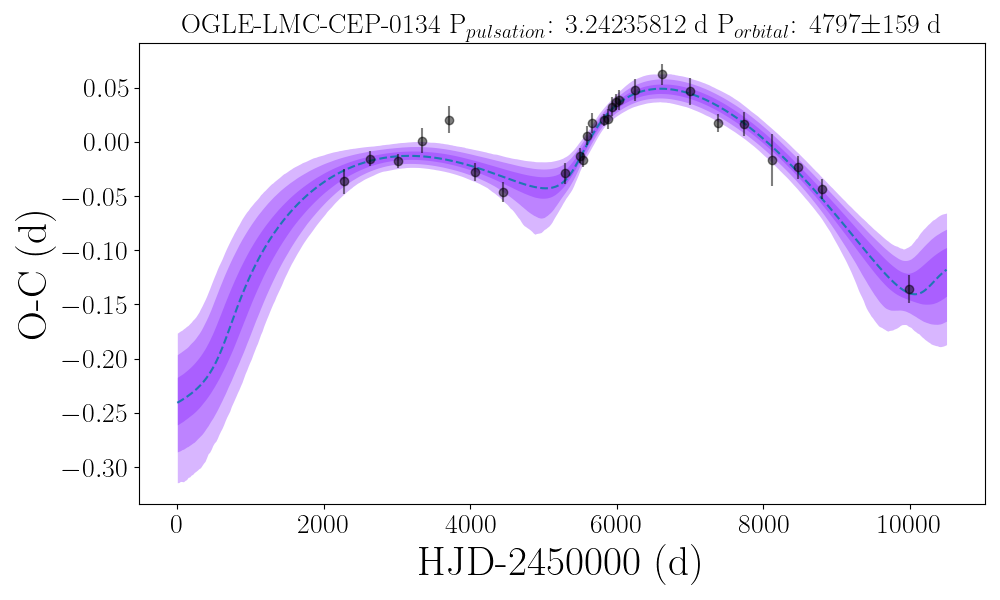}}
{\includegraphics[height=4.5cm,width=0.49\linewidth]{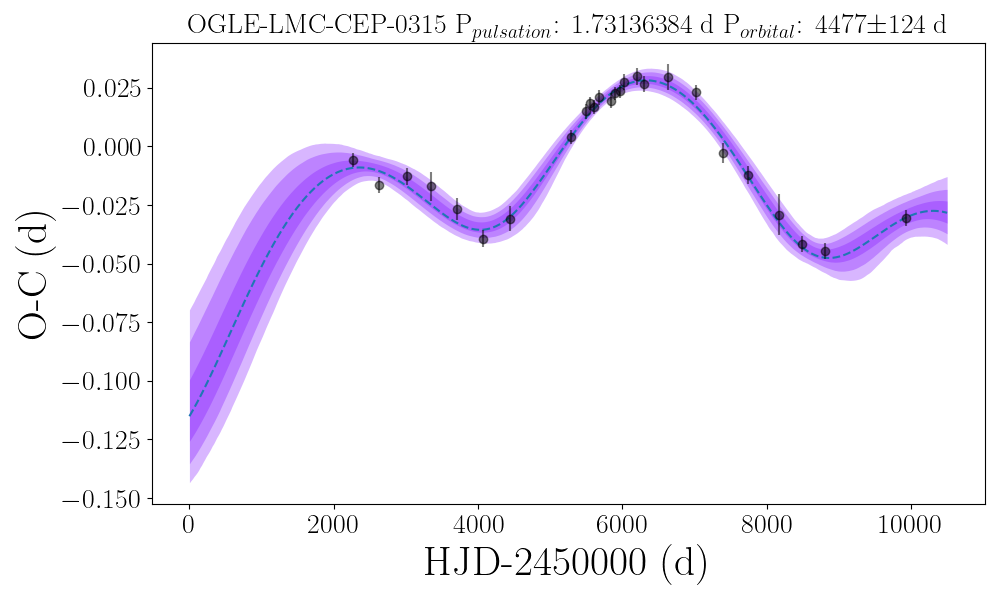}}
{\includegraphics[height=4.5cm,width=0.49\linewidth]{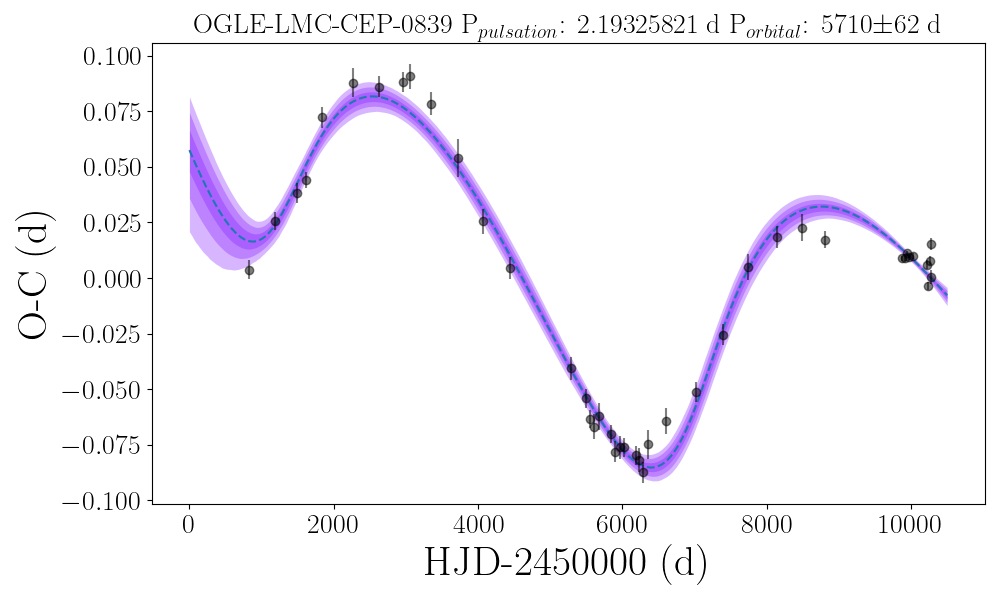}}
\caption{Remaining $O-C$ diagrams for LMC first-overtone mode Cepheid binary candidate sample}
\label{fig:appendix_ocplot_1Omode_LMC}
\end{center}
\end{figure*}

\begin{figure*}[ht!]
\ContinuedFloat
\begin{center}
{\includegraphics[height=4.5cm,width=0.49\linewidth]{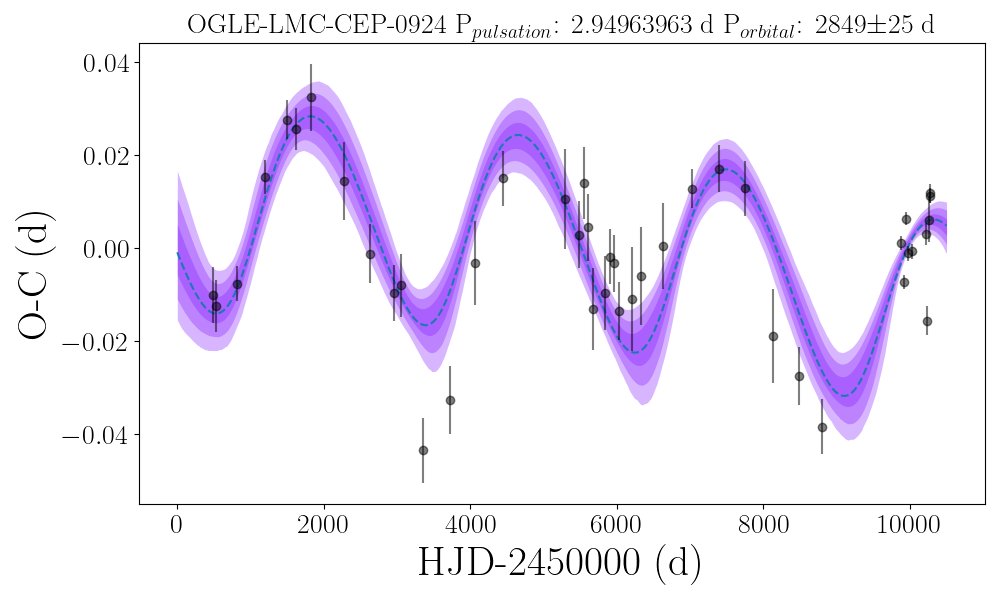}}
{\includegraphics[height=4.5cm,width=0.49\linewidth]{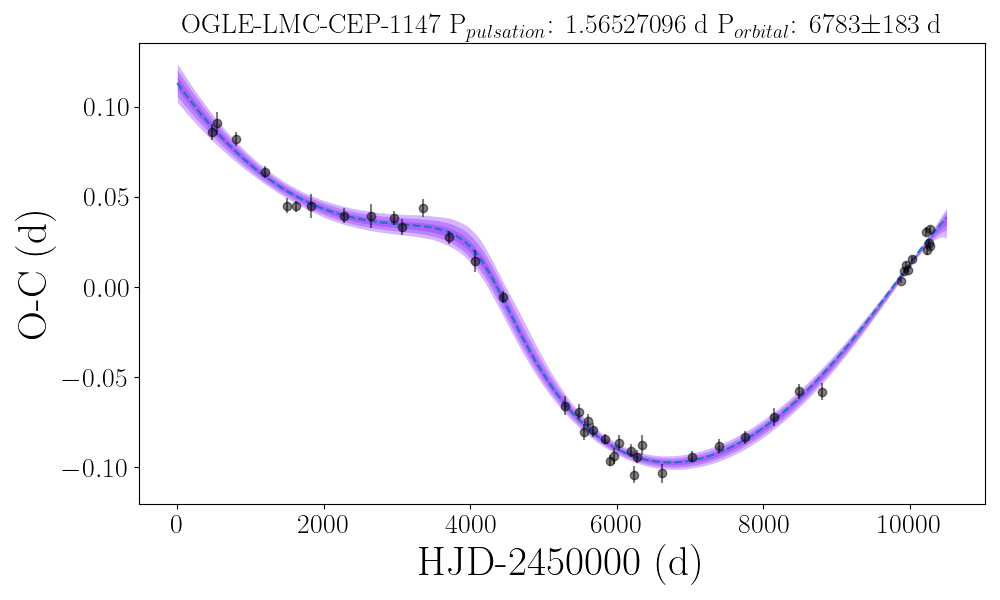}}
{\includegraphics[height=4.5cm,width=0.49\linewidth]{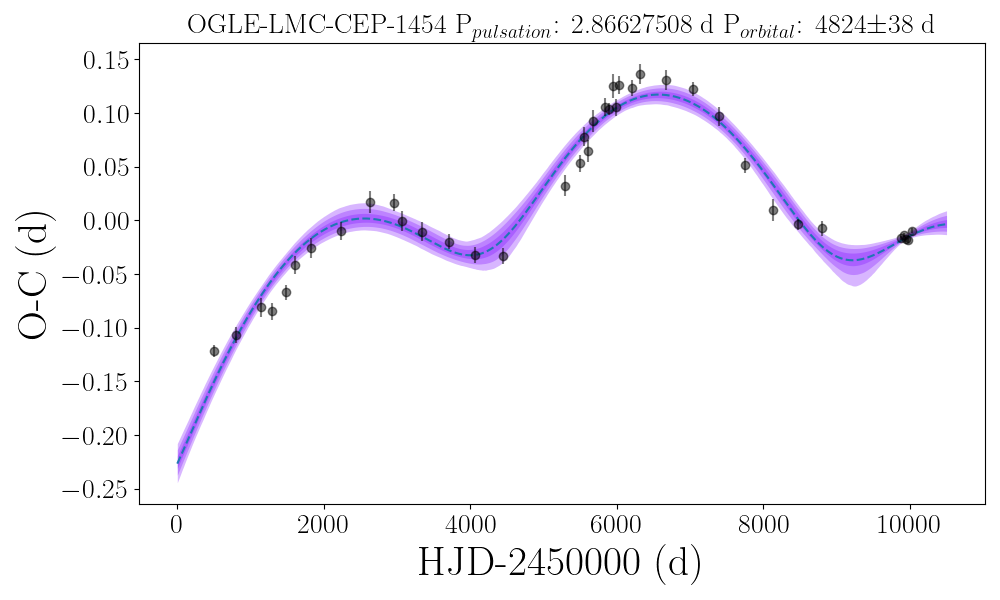}}
{\includegraphics[height=4.5cm,width=0.49\linewidth]{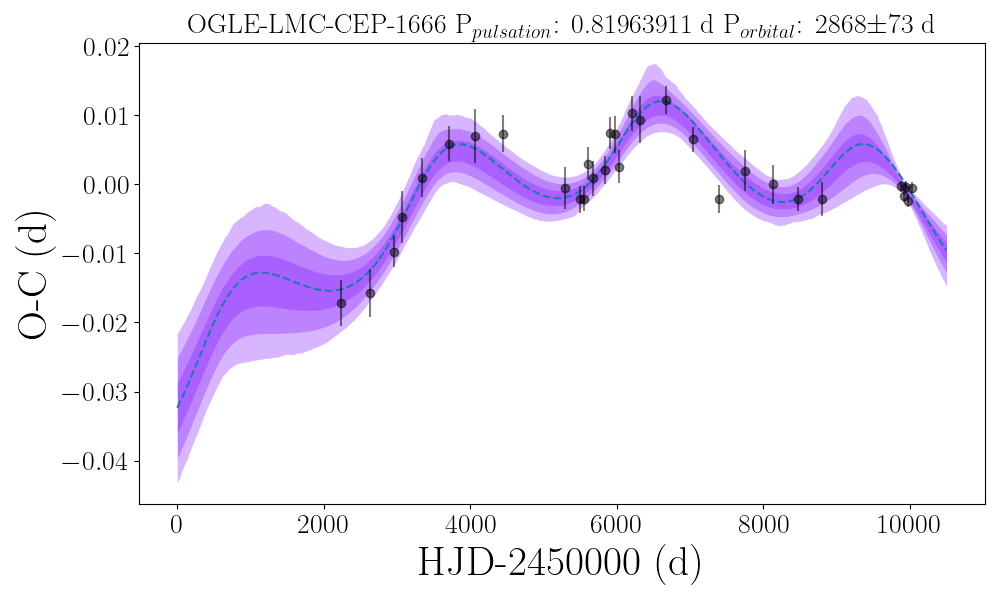}}
{\includegraphics[height=4.5cm,width=0.49\linewidth]{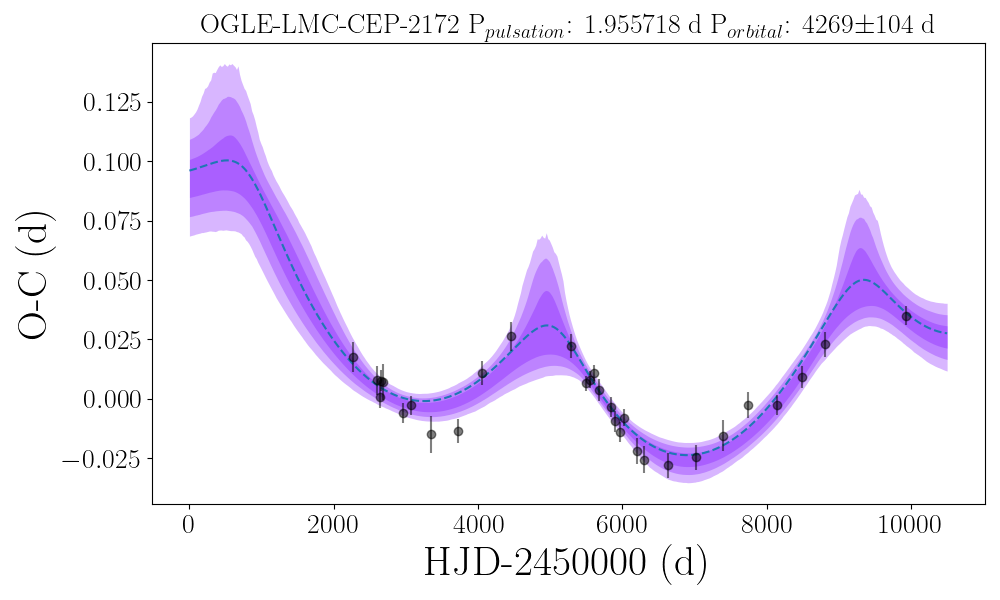}}
{\includegraphics[height=4.5cm,width=0.49\linewidth]{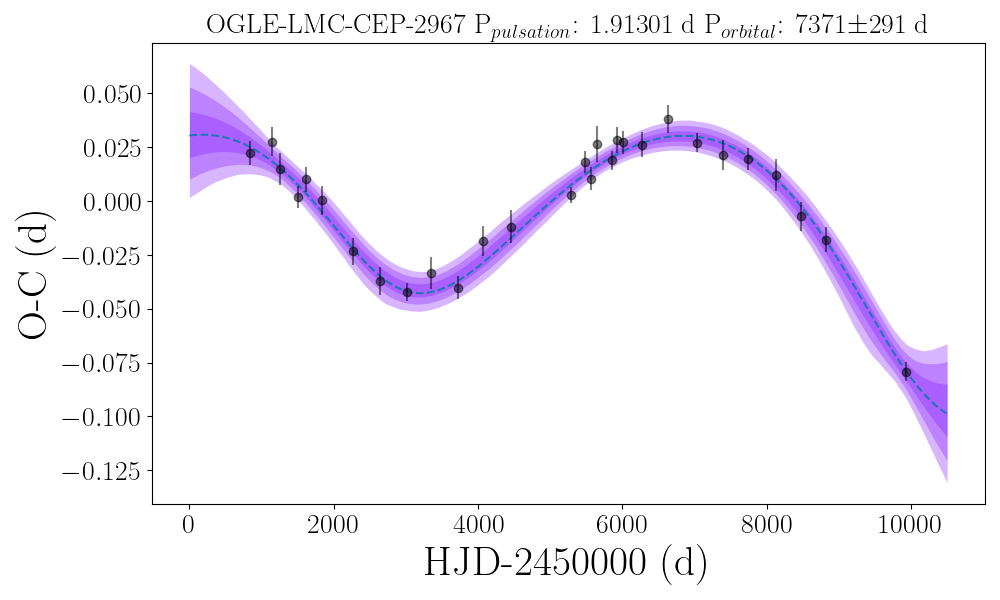}}
{\includegraphics[height=4.5cm,width=0.49\linewidth]{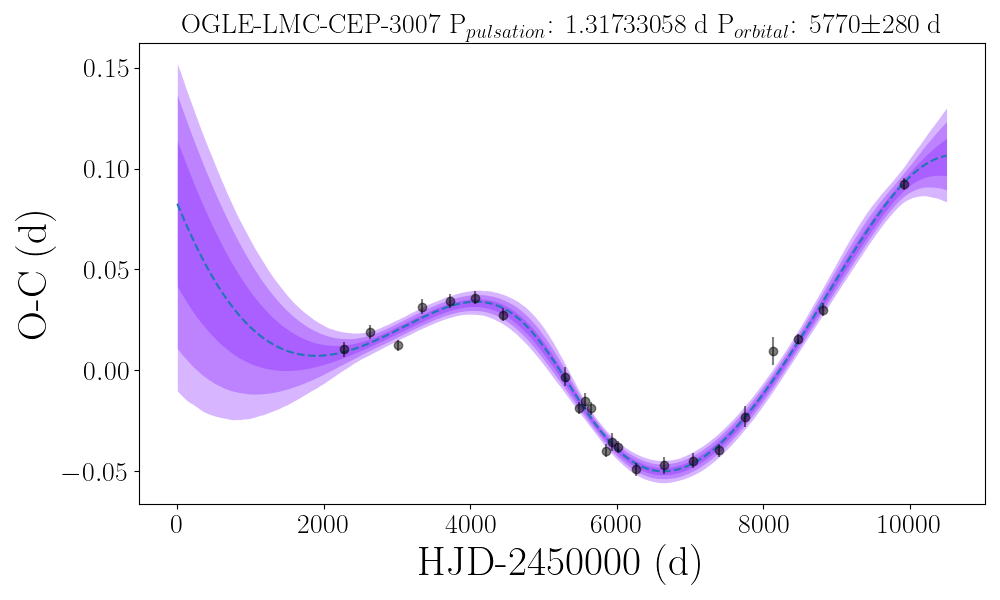}}
\caption{continued.}
\end{center}
\end{figure*}


\section{$O-C$ curves for SMC F mode}

\begin{figure*}[ht!]
\begin{center}
{\includegraphics[height=4.5cm,width=0.49\linewidth]{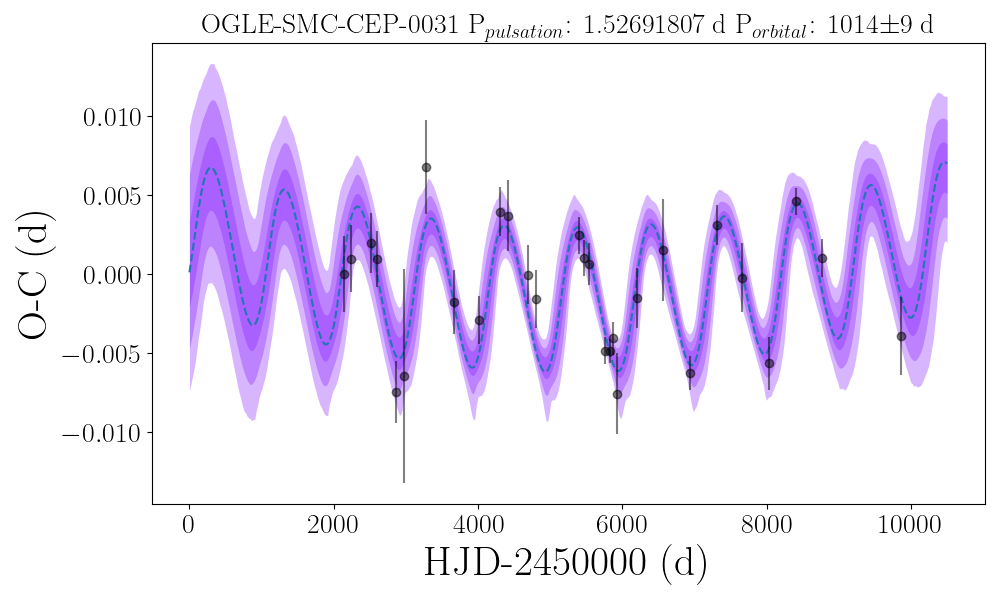}}
{\includegraphics[height=4.5cm,width=0.49\linewidth]{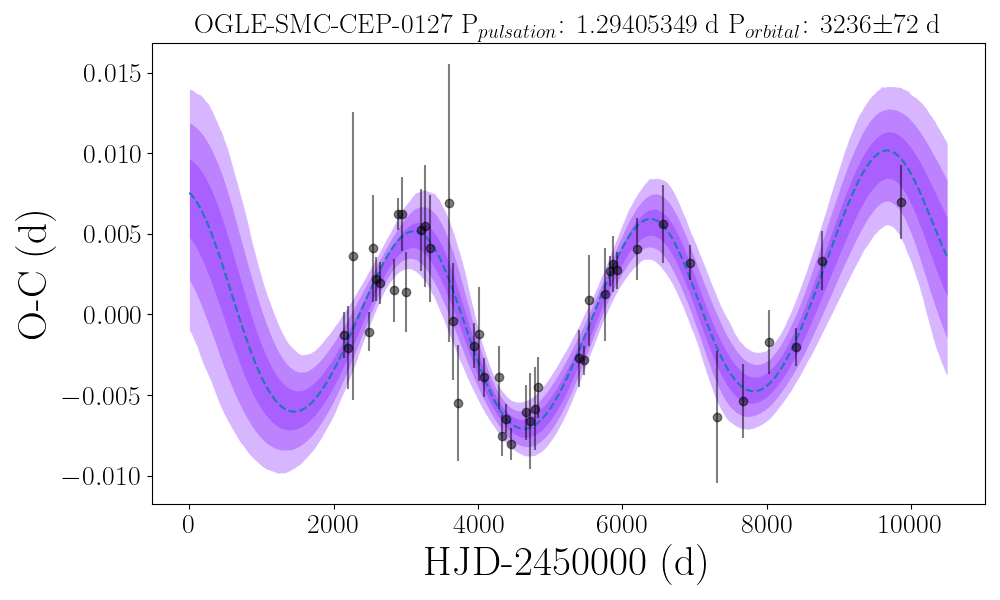}}
{\includegraphics[height=4.5cm,width=0.49\linewidth]{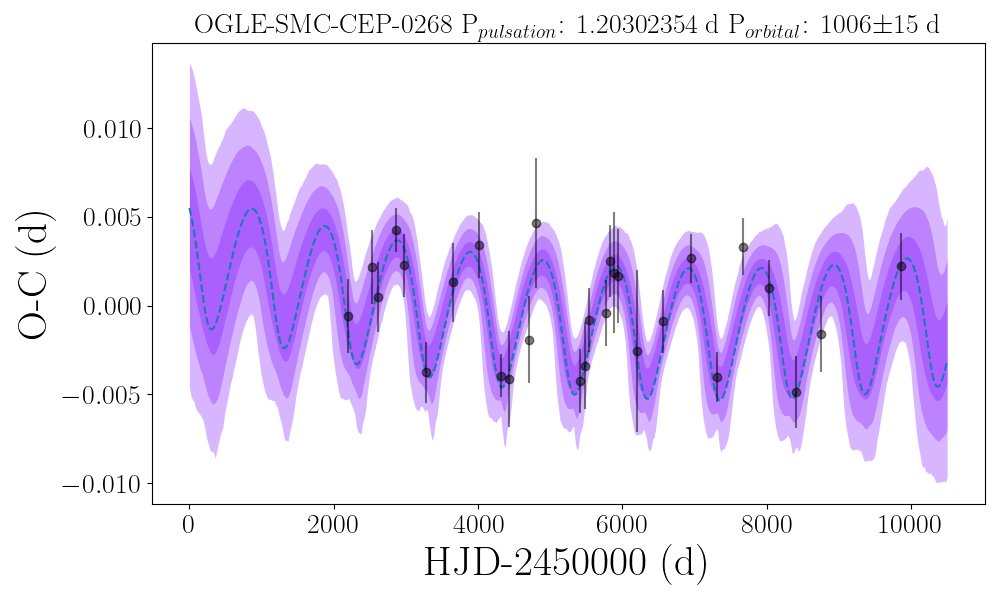}}
{\includegraphics[height=4.5cm,width=0.49\linewidth]{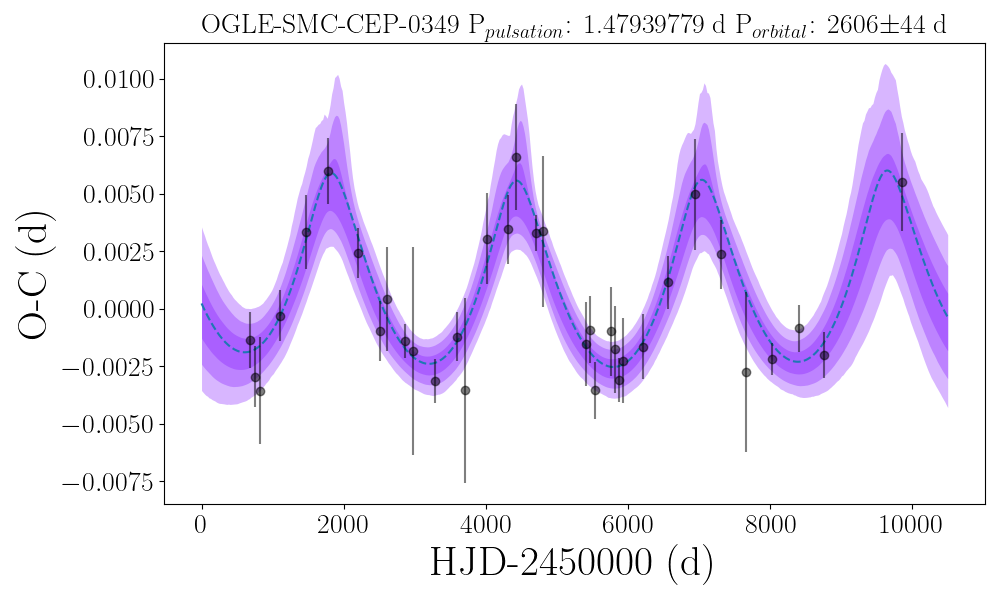}}
{\includegraphics[height=4.5cm,width=0.49\linewidth]{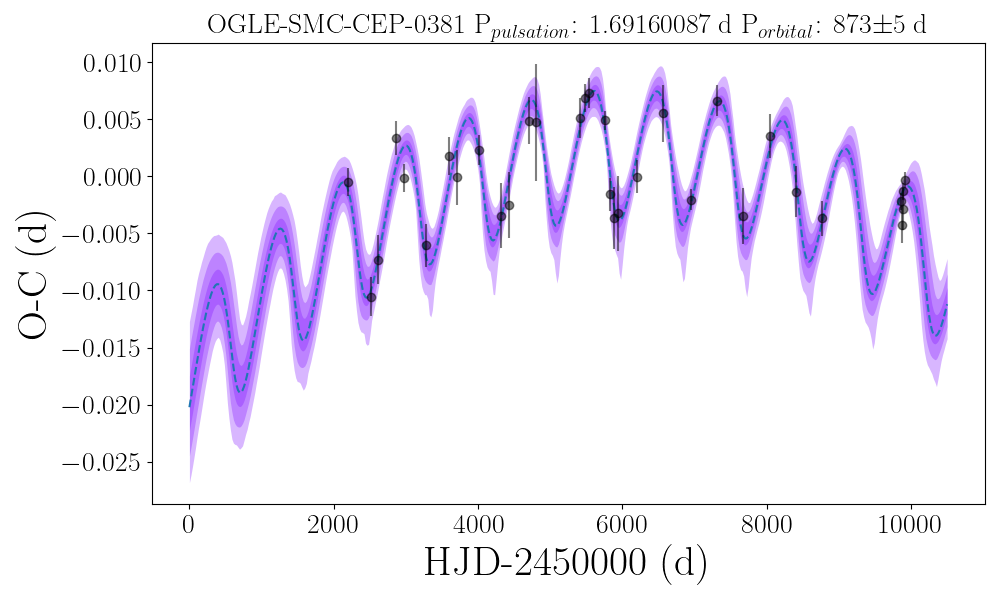}}
{\includegraphics[height=4.5cm,width=0.49\linewidth]{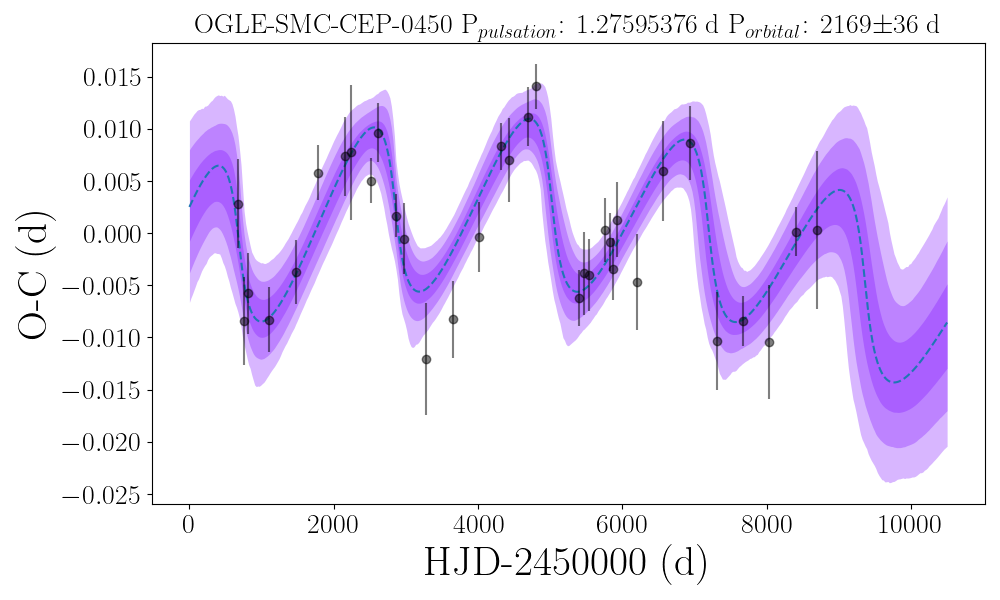}}
{\includegraphics[height=4.5cm,width=0.49\linewidth]{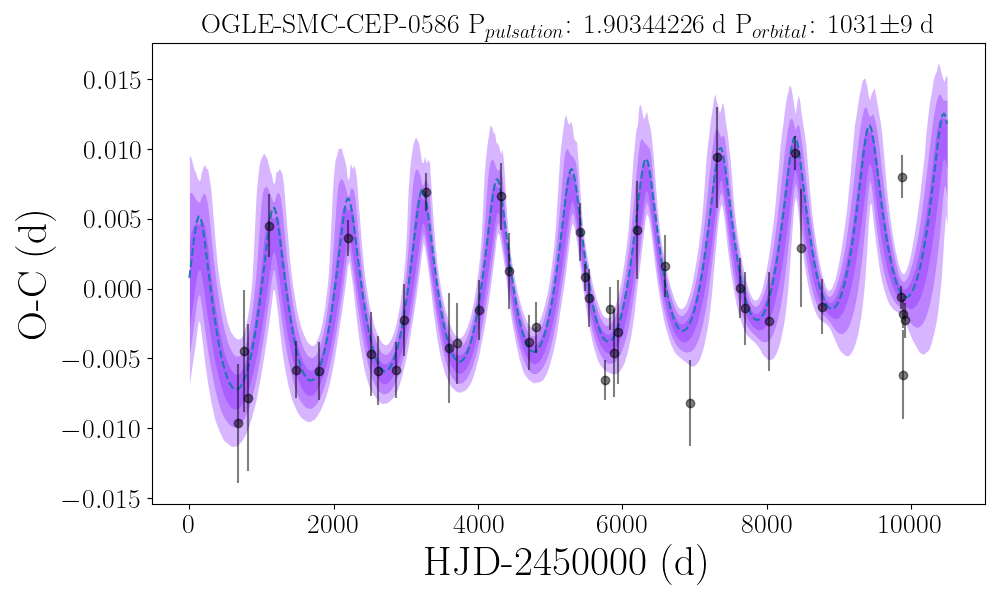}}
{\includegraphics[height=4.5cm,width=0.49\linewidth]{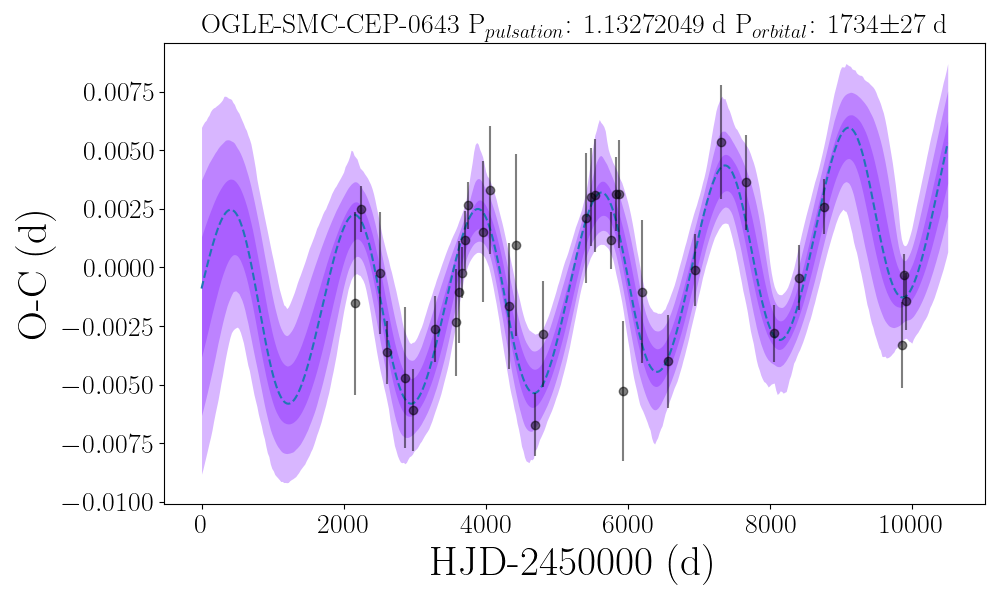}}
{\includegraphics[height=4.5cm,width=0.49\linewidth]{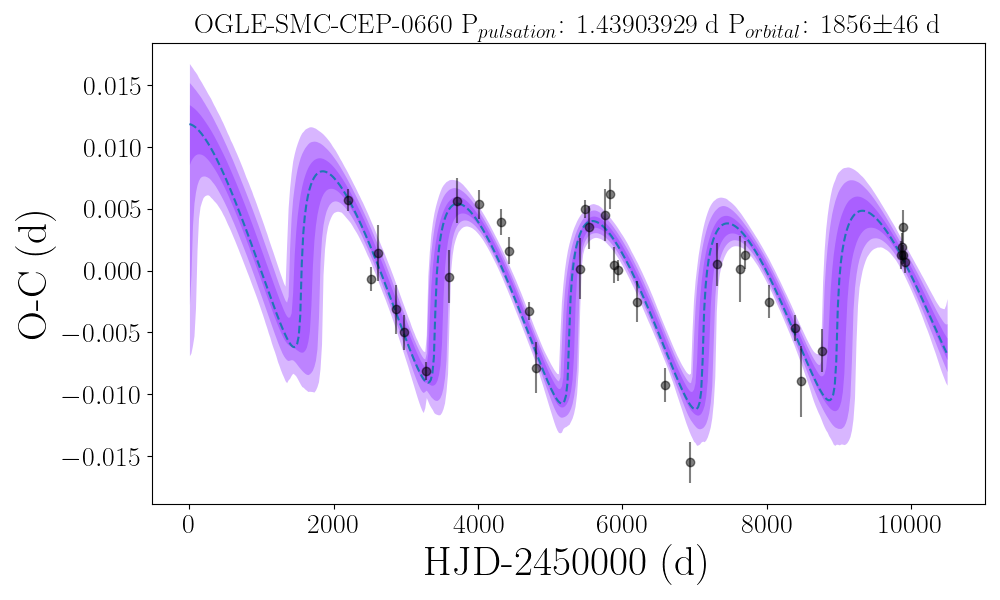}}
{\includegraphics[height=4.5cm,width=0.49\linewidth]{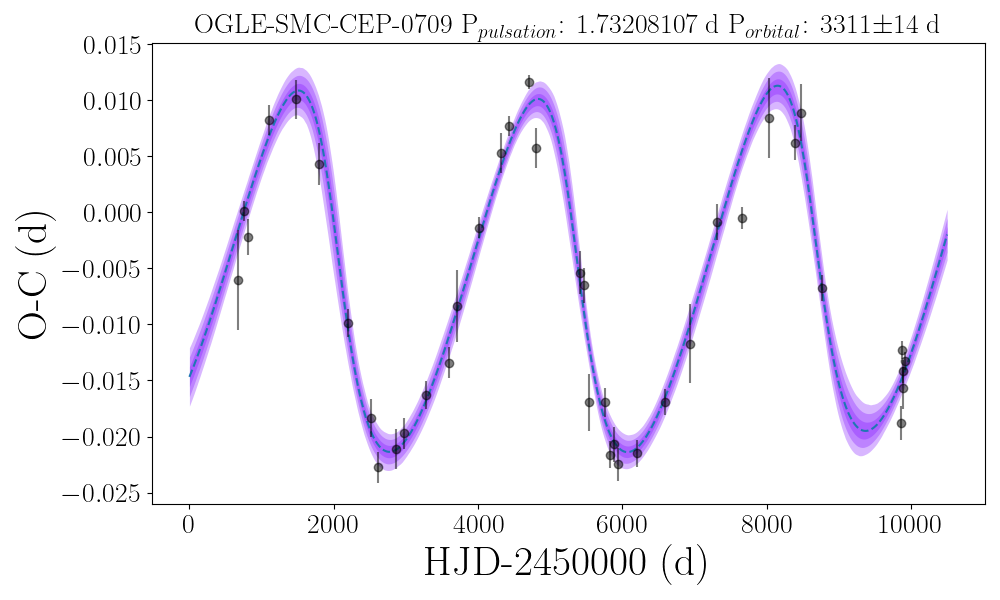}}
\caption{Remaining $O-C$ diagrams for SMC fundamental-mode Cepheid binary candidate sample}
\label{fig:appendix_ocplot_Fmode_SMC}
\end{center}
\end{figure*}

\begin{figure*}[ht!]
\ContinuedFloat
\begin{center}
{\includegraphics[height=4.5cm,width=0.49\linewidth]{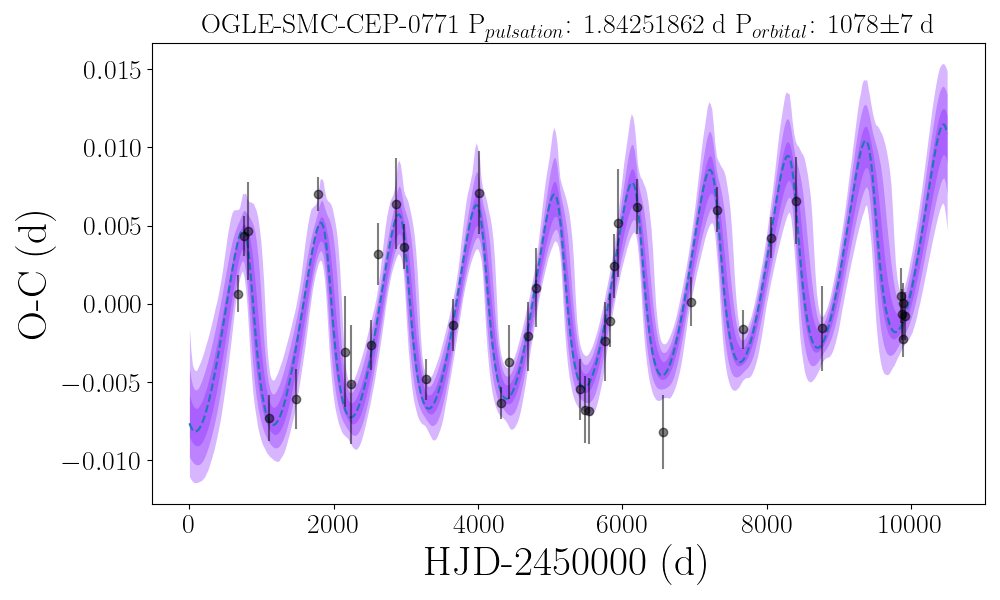}}
{\includegraphics[height=4.5cm,width=0.49\linewidth]{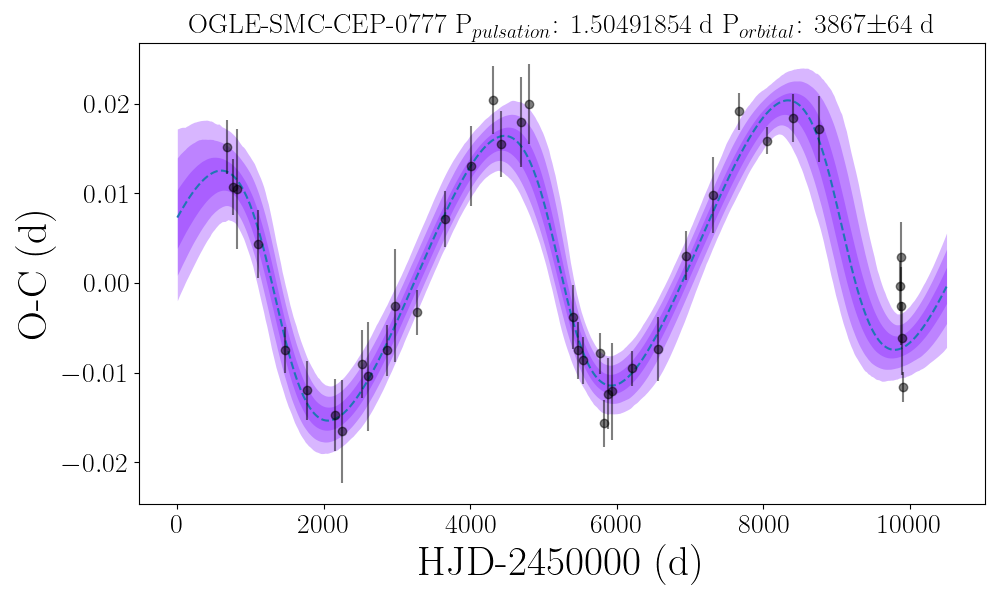}}
{\includegraphics[height=4.5cm,width=0.49\linewidth]{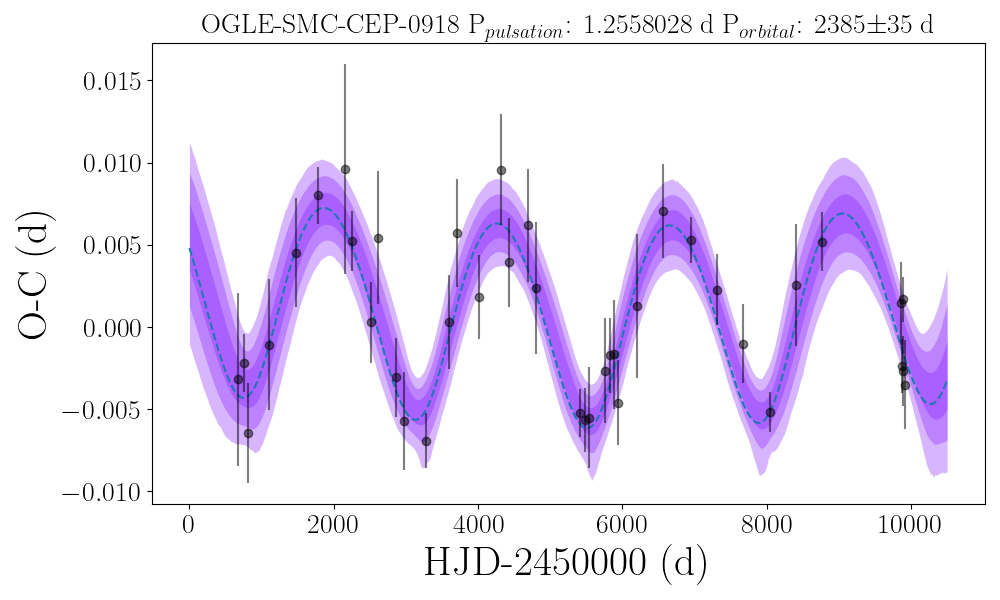}}
{\includegraphics[height=4.5cm,width=0.49\linewidth]{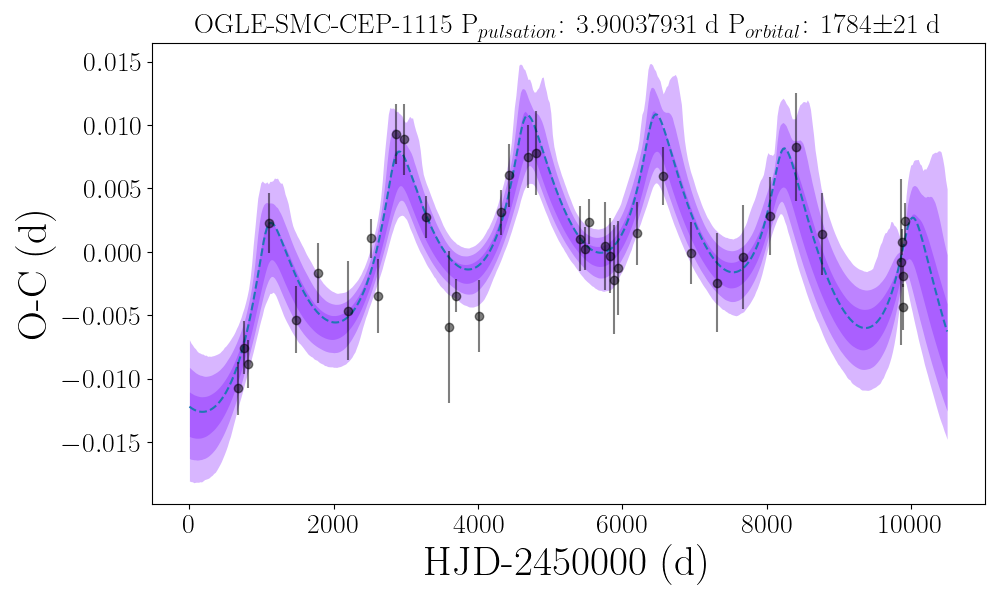}}
{\includegraphics[height=4.5cm,width=0.49\linewidth]{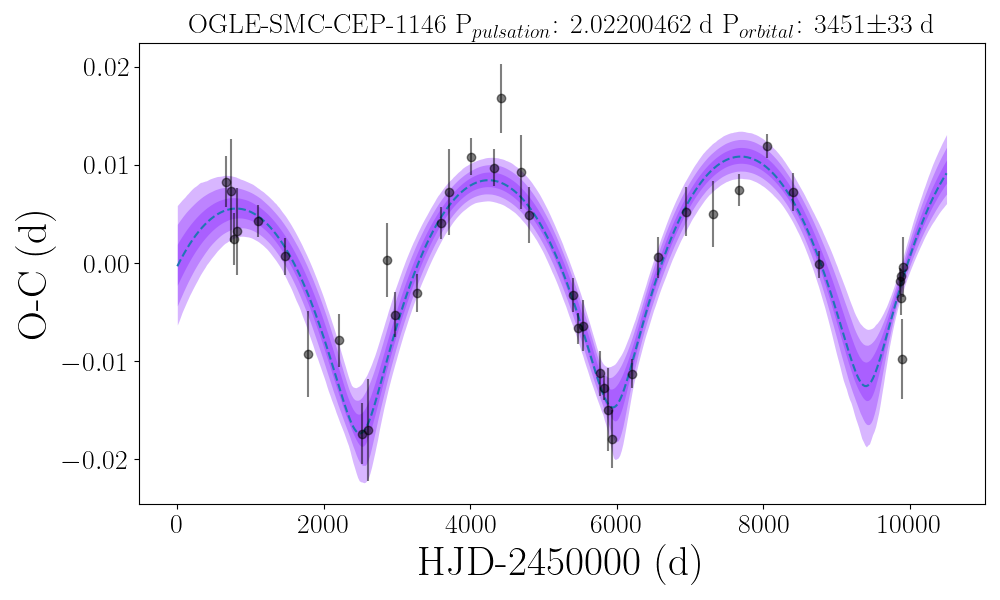}}
{\includegraphics[height=4.5cm,width=0.49\linewidth]{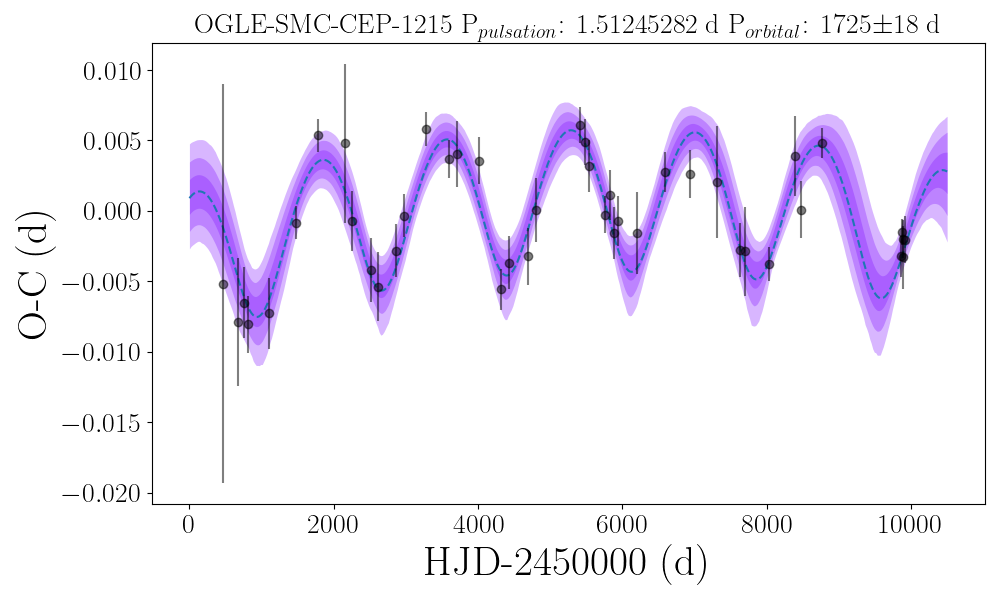}}
{\includegraphics[height=4.5cm,width=0.49\linewidth]{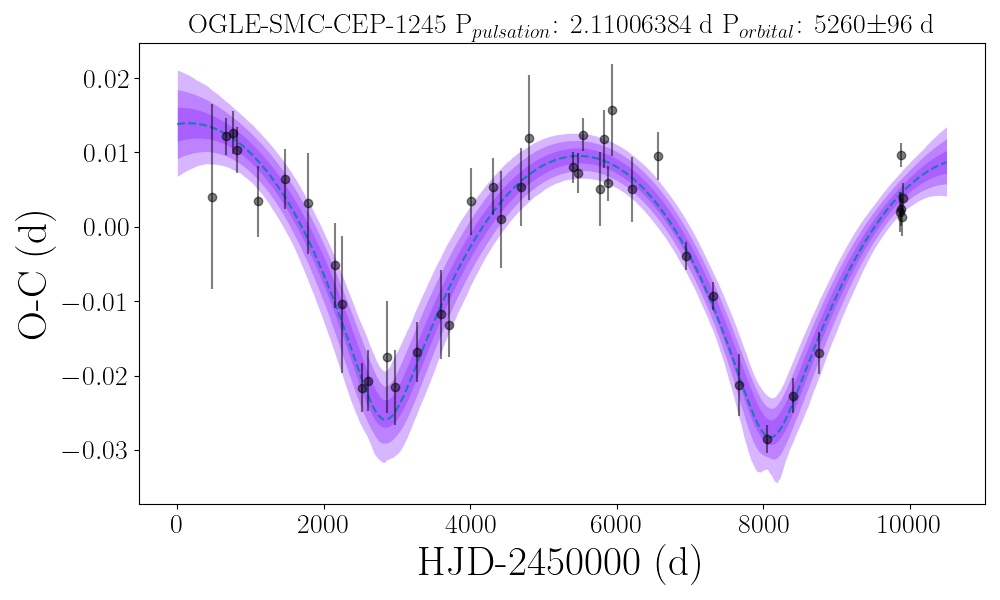}}
{\includegraphics[height=4.5cm,width=0.49\linewidth]{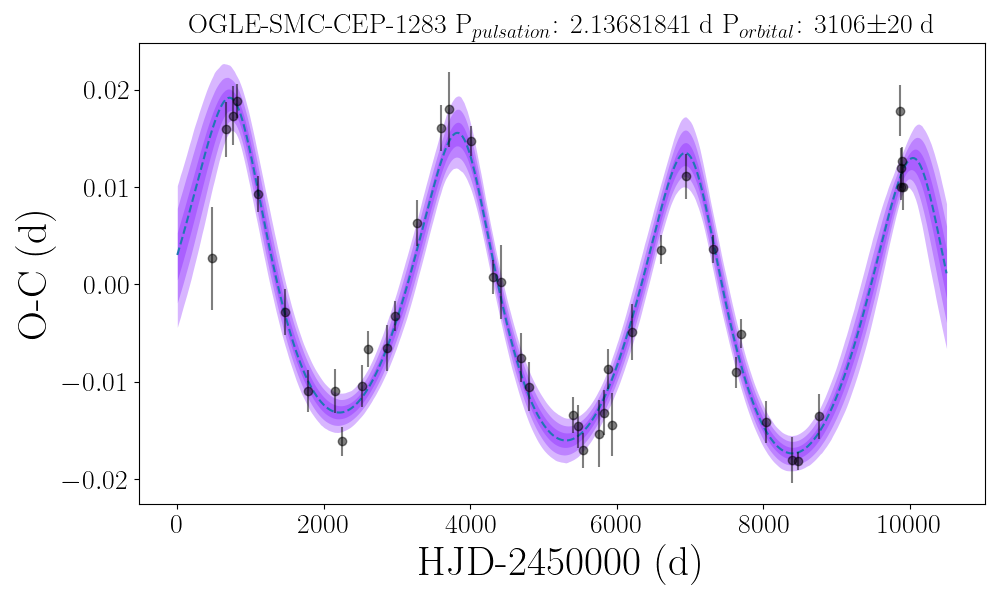}}
{\includegraphics[height=4.5cm,width=0.49\linewidth]{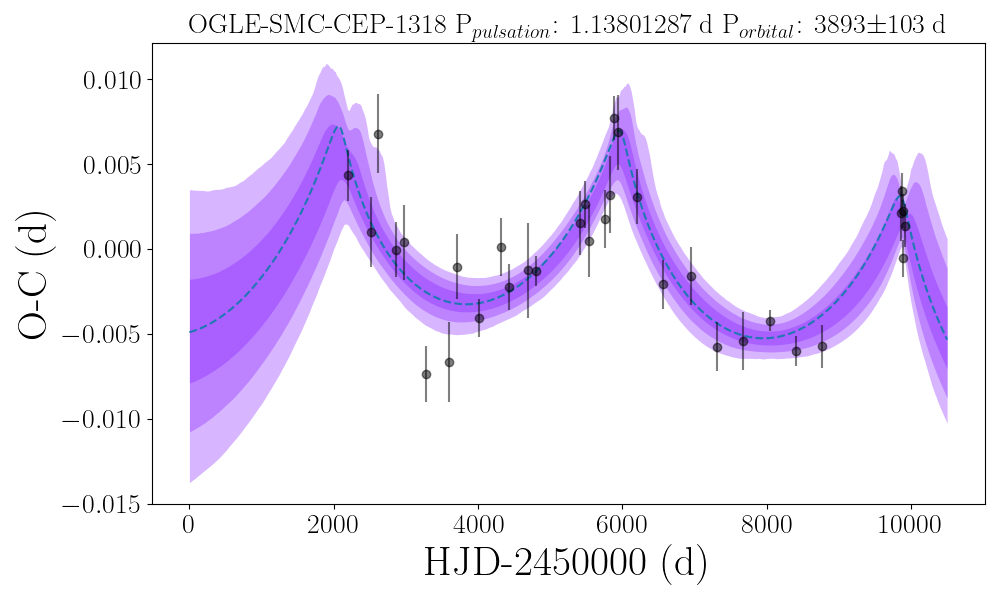}}
{\includegraphics[height=4.5cm,width=0.49\linewidth]{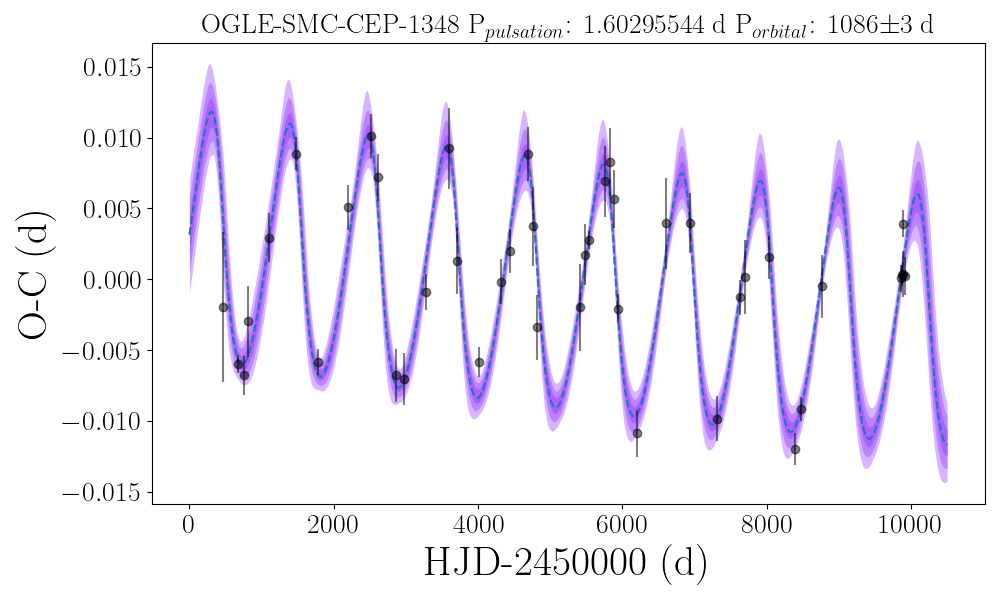}}
\caption{continued.}
\end{center}
\end{figure*}

\begin{figure*}[ht!]
\ContinuedFloat
\begin{center}
{\includegraphics[height=4.5cm,width=0.49\linewidth]{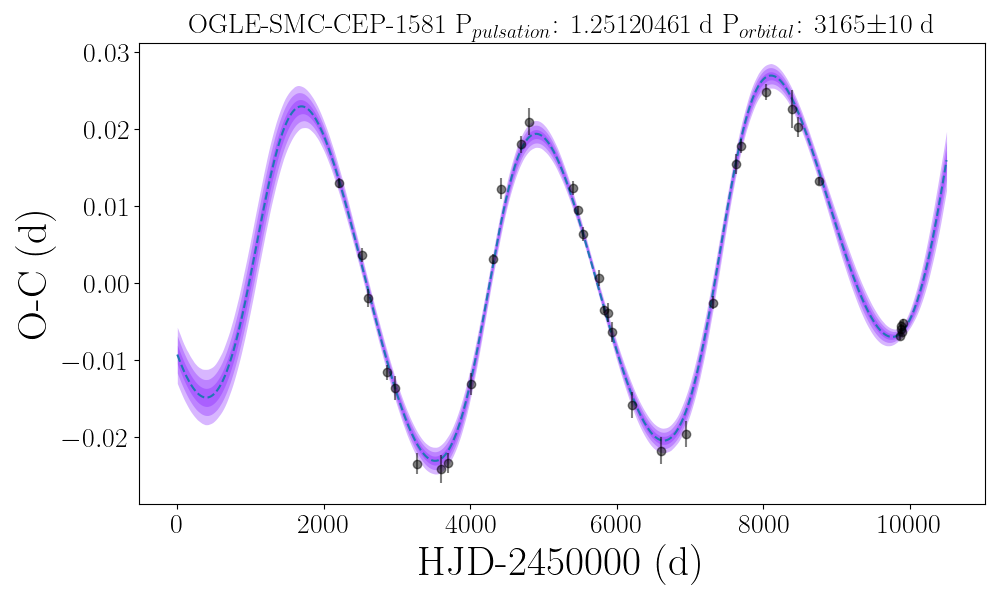}}
{\includegraphics[height=4.5cm,width=0.49\linewidth]{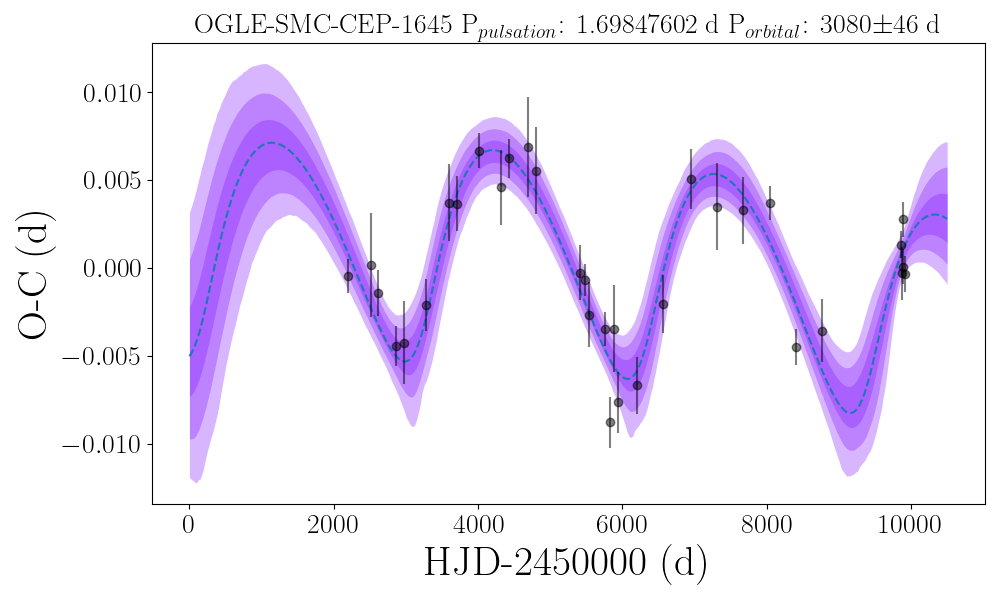}}
{\includegraphics[height=4.5cm,width=0.49\linewidth]{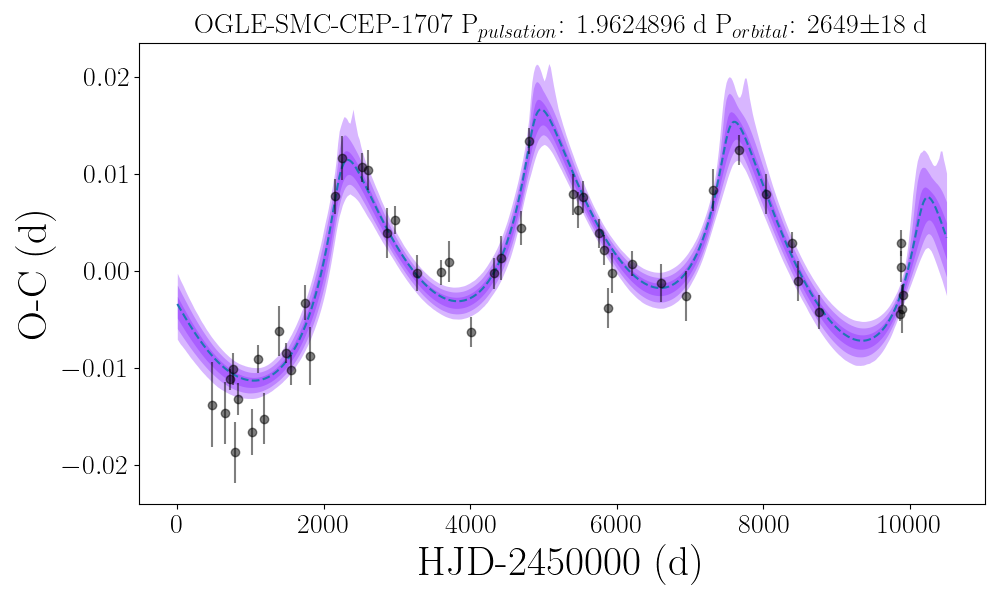}}
{\includegraphics[height=4.5cm,width=0.49\linewidth]{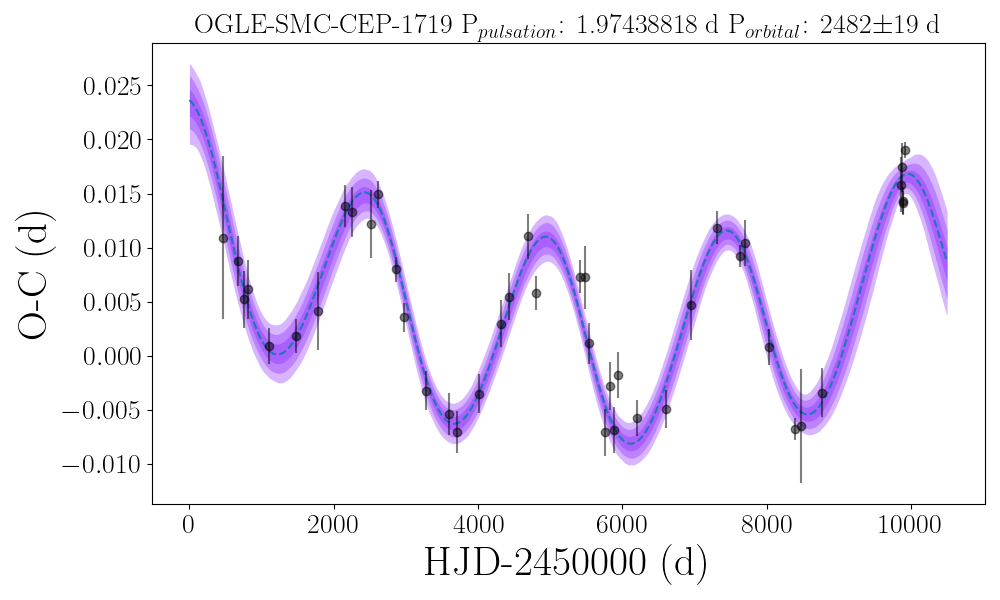}}
{\includegraphics[height=4.5cm,width=0.49\linewidth]{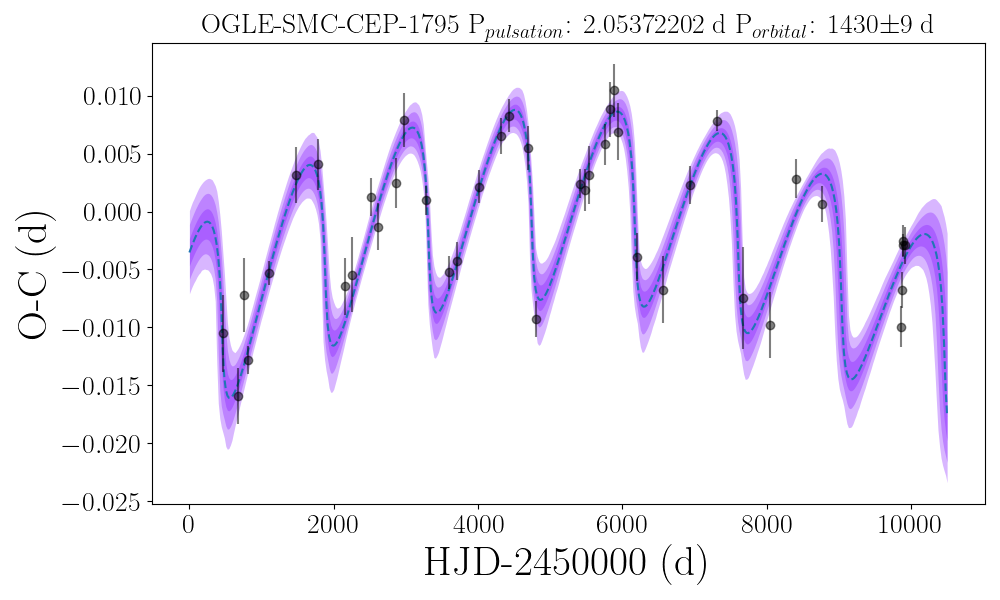}}
{\includegraphics[height=4.5cm,width=0.49\linewidth]{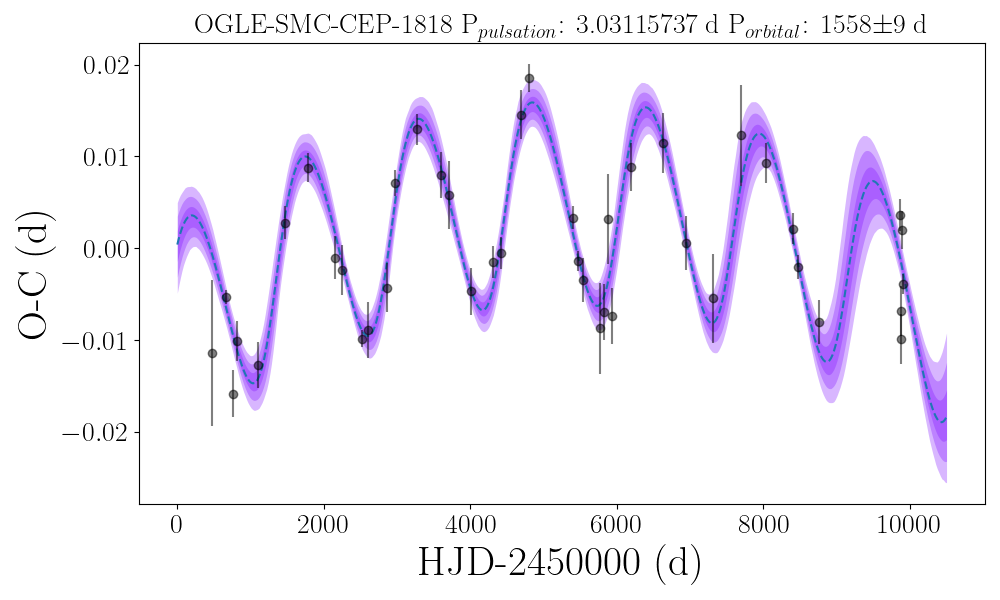}}
{\includegraphics[height=4.5cm,width=0.49\linewidth]{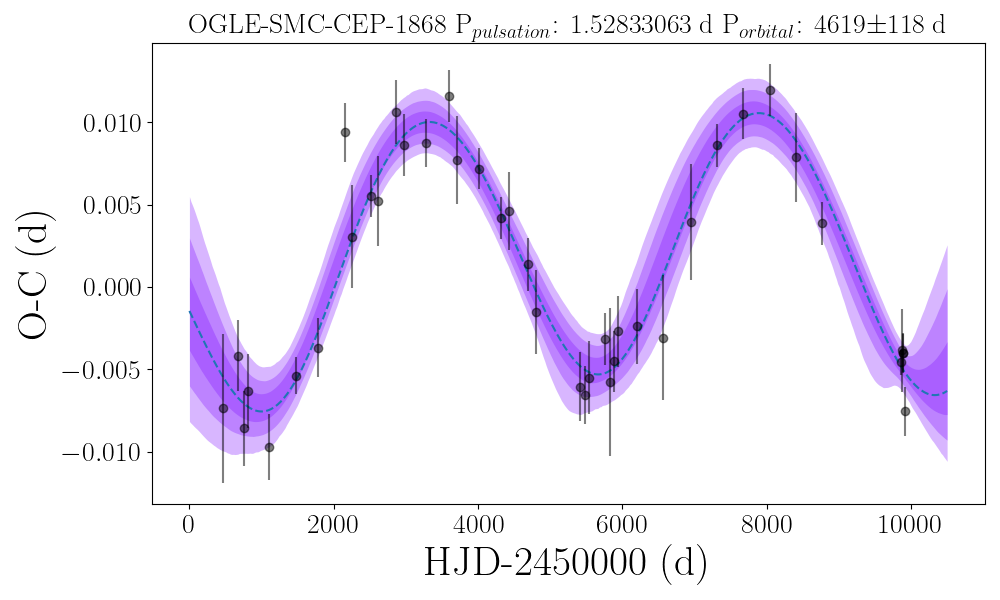}}
{\includegraphics[height=4.5cm,width=0.49\linewidth]{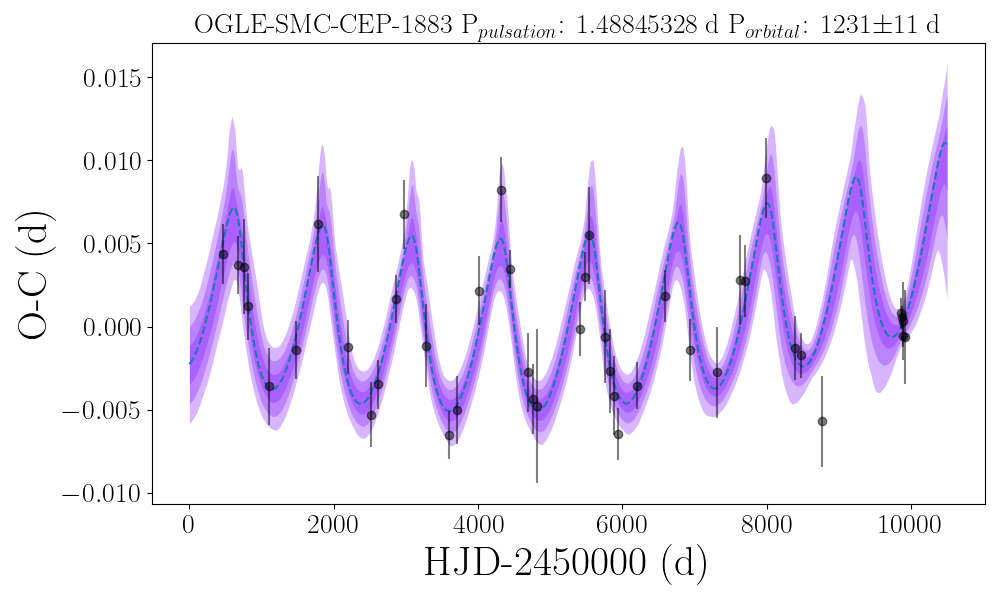}}
{\includegraphics[height=4.5cm,width=0.49\linewidth]{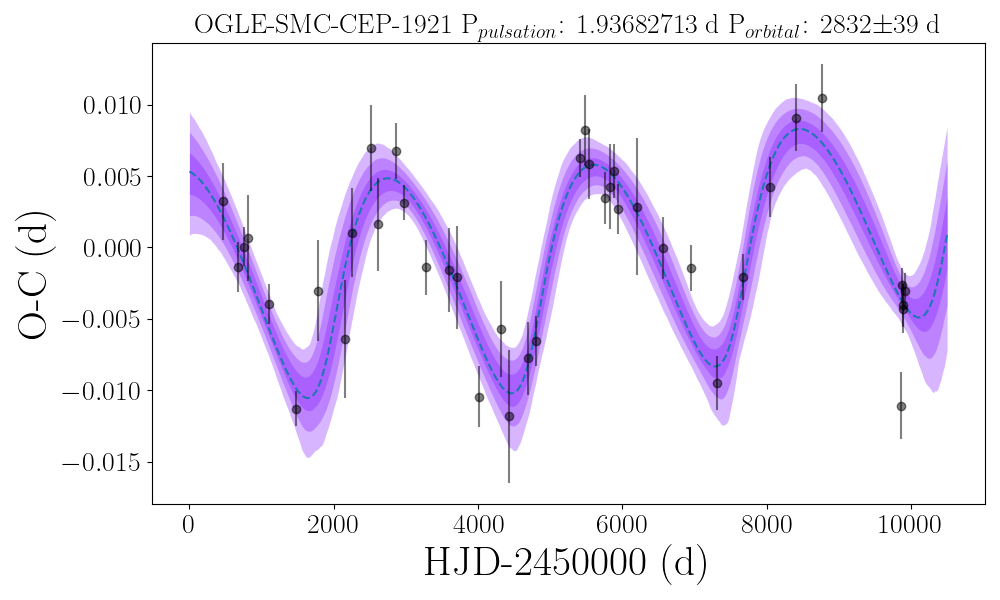}}
{\includegraphics[height=4.5cm,width=0.49\linewidth]{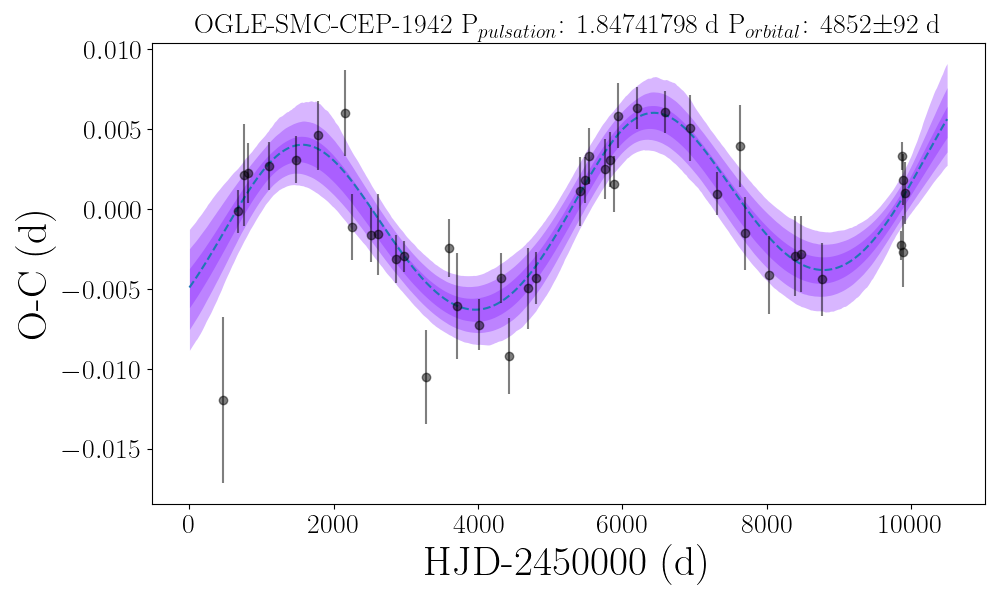}}
\caption{continued.}
\end{center}
\end{figure*}

\begin{figure*}[ht!]
\ContinuedFloat
\begin{center}
{\includegraphics[height=4.5cm,width=0.49\linewidth]{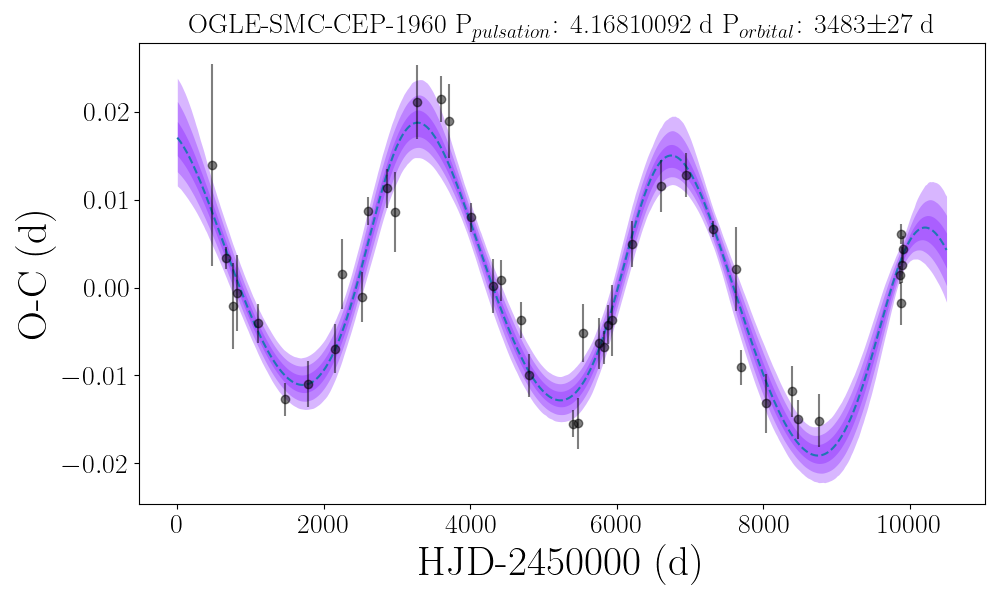}}
{\includegraphics[height=4.5cm,width=0.49\linewidth]{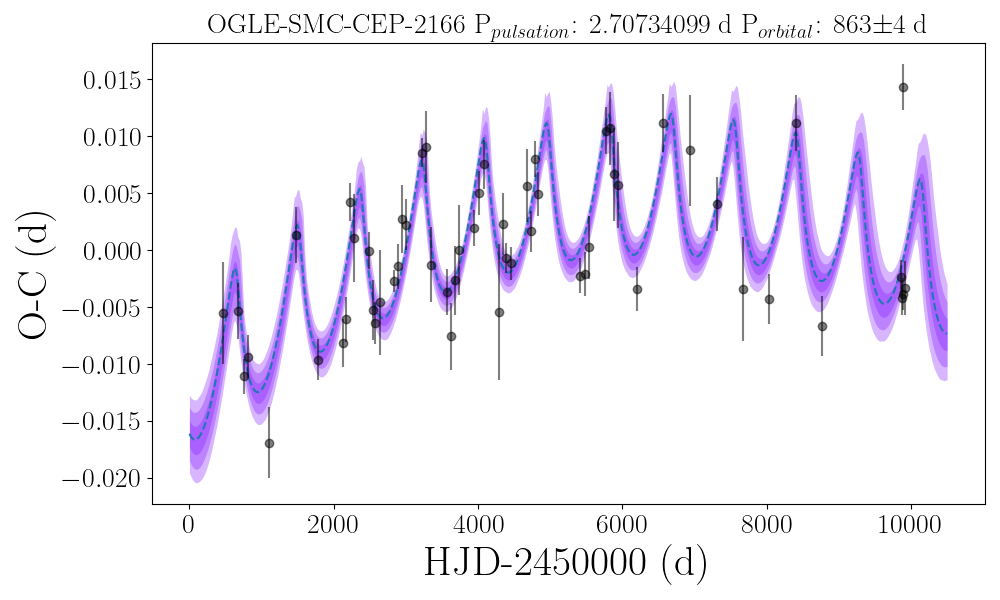}}
{\includegraphics[height=4.5cm,width=0.49\linewidth]{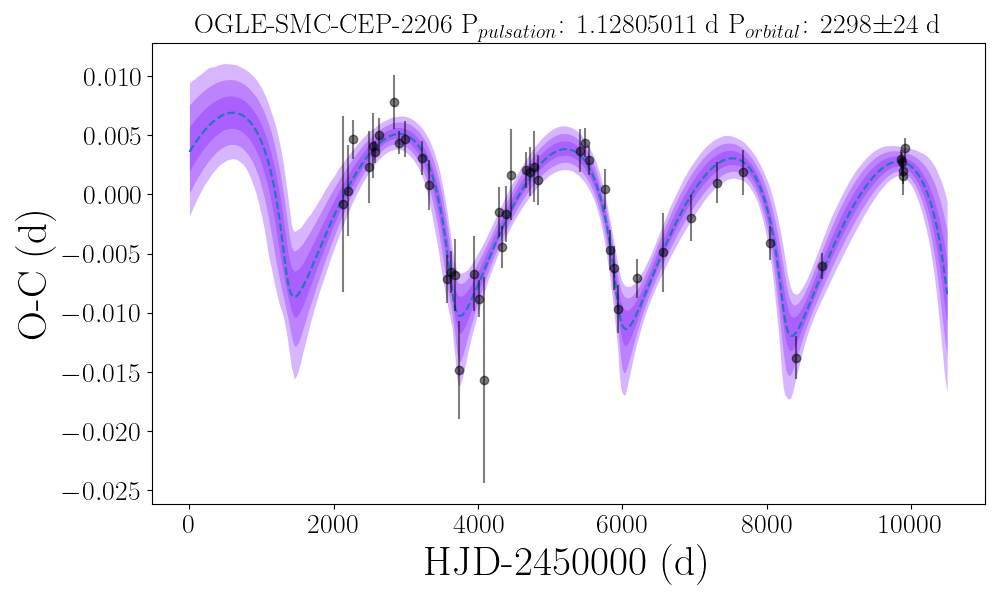}}
{\includegraphics[height=4.5cm,width=0.49\linewidth]{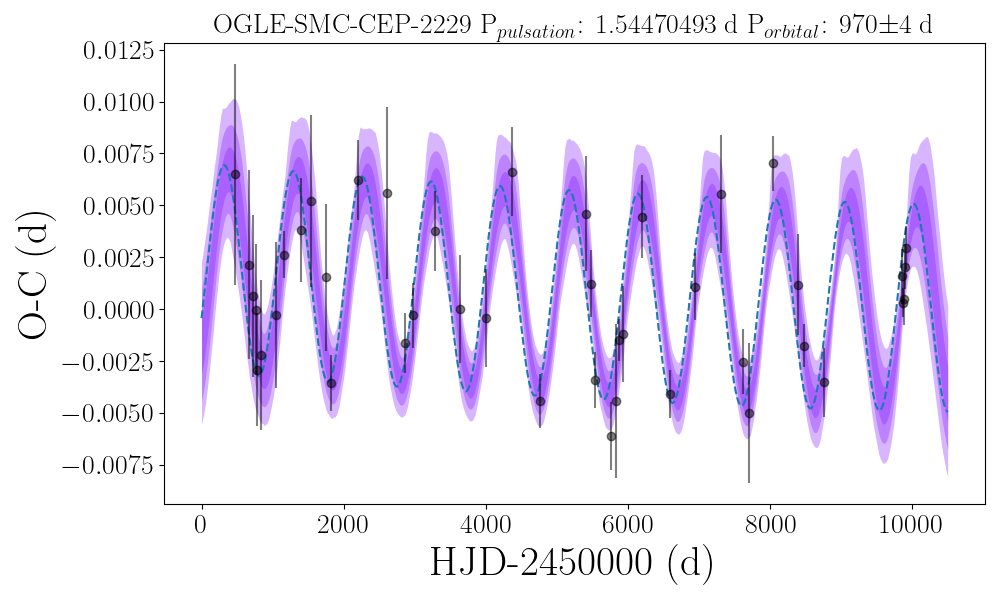}}
{\includegraphics[height=4.5cm,width=0.49\linewidth]{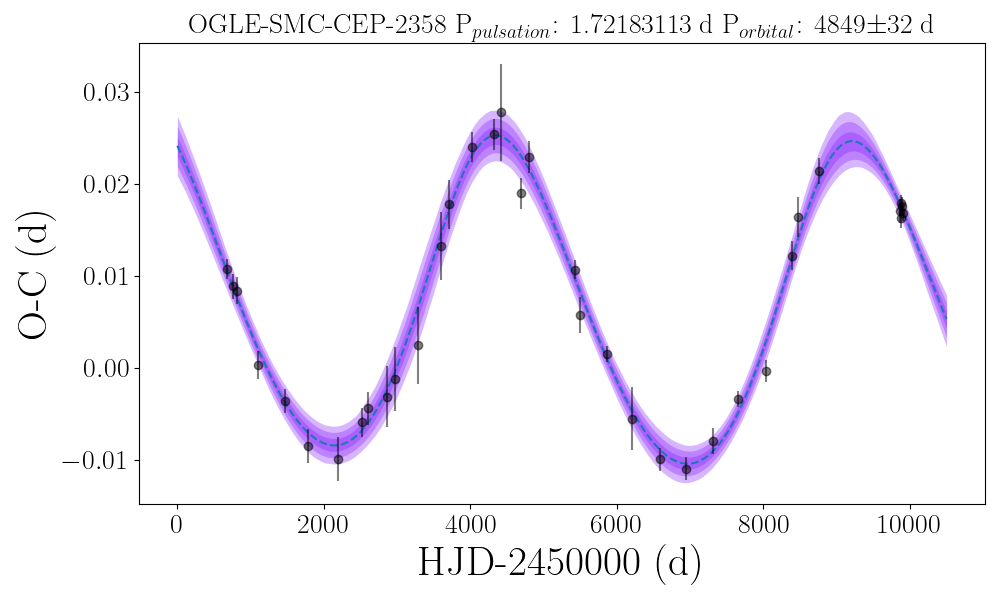}}
{\includegraphics[height=4.5cm,width=0.49\linewidth]{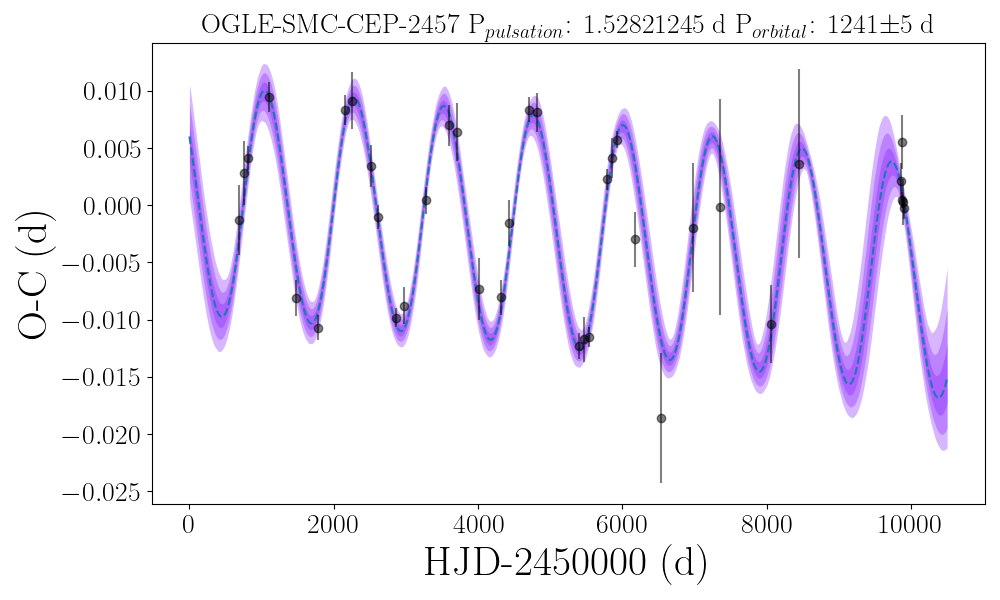}}
{\includegraphics[height=4.5cm,width=0.49\linewidth]{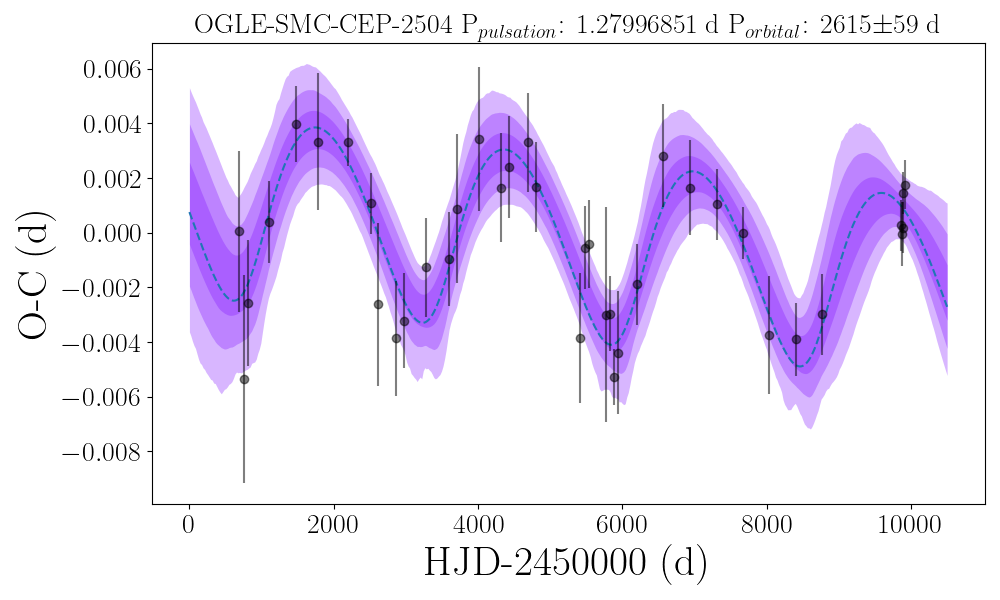}}
{\includegraphics[height=4.5cm,width=0.49\linewidth]{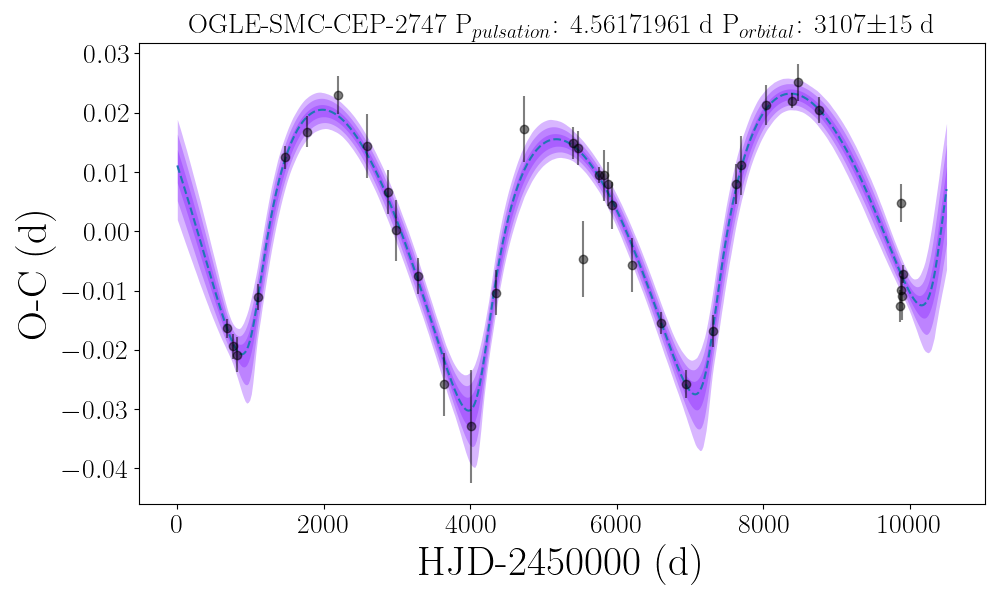}}
{\includegraphics[height=4.5cm,width=0.49\linewidth]{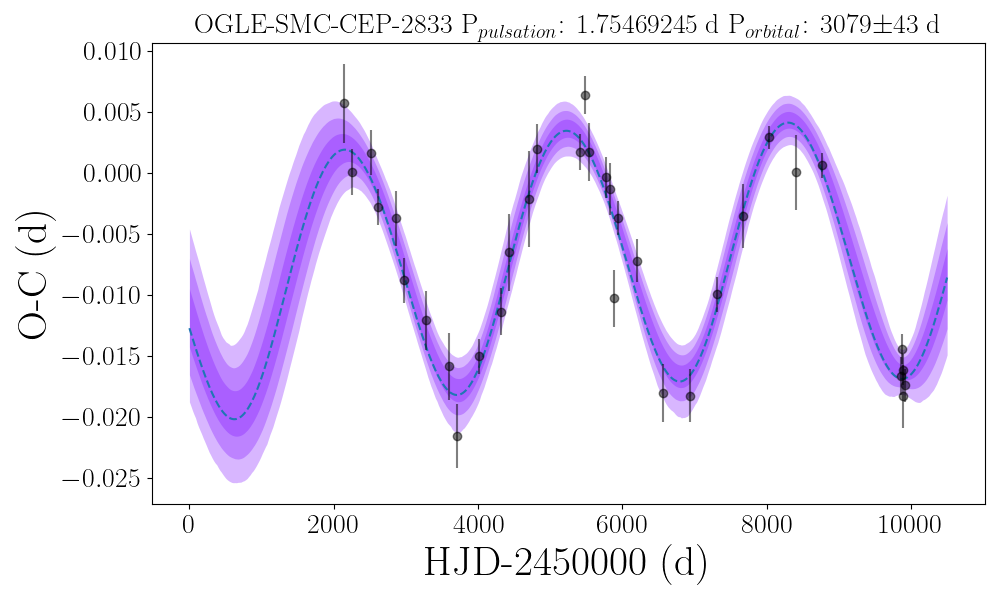}}
{\includegraphics[height=4.5cm,width=0.49\linewidth]{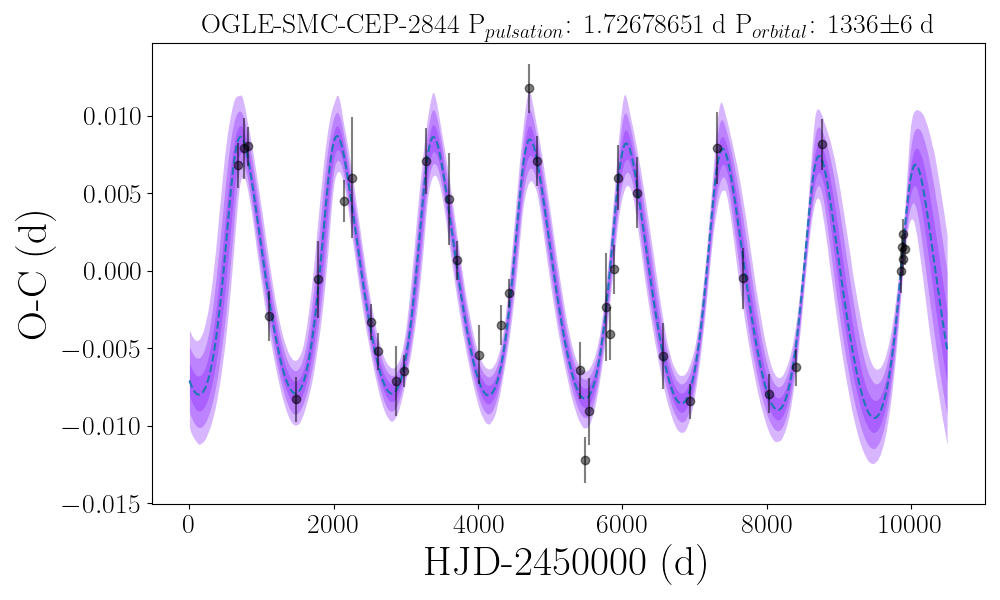}}
\caption{continued.}
\end{center}
\end{figure*}

\begin{figure*}[ht!]
\ContinuedFloat
\begin{center}
{\includegraphics[height=4.5cm,width=0.49\linewidth]{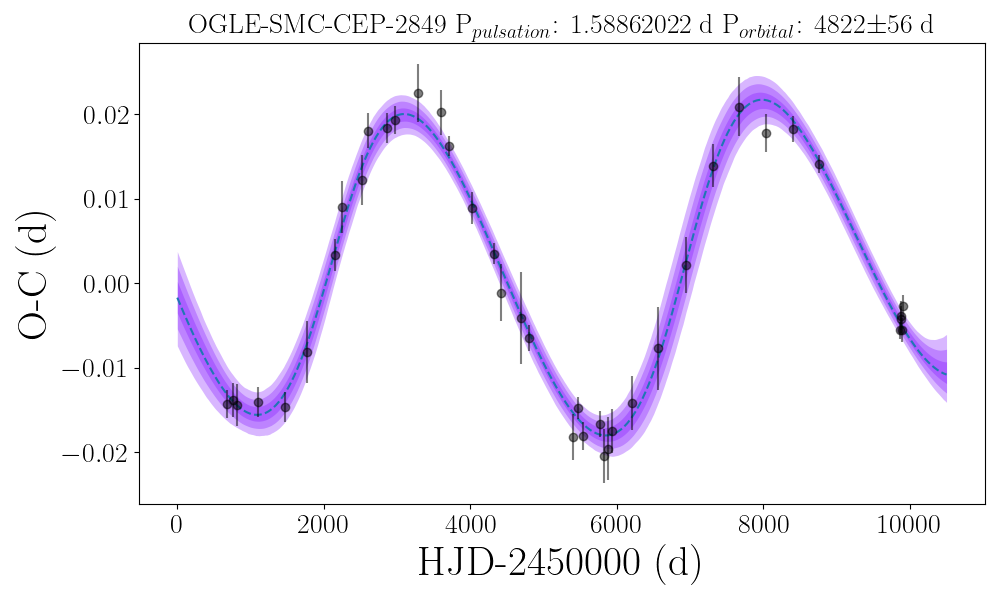}}
{\includegraphics[height=4.5cm,width=0.49\linewidth]{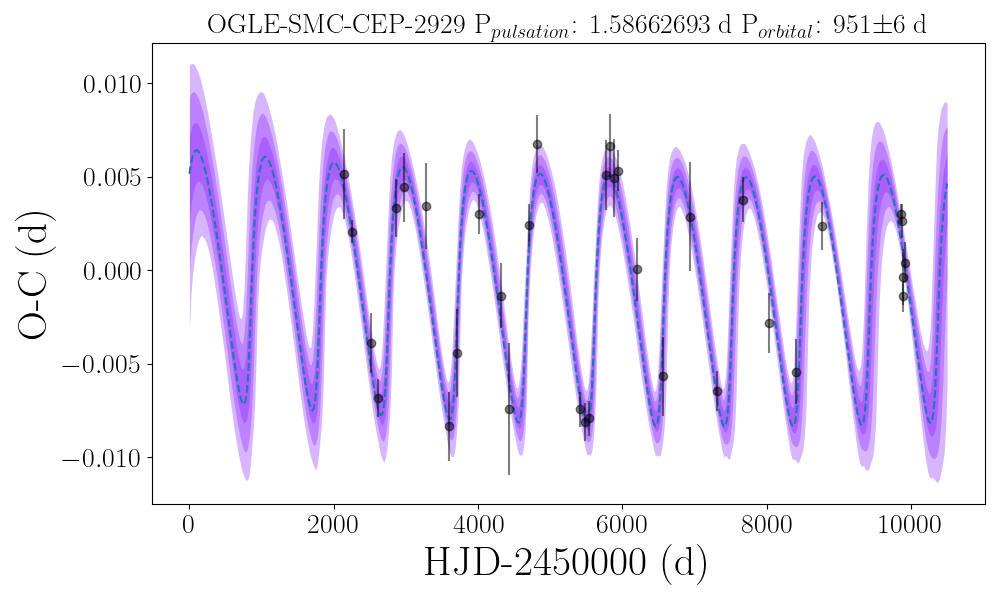}}
{\includegraphics[height=4.5cm,width=0.49\linewidth]{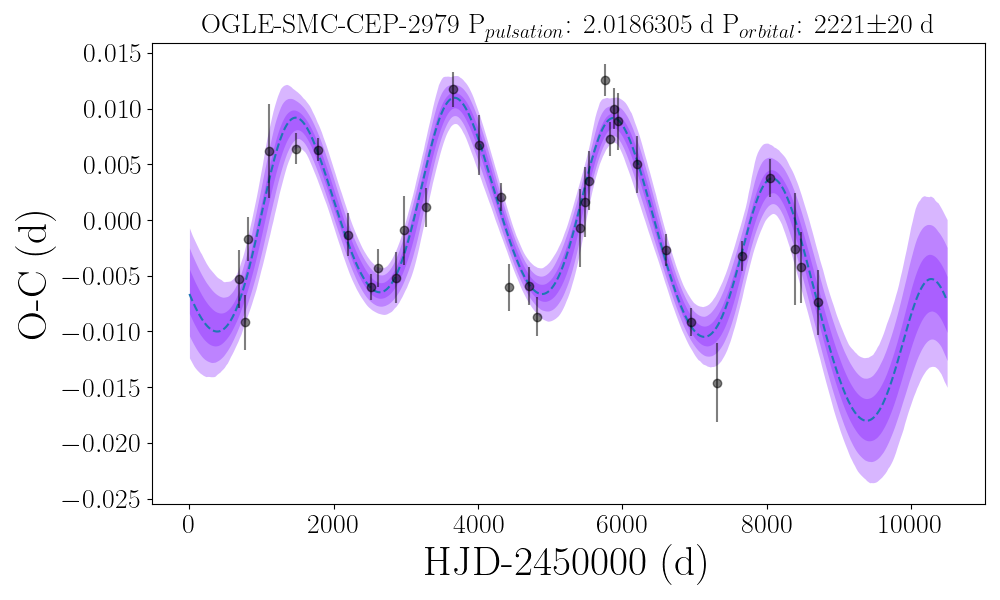}}
{\includegraphics[height=4.5cm,width=0.49\linewidth]{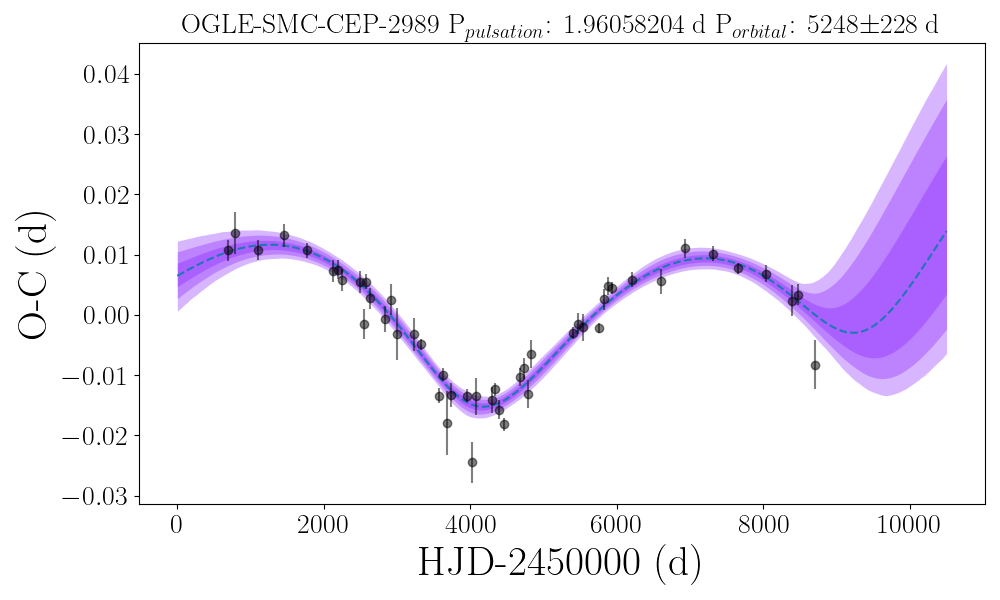}}
{\includegraphics[height=4.5cm,width=0.49\linewidth]{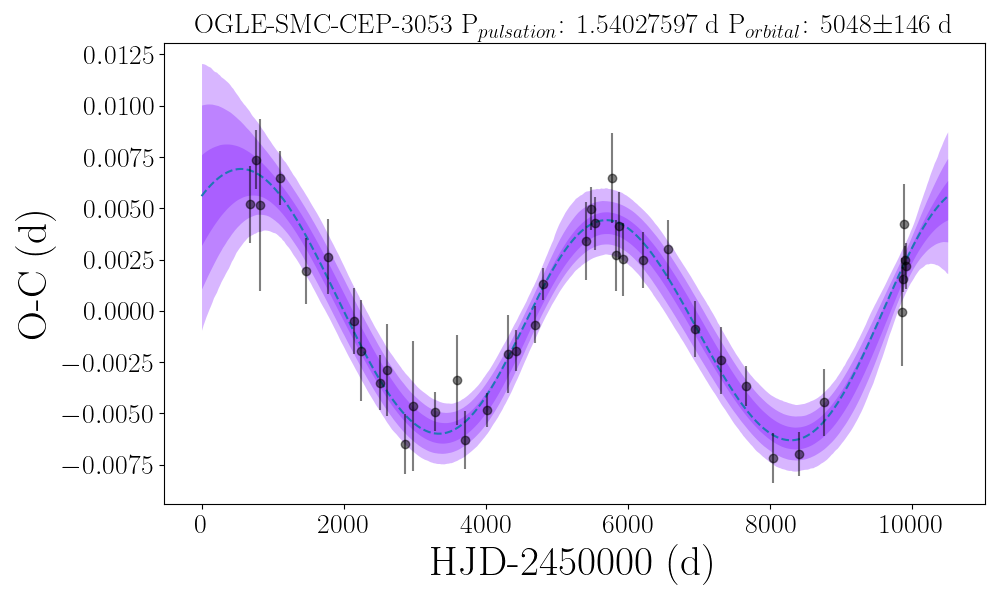}}
{\includegraphics[height=4.5cm,width=0.49\linewidth]{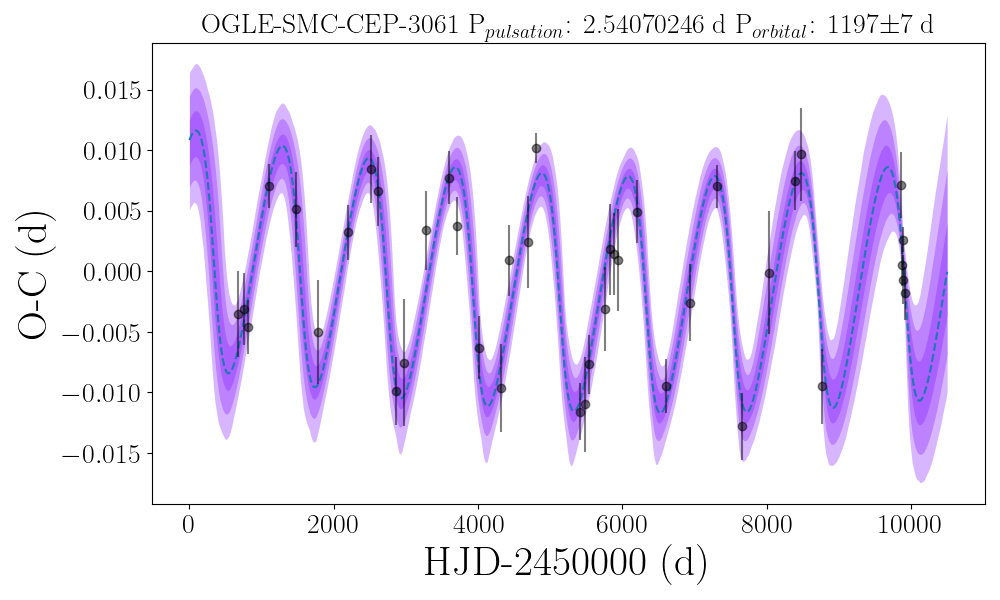}}
{\includegraphics[height=4.5cm,width=0.49\linewidth]{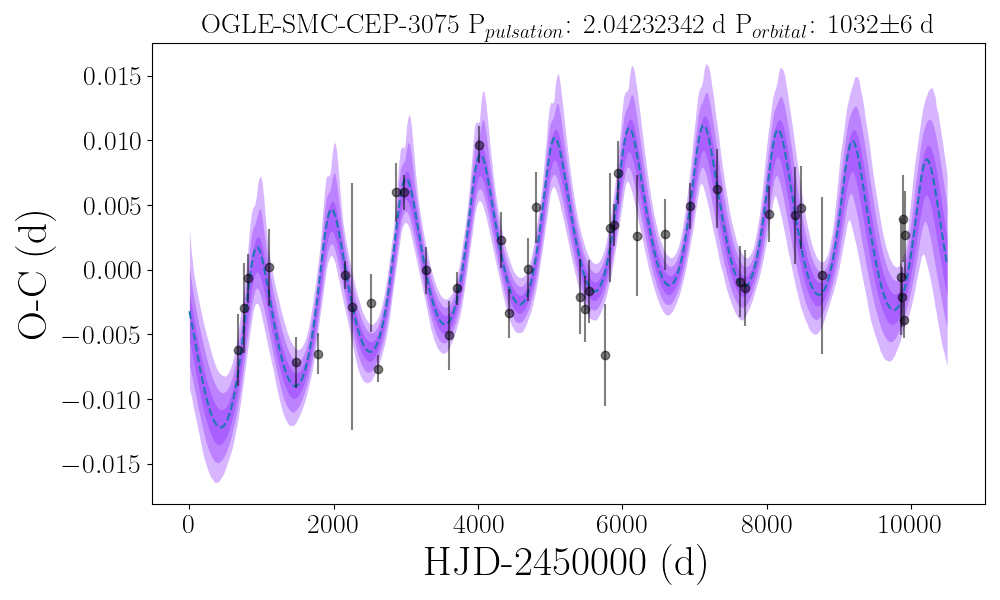}}
{\includegraphics[height=4.5cm,width=0.49\linewidth]{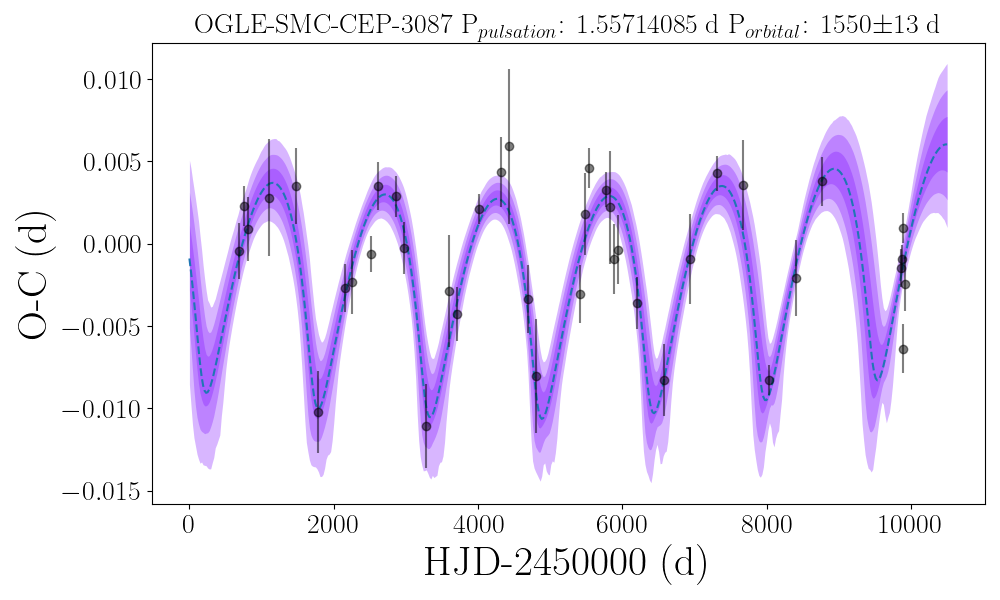}}
{\includegraphics[height=4.5cm,width=0.49\linewidth]{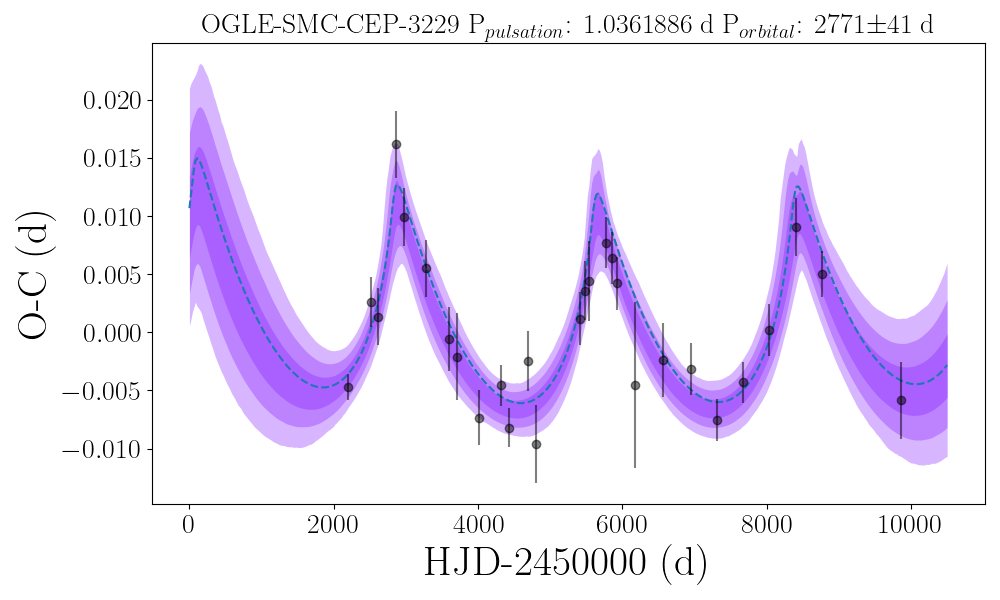}}
{\includegraphics[height=4.5cm,width=0.49\linewidth]{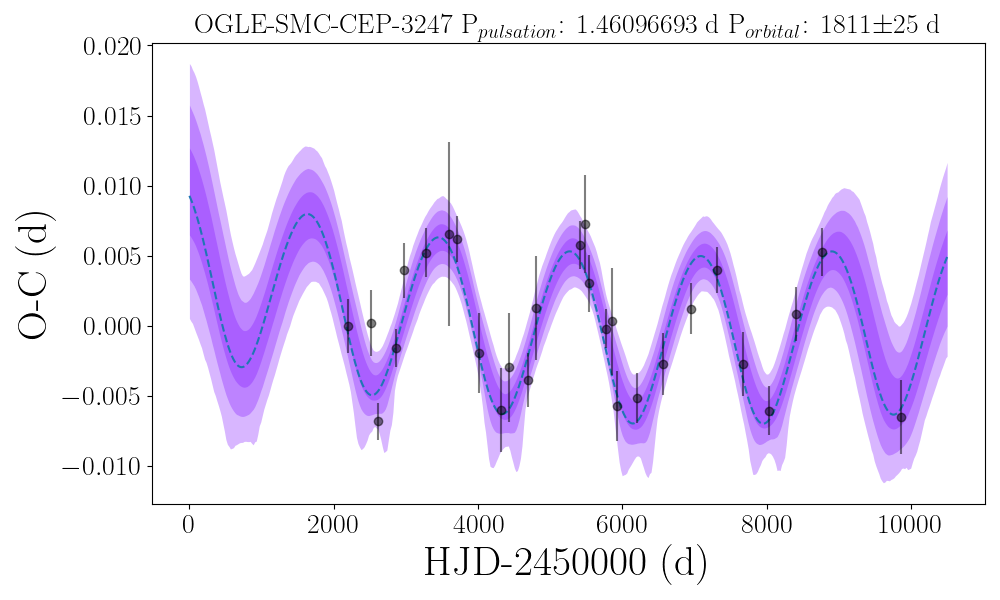}}
\caption{continued.}
\end{center}
\end{figure*}

\begin{figure*}[ht!]
\ContinuedFloat
\begin{center}
{\includegraphics[height=4.5cm,width=0.49\linewidth]{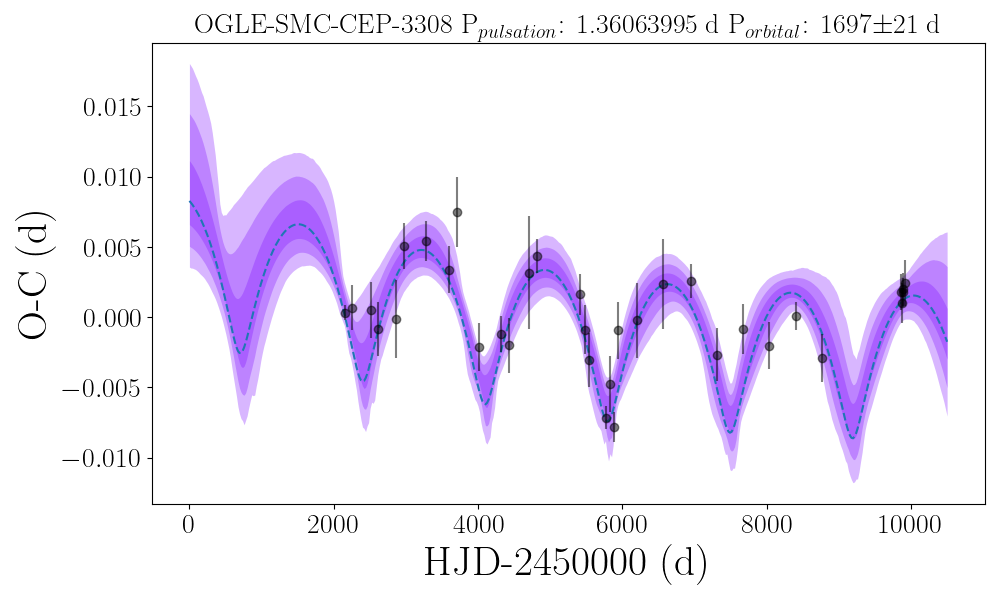}}
{\includegraphics[height=4.5cm,width=0.49\linewidth]{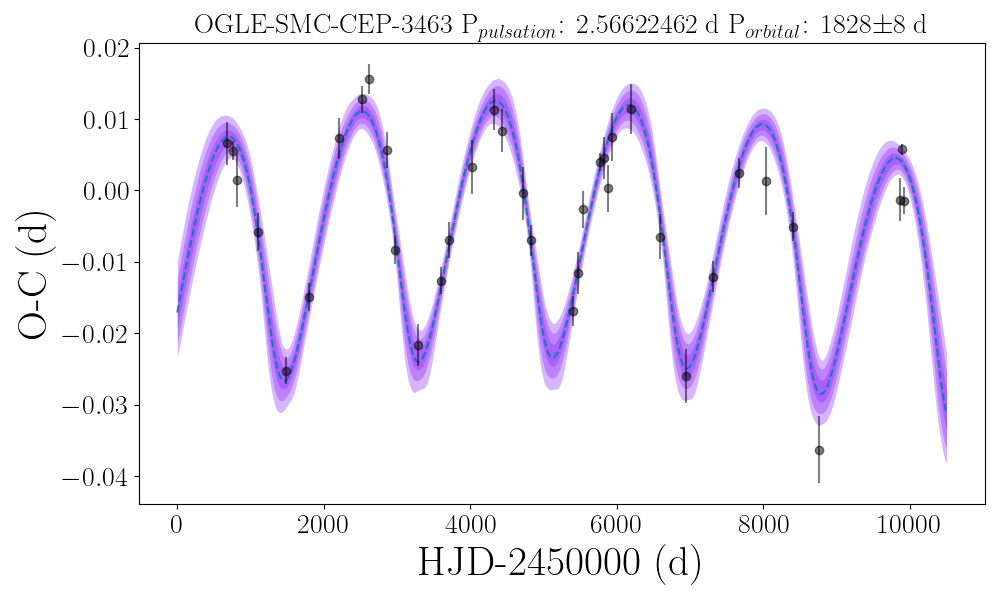}}
{\includegraphics[height=4.5cm,width=0.49\linewidth]{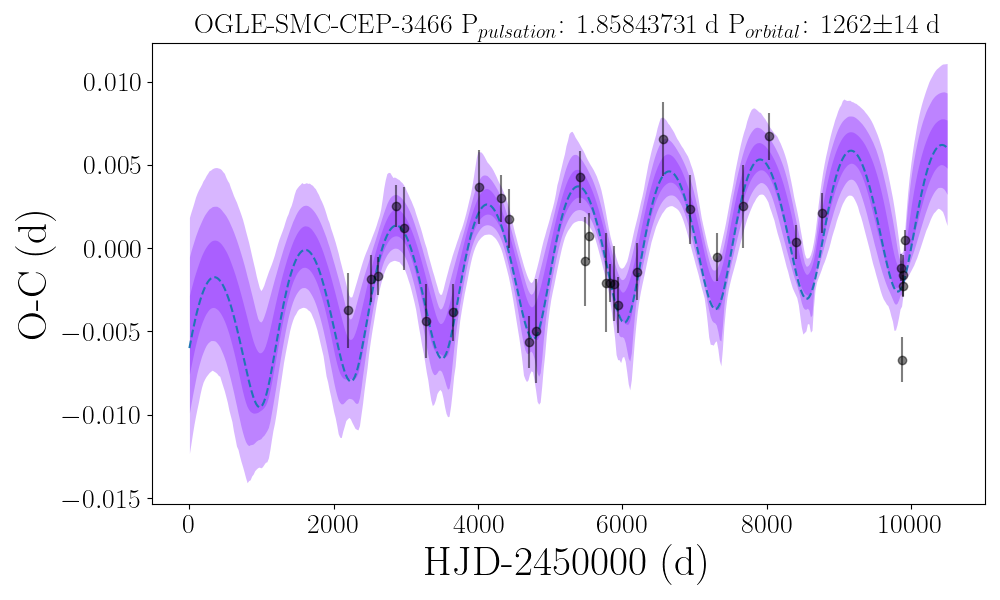}}
{\includegraphics[height=4.5cm,width=0.49\linewidth]{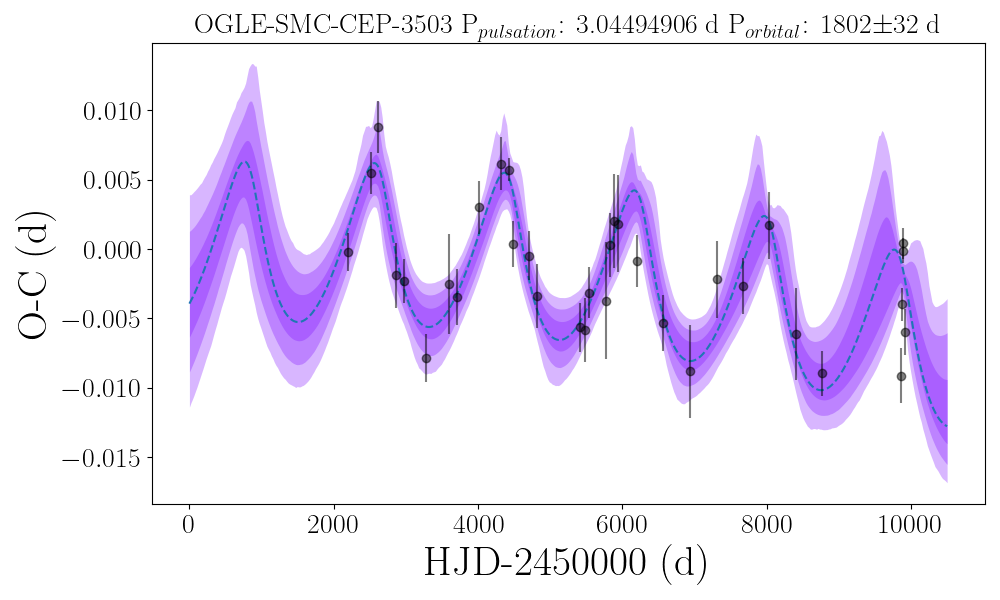}}
{\includegraphics[height=4.5cm,width=0.49\linewidth]{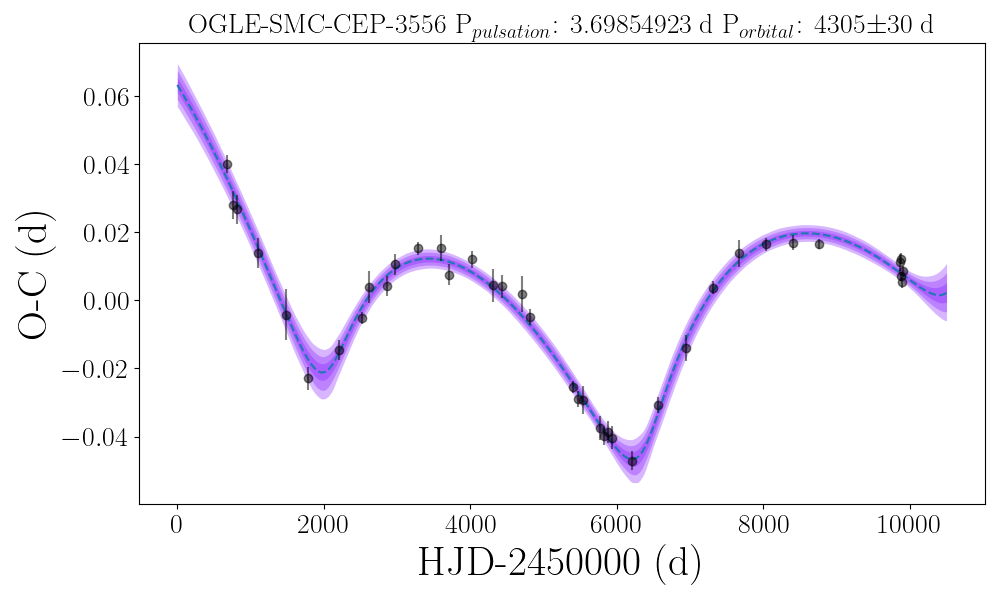}}
{\includegraphics[height=4.5cm,width=0.49\linewidth]{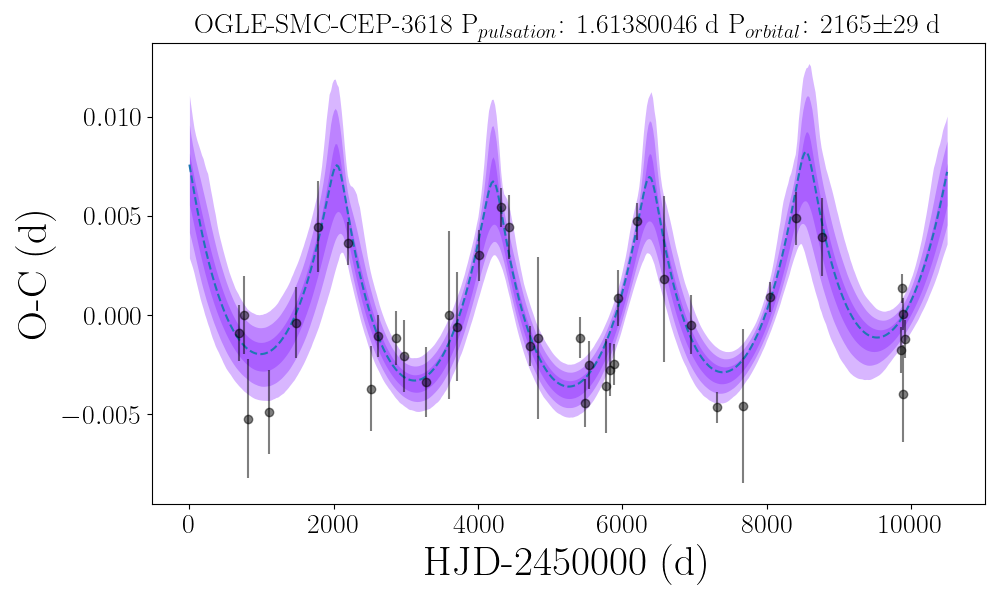}}
{\includegraphics[height=4.5cm,width=0.49\linewidth]{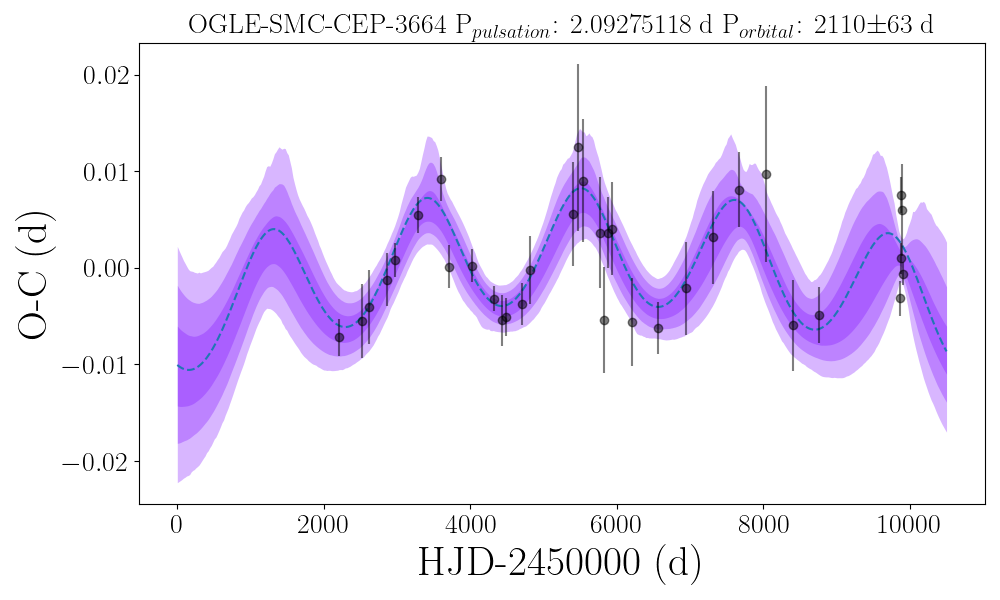}}
{\includegraphics[height=4.5cm,width=0.49\linewidth]{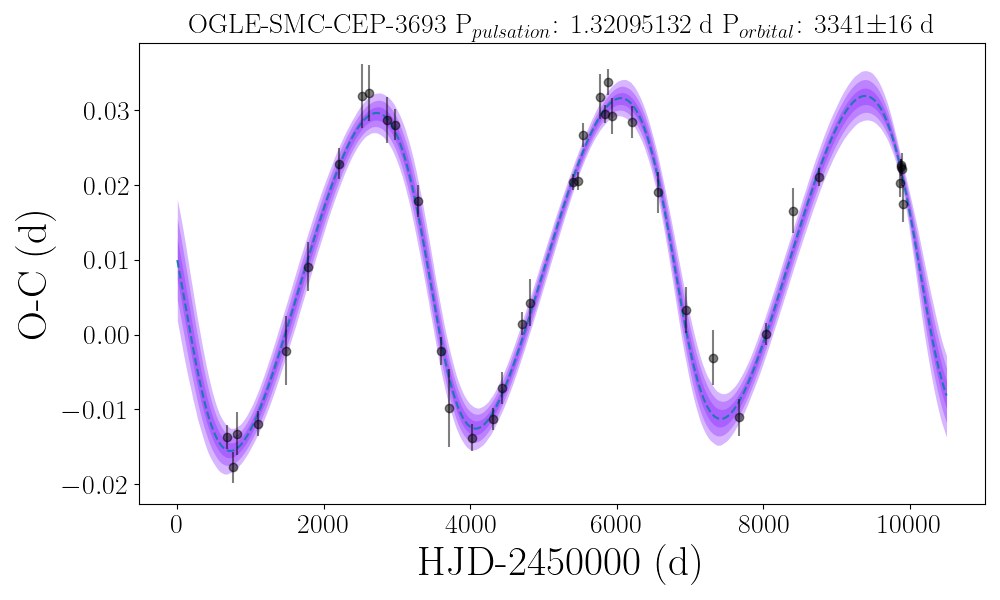}}
{\includegraphics[height=4.5cm,width=0.49\linewidth]{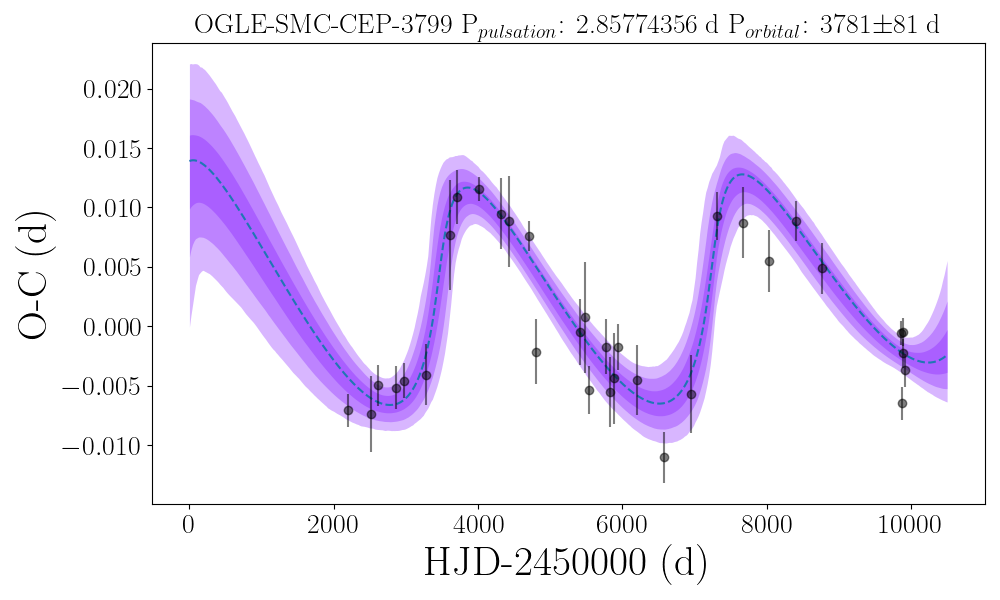}}
{\includegraphics[height=4.5cm,width=0.49\linewidth]{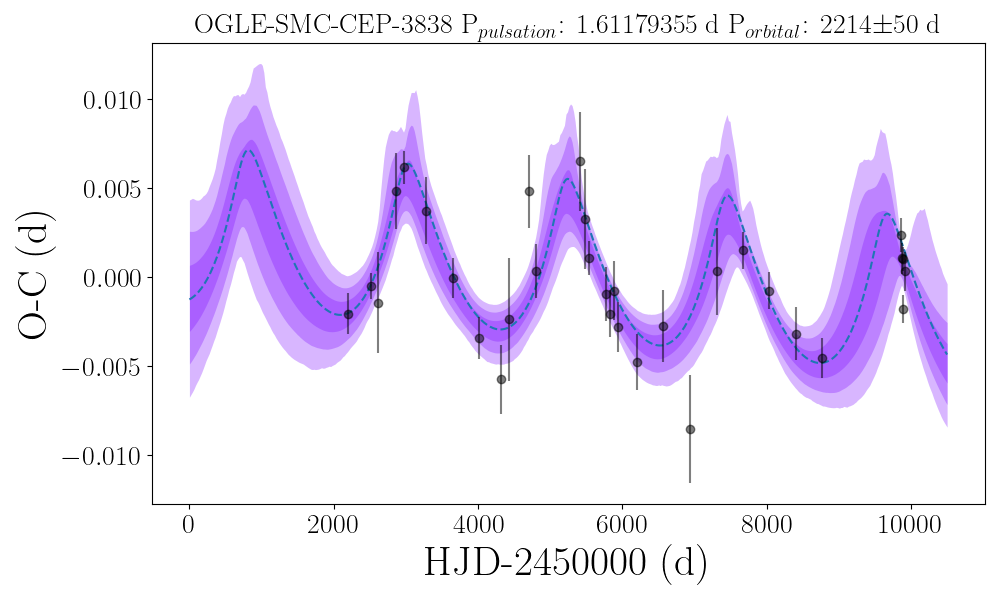}}
\caption{continued.}
\end{center}
\end{figure*}

\begin{figure*}[ht!]
\ContinuedFloat
\begin{center}
{\includegraphics[height=4.5cm,width=0.49\linewidth]{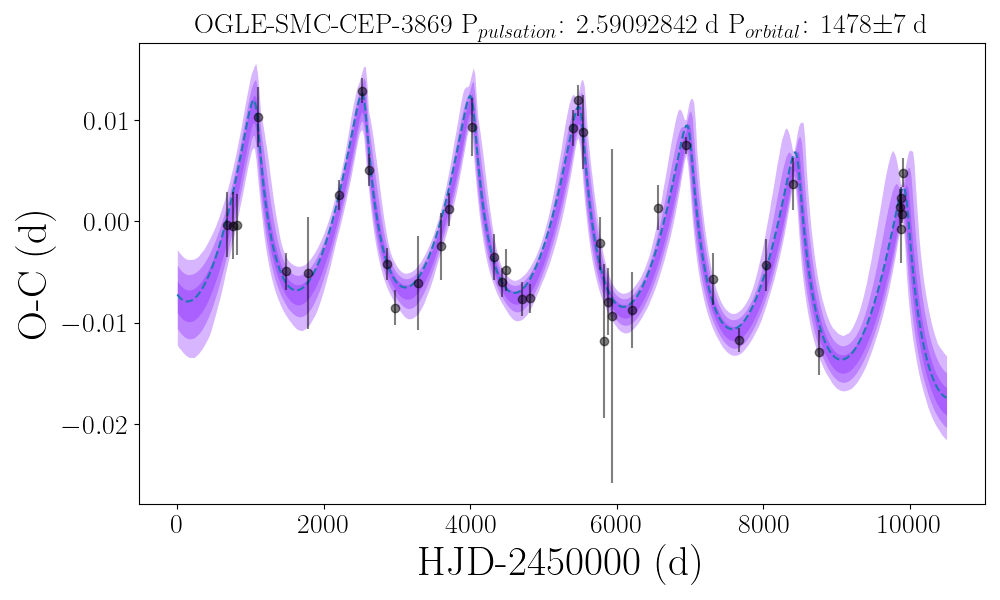}}
{\includegraphics[height=4.5cm,width=0.49\linewidth]{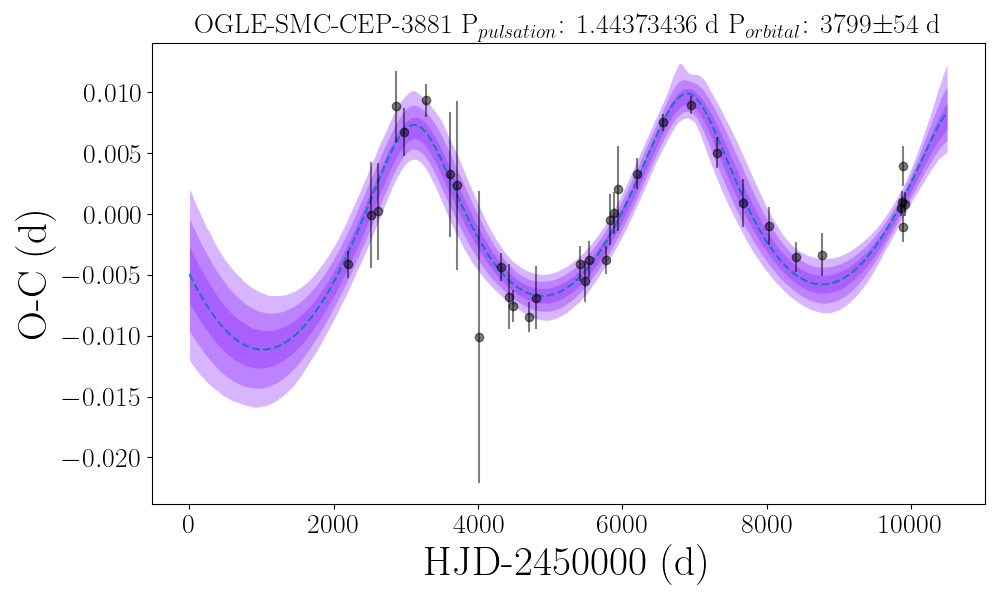}}
{\includegraphics[height=4.5cm,width=0.49\linewidth]{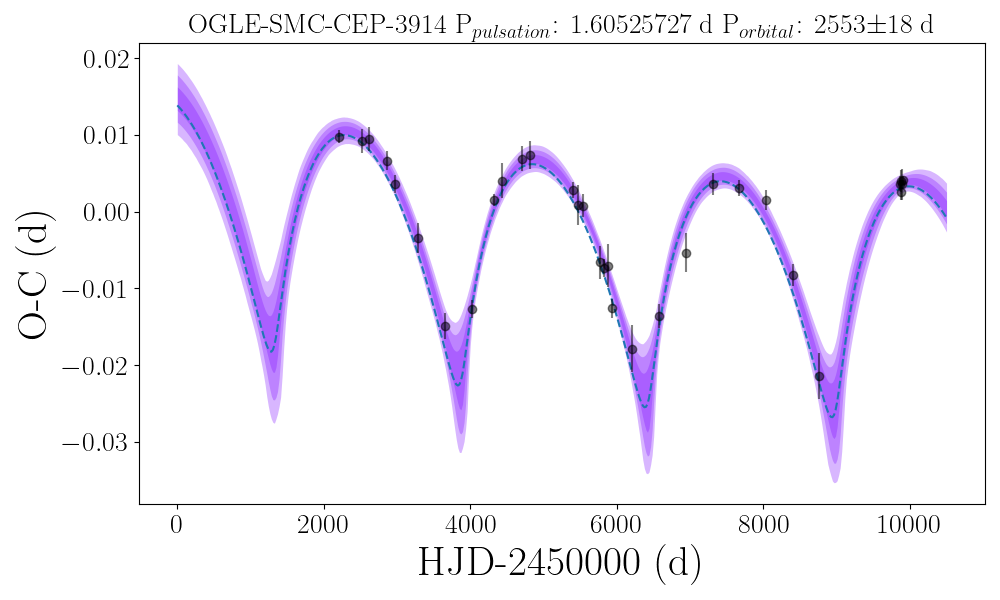}}
{\includegraphics[height=4.5cm,width=0.49\linewidth]{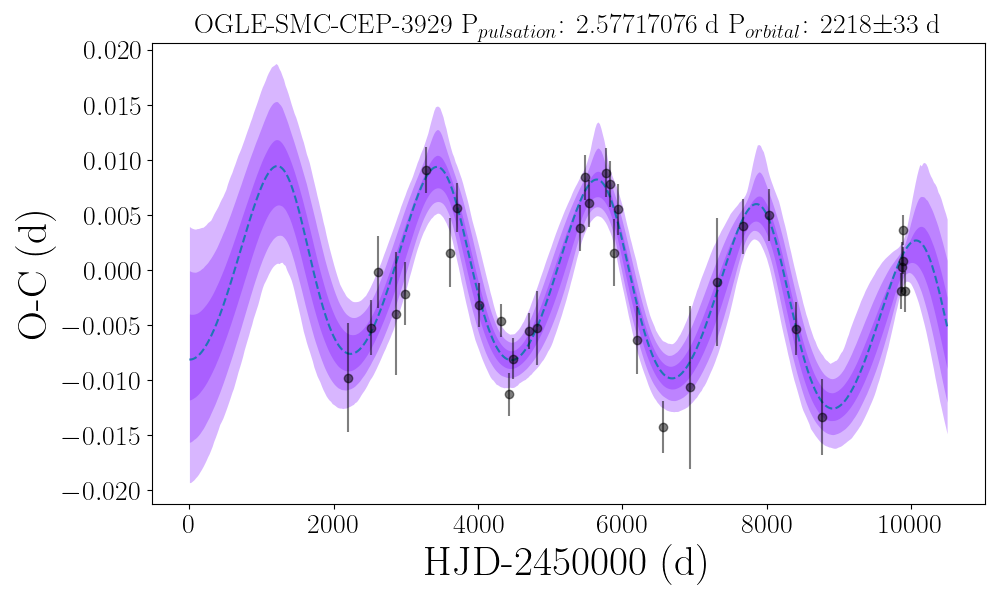}}
{\includegraphics[height=4.5cm,width=0.49\linewidth]{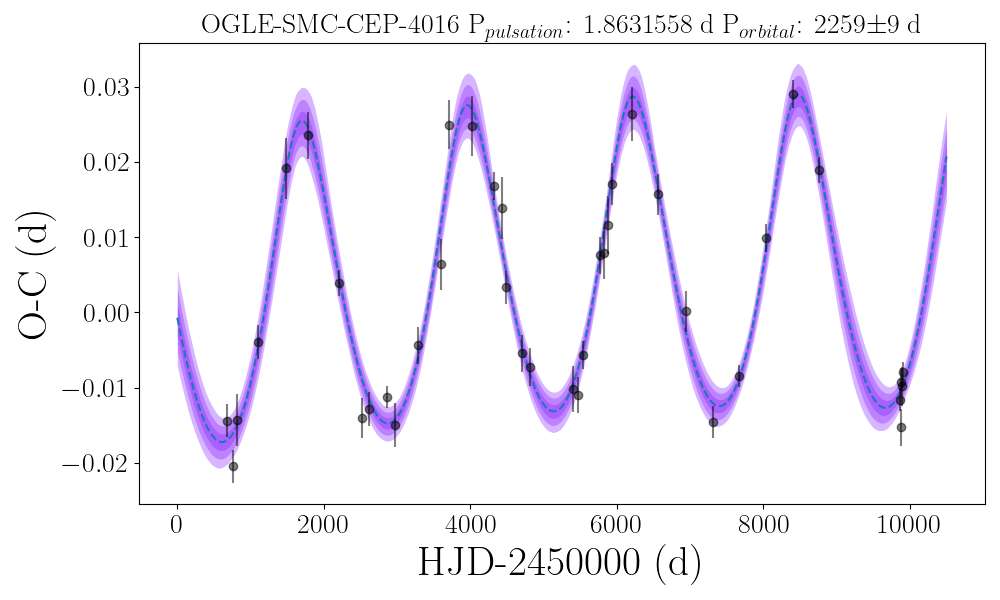}}
{\includegraphics[height=4.5cm,width=0.49\linewidth]{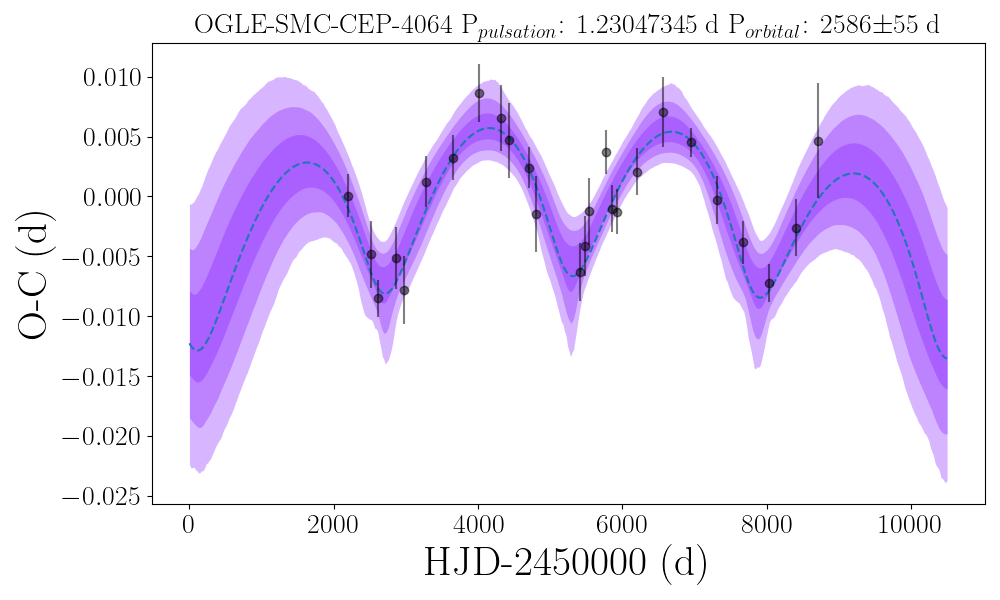}}
{\includegraphics[height=4.5cm,width=0.49\linewidth]{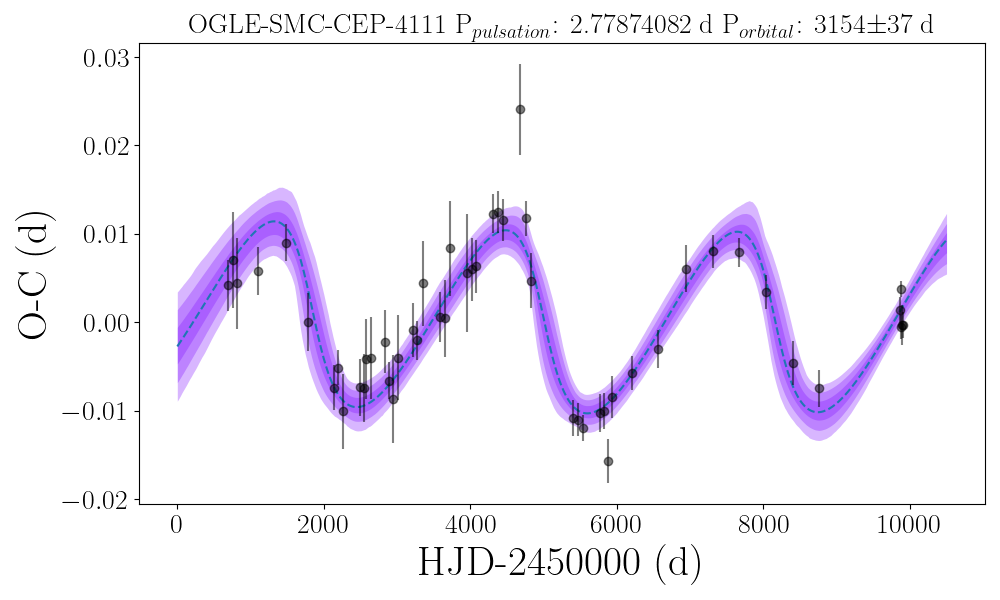}}
{\includegraphics[height=4.5cm,width=0.49\linewidth]{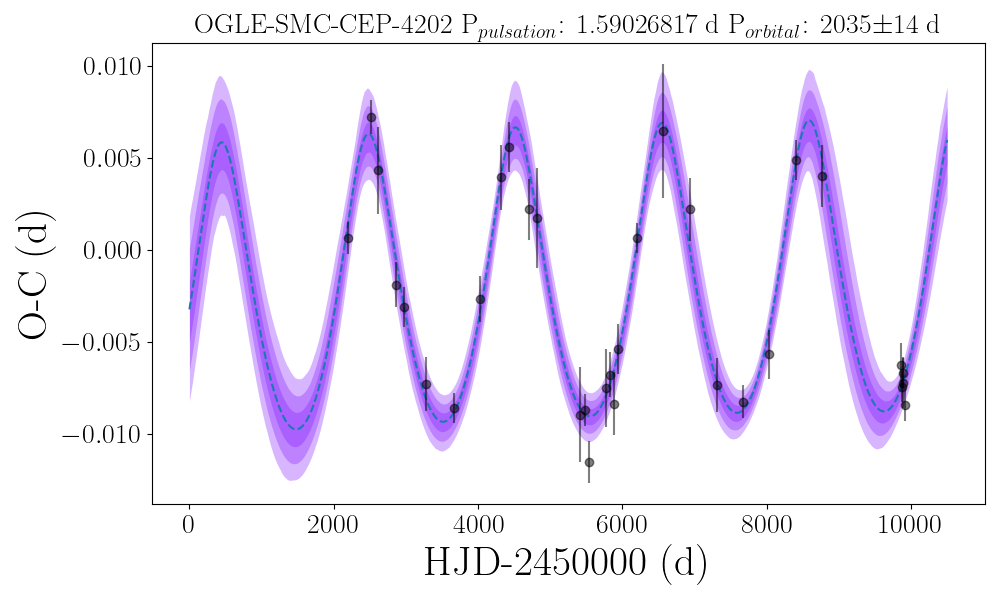}}
{\includegraphics[height=4.5cm,width=0.49\linewidth]{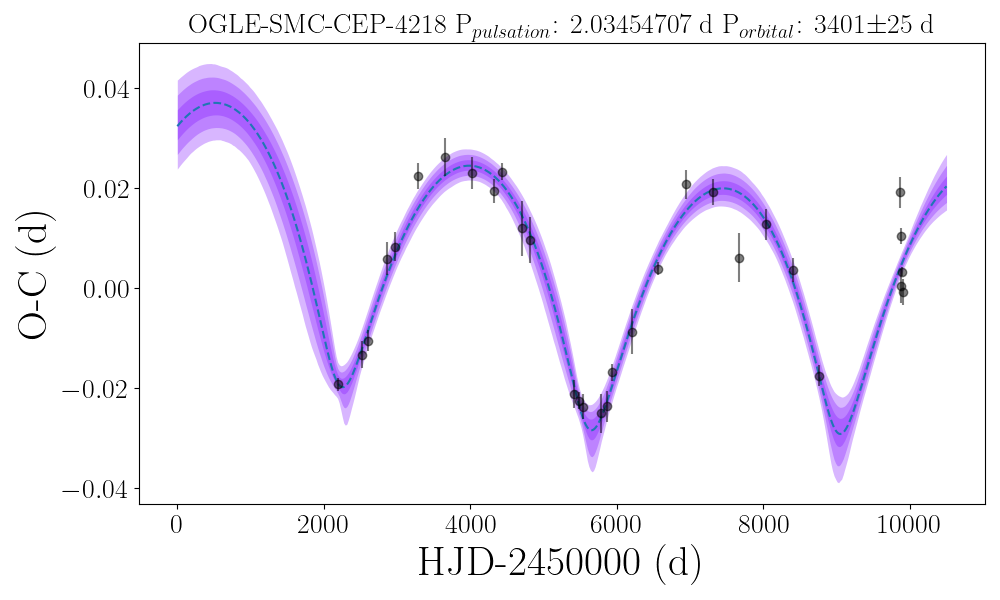}}
{\includegraphics[height=4.5cm,width=0.49\linewidth]{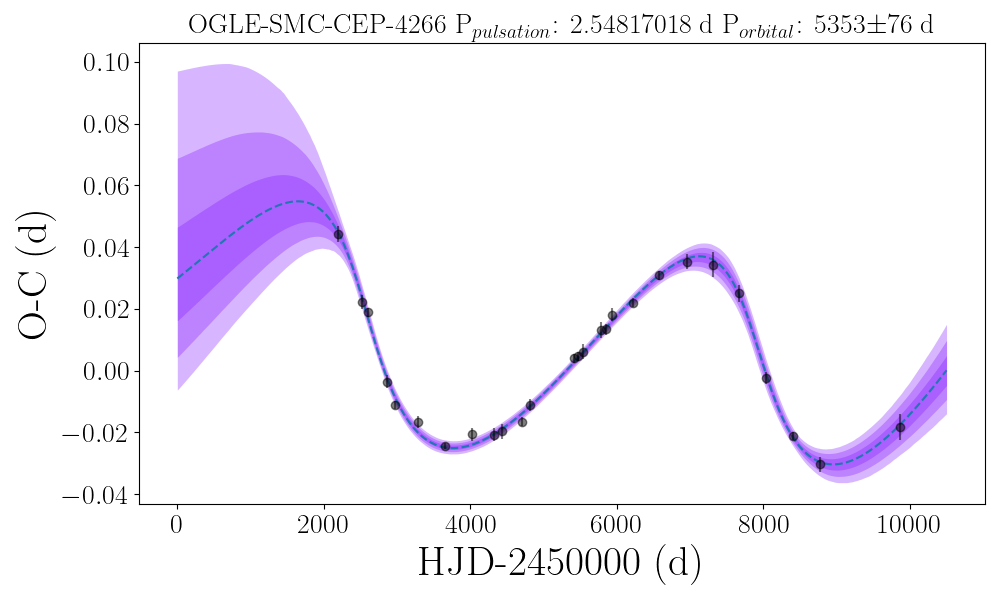}}
\caption{continued.}
\end{center}
\end{figure*}

\begin{figure*}[ht!]
\ContinuedFloat
\begin{center}
{\includegraphics[height=4.5cm,width=0.49\linewidth]{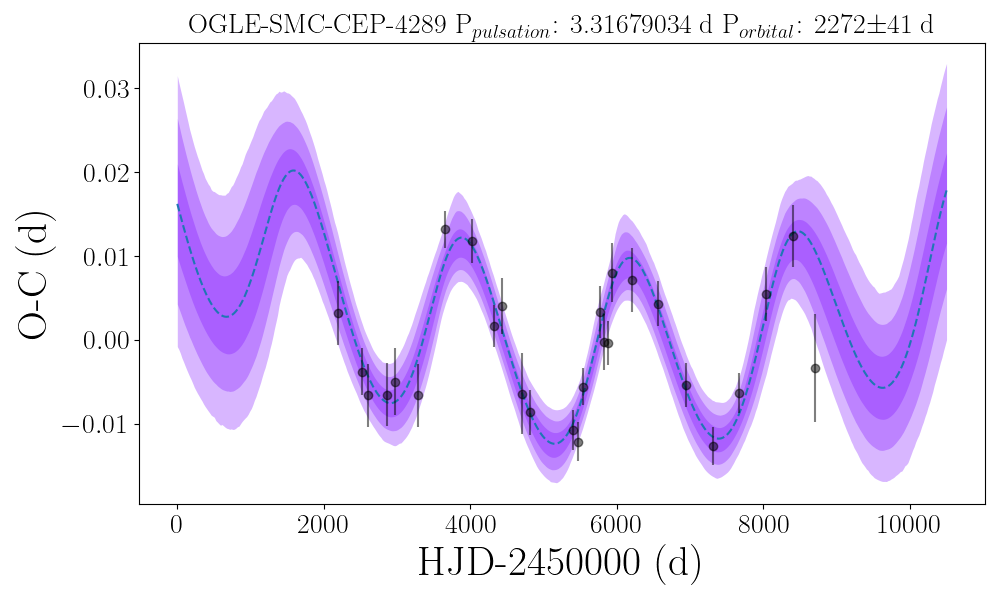}}
{\includegraphics[height=4.5cm,width=0.49\linewidth]{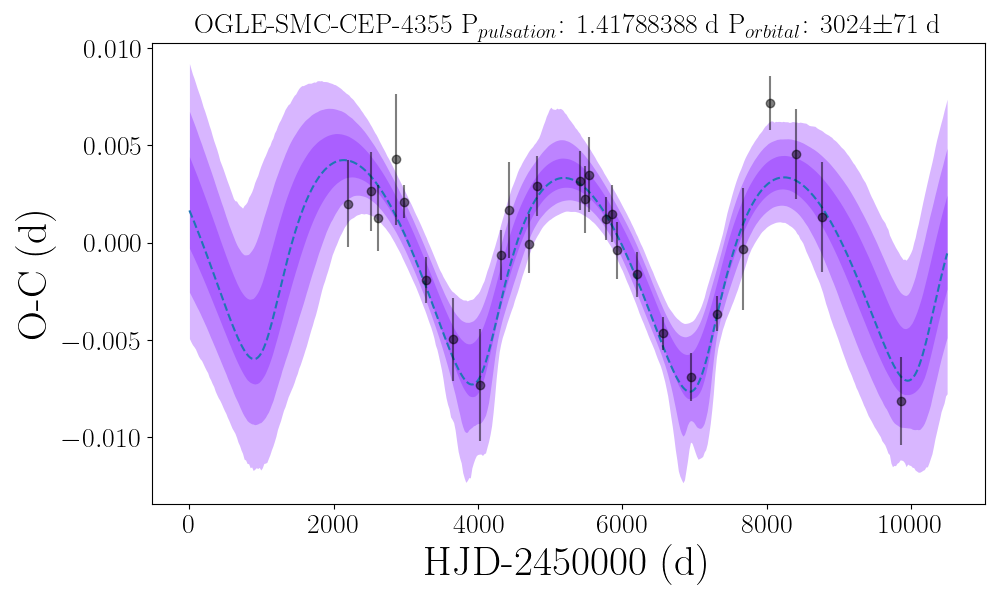}}
{\includegraphics[height=4.5cm,width=0.49\linewidth]{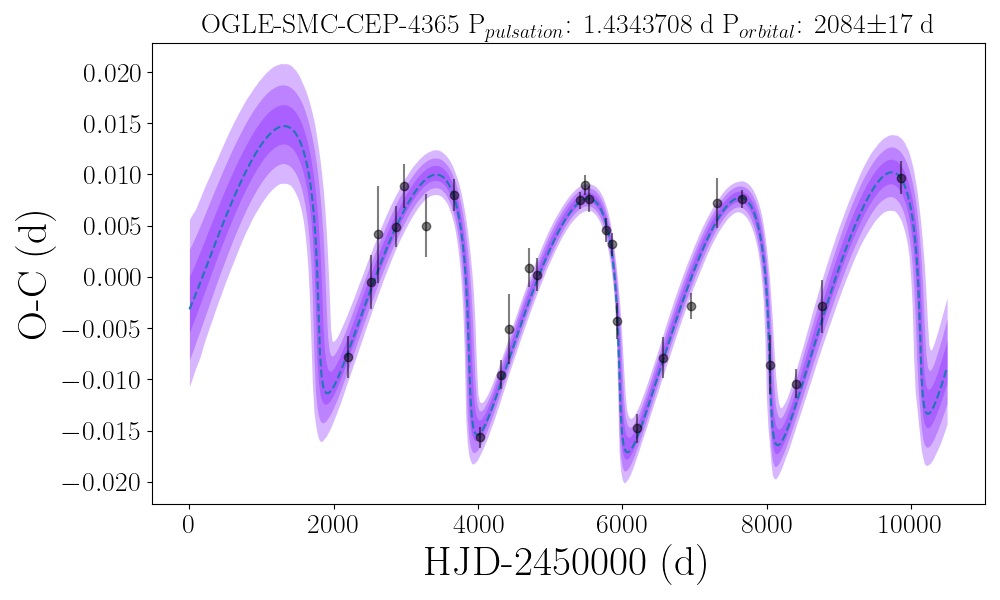}}
{\includegraphics[height=4.5cm,width=0.49\linewidth]{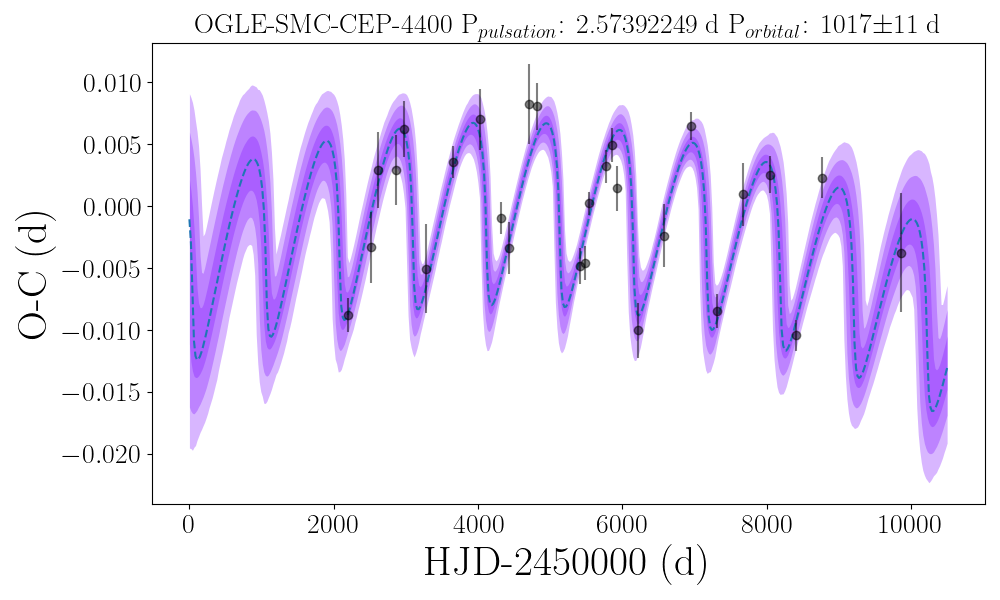}}
{\includegraphics[height=4.5cm,width=0.49\linewidth]{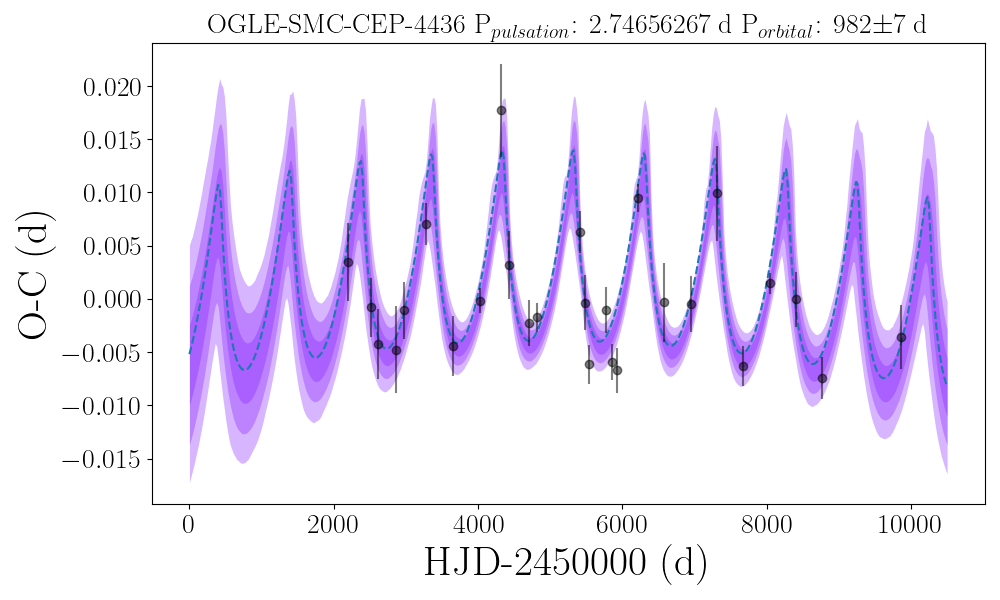}}
{\includegraphics[height=4.5cm,width=0.49\linewidth]{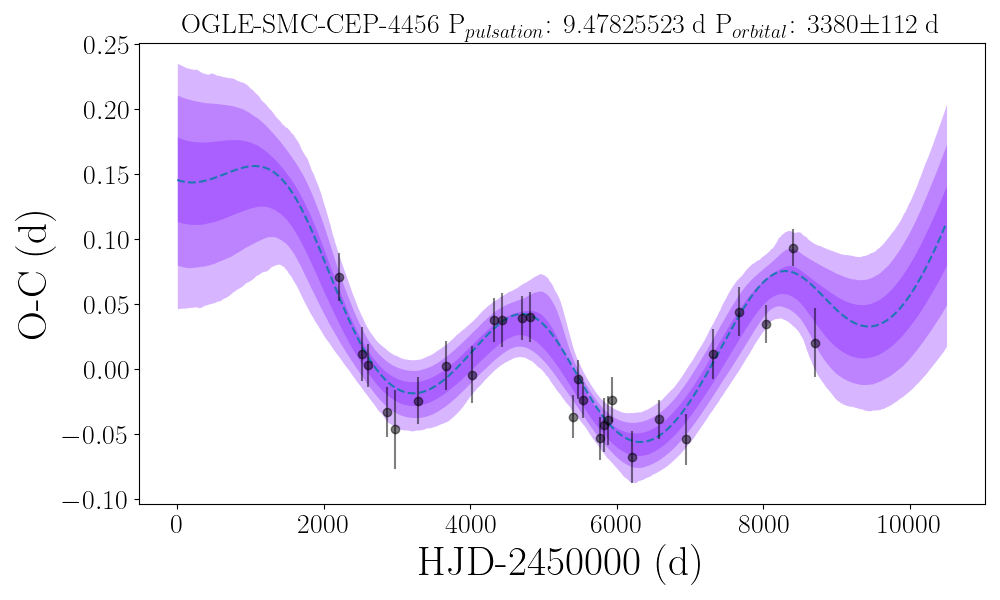}}
{\includegraphics[height=4.5cm,width=0.49\linewidth]{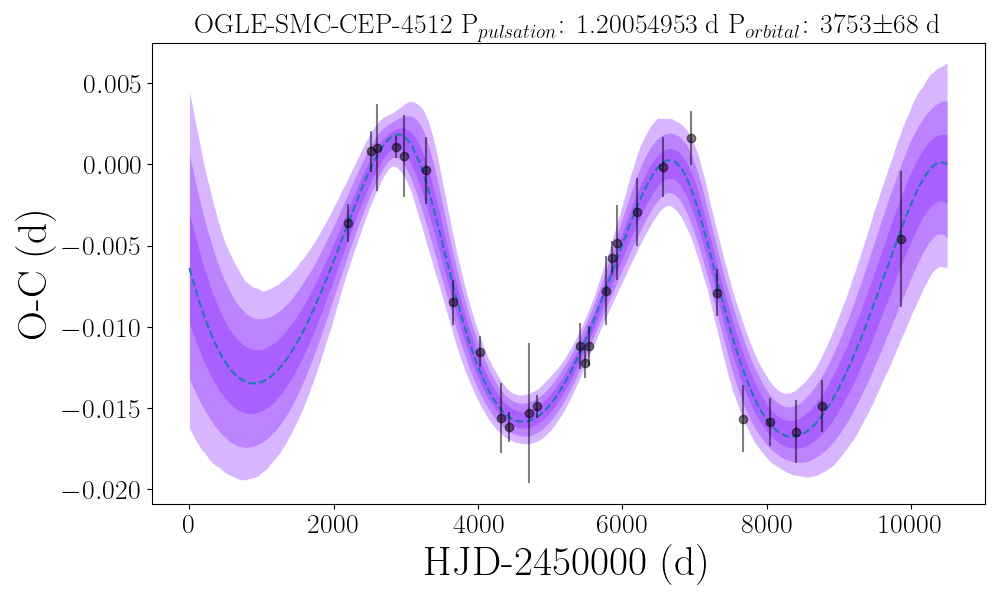}}
{\includegraphics[height=4.5cm,width=0.49\linewidth]{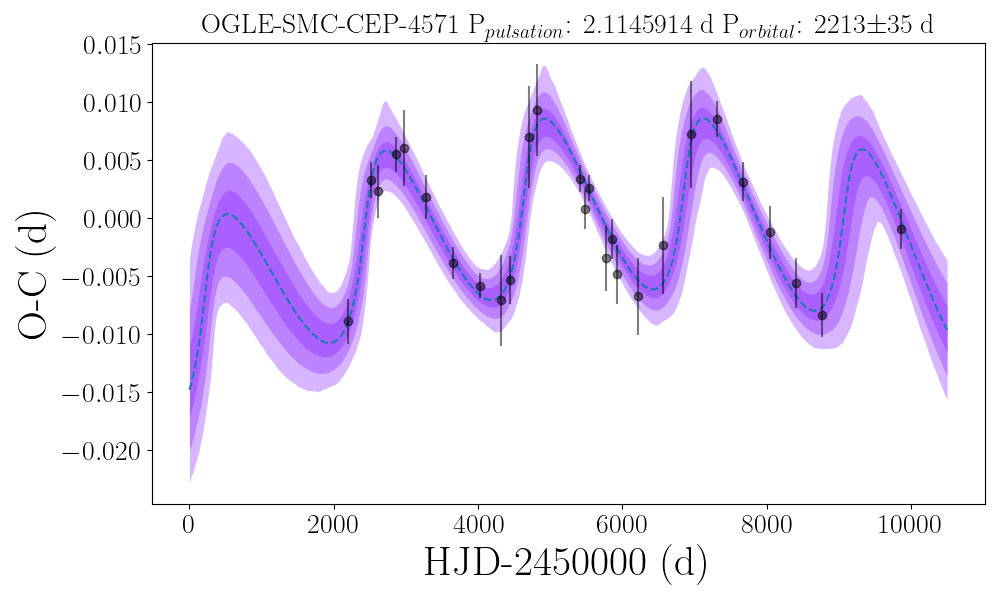}}
\caption{continued.}
\end{center}
\end{figure*}


\section{$O-C$ curves for SMC 1O mode}

\begin{figure*}[ht!]
\begin{center}
{\includegraphics[height=4.5cm,width=0.49\linewidth]{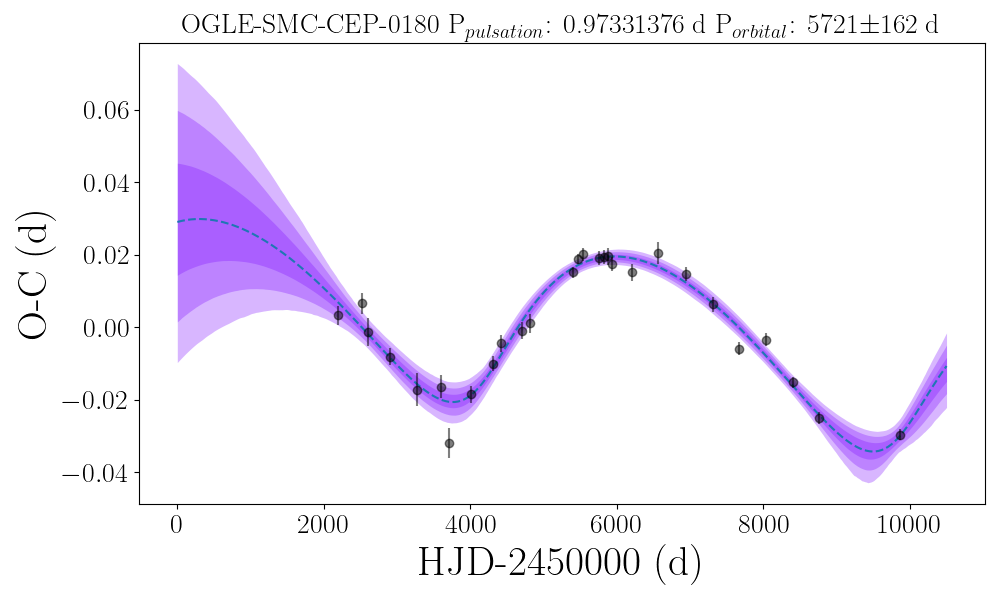}}
{\includegraphics[height=4.5cm,width=0.49\linewidth]{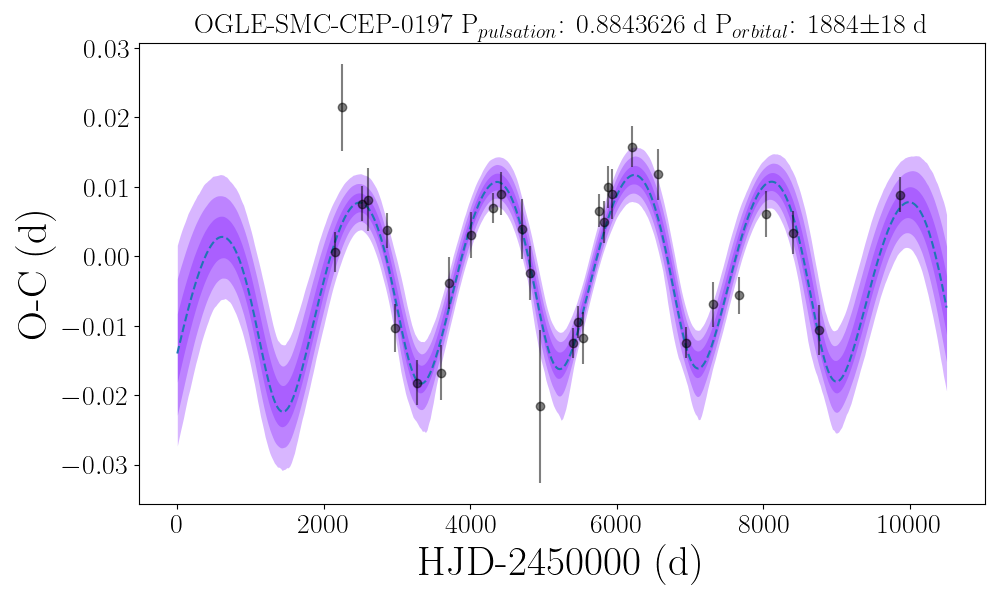}}
{\includegraphics[height=4.5cm,width=0.49\linewidth]{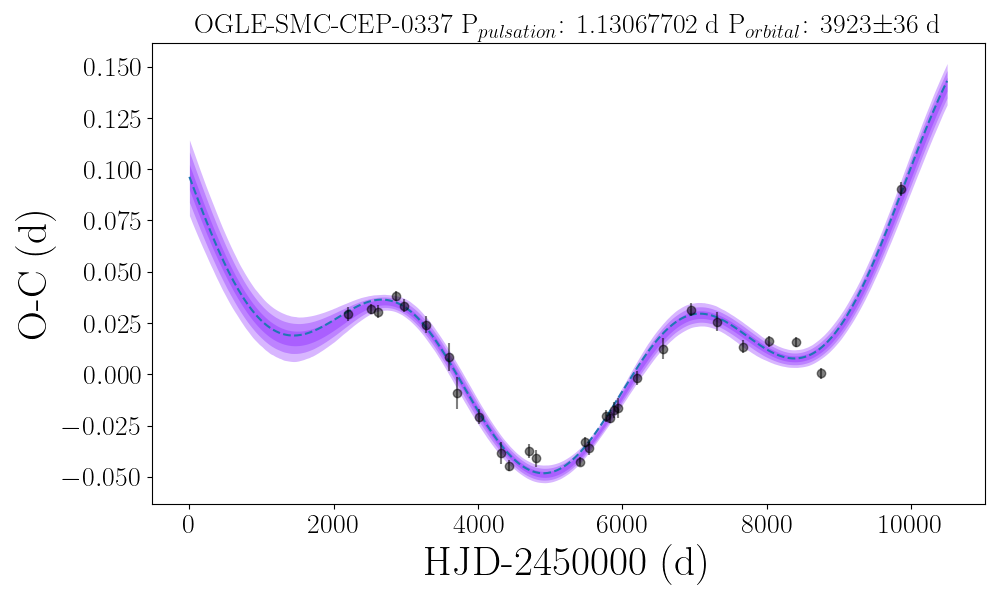}}
{\includegraphics[height=4.5cm,width=0.49\linewidth]{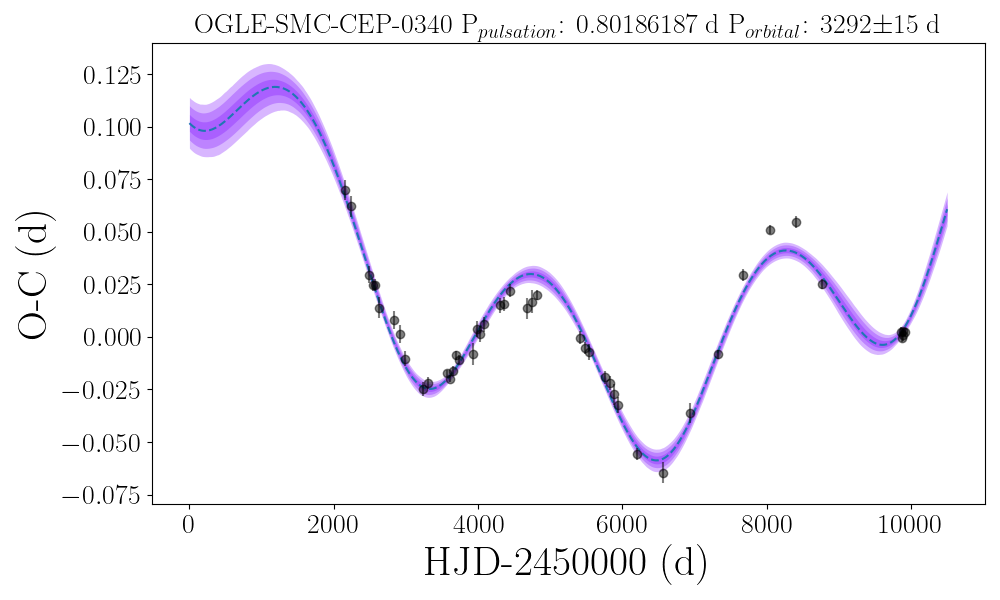}}
{\includegraphics[height=4.5cm,width=0.49\linewidth]{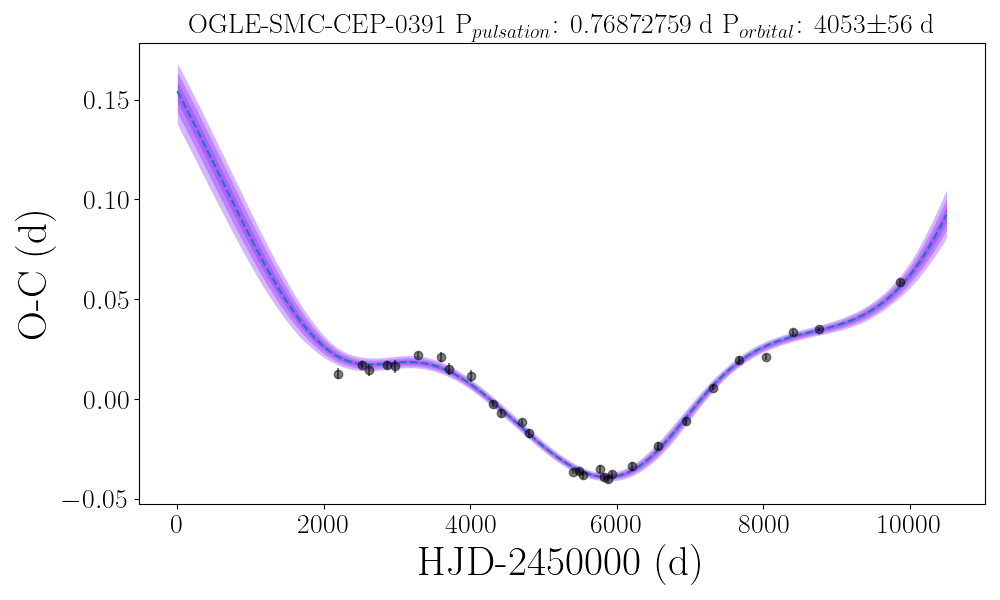}}
{\includegraphics[height=4.5cm,width=0.49\linewidth]{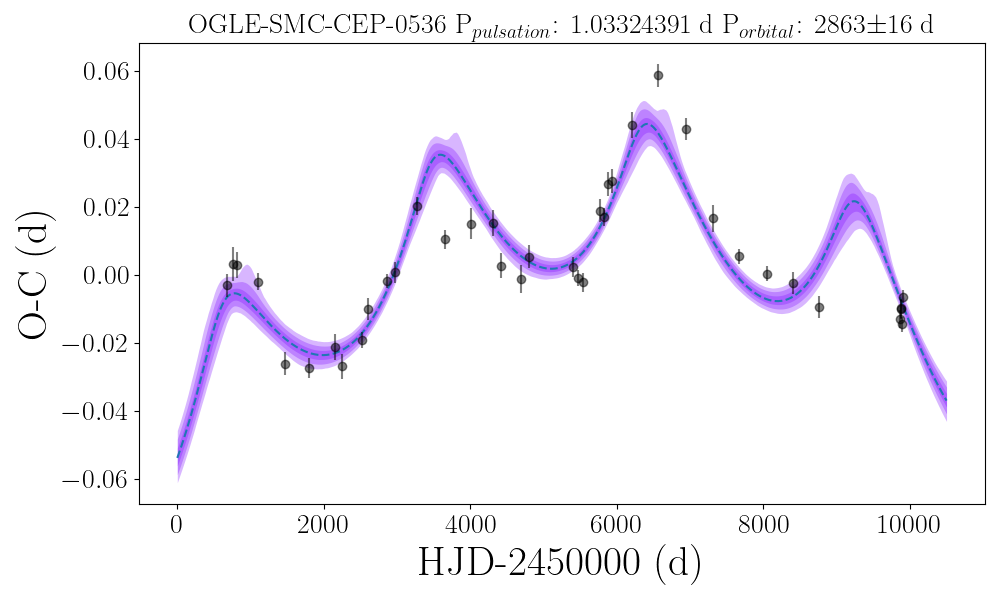}}
{\includegraphics[height=4.5cm,width=0.49\linewidth]{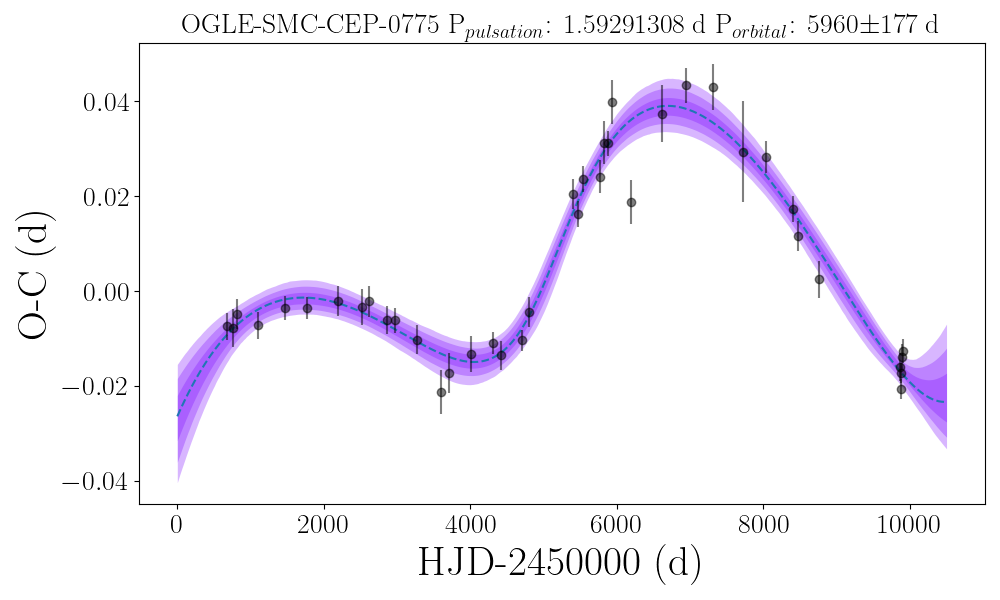}}
{\includegraphics[height=4.5cm,width=0.49\linewidth]{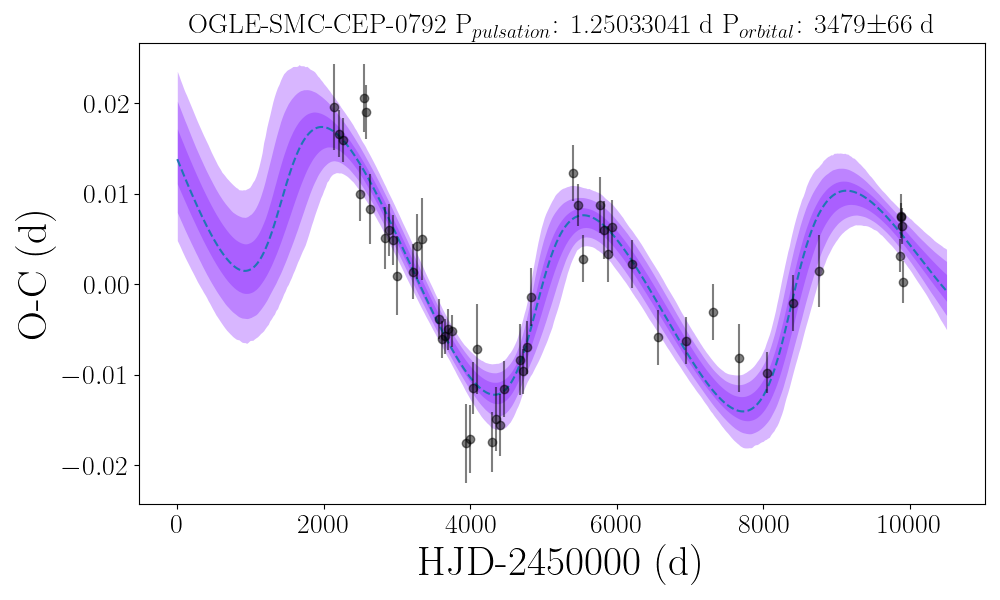}}
{\includegraphics[height=4.5cm,width=0.49\linewidth]{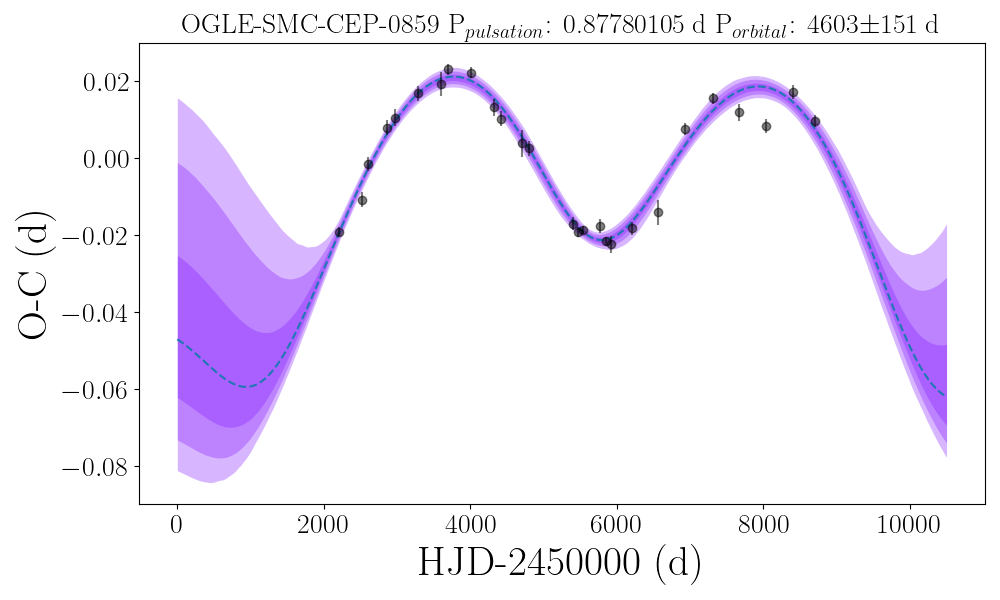}}
{\includegraphics[height=4.5cm,width=0.49\linewidth]{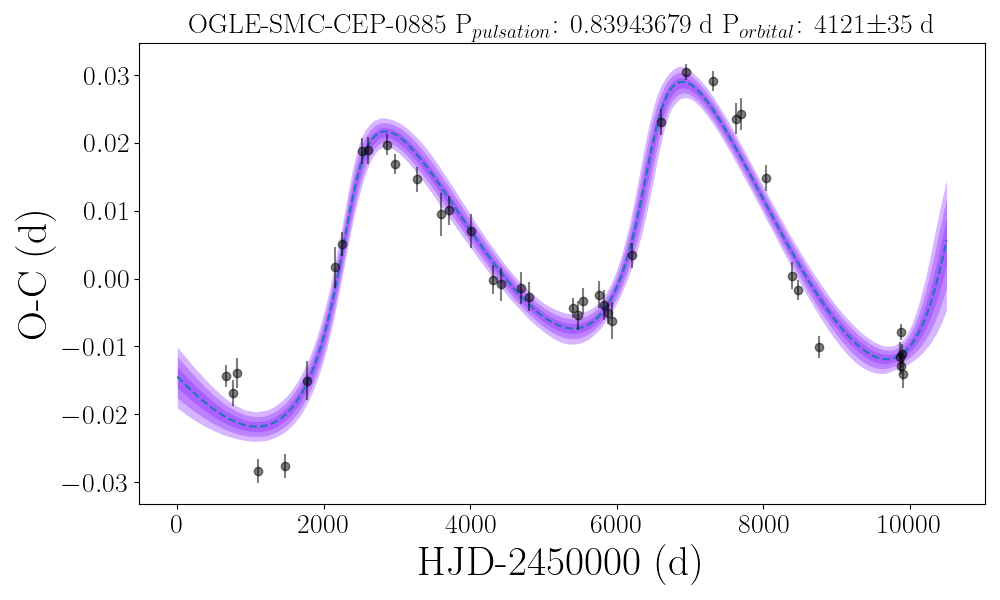}}
\caption{Remaining $O-C$ diagrams for SMC first-overtone mode Cepheid binary candidate sample}
\label{fig:appendix_ocplot_1Omode_SMC}
\end{center}
\end{figure*}

\begin{figure*}[ht!]
\ContinuedFloat
\begin{center}
{\includegraphics[height=4.5cm,width=0.49\linewidth]{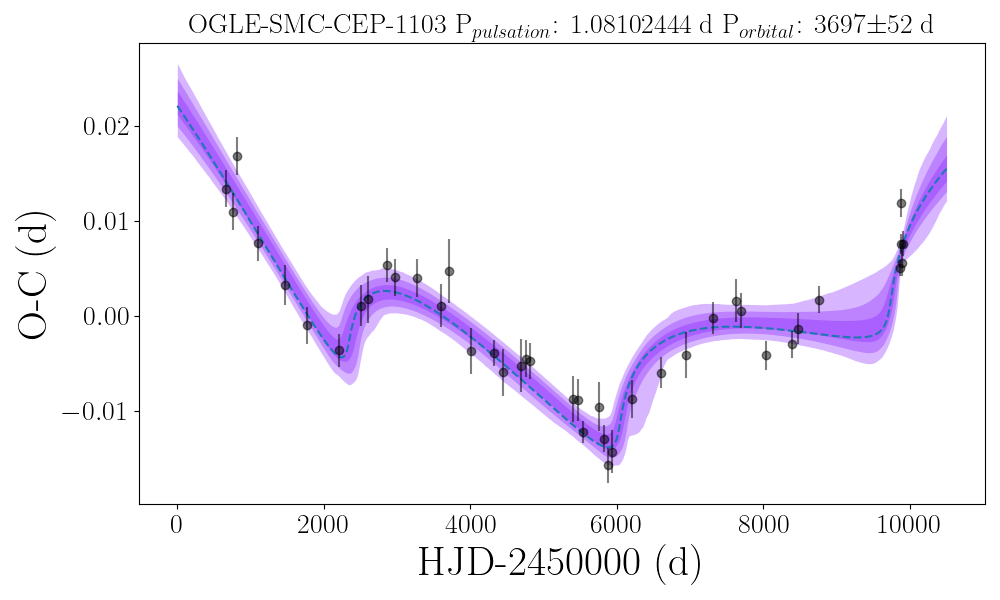}}
{\includegraphics[height=4.5cm,width=0.49\linewidth]{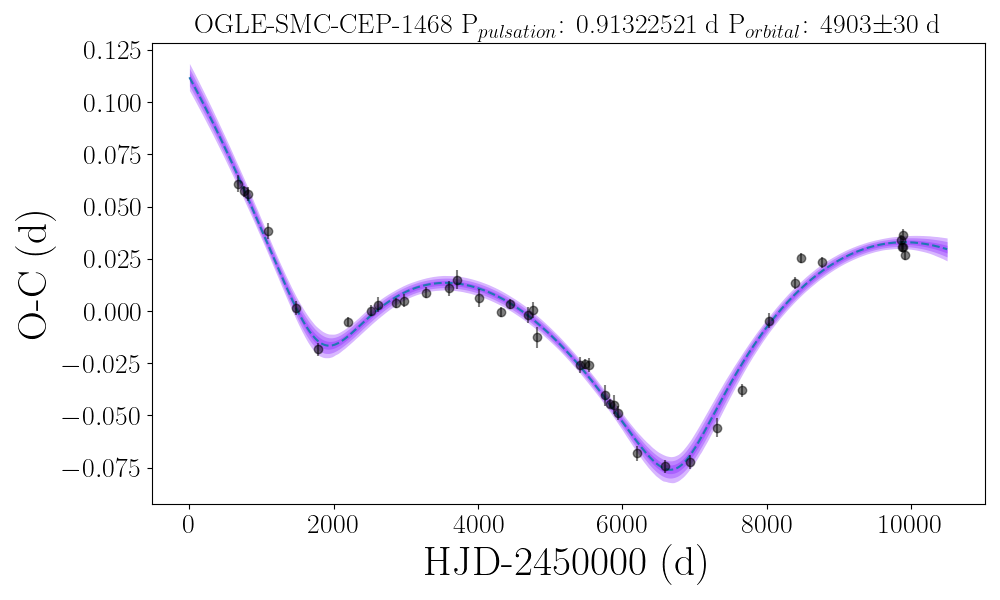}}
{\includegraphics[height=4.5cm,width=0.49\linewidth]{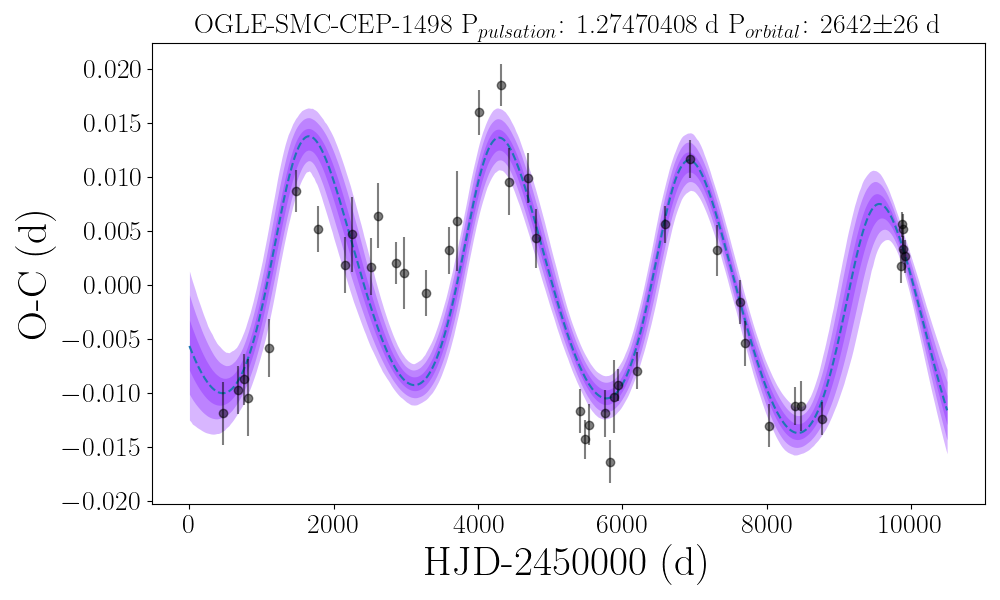}}
{\includegraphics[height=4.5cm,width=0.49\linewidth]{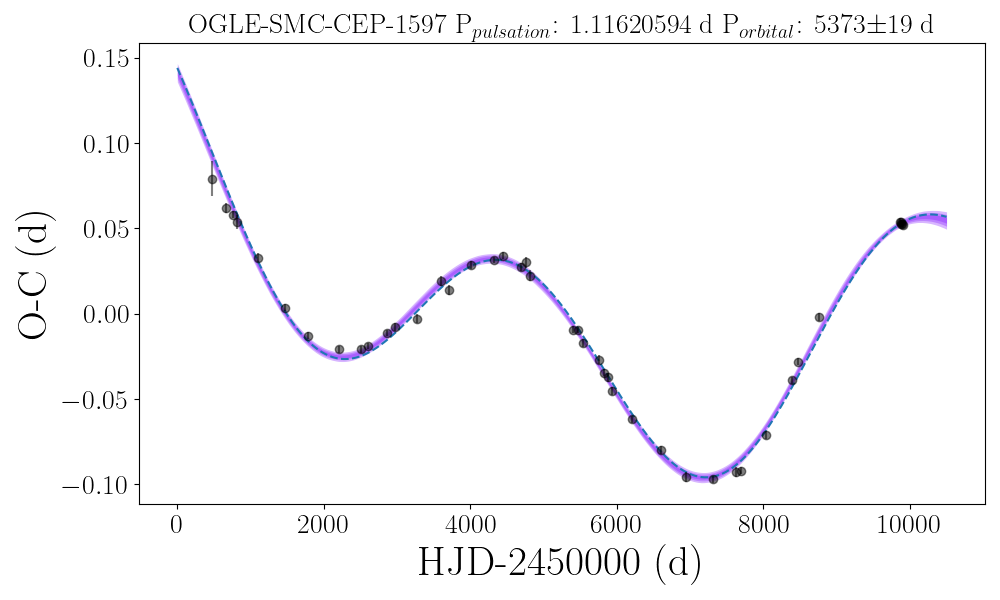}}
{\includegraphics[height=4.5cm,width=0.49\linewidth]{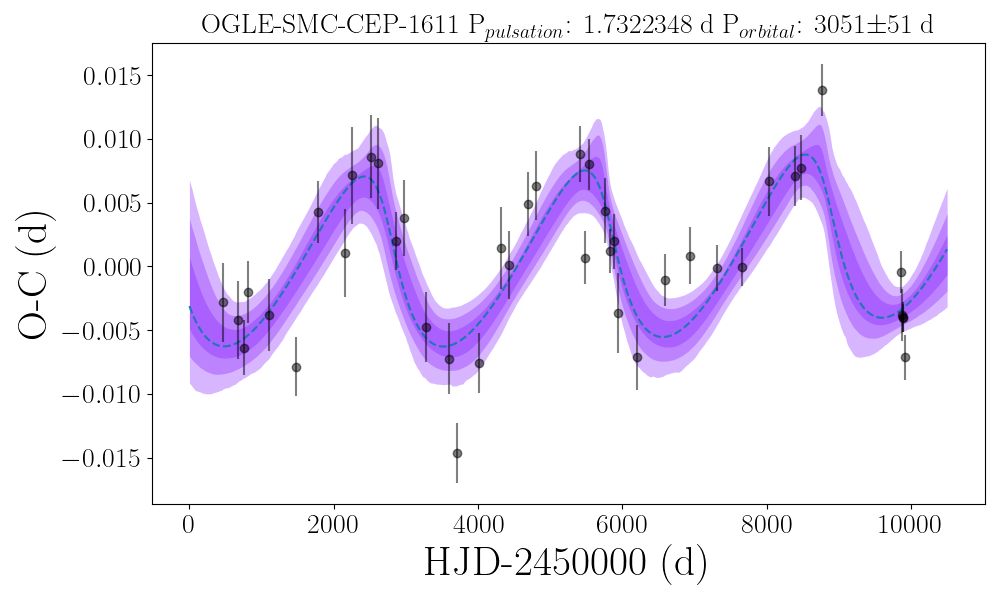}}
{\includegraphics[height=4.5cm,width=0.49\linewidth]{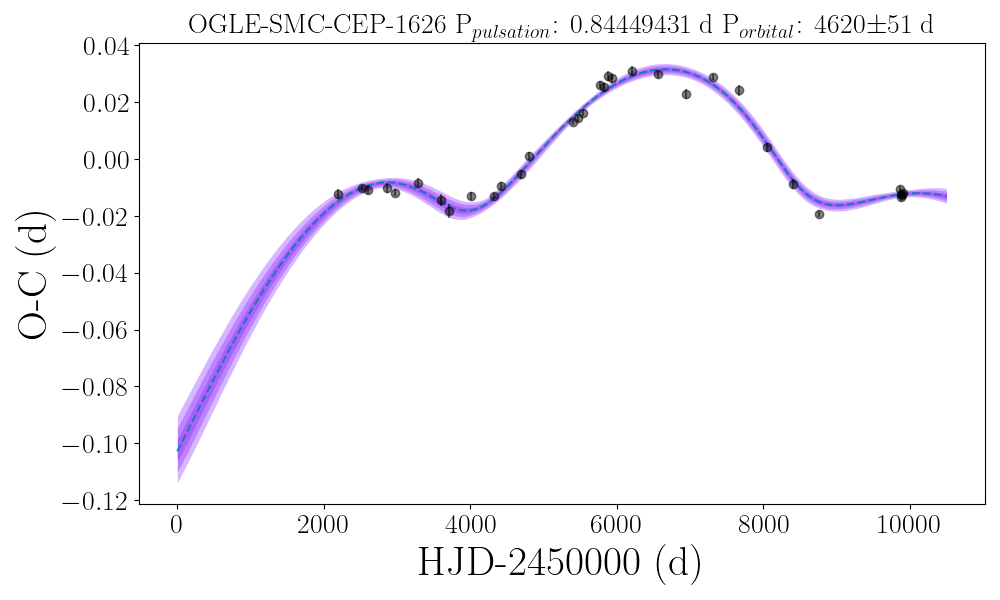}}
{\includegraphics[height=4.5cm,width=0.49\linewidth]{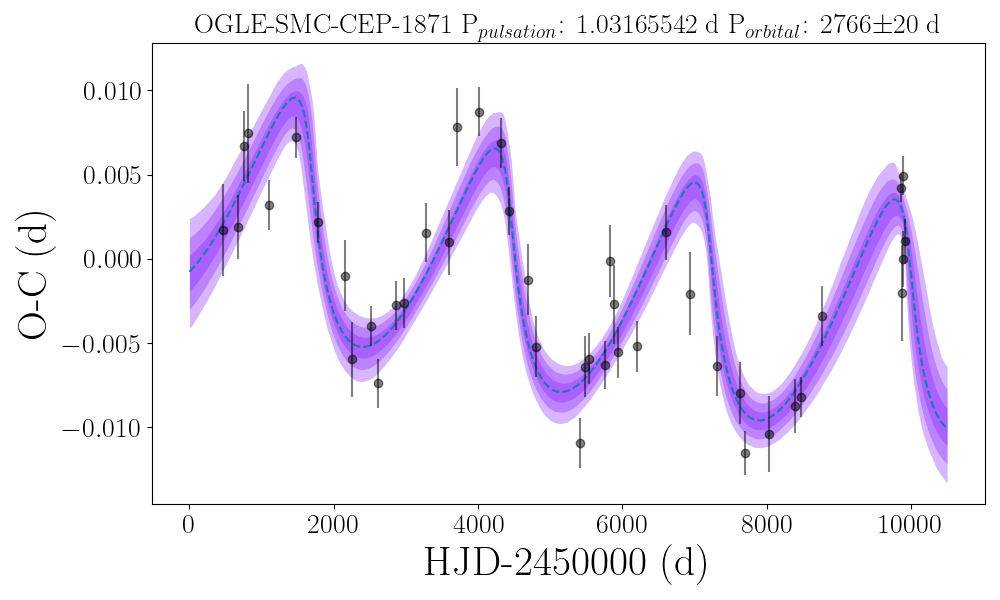}}
{\includegraphics[height=4.5cm,width=0.49\linewidth]{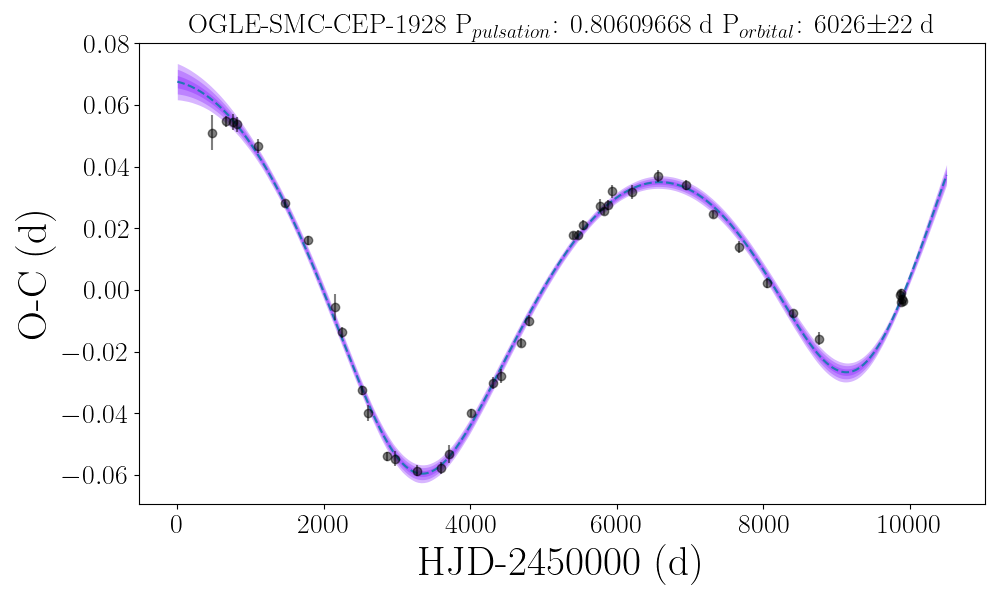}}
{\includegraphics[height=4.5cm,width=0.49\linewidth]{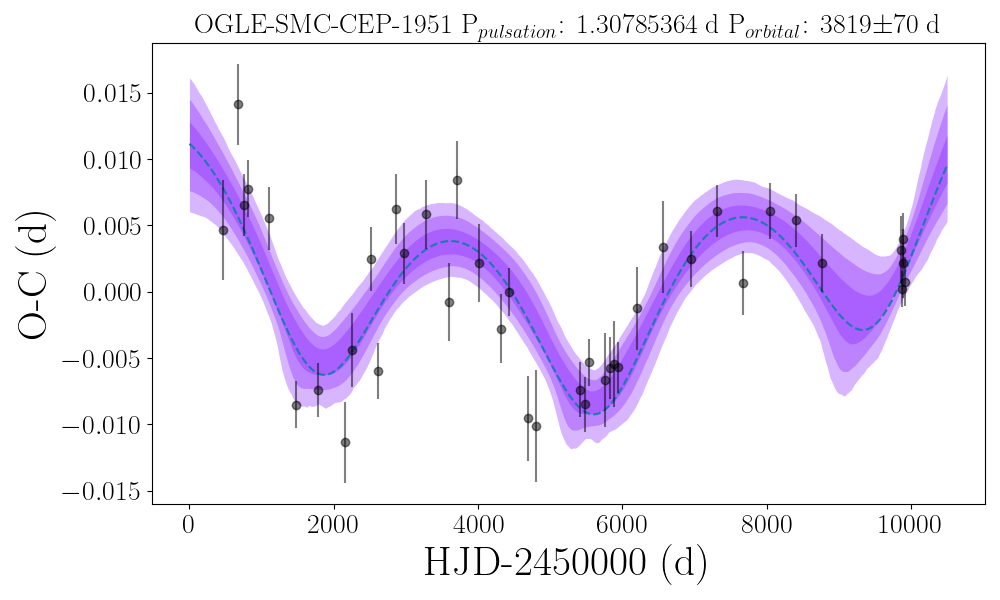}}
{\includegraphics[height=4.5cm,width=0.49\linewidth]{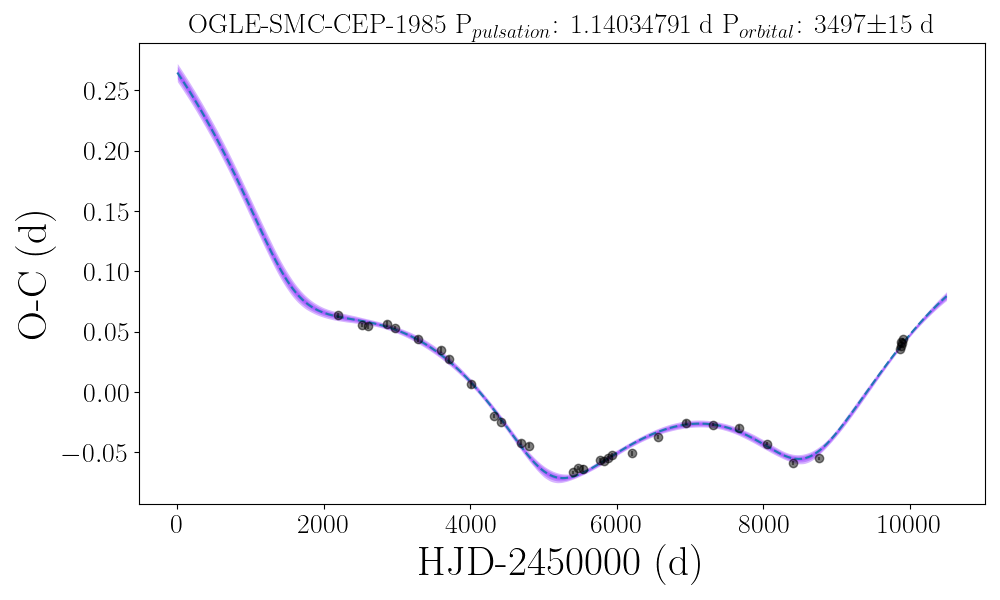}}
\caption{continued.}
\end{center}
\end{figure*}

\begin{figure*}[ht!]
\ContinuedFloat
\begin{center}
{\includegraphics[height=4.5cm,width=0.49\linewidth]{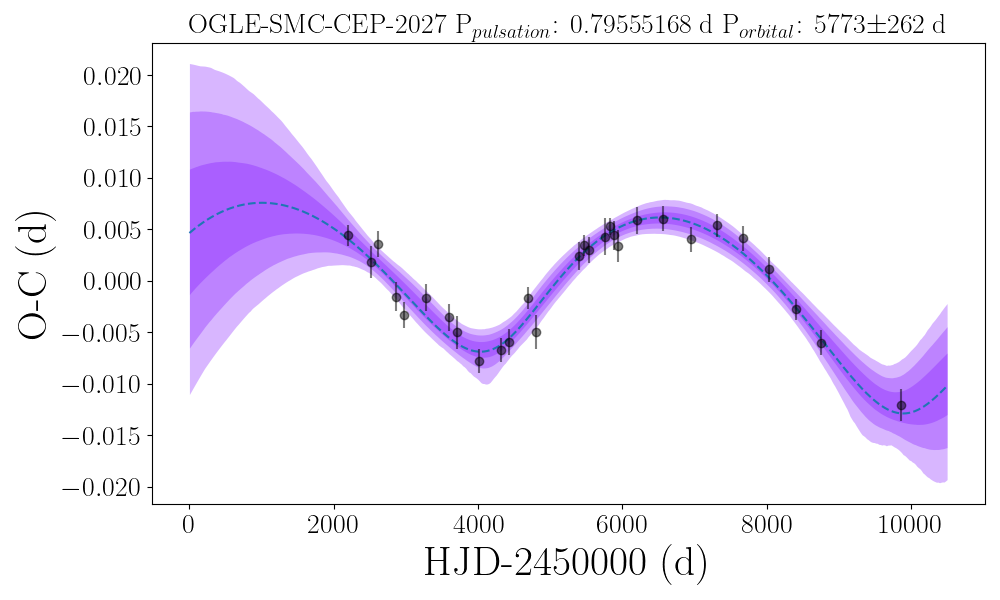}}
{\includegraphics[height=4.5cm,width=0.49\linewidth]{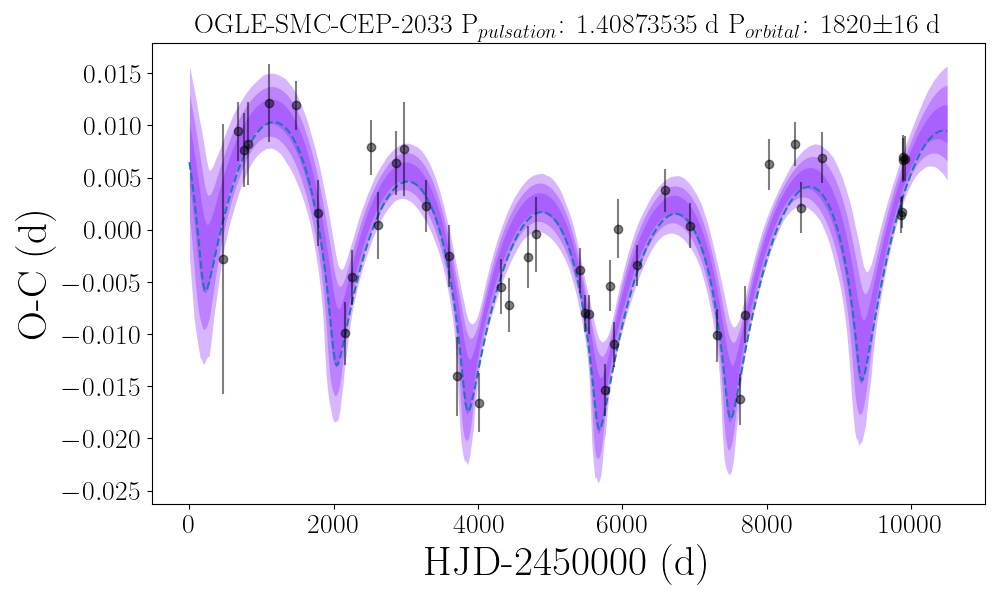}}
{\includegraphics[height=4.5cm,width=0.49\linewidth]{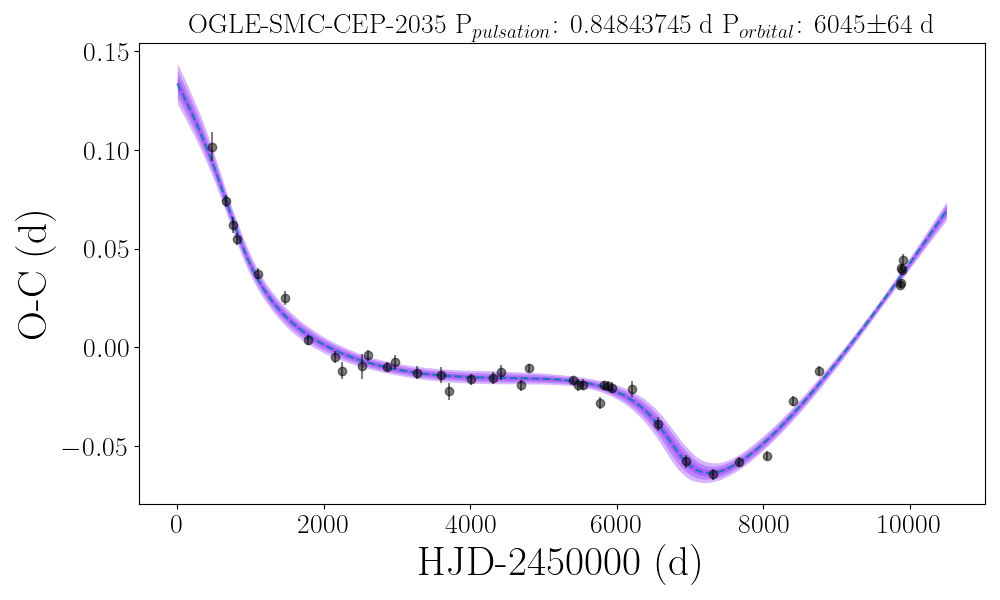}}
{\includegraphics[height=4.5cm,width=0.49\linewidth]{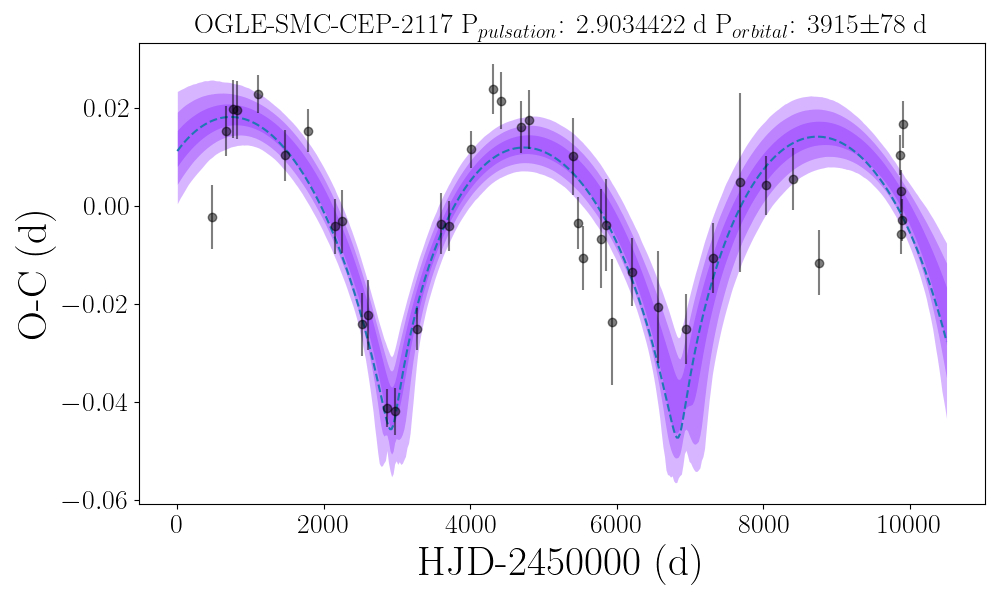}}
{\includegraphics[height=4.5cm,width=0.49\linewidth]{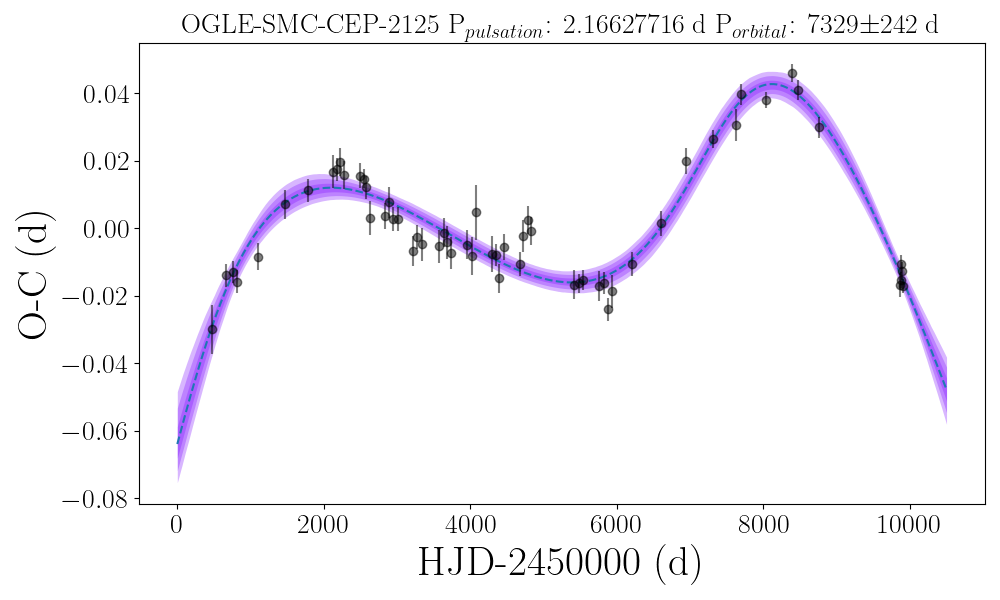}}
{\includegraphics[height=4.5cm,width=0.49\linewidth]{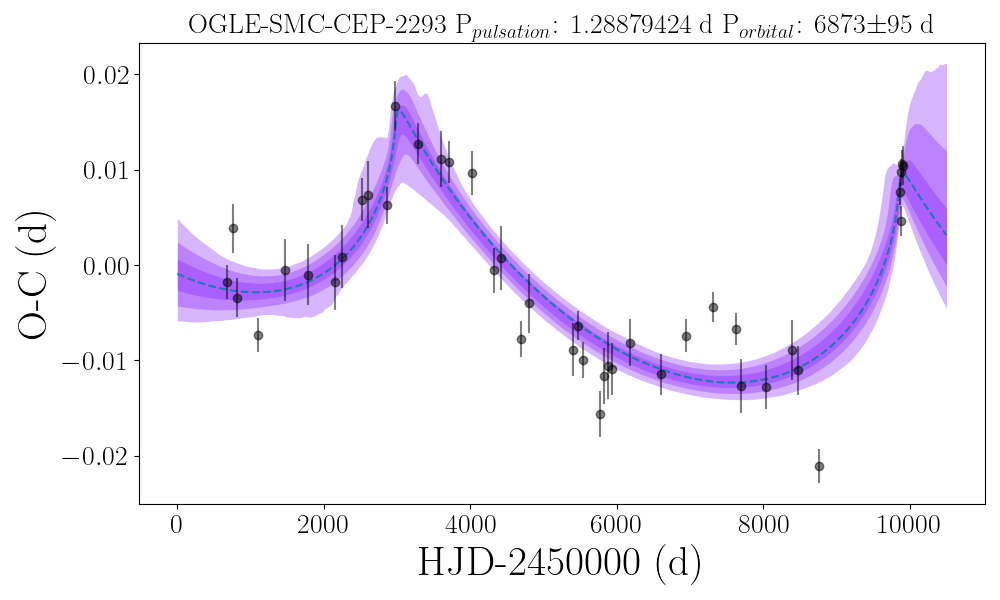}}
{\includegraphics[height=4.5cm,width=0.49\linewidth]{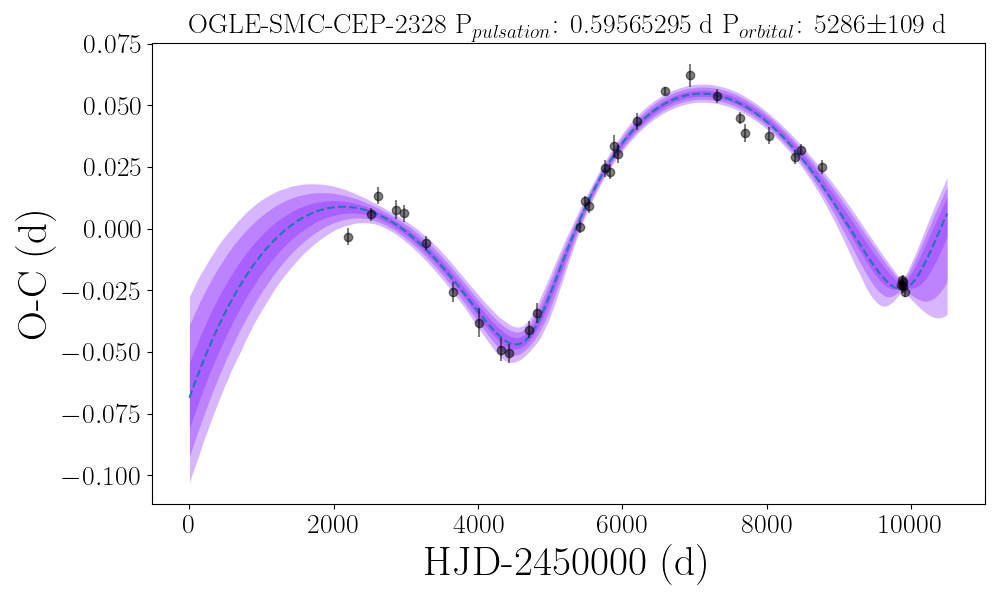}}
{\includegraphics[height=4.5cm,width=0.49\linewidth]{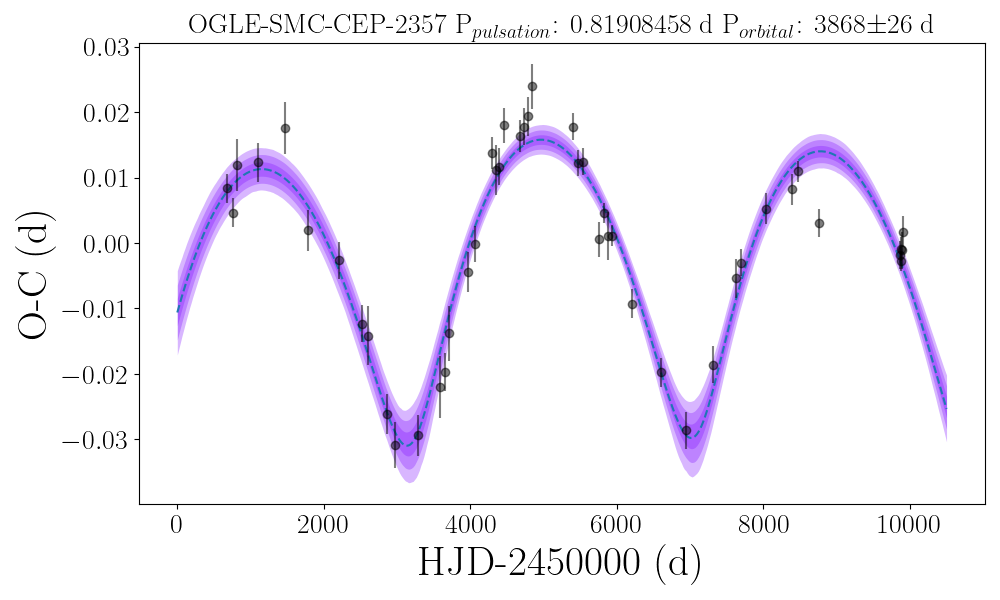}}
{\includegraphics[height=4.5cm,width=0.49\linewidth]{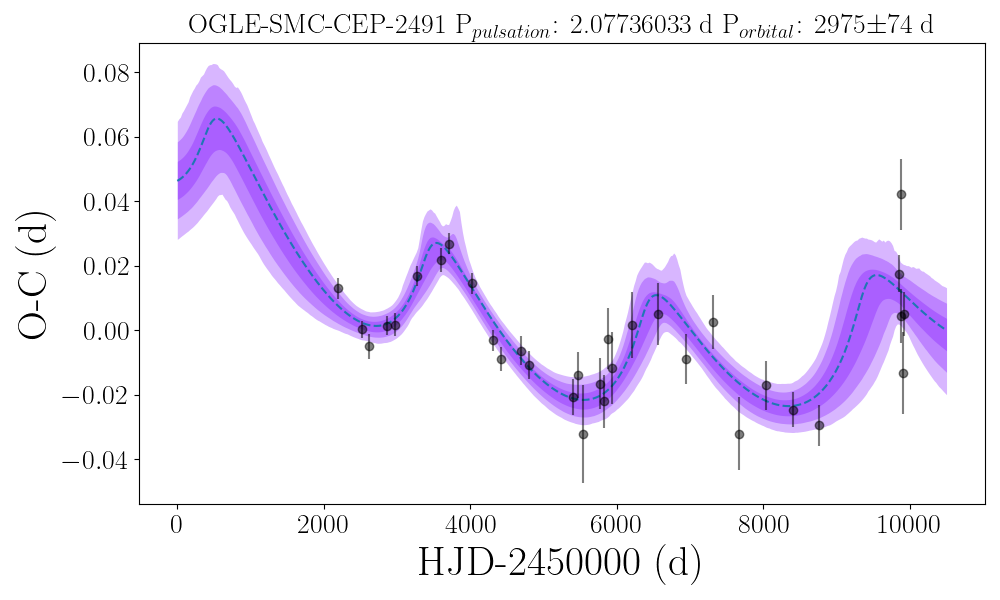}}
{\includegraphics[height=4.5cm,width=0.49\linewidth]{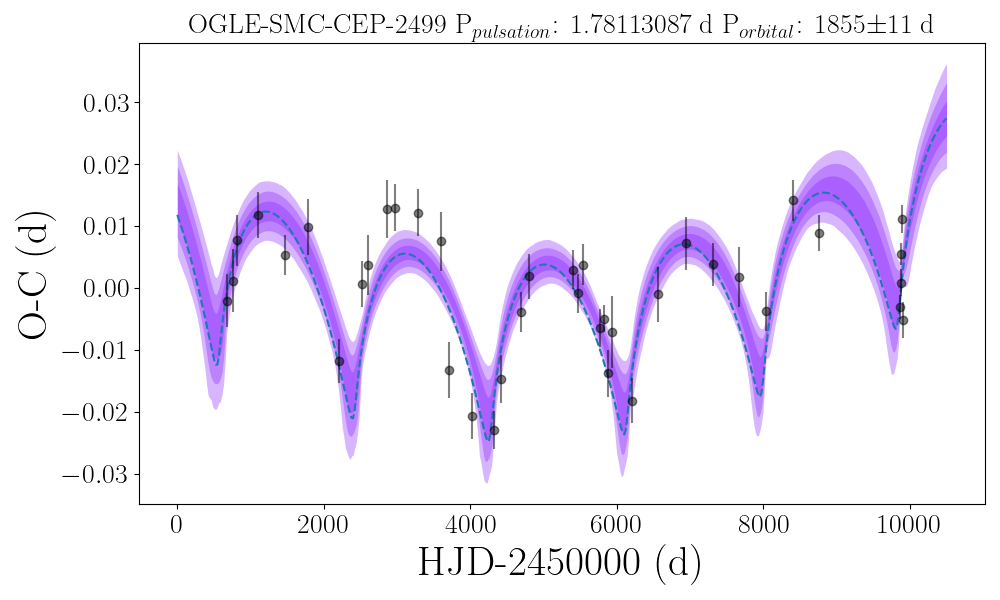}}
\caption{continued.}
\end{center}
\end{figure*}

\begin{figure*}[ht!]
\ContinuedFloat
\begin{center}
{\includegraphics[height=4.5cm,width=0.49\linewidth]{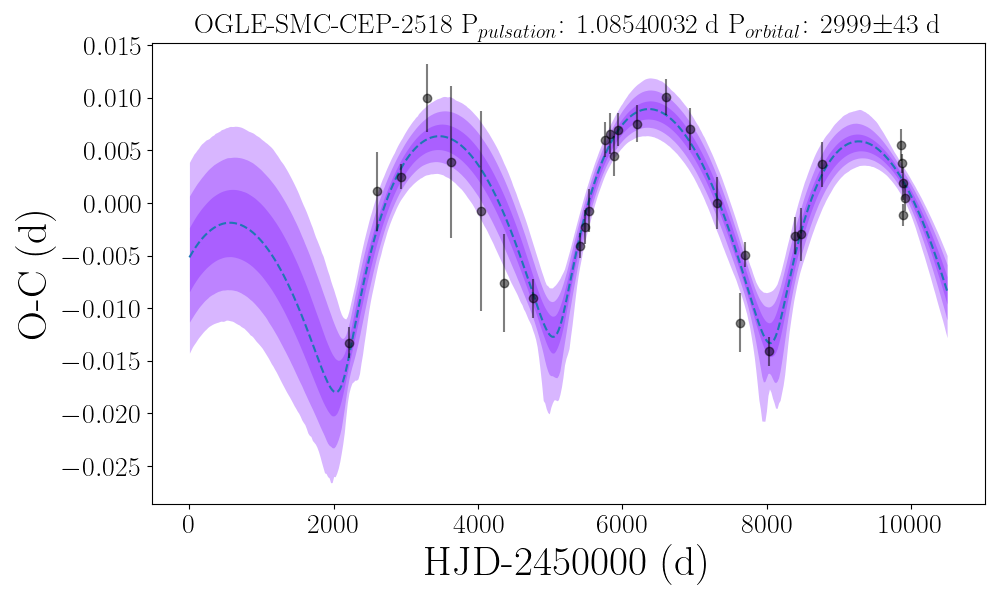}}
{\includegraphics[height=4.5cm,width=0.49\linewidth]{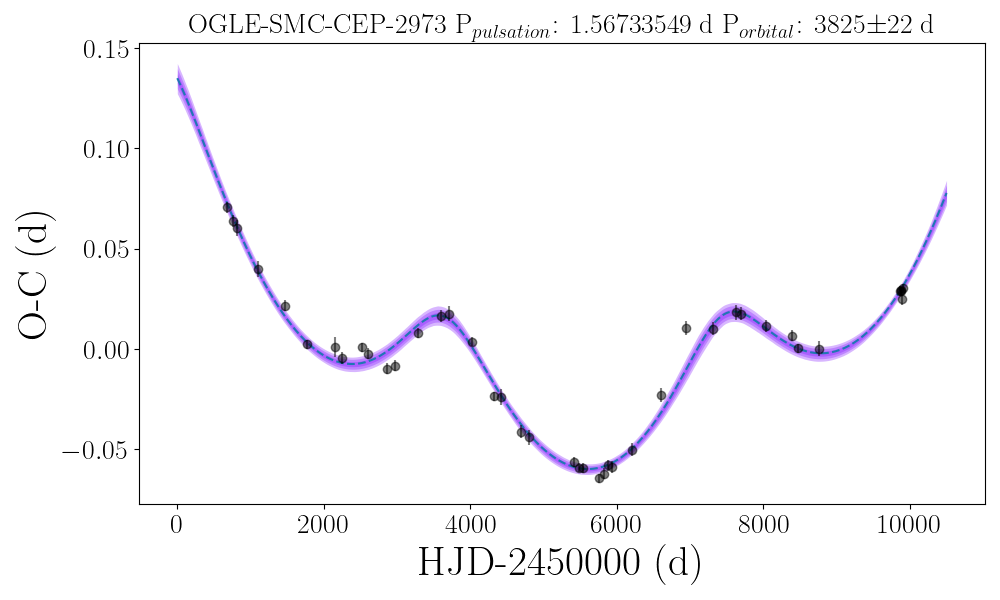}}
{\includegraphics[height=4.5cm,width=0.49\linewidth]{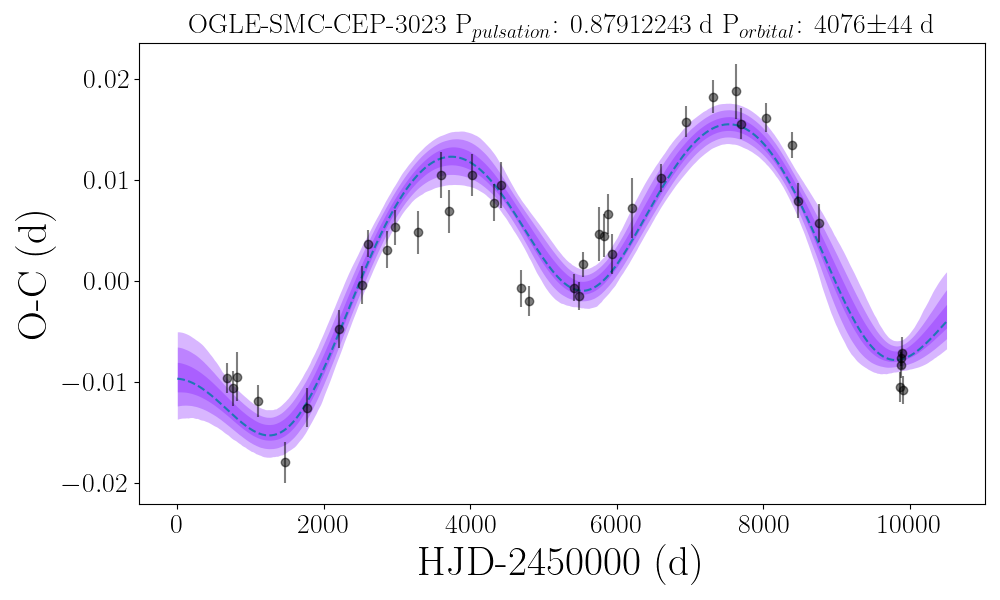}}
{\includegraphics[height=4.5cm,width=0.49\linewidth]{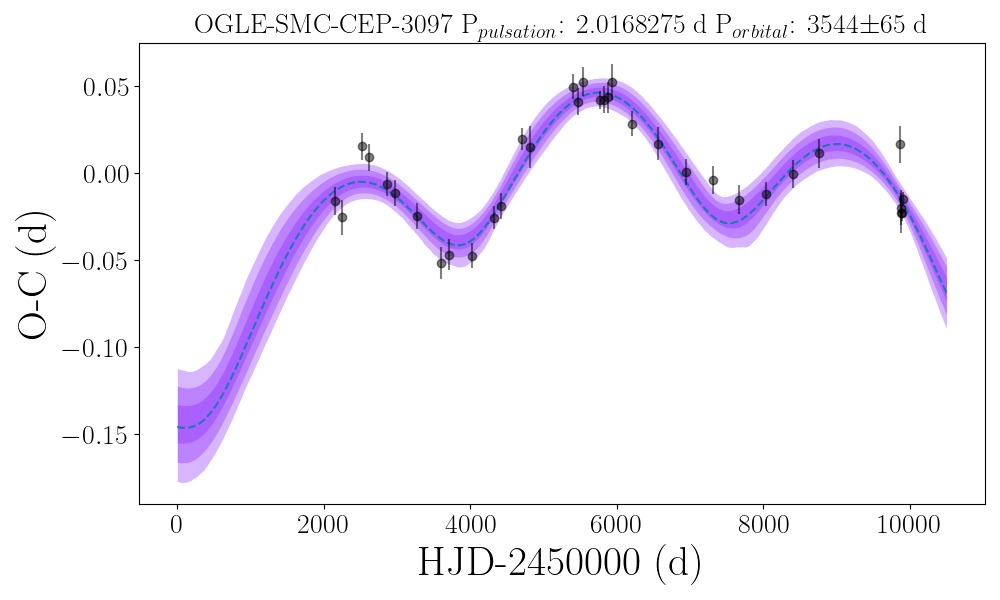}}
{\includegraphics[height=4.5cm,width=0.49\linewidth]{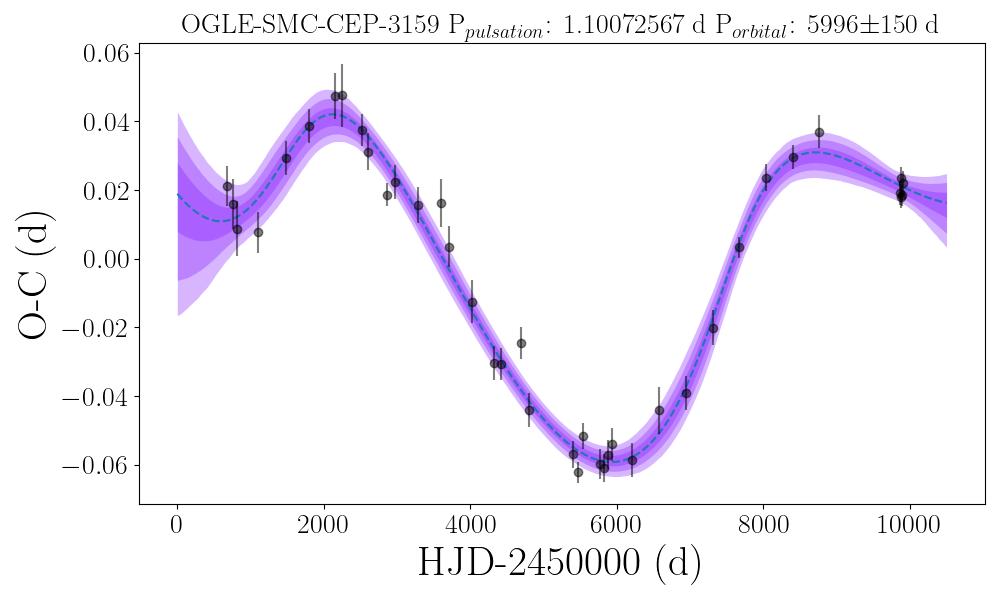}}
{\includegraphics[height=4.5cm,width=0.49\linewidth]{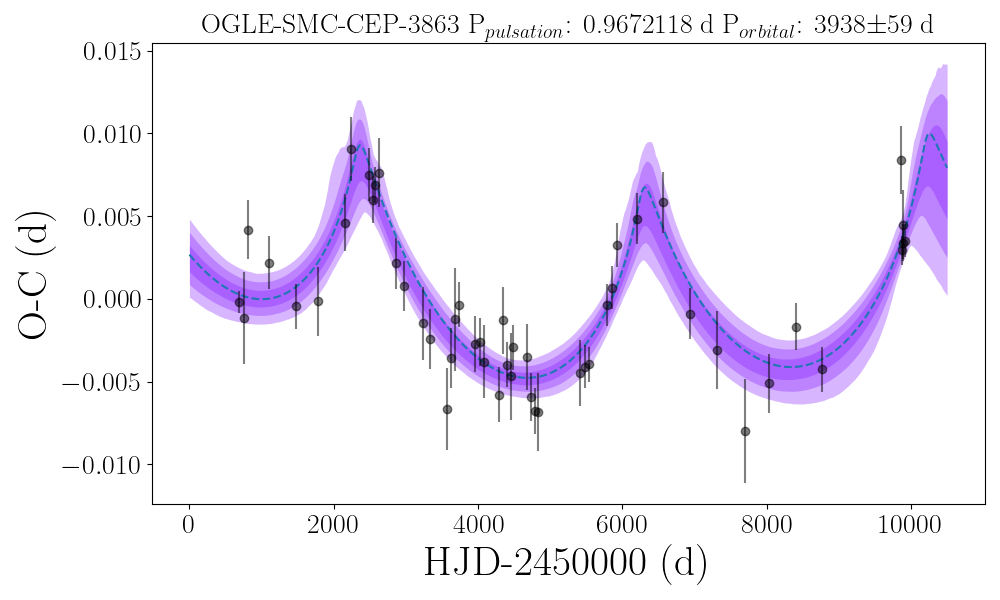}}
{\includegraphics[height=4.5cm,width=0.49\linewidth]{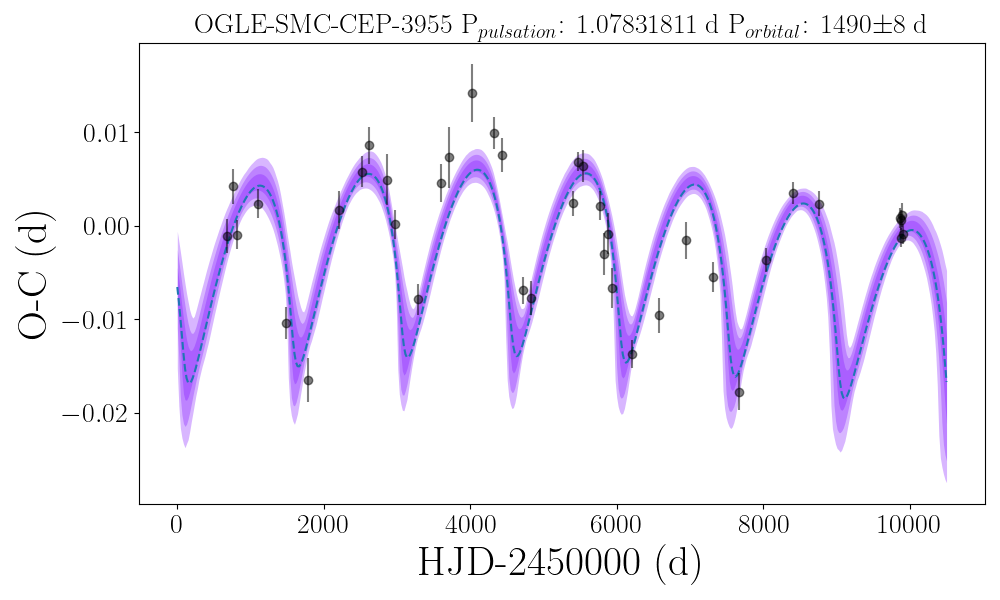}}
{\includegraphics[height=4.5cm,width=0.49\linewidth]{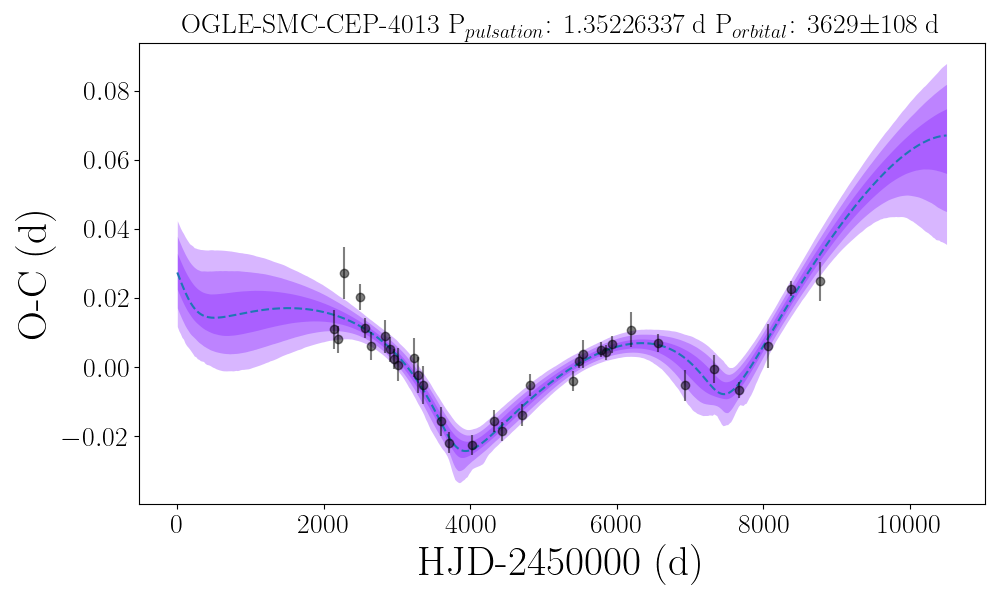}}
{\includegraphics[height=4.5cm,width=0.49\linewidth]{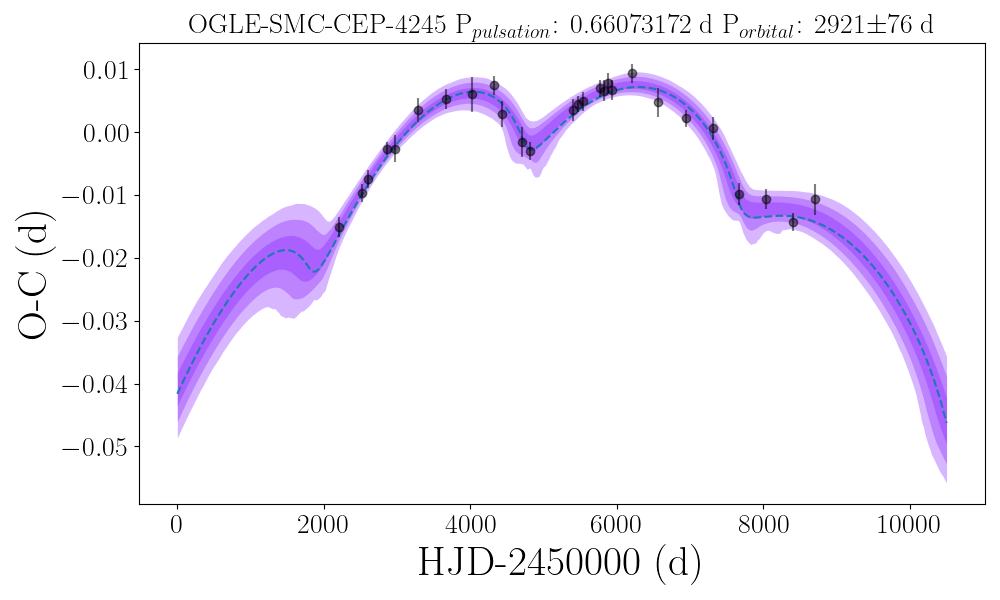}}
{\includegraphics[height=4.5cm,width=0.49\linewidth]{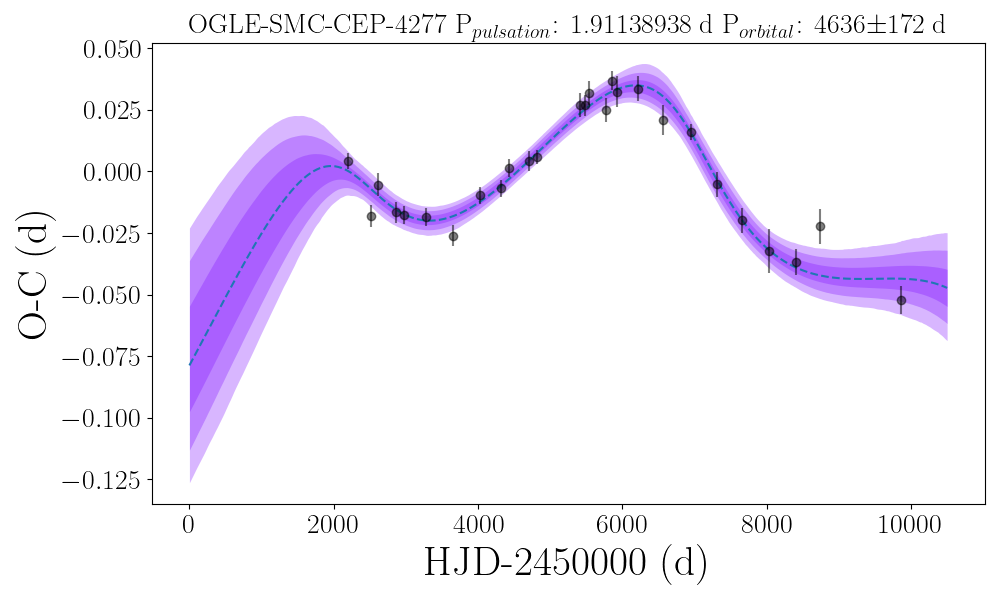}}
\caption{continued.}
\end{center}
\end{figure*}

\begin{figure*}[ht!]
\ContinuedFloat
\begin{center}
{\includegraphics[height=4.5cm,width=0.49\linewidth]{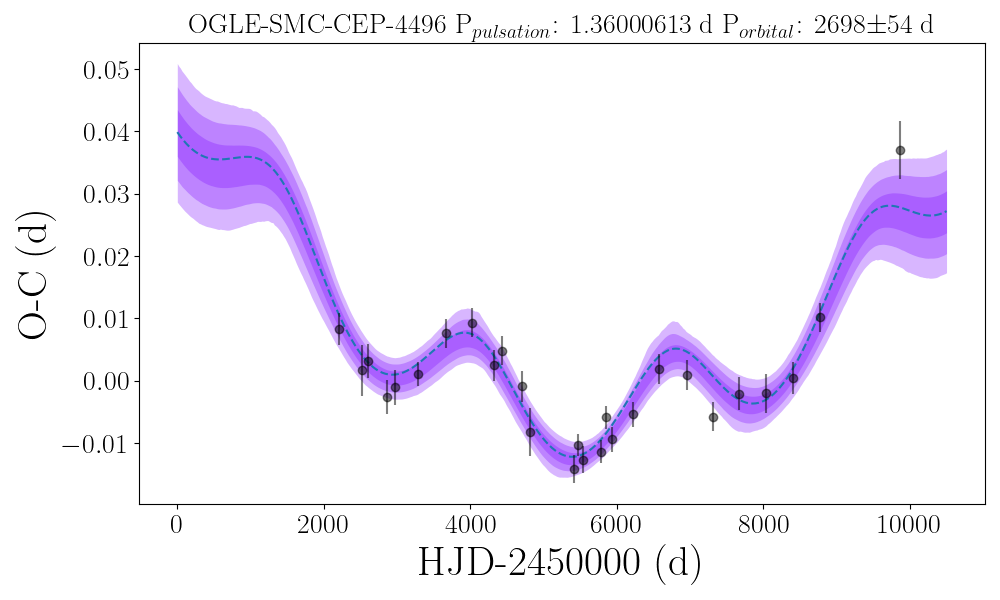}}
{\includegraphics[height=4.5cm,width=0.49\linewidth]{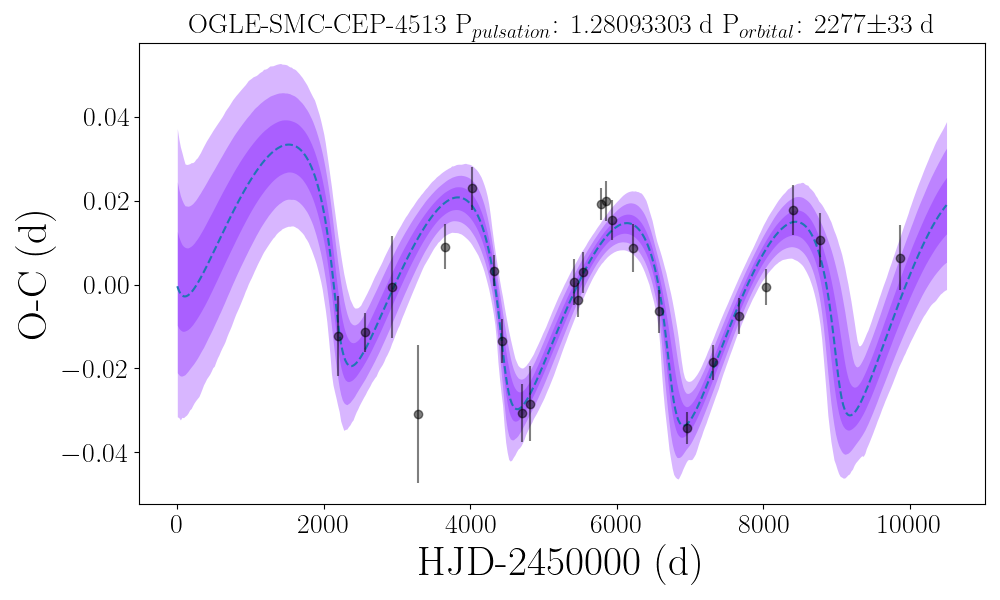}}
{\includegraphics[height=4.5cm,width=0.49\linewidth]{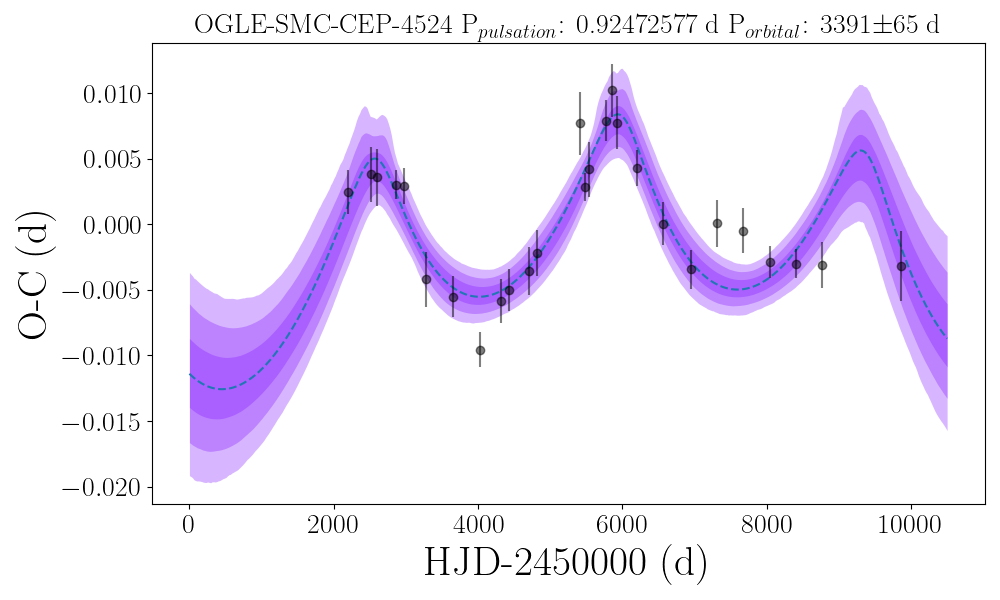}}
\caption{continued.}
\end{center}
\end{figure*}


\section{$O-C$ curves for high-mass-function candidates}

\begin{figure*}[ht!]
\begin{center}
{\includegraphics[height=4.5cm,width=0.49\linewidth]{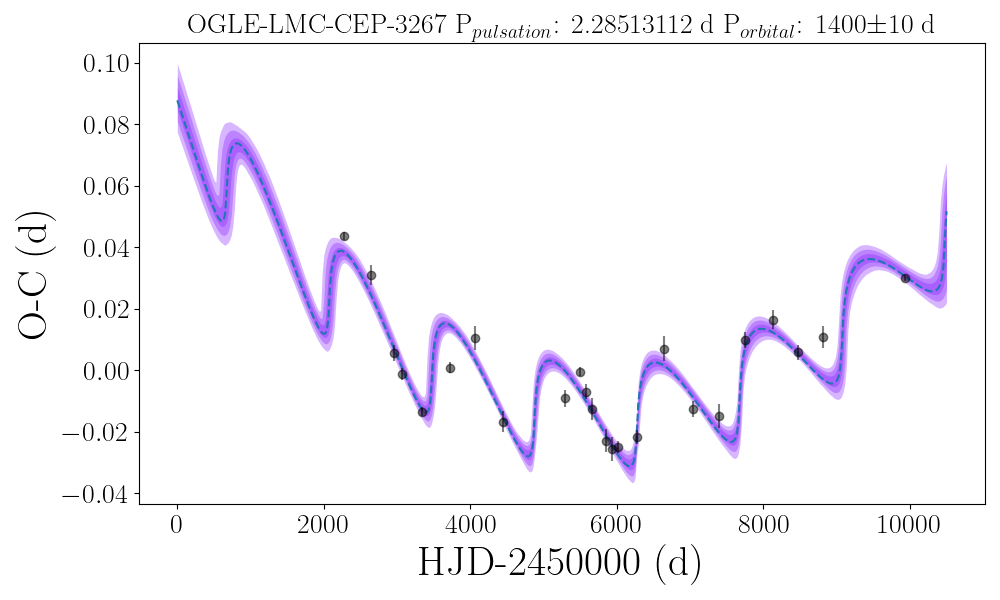}}
{\includegraphics[height=4.5cm,width=0.49\linewidth]{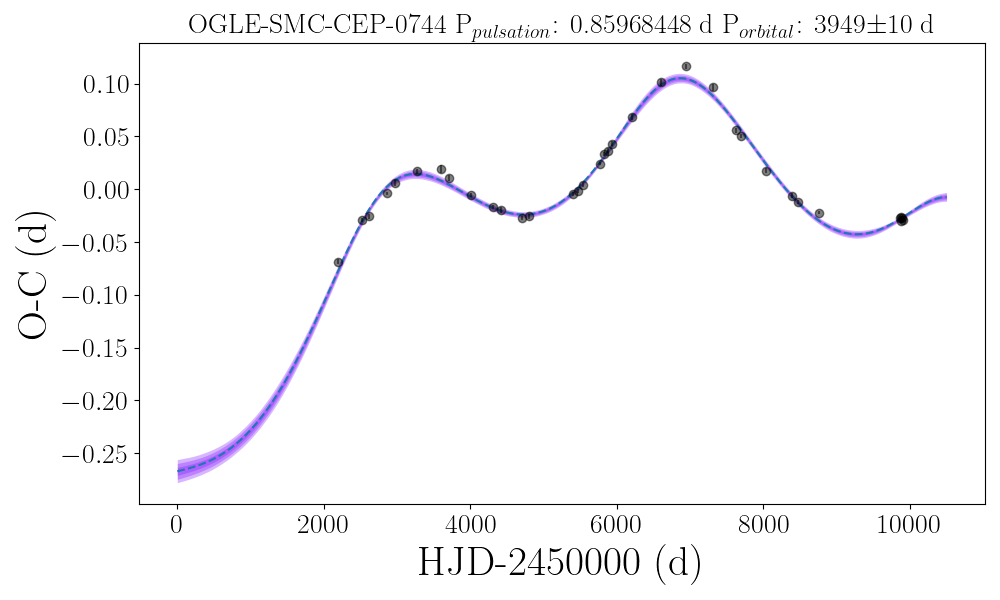}}
{\includegraphics[height=4.5cm,width=0.49\linewidth]{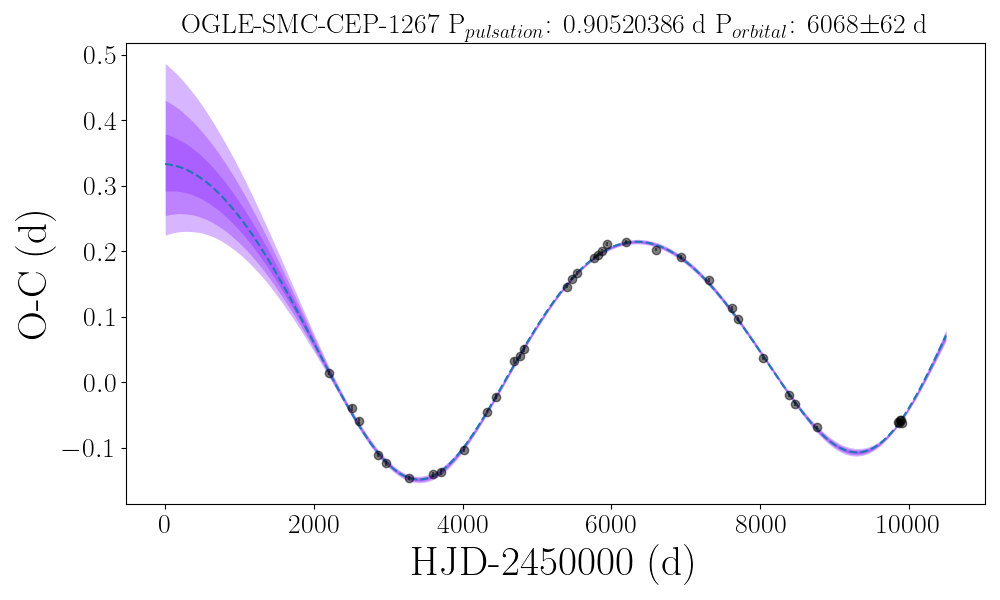}}
{\includegraphics[height=4.5cm,width=0.49\linewidth]{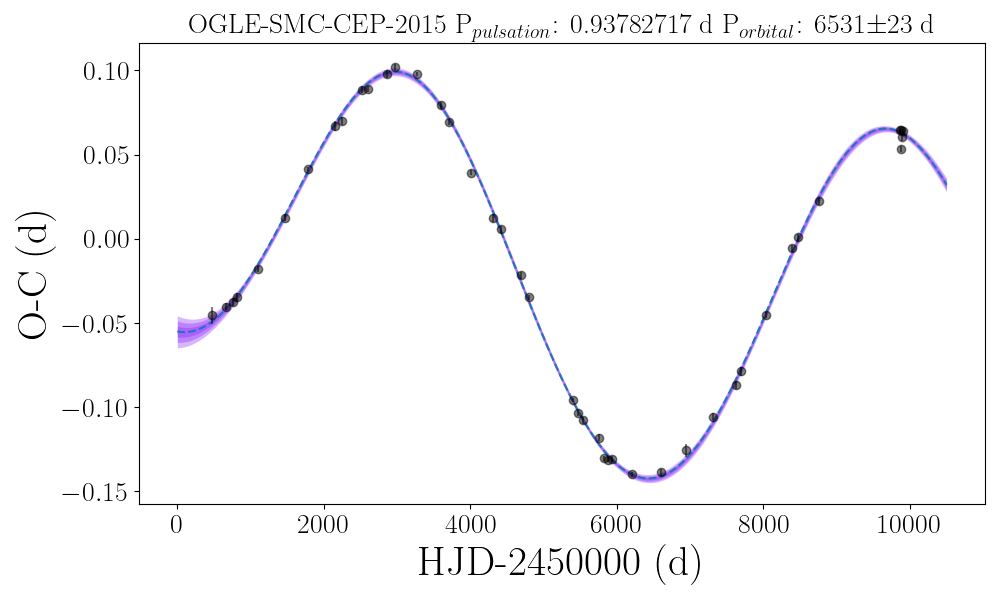}}

\caption{Remaining $O-C$ diagram for peculiar high mass-function candidates as described in section \ref{subsec: High Mass function candidates}}
\label{fig:appendix_Peculiar mass-function candidates}
\end{center}
\end{figure*}

\begin{figure*}[ht!]
\ContinuedFloat
\begin{center}
{\includegraphics[height=4.5cm,width=0.49\linewidth]{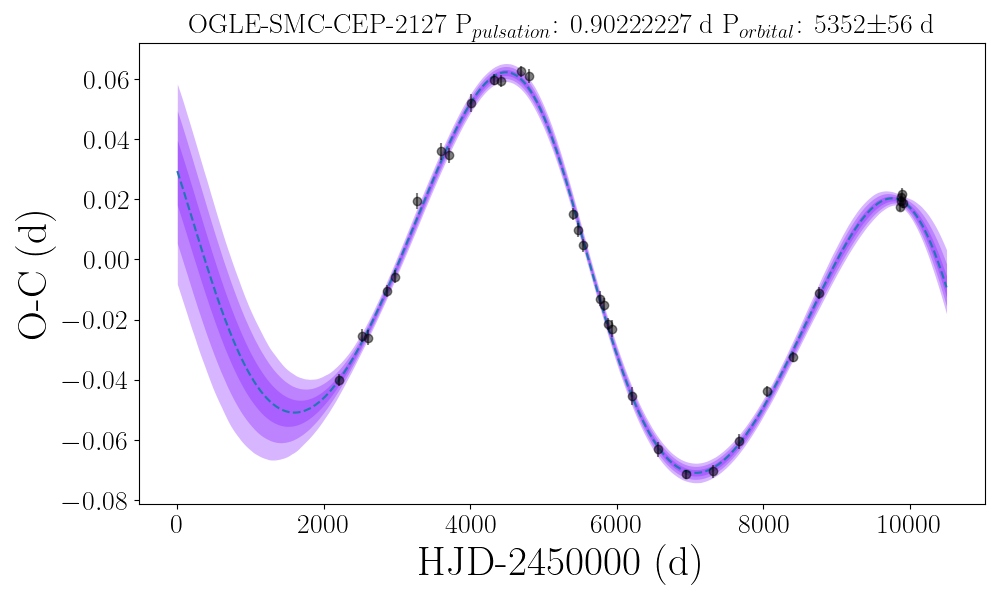}}
{\includegraphics[height=4.5cm,width=0.49\linewidth]{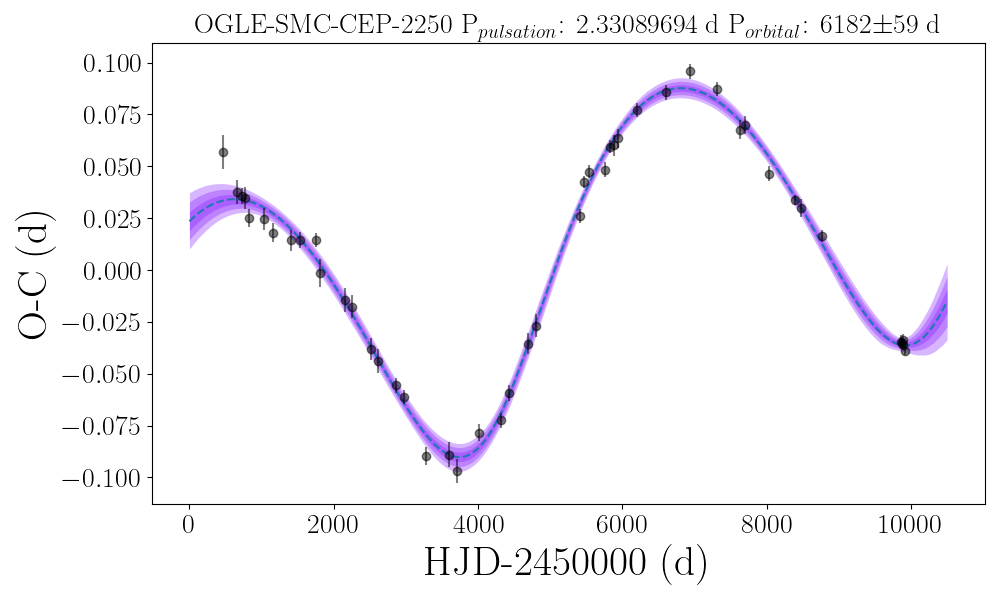}}
{\includegraphics[height=4.5cm,width=0.49\linewidth]{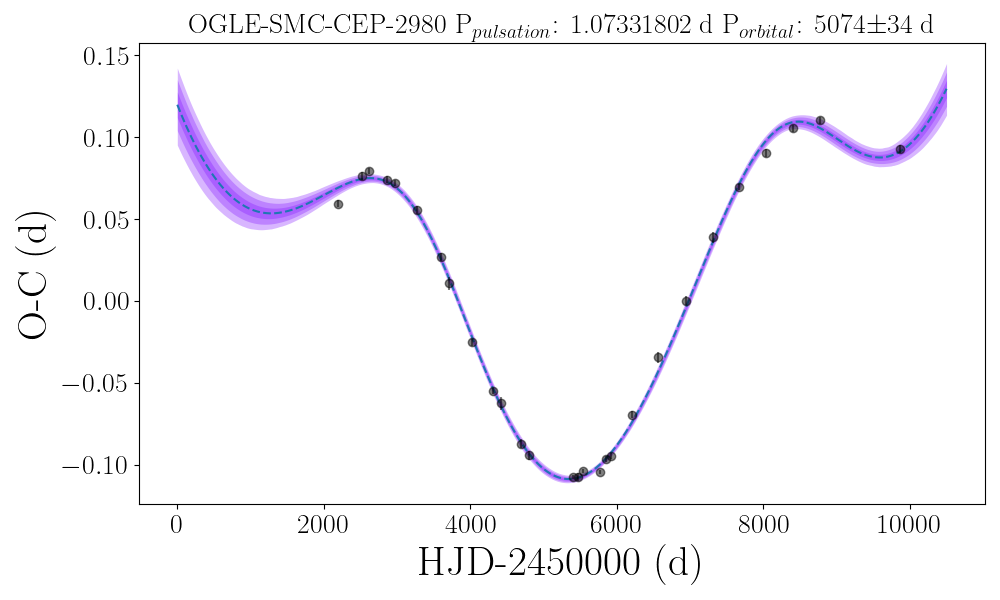}}
{\includegraphics[height=4.5cm,width=0.49\linewidth]{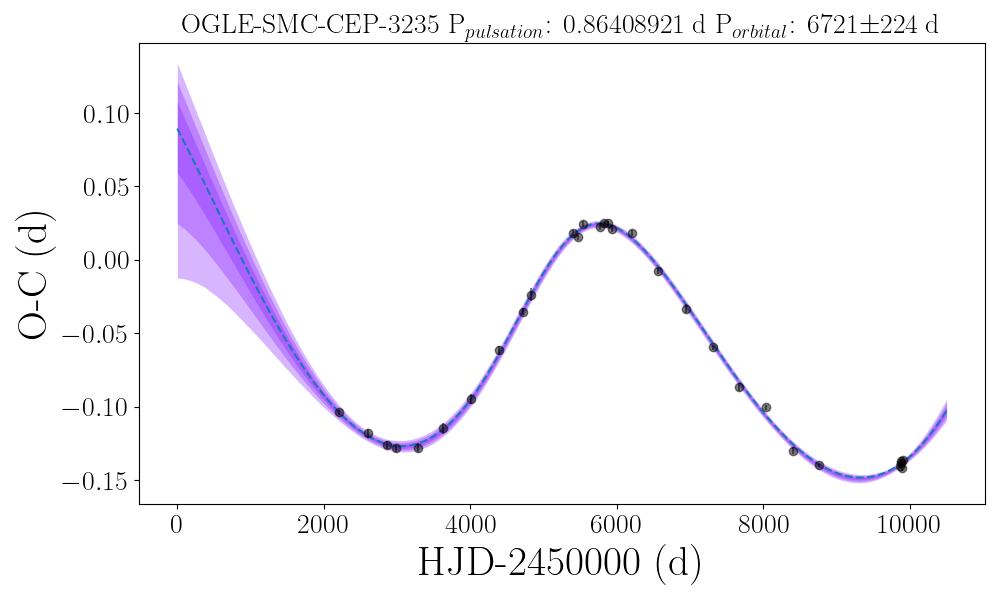}}
{\includegraphics[height=4.5cm,width=0.49\linewidth]{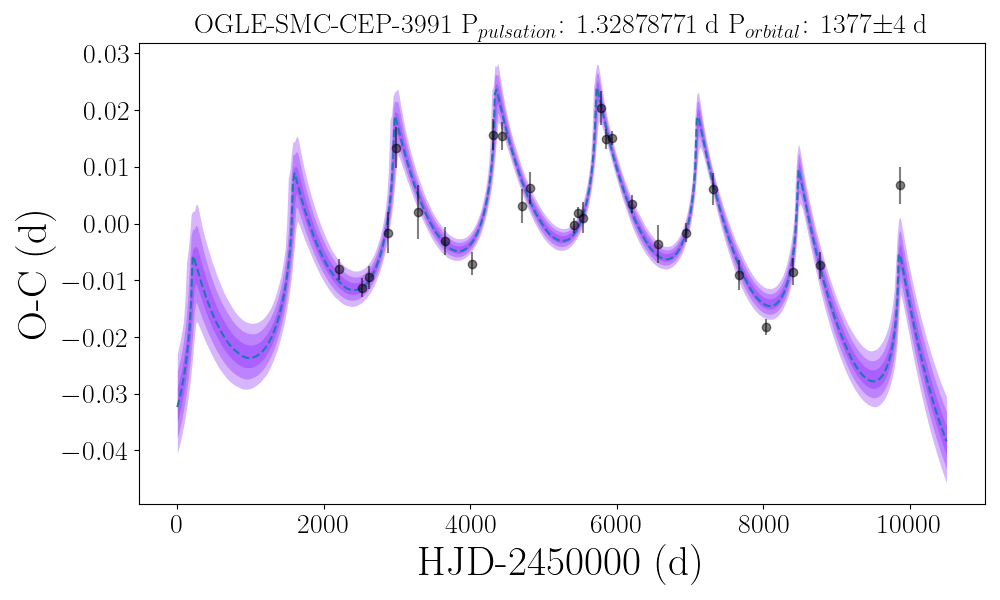}}
{\includegraphics[height=4.5cm,width=0.49\linewidth]{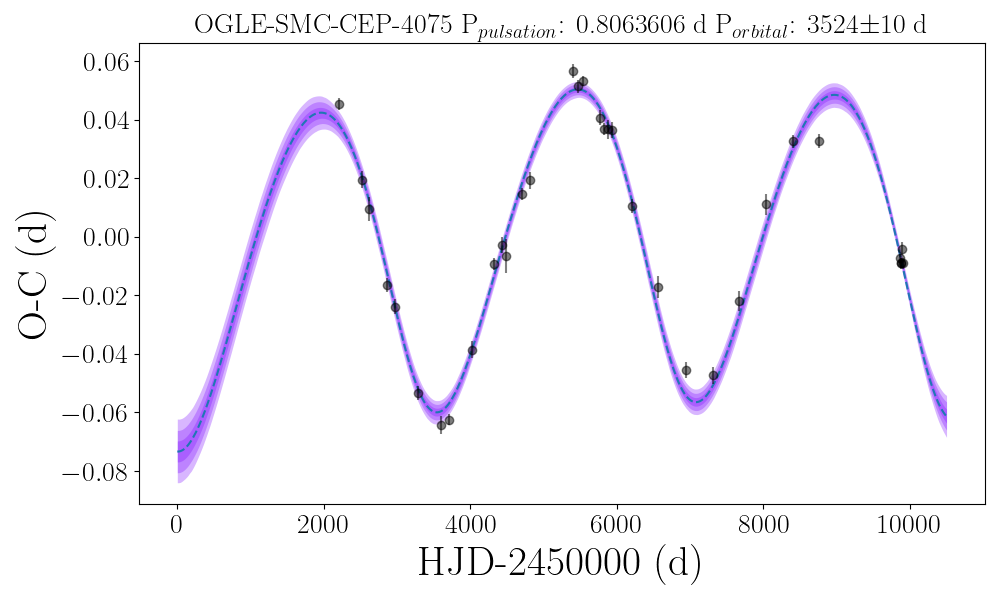}}
{\includegraphics[height=4.5cm,width=0.49\linewidth]{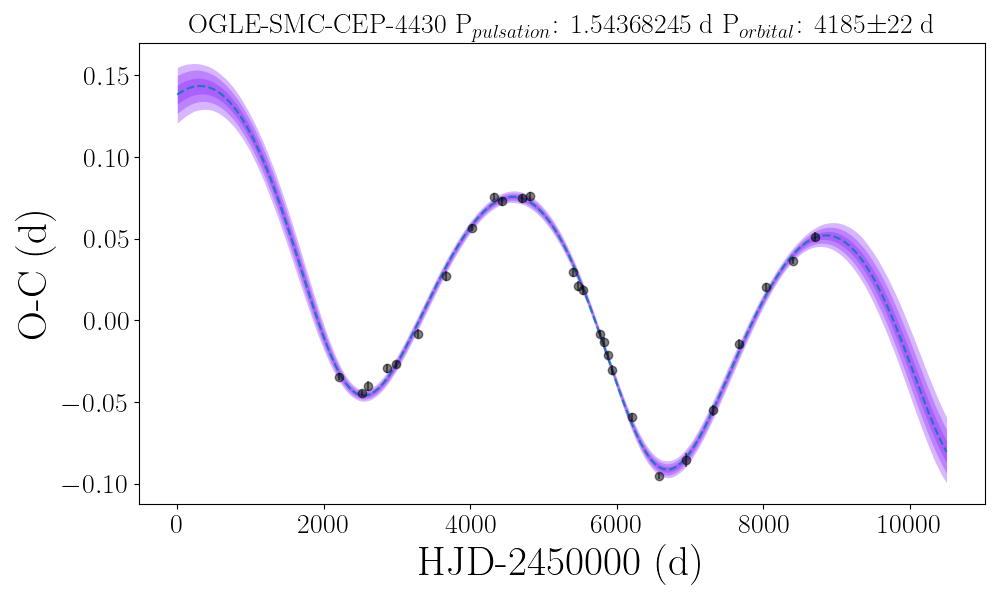}}
\caption{continued.}
\end{center}
\end{figure*}

\end{appendix}

\end{document}